\documentclass[twocolumn]{aastex63}

\usepackage{natbib}
\usepackage[caption=false]{subfig}

\graphicspath{{./}{figures/}}
\newcommand{\Msun}{\ensuremath{\textrm{M}_{\odot}}\,}

\submitjournal{ApJ}

\shorttitle{Calcium rich gap transients from ZTF}
\shortauthors{K. De et al.}
\begin{document}

\title{The Zwicky Transient Facility Census of the Local Universe I: Systematic search for Calcium rich gap transients reveal three related spectroscopic sub-classes}

\correspondingauthor{Kishalay De}
\email{kde@astro.caltech.edu}

\author[0000-0002-8989-0542]{Kishalay De}
\affil{Cahill Center for Astrophysics, California Institute of Technology, 1200 E. California Blvd. Pasadena, CA 91125, USA}

\author{Mansi M. Kasliwal}
\affil{Cahill Center for Astrophysics, California Institute of Technology, 1200 E. California Blvd. Pasadena, CA 91125, USA}

\author{Anastasios Tzanidakis} 
\affil{Cahill Center for Astrophysics, California Institute of Technology, 1200 E. California Blvd. Pasadena, CA 91125, USA}

\author{U. Christoffer Fremling}
\affil{Cahill Center for Astrophysics, California Institute of Technology, 1200 E. California Blvd. Pasadena, CA 91125, USA}

\author{Scott Adams} 
\affil{Cahill Center for Astrophysics, California Institute of Technology, 1200 E. California Blvd. Pasadena, CA 91125, USA}

\author[0000-0002-8977-1498]{Igor Andreoni} 
\affil{Cahill Center for Astrophysics, California Institute of Technology, 1200 E. California Blvd. Pasadena, CA 91125, USA}

\author{Ashot Bagdasaryan} 
\affil{Cahill Center for Astrophysics, California Institute of Technology, 1200 E. California Blvd. Pasadena, CA 91125, USA}

\author[0000-0001-8018-5348]{Eric C. Bellm}
\affiliation{DIRAC Institute, Department of Astronomy, University of Washington, 3910 15th Avenue NE, Seattle, WA 98195, USA}

\author{Lars Bildsten}
\affil{Department of Physics, University of California, Santa Barbara, CA 93106, USA}
\affil{Kavli Institute for Theoretical Physics, University of California, Santa Barbara, CA 93106, USA}

\author{Christopher Cannella}
\affil{Cahill Center for Astrophysics, California Institute of Technology, 1200 E. California Blvd. Pasadena, CA 91125, USA}

\author[0000-0002-6877-7655]{David O. Cook}
\affil{Caltech / IPAC, 1200 E. California Blvd, Pasadena, CA 91125, USA}

\author{Alexandre Delacroix}
\affil{Caltech Optical Observatories, California Institute of Technology, Pasadena, CA 91125, USA}

\author{Andrew Drake}
\affil{Cahill Center for Astrophysics, California Institute of Technology, 1200 E. California Blvd. Pasadena, CA 91125, USA}

\author[0000-0001-5060-8733]{Dmitry Duev}
\affil{Cahill Center for Astrophysics, California Institute of Technology, 1200 E. California Blvd. Pasadena, CA 91125, USA}

\author[0000-0001-7344-0208]{Alison Dugas} 
\affil{Cahill Center for Astrophysics, California Institute of Technology, 1200 E. California Blvd. Pasadena, CA 91125, USA}
\affil{Institute for Astronomy, University of Hawai'i, 2680 Woodlawn Drive, Honolulu, HI 96822, USA}

\author{Sara Frederick}
\affil{Department of Astronomy, University of Maryland, College Park, MD 20742, USA}

\author{Avishay Gal-Yam}
\affil{Benoziyo Center for Astrophysics, The Weizmann Institute of Science, Rehovot 76100, Israel}

\author{Daniel Goldstein}
\affil{Cahill Center for Astrophysics, California Institute of Technology, 1200 E. California Blvd. Pasadena, CA 91125, USA}

\author[0000-0001-8205-2506]{V. Zach Golkhou}
\affiliation{DIRAC Institute, Department of Astronomy, University of Washington, 3910 15th Avenue NE, Seattle, WA 98195, USA} 
\affiliation{The eScience Institute, University of Washington, Seattle, WA 98195, USA}
\altaffiliation{Moore-Sloan, WRF Innovation in Data Science, and DIRAC Fellow}

\author{Matthew J. Graham}
\affil{Cahill Center for Astrophysics, California Institute of Technology, 1200 E. California Blvd. Pasadena, CA 91125, USA}

\author{David Hale}
\affil{Caltech Optical Observatories, California Institute of Technology, Pasadena, CA 91125, USA}

\author{Matthew Hankins}
\affil{Cahill Center for Astrophysics, California Institute of Technology, 1200 E. California Blvd. Pasadena, CA 91125, USA}

\author[0000-0003-3367-3415]{George Helou}
\affiliation{IPAC, California Institute of Technology, 1200 E. California Blvd, Pasadena, CA 91125, USA}

\author[0000-0002-9017-3567]{Anna Y. Q. Ho}
\affil{Cahill Center for Astrophysics, California Institute of Technology, 1200 E. California Blvd. Pasadena, CA 91125, USA}

\author[0000-0002-7996-8780]{Ido Irani}
\affil{Department of Particle Physics and Astrophysics, Weizmann
Institute of Science, Rehovot 7610001, Israel}

\author[0000-0001-5754-4007]{Jacob E. Jencson}
\affiliation{Steward Observatory, University of Arizona, 933 North Cherry Avenue, Tucson, AZ 85721-0065, USA}

\author{Stephen Kaye}
\affil{Caltech Optical Observatories, California Institute of Technology, Pasadena, CA 91125, USA}

\author{S. R. Kulkarni}
\affil{Cahill Center for Astrophysics, California Institute of Technology, 1200 E. California Blvd. Pasadena, CA 91125, USA}

\author[0000-0002-6540-1484]{Thomas Kupfer}
\affiliation{Kavli Institute for Theoretical Physics, University of California, Santa Barbara, CA 93106, USA}

\author[0000-0003-2451-5482]{Russ R. Laher}
\affiliation{IPAC, California Institute of Technology, 1200 E. California Blvd, Pasadena, CA 91125, USA}

\author{Robin Leadbeater}
\affil{Three Hills Observatory, The Birches CA71JF, UK}

\author[0000-0001-9454-4639]{Ragnhild Lunnan}
\affiliation{Department of Astronomy, The Oskar Klein Center, Stockholm University, AlbaNova, 10691 Stockholm, Sweden}

\author[0000-0002-8532-9395]{Frank J. Masci}
\affiliation{IPAC, California Institute of Technology, 1200 E. CaliforniaBlvd, Pasadena, CA 91125, USA}

\author{Adam A. Miller}
\affiliation{Center for Interdisciplinary Exploration and Research in Astrophysics (CIERA) and Department of Physics and Astronomy,
Northwestern University, 2145 Sheridan Road, Evanston, IL 60208, USA}
\affiliation{The Adler Planetarium, Chicago, IL 60605, USA}

\author{James D. Neill}
\affil{Cahill Center for Astrophysics, California Institute of Technology, 1200 E. California Blvd. Pasadena, CA 91125, USA}

\author{Eran O. Ofek}
\affil{Benoziyo Center for Astrophysics, The Weizmann Institute of Science, Rehovot 76100, Israel}

\author{Daniel A. Perley}
\affiliation{Astrophysics Research Institute, Liverpool John Moores University, IC2, Liverpool Science Park, 146 Brownlow Hill, Liverpool L3 5RF, UK}

\author{Abigail Polin}
\affil{Department of Astronomy, University of California, Berkeley, CA, 94720-3411, USA}
\affil{Lawrence Berkeley National Laboratory, Berkeley, California 94720, USA}

\author{Thomas A. Prince}
\affil{Cahill Center for Astrophysics, California Institute of Technology, 1200 E. California Blvd. Pasadena, CA 91125, USA}

\author{Eliot Quataert}
\affil{Department of Astronomy, University of California, Berkeley, CA, 94720-3411, USA}

\author{Dan Reiley}
\affil{Caltech Optical Observatories, California Institute of Technology, Pasadena, CA 91125, USA}

\author{Reed L. Riddle}
\affiliation{Caltech Optical Observatories, California Institute of Technology, Pasadena, CA 91125, USA}

\author[0000-0001-7648-4142]{Ben Rusholme}
\affiliation{IPAC, California Institute of Technology, 1200 E. California Blvd, Pasadena, CA 91125, USA}

\author[0000-0003-4531-1745]{Yashvi Sharma}
\affil{Cahill Center for Astrophysics, California Institute of Technology, 1200 E. California Blvd. Pasadena, CA 91125, USA}

\author[0000-0003-4401-0430]{David L. Shupe}
\affiliation{IPAC, California Institute of Technology, 1200 E. California Blvd, Pasadena, CA 91125, USA}
             
\author[0000-0003-1546-6615]{Jesper Sollerman}
\affiliation{Department of Astronomy, The Oskar Klein Center, Stockholm University, AlbaNova, 10691 Stockholm, Sweden}

\author[0000-0003-3433-1492]{Leonardo Tartaglia}
\affiliation{Department of Astronomy, The Oskar Klein Center, Stockholm University, AlbaNova, 10691 Stockholm, Sweden}

\author{Richard Walters}
\affil{Caltech Optical Observatories, California Institute of Technology, Pasadena, CA 91125, USA}

\author{Lin Yan}
\affil{Cahill Center for Astrophysics, California Institute of Technology, 1200 E. California Blvd. Pasadena, CA 91125, USA}

\author[0000-0001-6747-8509]{Yuhan Yao}
\affil{Cahill Center for Astrophysics, California Institute of Technology, 1200 E. California Blvd. Pasadena, CA 91125, USA}

\begin{abstract}
Using the Zwicky Transient Facility alert stream, we are conducting a large campaign to spectroscopically classify all transients occurring in galaxies in the Census of the Local Universe (CLU) catalog. The aim of the experiment is to construct a spectroscopically complete, volume-limited sample of transients coincident within 100\arcsec\, of CLU galaxies out to 200\,Mpc, and to a depth of 20\,mag. We describe the survey design and spectroscopic completeness from the first 16 months of operations. We present results from a systematic search for Calcium rich gap transients in the sample of 22 low luminosity (peak absolute magnitude $M > -17$), hydrogen poor events found in the experiment (out of 754 spectroscopically classified SNe). We report the detection of eight Calcium rich gap transients, and place a lower limit on the volumetric rate of these events to be $\approx 15\pm5$\% of the SN\,Ia rate. Combining this sample with ten events from the literature, we find evidence of a likely continuum of spectroscopic properties ranging from events with SN\,Ia-like features (Ca-Ia objects) to SN\,Ib/c-like features (Ca-Ib/c objects) at peak light. Within the Ca-Ib/c events, we find two populations of events distinguished by their red ($g - r \approx 1.5$\,mag) or green ($g - r \approx 0.5$\,mag) spectral colors at $r$-band peak. Red Ca-Ib/c events are characterized by strongly line-blanketed spectra, systematically slower light curves, weaker and slower (by $\approx 3000$ km s$^{-1}$) He lines and lower [Ca II]/[O I] flux ratio (by a factor of $\approx 2$) in the nebular phase compared to the green Ca-Ib/c objects. Ca-Ia objects show typically more luminous light curves ($M_r \approx -16.9\,\rm{mag}$ at peak) than Ca-Ib/c objects, strong line blanketing signatures with intermediate red colors ($g-r\approx 1\,\rm{mag}$ at $r$-band peak), slow light curves, and weak or no [O I] in the nebular phase. Together, we find that the continuum of spectroscopic properties and the correlations thereof, the volumetric rates and striking old environments of these events are consistent with progenitor channels involving the explosive burning of He shells on low mass white dwarfs over a range of shell and core masses. We posit that the Ca-Ia and red Ca-Ib/c objects are broadly consistent with scenarios invoking the double detonation of He shells on white dwarfs with high He burning efficiency, while green Ca-Ib/c objects could arise from He shell explosions with lower He burning efficiency such as detonations in lower density He shells or He shell deflagrations.
\end{abstract}


\section{Introduction}
Calcium rich gap transients represent an emerging population of faint and fast evolving supernovae identified by their conspicuous [Ca II] emission in nebular phase spectra \citep{Perets2010, Sullivan2011, Kasliwal2012a, Valenti2014, Lunnan2017, De2018b}. Their photometric evolution is characterized by timescales and peak luminosities faster and fainter than those of typical core-collapse and thermonuclear supernovae (SNe), while their photospheric phase velocities are largely similar to normal Type Ib/c SNe ($\sim 8000$ km\,s$^{-1}$; see \citealt{Filippenko1997, Gal-Yam2017} for a review).  Yet, their most striking feature remains their preference for remote locations in the far outskirts of galaxies in old quiescent environments, in stark contrast to normal stripped envelope SNe which are found close to star formation \citep{Perets2010, Lunnan2017}. Together with the non-detection of any parent stellar populations in late-time imaging of the locations of these objects, their remote locations suggest that these transients arise from very old progenitors that may have traveled far away from their parent stellar population or were possibly formed in these remote locations \citep{Yuan2013,Lyman2014,Perets2014, Lyman2016b,Lunnan2017}. Their host offset distribution has been shown to be more skewed towards larger offsets than Type Ia SNe and even short gamma-ray bursts \citep{Lunnan2017}, while their hosts are preferentially found in group and cluster environments \citep{Mulchaey2014,Foley2015,Lunnan2017}. \citet{Shen2019} show that the radial offset distribution of the sample may be consistent with that of globular clusters, and potentially indicative of a progenitor population that has been kicked out of nearby globular clusters.

The progenitors of Calcium rich gap transients remain unknown to date, and are currently only constrained with circumstantial evidence. Specifically, their remote locations and old host environments point to old progenitors involving white dwarfs (WDs) in binary systems. Suggested channels include helium shell detonations on WDs \citep{Bildsten2007, Shen2010, Waldman2011, Sim2012, Dessart2015a, Meng2015}, double detonations of He shells on the surface of WDs \citep{Sim2012, Polin2019a, Polin2019b}, mergers of WDs with neutron stars \citep{Metzger2012, Margalit2016, Toonen2018, Zenati2019}, tidal disruptions of WDs by intermediate mass black holes \citep{Rosswog2008, Macleod2014, Sell2015, Macleod2016, Sell2018, Kawana2020} and even extreme core-collapse SNe from highly stripped massive stars \citep{Tauris2015, Moriya2017}; however their old environments make core-collapse SNe unlikely \citep{Perets2011}. If they arise from binary WD systems, \citet{Meng2015} show that the old environments and consequently long delay times constrain the progenitor binary to consist of low mass CO ($\lesssim 0.6$\,\Msun) and He ($\lesssim 0.25$\,\Msun) WDs.

Constraining their progenitors and rates is not only important for our understanding of these potentially common types of transients, but also to shed light on a likely common end point in binary stellar evolution involving white dwarfs in binary systems, and their possible significant contribution to the enrichment of the intergalactic medium with Ca \citep{Mulchaey2014, Mernier2016}. Estimates of the volumetric rates of this population from previous transient experiments include an estimate of $7 \pm 5$\% of the SN Ia rate from the Lick Observatory Supernova Search (LOSS; \citealt{Li2011,Perets2010}) and a lower limit of $\approx 3$\% of the SN Ia rate from the Palomar Transient Factory \citep{Kasliwal2012a}. A later estimate based on post-facto simulations of the detection efficiency and survey cadence of the Palomar Transient Factory suggests that their rates may be as high as $\approx 30$\% of the SN Ia rate \citep{Frohmaier2018}. However, the known sample of objects were found largely by follow-up of isolated events outside of systematic SN classification efforts, leaving considerable uncertainty on the rates of the class.

The number of reported Ca-rich gap transients  in the literature, as well as the diversity in their observed properties, have risen substantially in the last decade with large scale optical transient surveys. As per the name of the class, the detection of strong [Ca II] emission in the nebular phase spectra with high [Ca II]/[O I] ratio\footnote{Throughout this paper, we refer to the flux ratio of the forbidden [Ca II] $\lambda \lambda 7291, 7324$ to [O I] $\lambda 6300, 6364$ lines as [Ca II]/[O I]} is the primary criterion used to relate objects to the class of Ca-rich transients \citep{Valenti2014, Milisavljevic2017, De2018b}. In addition, \citet{Kasliwal2012a} defined this class of `gap' transients by their i) faint peak luminosity, ii) fast photometric evolution, iii) photospheric phase velocities similar to normal hydrogen-poor SNe ($\sim 8000$ km s$^{-1}$) and iv) early evolution to the nebular phase, notably without any constraints on the photospheric phase spectra of the transient. These criteria are consistent with the prototype event of the class SN\,2005E \citep{Perets2010}, which was also spectroscopically similar to Type Ib SNe near peak light.

There are seven other events in the class of Ca-rich gap transients that are spectroscopically similar to Type Ib/c SNe near peak light -- SN\,2007ke, SN\,2010et, PTF\,11bij \citep{Kasliwal2012a}, SN\,2012hn \citep{Valenti2014}, PTF\,11kmb, PTF\,12bho \citep{Lunnan2017} and SN\,2016hgs \citep{De2018b}. Other Type Ic SNe which show evidence of strong [Ca II] emission in the early nebular phase include iPTF\,14gqr \citep{De2018a}, SN\,2018kzr \citep{McBrien2019} and SN\,2019bkc \citep{Chen2020, Prentice2020}, although they were more luminous ($M_{peak} \lesssim -17.5$) than the typical Ca-rich gap transient. However, spectroscopic similarity to Type Ib/c SNe does not appear to be a defining characteristic of the class. Notable exceptions include PTF\,09dav \citep{Sullivan2011, Kasliwal2012a} and SN\,2016hnk \citep{Galbany2019, Jacobson-Galan2019}, which exhibit similarities to sub-luminous SN1991bg-like Type Ia SNe \citep{Filippenko1992b} near peak light. Similarly, iPTF\,15eqv \citep{Milisavljevic2017} was a peculiar hydrogen-rich SN\,IIb which exhibited high [Ca II]/[O I] ratio in nebular phase spectra, noting that it was luminous at peak and consistent with a core-collapse SN in a star forming environment. Another potential menmber of the class, SN\,2005cz \citep{Kawabata2010, Perets2011} exhibited high [Ca II]/[O I] ratio in its nebular phase spectrum; yet the lack of photometry around peak light precludes a confirmed association with this class of faint and fast evolving transients, as is the case for several candidates presented in \citet{Fillipenko2003}. 

The large heterogeneity in the peak luminosity and spectroscopic appearance of objects likely points to a diversity in explosions that produce high [Ca II]/[O I] in their nebular phase spectra. Yet, the small number of total reported events ($\lesssim 10$) has prevented a holistic analysis of the spectroscopic and photometric properties of this class. Most previous studies have focused on one or two events, each of which have been suggested to be unique members of this emerging population which remains poorly understood. In particular, previous works have not characterized the nebular phase behavior of a systematically selected sample of low-luminosity hydrogen poor transients to be able to quantitatively place the photometric and spectroscopic properties of the class of Ca-rich gap transients in a broader context. Such an analysis with a large sample can yield vital clues to trends within the population and shed light on the underlying explosions. The aim of this paper is to systematically uncover and analyze this population of faint and fast evolving hydrogen-poor events that exhibit high [Ca II]/[O I] in nebular phase spectra.

While galaxy-targeted supernova surveys are sensitive to transients occurring close to their host galaxies, the known preference of these transients for large host offsets necessitates a wide-field search approach that is sensitive to transients at large projected offsets from their host galaxies. Given the faint peak luminosity ($M_r \approx -16$) and relatively low volumetric rates ($\sim 10$\% of SN\,Ia rate) of these events, finding a large sample of events thus requires a sufficiently deep (depth $r \gtrsim 20$ mag to find events out to $\approx 150$\,Mpc) optical all sky survey with a cadence of $\lesssim 4\,\rm{d}$ to detect these short lived events.  At the same time, due to the large rate of higher redshift Type Ia SNe at this depth ($\sim 8500$ yr$^{-1}$ down to $r = 20\,\rm{mag}$ limiting magnitude; \citealt{Feindt2019}), finding a systematic sample of these local universe events requires a targeted approach to classify transients in the local universe by cross-matching transients to known nearby galaxies. Such an approach is now possible with large catalogs of galaxies with known spectroscopic redshifts like the Census of the Local Universe (CLU; \citealt{Cook2019}) catalog and GLADE \citep{Dalya2018}.

This paper presents the first in a series of publications from the Census of the Local Universe experiment of the Zwicky Transient Facility (ZTF; \citealt{Bellm2019a, Graham2019}). This paper provides an overview of the sample selection and spectroscopic completeness of this volume-limited experiment. Here, we focus on the identification of Ca-rich gap transients, specifically on the class of faint and hydrogen-poor transients that exhibit Ca-rich spectra in the nebular phase. We briefly describe the design of the experiment and sample selection in Section \ref{sec:observations}. Section \ref{sec:analysis} presents an analysis of the photometric and spectroscopic properties of the combined sample of transients from ZTF and the literature, specifically noting the presence of two spectroscopic classes and a continuum of properties across these classes. Using the controlled selection criteria of the experiment, we present an analysis of the host environments of these transients in Section \ref{sec:environments} while Section \ref{sec:rates} presents a discussion on the estimated volumetric rates of these events. In Section \ref{sec:discussion}, we combine all of the results to constrain the progenitors of this class and summarize our conclusions in Section \ref{sec:summary}. Calculations in this paper assume WMAP9 flat $\Lambda$CDM cosmology with $H_0 = 69.3$ km s$^{-1}$ Mpc$^{-1}$ and $\Omega_M = 0.286$ \citep{Hinshaw2013}. We use the median redshift independent distance estimates from the NASA Extragalactic Database (NED) for transients hosted in galaxies that have such measurements, and redshift derived distance estimates otherwise. For the redshift derived distance estimates in this local universe sample ($z < 0.05$), the typical uncertainty in the luminosity distance and projected offsets is $\lesssim 5$\% for peculiar velocities $\lesssim 300$\,km s$^{-1}$. Times reported are in UT throughout this paper.

\section{Observations}
\label{sec:observations}
\subsection{The Census of the Local Universe Experiment}

The Zwicky Transient Facility is a wide-field optical time domain survey running out of the 48-inch Schmidt telescope (P48) at Palomar observatory \citep{Bellm2019a, Graham2019}. With a field of view of 47 square degrees, the instrument achieves median limiting magnitude of $r \approx 20.5\,\rm{mag}$ in 30 s exposures of the sky and a survey speed of $\approx 3750$ square degrees per hour \citep{Dekany2016}. ZTF observing time is divided into a public component (40\%), a collaboration component (40\%) and a Caltech component (20\%). \citet{Bellm2019b} provide an overview of the various ZTF surveys undertaken in the first year of operations, and the survey scheduling system designed to carry out operations to maximize volumetric survey speed. The public component is a 3-day cadence $g$ + $r$ survey of the entire northern sky ($\approx 34$\% of P48 time) together with a 1-day cadence $g$ + $r$ survey of the Galactic plane ($\approx 6$\% of P48 time). The collaboration time was dedicated to high cadence (3$g$ + 3$r$ per night) observations of $\approx 2500$ square degrees and a slower cadence ($\approx 4\,\rm{d}$) $i$-band survey. The Caltech time was dedicated to a one night cadence $g$ + $r$ survey of $\approx 3000$ square degrees. Transients in the difference imaging pipeline (based on the ZOGY subtraction algorithm; \citealt{Zackay2016}) of ZTF \citep{Masci2019} are reported and distributed in Avro alert packets\footnote{\url{https://avro.apache.org}} \citep{Patterson2019}, including photometry and metadata for the detected transient, as well as a 30-day history for the previous detections and non-detections.

The ZTF Census of the Local Universe (CLU) experiment has been designed to build up a spectroscopically classified sample of transients in the local universe (within 200 Mpc) by classifying all transients found coincident with galaxies in the Census of the Local Universe (CLU; \citealt{Cook2019}) catalog. The CLU catalog consists of $\sim 234500$ galaxies with previously known redshifts compiled from several previous spectroscopic surveys (called CLU-compiled; see \citealt{Cook2019}), along with additional nearby galaxies found in a wide, narrow-band (Halpha) survey covering 3$\pi$ of the northern sky with the Palomar 48-inch telescope \citep{Cook2019}. The initial filter for the experiment used the CLU-compiled catalog together with $\sim 1000$ of the highest significance ($> 25\sigma$) candidates from the  H$\alpha$ survey. Starting from April 2019, we initiated the use of the next data release which included a larger sample ($\approx 38000$ candidates with significance $> 3\sigma$) of high confidence nearby galaxies from the CLU H$\alpha$ survey (see \citealt{Cook2019} for a description). Based on the transient sample found in this experiment, we find that $\approx 1$\% of transients were hosted in the CLU H$\alpha$ survey galaxies before the April 2019 update, while $\approx 10\%$ of transients were hosted in galaxies from the H$\alpha$ catalog following the inclusion of the next data release.

The CLU experiment was initiated on 2018 June 01 and we restrict the sample of transients in this paper to events saved until 2019 September 30. The sample selection for the transients was implemented as a part of a custom filter implemented on the GROWTH Marshal \citep{Kasliwal2019}, which is a web-portal for vetting and coordinating follow-up of transients. The selection criteria for ZTF alerts to be saved in the CLU experiment are as follows:
\begin{enumerate}
    \item Each alert packet is spatially cross-matched to the CLU catalog of galaxies. The size of the spatial cross-match is set to $3 \times D_{25}$ where $D_{25}$ is the isophotal major axis containing 25\% of the total light of the galaxy, as contained in the CLU catalog. If a $D_{25}$ radius is not available for the galaxy, a default cross-match radius of 280\,\arcsec\, was used\footnote{280\arcsec is $3\times$ the median $D_{25}$ value of all galaxies in the CLU catalog.}. The cross-matching was performed on the dedicated time domain astronomy server called \texttt{kowalski} at Caltech \citep{Duev2019}. 
    \item The alert candidate was produced as a positive candidate in the subtraction, i.e. the source flux has increased from the reference image.
    \item The alert candidate has a real-bogus score (rbscore) of $> 0.3$ as classified by the ZTF machine learning algorithm \citep{Mahabal2019}. This choice produces a false negative rate of $< 3$\% \citep{Duev2019}.
    \item The alert candidate is at least 20\,\arcsec away from a star brighter than 15.0 mag.
    \item Alert candidates within 1\arcsec of a known star in PS1 \citep{Chambers2016} are rejected. The identification of stars is based on the machine learning based star/galaxy classification score presented in \citet{Tachibana2018}, which is available for the three nearest sources in the ZTF alert packets. We use a \texttt{sgscore} threshold of 0.6, i.e. candidates within 1\arcsec\, of a PS1 source with \texttt{sgscore} $> 0.6$ are rejected.
    \item The alert candidate was at least 2\,\arcsec away from the nearest solar system object and was detected at least twice in the survey separated by 50 mins. The former criterion removes known asteroids in the ZTF alert stream while the latter removes unknown solar system objects.
\end{enumerate}{}
No further magnitude cut was applied to this transient stream. This filter produces typically $\approx 100$ sources to be vetted by on-duty astronomers every day, which involves an inspection of the science, reference and difference image cutouts contained in the ZTF alert packets. The human vetting is required to remove alerts from remaining stellar sources that pass the filter, remove variability from known AGN and identify remaining bogus sources before assigning appropriate spectroscopic follow-up. For 2018 and 2019, spectroscopic follow-up was exclusively limited to transients within 100\arcsec of the nearest CLU galaxy to remove the large amount of contamination of SNe in background galaxies. Transients coincident with known background galaxies (with known spectroscopic redshifts or photo-z $> 0.1$) were excluded from the sample in order to avoid the large number of false positives from background AGN and Type Ia SNe. As such, the experiment may be incomplete to transients occurring at very large projected offsets of $> 100$\arcsec\, from their parent galaxies, corresponding to a physical projected distance of 100\,kpc at 200 Mpc, 50\,kpc at 100 Mpc and 25\,kpc at 50 Mpc. In addition, sources coincident with the nuclei of known Active Galactic Nuclei (AGN) and with long term history of variability are not assigned for follow-up. On average, 10 - 15 SNe are saved for spectroscopic follow-up every week, which were coordinated via source pages on the GROWTH Marshal. Starting from November 2019, sources from the public ZTF data stream are reported to the Transient Name Server (TNS\footnote{\url{https://wis-tns.weizmann.ac.il/}}) as soon as they are saved by a human scanner \citep{De2019c}.

All transients saved after this vetting process are systematically assigned for spectroscopic follow-up. We prioritized follow-up of sources that were brighter than or were going to peak at brighter than 20 mag (in either $r_{\text{ZTF}}$ or $g_{\text{ZTF}}$; see \citealt{Bellm2019a}), which was selected to be the target limiting magnitude for the experiment. Given the typical ZTF limiting magnitude of $20.5\,\rm{mag}$, we did not apply any magnitude cuts to the filter to be able to track the photometric evolution and assign follow-up for transients peaking around 20 mag. Spectroscopic classifications were performed using a multi-tiered approach -- i) sources brighter than 19 mag were assigned for spectroscopic follow-up on the robotic Spectral Energy Distribution Machine (SEDM; \citealt{Blagorodnova2018}) and ii) sources between 19 and 20 mag were assigned for spectroscopic follow-up on the Double Beam Spectrograph (DBSP; \citealt{Oke1982}) on the 200-inch Hale telescope (P200) at Palomar observatory and iii) sources fainter than 20 mag were assigned for lower priority follow-up on P200 + DBSP. Sources assigned to the SEDM queue remain as follow-up targets for a duration of 7 days after which they are re-assigned to SEDM (if still brighter than 19 mag) or to P200 otherwise.  Spectroscopic follow-up on P200 was required for bright sources ($< 19$\,mag) when coincident with the nuclei of nearby galaxies, where host galaxy contamination was difficult to remove in the SEDM spectra. In cases where P200 classifications were not possible due to poor weather or due to large host contamination in P200 data for faint targets, we also used the Low Resolution Imaging Spectrometer (LRIS; \citealt{Oke1995}) on the Keck-I telescope for spectroscopic classifications.

The spectroscopic follow-up effort for bright sources ($m<19$\,mag) was coordinated with the Bright Transient Survey (BTS; \citealt{Fremling2019}) experiment which aims to classify all transients brighter than $19$ mag in the ZTF public alert stream. Community follow-up for bright transients overlapping with BTS have also aided in the spectroscopic classification completeness of the sample. Classifications are done with the \texttt{SuperNova IDentification} \citep{Blondin2007} (SNID) code by automatic execution on spectra produced by the SEDM automated pipeline \citep{Rigault2019} and manual execution for all other instruments. For spectra contaminated by the underlying host light, we used \texttt{superfit} \citep{Howell2005} to attempt host subtraction and derive a classification. The final classification is made by human inspection of the best-fit templates matched from SNID or \texttt{superfit}.

\subsection{Spectroscopic completeness}

The aim of the tiered approach to spectroscopic classification was to obtain high spectroscopic completeness for transients that peaked brighter than 20\, mag in galaxies within the local universe. This magnitude limit corresponds to a luminosity completeness of $M < -16.5\,\rm{mag}$ for all galaxies in the 200\,Mpc volume of the CLU catalog and $M < -15\,\rm{mag}$ for galaxies within 100\,Mpc. However, given the galaxy targeted nature of the target selection, the experiment is not sensitive to transients that occur in nearby galaxies with previously unknown spectroscopic redshifts. Based on results from the ZTF BTS, the completeness (in terms of galaxy counts) of the compiled catalog is $\approx 80$\% at the lowest redshifts and decreases to $\approx 50$\% at the edge of the 200 Mpc volume \citep{Fremling2019}. While effort was made to have complete spectroscopic classifications to a depth of $\approx 20$\,mag, classifications were not always possible due to several reasons. These include the difficulty of following up transients found close to the sun, classifying transients on the nuclei of bright galaxies (where low resolution SEDM spectra are dominated by host galaxy light) that faded before a scheduled run on the P200 / Keck, and due to loss of P200 / Keck time in periods of bad weather. We thus evaluate the spectroscopic completeness of the experiment in the first year of operations. 

\begin{figure}
    \centering
    \includegraphics[width=\columnwidth]{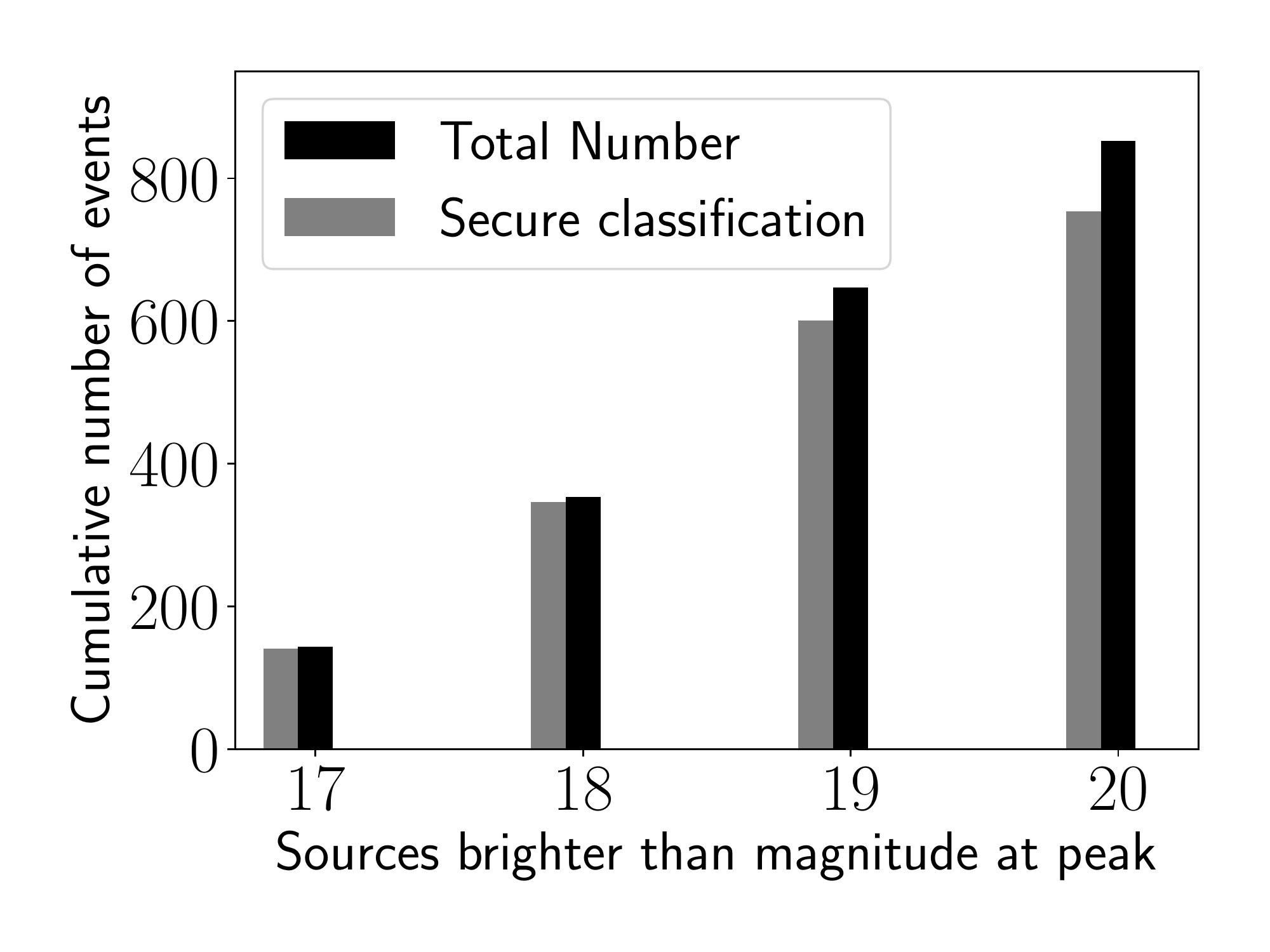}
    \caption{Cumulative number of events that were saved (in black) and classified (in grey) as a function of the peak magnitude in the CLU sample of events. }
    \label{fig:cumcomplete}
\end{figure}

We restrict this sample to transients that were saved between\footnote{Accounting for observation gaps due to instrument maintenance and poor weather, this period contains a total of 390 nights of full or partial ZTF operations.} 2018 June 01 and 2019 September 30, and to sources detected in any of the public or internal collaboration surveys (note that the public survey has the largest footprint on the sky). We define our primary sample of sources such that they are detected at least two times, and peaked at a magnitude brighter than 20 mag in either the $r_{\text{ZTF}}$ or $g_{\text{ZTF}}$ filters. A total of 852 candidate SNe were saved during this time period, out of which 563 were also included in the BTS program. 754 out of these 852 events were reliably classified, while 98 were unclassified either due to the lack of follow-up spectroscopy or due to ambiguous classifications from spectroscopic data. As such, the spectroscopic completeness of the complete acquired sample is 88.5\% for all transients that had at least one detection brighter than $20$\,mag. The corresponding classification completeness for $m_\textrm{peak} < 19$\,mag is 92.9\%, $m_\textrm{peak} < 18$\,mag is 98.0\% and $m_\textrm{peak} < 17$\,mag is 98.6\% for the entire experiment duration mentioned above. We show a cumulative plot of the number of sources saved and classified as a function of the peak magnitude in Figure \ref{fig:cumcomplete}.

\subsection{Sample of Ca-rich gap transients}
\label{sec:cand_select}

\begin{table*}
    \centering
    \scriptsize
    \caption{Summary of the properties of the Ca-rich gap transients presented in this paper, together with the sample of events that passed the selection criteria based on peak light photometry and spectroscopy only, but did not exhibit high [Ca II]/[O I] in their nebular phase spectra. The column Spec Type only refers to the spectroscopic appearance of the object near peak light. The Ca-rich objects are indicated with Ca in the Spec Type column. For cases where the object did not turn nebular even at the latest phases of spectroscopic follow-up, we indicate the [Ca II]/[O I] flux ratio with NN and indicate the phase of the latest available spectrum. In the case of SN\,2018gwo (indicated with *), the object did not pass the primary selection criteria but is a likely Ca-rich gap transient when combined with publicly available photometry and nebular phase follow-up from our campaign. Details on the objects that did not pass the nebular phase criterion are discussed in Appendix \ref{sec:control}, highlighting why each object was excluded from the Ca-rich sample. For one event (SN\,2019gau), we do not detect the nebular emission features and hence denote the [Ca II]/[O I] ratio with --.}
    \begin{tabular}{lccccccc}
    \hline
    \hline
         Object & RA & Dec & Spec Type & Peak $r$ mag & [Ca II] / [O I] flux & Redshift & Host Offset \\
         & J2000 & J2000 & & (Abs. Mag) & Value / Phase (days) & & (\arcsec / kpc)\\
    \hline
         ZTF\,18aayhylv / SN\,2018ckd & 14h06m11.94s & +09$^\circ$20\arcmin$39\farcs33$ & Ca-Ib & $-16.20$ & $> 3.38$ / +58 & 0.024 & 39.03 / 19.08 \\
         ZTF\,18abmxelh / SN\,2018lqo & 16h28m43.26s & +41$^\circ$07\arcmin$58\farcs66$ & Ca-Ib & $-16.21$ & $>12.5$ / +49 & 0.033 & 23.25 / 15.46 \\
         ZTF\,18abttsrb / SN\,2018lqu & 15h54m11.47s &  +13$^\circ$30\arcmin$50\farcs87$& Ca-Ib  & $-16.44$ & $>8.38$ / +31 & 0.036 & 37.12 / 26.70\\
        ZTF\,18acbwazl / SN\,2018gwo* & 12h08m38.82s & +68$^\circ$46\arcmin$44\farcs42$ & Ca-Ic & $<-16.0$ & $5.16$ / +53 & 0.008 & 54.20 / 8.56 \\
        ZTF\,18acsodbf / SN\,2018kjy  & 06h47m17.96s & +74$^\circ$14\arcmin$05\farcs90$  & Ca-Ib & $-15.62$ & $4.44$ / +111 & 0.018 & 17.18 / 6.35 \\
        ZTF\,19aaznwze / SN\,2019hty  & 12h55m33.03s & +32$^\circ$12\arcmin$21\farcs70$  & Ca-Ib  & $-16.29$ & $>3.27$ / +38 & 0.023 & 18.74 / 8.73\\
        ZTF\,19abrdxbh / SN\,2019ofm  & 14h50m54.65s & +27$^\circ$34\arcmin$57\farcs59$ & Ca-Ia & $-16.84$ & $>2.13$ / +175 & 0.030 & 18.24 / 11.16\\
        ZTF\,19abwtqsk / SN\,2019pxu  &05h10m12.60s & -00$^\circ$46\arcmin$38\farcs63$ & Ca-Ib & $-16.72$ & $>8.30$ / +146 & 0.028 & 30.93 / 17.56 \\
        ZTF\,19aamfupk / SN\,2019ccm & 04h41m05.36s & +73$^\circ$40\arcmin$23\farcs10$ & SN\,Ib & $-16.40$ & 1.18 / +207 & 0.015 & 14.73 / 4.54 \\
        ZTF\,19aanfsmc / SN\,2019txl & 09h32m59.36s &  +27$^\circ$30\arcmin$07\farcs80$ & SN\,Ib & $-16.21$ & 0.87 / +330 & 0.034 & 11.68 / 7.96 \\
        ZTF\,19aasqseq / SN\,2019txt & 09h59m06.38s & +17$^\circ$49\arcmin$09\farcs99$ & SN\,Ib & $-15.90$ &  1.34 / +180 & 0.026 & 19.94 / 10.69 \\
        ZTF\,19abgqruu / SN\,2019mjo & 00h06m59.83s & +03$^\circ$27\arcmin$39\farcs70$ & SN\,Ib-pec & $-16.62$ & NN / +180 & 0.041 & 12.38 / 10.06\\
        ZTF\,18abdffeo / SN\,2018dbg & 14h17m58.86s & +26$^\circ$24\arcmin$44\farcs59$ & SN\,Ib/c & $-16.65$ & NN / +22 & 0.015 & 2.01 / 0.61\\
        ZTF\,19aarrdoz / SN\,2019txr & 08h42m31.91s & +56$^\circ$17\arcmin$42\farcs19$ & SN\,Ib/c & $-16.73$ & $< 1$ / +270 & 0.044 & 1.95 / 1.71\\
        ZTF\,18aboabxv / SN\,2018fob & 15h13m07.23s & +41$^\circ$16\arcmin$11\farcs07$ & SN\,Ic & $-16.93$ & 0.87 / +212 & 0.029 & 18.47 / 10.84 \\
        ZTF\,19aadttht / SN\,2019yz & 15h41m57.30s & +00$^\circ$42\arcmin$39\farcs41$ & SN\,Ic & $-16.63$ & 0.59 / +242 & 0.006 & 34.17 / 4.54\\
        ZTF\,19aadwtoe / SN\,2019abb & 07h54m17.26s & +14$^\circ$16\arcmin$22\farcs42$ & SN\,Ic & $-16.56$ & 0.79 / +357 & 0.015 & 4.26 / 1.34\\
        ZTF\,19aailcgs / SN\,2019ape & 10h51m42.55s & +18$^\circ$28\arcmin$52\farcs62$ & SN\,Ic & $-16.62$ & 0.87 / +180 & 0.021 & 11.53 / 4.79 \\
        ZTF\,19abhhdwf / SN\,2019ouq & 17h01m41.94s & +30$^\circ$06\arcmin$34\farcs43$ & SN\,Ic & $-16.68$ & $< 1$ / +170 & 0.036 & 8.71 / 6.27\\
        ZTF\,18acushie / SN\,2018kqr & 08h50m03.60s & +55$^\circ$10\arcmin$09\farcs54$ & SN\,Ic-BL & $-16.75$ & NN / +16 & 0.045 & 3.15 / 2.82\\
        ZTF\,19aavlfvn / SN\,2019gau & 14h38m10.42s & +10$^\circ$08\arcmin$04\farcs93$ & SN\,Ia & $-16.75$ & -- / +260 & 0.028 & 1.30 / 0.73 \\
        ZTF\,19aawhlcn / SN\,2019gsc & 14h37m45.25s & +52$^\circ$43\arcmin$36\farcs28$ & SN\,Ia\,02cx & $-13.90$ & NN / +60 & 0.011 & 10.99 / 2.50\\
        ZTF\,19abalbim / SN\,2019ttf & 18h42m15.87s & +24$^\circ$53\arcmin$48\farcs99$ & SN\,Ia\,02cx & $-13.99$ & NN / +230 & 0.011 & 11.07 / 2.52\\
        \hline
    \end{tabular}
    \label{tab:summary}
\end{table*}

\begin{figure*}
\centering
\includegraphics[width=\textwidth]{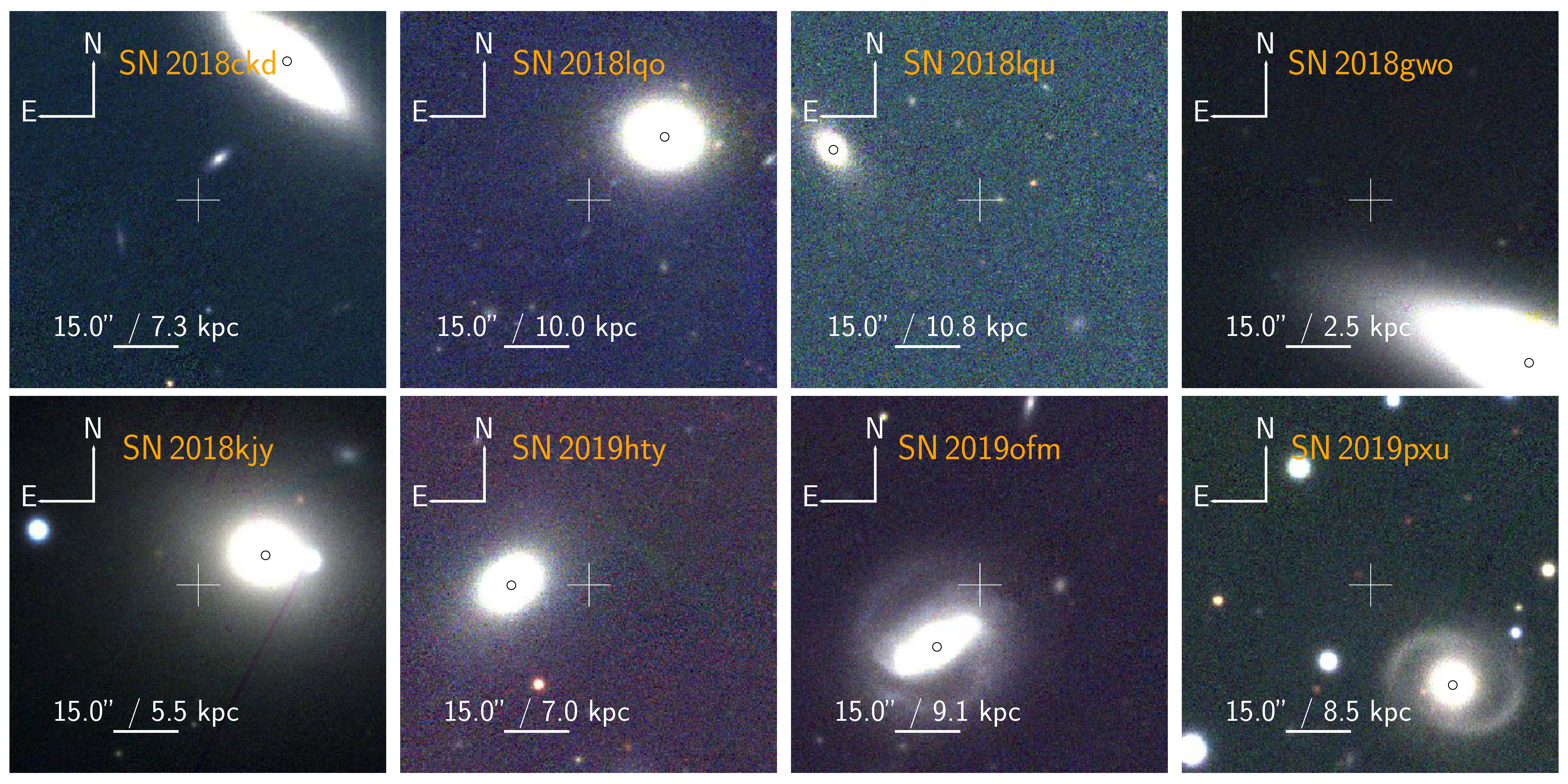}
\caption{RGB composite archival images of the locations of the 8 Ca-rich gap transients in the ZTF CLU sample. Images were taken from the PS1 survey \citep{Chambers2016}. In each panel, the white cross-hair shows the location of the detected transient and the scale shows the projected angular size and physical scale at the redshift of the host galaxy. The apparent host galaxy is marked with a black circle at its core in each panel.}
\label{fig:cutouts}
\end{figure*}

This work focuses on the sample of Ca-rich gap transients identified in this experiment in the aforementioned period of operations (i.e. between 2018 June 01 to 2019 September 30). Given the large number of supernovae that are classified as a part of the experiment, it is not possible to perform nebular phase spectroscopic follow-up of all events. We thus identified candidate Ca-rich gap transients with a simple selection criteria using photometry from the ZTF survey and peak-light spectroscopic properties (from the classification effort), focusing on the population of hydrogen-poor low luminosity events in the sample. Although the criteria were motivated by the known properties of the previous sample of events, we deliberately kept these minimal for candidate selection due to the uncertainties on the intrinsic properties of the class and the small number of previously reported events. The candidate selection criteria were:
\begin{enumerate}
    \item The transient should have a peak luminosity (in $g_{\text{ZTF}}$ or $r_{\text{ZTF}}$ filter) fainter than $M = -17\,\rm{mag}$ at the known redshift of the host galaxy (after correcting for Galactic extinction, but not possible host galaxy extinction). We require at least two detections on the rise of the light curve in the same filter to be able to constrain the peak luminosity.
    \item The spectroscopic properties of the source should be consistent with a hydrogen-poor SN near peak light. We do not include any events that exhibit broad hydrogen features in their spectra (i.e. Balmer lines that are not emanating from the underlying host galaxy). We obtained at least one epoch of late-time (at $\gtrsim 30\,\rm{d}$ after peak light) spectroscopy with DBSP and LRIS on the Keck-I telescope for events that passed (1) and this criterion.
    \item We require that the sources exhibit an early transition to the nebular phase, which we confirm by either the appearance of nebular emission lines and a fading continuum starting at +$30\,\rm{d}$ from peak light or a complete transition to the nebular phase by $\approx 150\,\rm{d}$ after peak light.
    \item As strong [Ca II] emission is the hallmark feature of this class, we require that the nebular phase spectrum should exhibit [Ca II]/[O I] $> 2$ \citep{Milisavljevic2017, De2018b} at any phase where the spectrum exhibits nebular emission lines.
\end{enumerate}

Comparing the selection criteria to \citet{Kasliwal2012a}, we note that we do not select candidates based on the fast photometric evolution or photospheric phase velocities. This choice makes us more sensitive to events with larger diversity in ejecta masses and velocities. Applying selection criteria (1) and (2) to the sample of events in the volume limited experiment, we were left with 22 events which were followed up with spectroscopy in the nebular phase. Seven of these 22 sources were found to qualify the criterion for [Ca II]/[O I] $> 2$ in the nebular phase, which defines the primary sample for this paper. In addition, we present observations of SN\,2018gwo, a nearby Type Ic SN which was not detected in ZTF data before peak (due to a maintenance break in October 2018), but was recovered on its radioactive decline tail. Combined with publicly available photometry and spectroscopy, we show that SN\,2018gwo is a likely Ca-rich gap transient at a distance of 28 Mpc. We apply the same selection criteria to the published literature sample of Ca-rich gap transients to include as our comparison sample. The set of ten literature events satisfying our cuts are SN\,2005E, SN\,2007ke, PTF\,09dav, SN\,2010et, PTF\,11bij, SN\,2012hn, PTF\,11kmb, PTF\,12bho and SN\,2016hgs and SN\,2016hnk. We describe the initial detection to final classification of each of the individual objects in the sample of Ca-rich gap transients presented in this paper. We present a discussion of the properties of the remaining objects that passed the light curve criteria but did not pass the nebular phase criterion in Appendix \ref{sec:control}, specifically highlighting how we rule out the Ca-rich classification for each event. Henceforth, we refer all phases with respect to time of $r$-band peak (see Section \ref{sec:lcfits}).

\subsubsection{SN\,2018ckd / ZTF\,18aayhylv}
ZTF\,18aayhylv (= SN\,2018ckd) was first detected in the ZTF difference imaging pipeline on 2018-06-07.19 ($\rm{MJD}=58276.19$) at J2000 coordinates $\alpha = $ 14:06:11.95 and $\delta =$ 09:20:39.3, at a magnitude of $r = 19.39 \pm 0.09\,\rm{mag}$. The transient passed the filter on the GROWTH Marshal on 2019-06-10 (second detection) and was saved for spectroscopic follow-up. The transient was detected by the Catalina Realtime Transient Survey (CRTS; \citealt{Drake2009}) on 2019-06-12 and was reported to the TNS on the same date \citep{SN2018ckd}, and was assigned the IAU name AT\,2018ckd. The transient was not detected in the ZTF alert production pipeline on 2018-06-01.31 to 5$\sigma$ limit of $r\approx 19.82\,\rm{mag}$. However, the transient is detected at $\approx 4\sigma$ significance with $r\approx 20.25$\,mag with forced photometry (see Section \ref{sec:photometry}) at the transient location in the ZTF difference images \citep{Masci2019}. The last non-detection of the source in forced photometry was at 2018-05-29.31 down to a 5$\sigma$ limit of $r \approx 19.65\,\rm{mag}$. 

The transient was found in the outskirts of NGC 5463, a S0 galaxy at $z = 0.024$ (Figure \ref{fig:cutouts}), at a projected offset of $\approx 39$\arcsec from the host center, corresponding to a physical separation of $\approx 19$\,kpc. We obtained a spectrum of ZTF\,18aayhylv with DBSP on 2018-06-12, which exhibited P-Cygni features of He I and Ca II similar to Type Ib SNe. Subsequent photometry from ZTF and follow-up with the P60 + SEDM confirmed a faint and fast evolving (rise time $\lesssim 15\,\rm{d}$) light curve peaking at an absolute magnitude of $M\approx-16.0\,\rm{mag}$ (Figure \ref{fig:photometry}). A nebular phase spectrum of the source at $\approx +60\,\rm{d}$ from $r$-band peak with LRIS on the Keck-I telescope showed strong [Ca II] emission lines with weak [O I] emission, confirming an early transition to the nebular phase and the classification of this source as a Ca-rich gap transient.

\subsubsection{SN\,2018lqo / ZTF\,18abmxelh}

ZTF\,18abmxelh (= SN\,2018lqo) was first detected in the ZTF difference imaging pipeline on 2018-08-10.18 ($\rm{MJD}=58340.18$) at J2000 coordinates $\alpha = $ 16:28:43.26 and $\delta =$ 41:07:58.7, at a magnitude of $r = 20.11 \pm 0.17\,\rm{mag}$. The transient passed machine learning thresholds on the GROWTH Marshal on 2019-08-16 and was saved for spectroscopic follow-up. We reported the transient to the TNS on 2019-10-28 \citep{SN2018lqo}, leading to its IAU name of AT\,2018lqo. With forced photometry on the ZTF difference images, we find that the transient was not detected on 2018-08-07.18 ($\rm{MJD}= 58337.18$) down to a $5\sigma$ limit of $r \approx 21.16\,\rm{mag}$. The transient exhibited an initial fading of $\approx 0.7$\,mag in $\approx 3\,\rm{d}$ following the first detection, followed by a rise to a peak $\approx 10\,\rm{d}$ later. 

The transient was found in the outskirts of CGCG 224-043, an E-type galaxy at $z = 0.032$ (Figure \ref{fig:cutouts}), at a projected offset of $\approx 23\farcs2$, corresponding to a physical separation of $\approx 14.9$ kpc. We obtained a spectrum of ZTF\,18abmxelh with DBSP on 2018-08-21, which exhibited P-Cygni features of He I and Ca II similar to Type Ib SNe. Subsequent photometry from ZTF and follow-up with the P60 + SEDM indicated a faint and fast evolving (rise time $\lesssim 15\,\rm{d}$) light curve peaking at an absolute magnitude of $M\approx-16.1\,\rm{mag}$ (Figure \ref{fig:photometry}). A nebular phase spectrum of the source at $\approx +50\,\rm{d}$ from $r$-band peak with LRIS on the Keck-I telescope showed strong [Ca II] emission lines, confirming a fast nebular phase transition and the classification of this source as a Ca-rich gap transient.

\subsubsection{SN\,2018lqu / ZTF\,18abttsrb}
ZTF\,18abttsrb (= SN\,2018lqu) was first detected in the ZTF difference imaging pipeline on 2018-09-03.13 ($\rm{MJD}= 58364.13$) at J2000 coordinates $\alpha = $ 15:54:11.48 and $\delta =$ +13:30:50.9, at a magnitude of $r = 20.14 \pm 0.33\,\rm{mag}$. The transient was saved as a candidate supernova on its second detection on 2018-09-07, and assigned spectroscopic follow-up. We reported the transient to the TNS on 2019-11-06 \citep{SN2018lqu}, leading to its IAU name of AT\,2018lqu. With forced photometry on the ZTF difference images, we find that the transient was not detected on 2018-08-16.17 ($\rm{MJD}= 58346.17$) down to a $5\sigma$ limit of $r \approx 21.16\,\rm{mag}$.

The transient was found in the outskirts of WISEA J155413.91+133102.4, a E-type galaxy at $z = 0.035$ (Figure \ref{fig:cutouts}), at a projected offset of $\approx 37\farcs1$, corresponding to a physical separation of $\approx 26.0$ kpc. We obtained a spectrum of ZTF\,18abttsrb with DBSP on 2018-09-12, which exhibited P-Cygni features of He I and Ca II similar to Type Ib SNe. Subsequent photometry from ZTF and follow-up with the P60 + SEDM indicated a faint and fast evolving (rise time $\lesssim 15\,\rm{d}$) light curve peaking at an absolute magnitude of $M\approx-16.4\,\rm{mag}$ (Figure \ref{fig:photometry}). A nebular phase spectrum of the source at $\approx +30\,\rm{d}$ from $r$-band peak with LRIS on the Keck-I telescope showed strong [Ca II] and weak [O I] emission lines, confirming an early transition to the nebular phase and the classification of this source as a Ca-rich gap transient.

\subsubsection{SN\,2018gwo / ZTF\,18acbwazl / Gaia\,18dfp / PS\,19lf}
ZTF\,18acbwazl (= SN\,2018gwo) was first detected in the ZTF difference imaging pipeline on 2018-10-31.49 ($\rm{MJD}= 58422.49$) at J2000 coordinates $\alpha = $ 12:08:38.83 and $\delta =$ +68:46:44.4, at a magnitude of $g = 19.20 \pm 0.17\,\rm{mag}$. Since the source was detected multiple times in the same night as part of the collaboration high cadence survey, the source was saved to the GROWTH Marshal on 2018-10-31. The source was detected by ZTF after a month-long gap in survey operations due to maintenance on the P48 camera. The source was first detected on 2018-09-28 by \citet{SN2018gwo} at $16.4\,\rm{mag}$ (clear filter) and reported to the TNS with the IAU Name AT\,2018gwo, shortly after the sky region emerged from solar conjunction. An upper limit of 17 mag was reported on the previous night. A low resolution spectrum from the Three Hills Observatory was reported to the TNS on 2018-09-30 and 2018-10-06 by \citet{Leadbeater2018}, consistent with a Type Ib/c SN with a reddened continuum near peak light, renaming this source to SN\,2018gwo. Subsequent ZTF photometry showed that the source was detected by ZTF on its post peak decline tail. We obtained a spectrum of the source with P60 + SEDM on 2018-11-06, which exhibited a weak continuum with emerging broad [Ca II] and Ca II lines.

The transient was found in the outskirts of NGC 4128, a S0 galaxy at a distance (median reported in NED) of 28.6 Mpc (Figure \ref{fig:cutouts}), at a projected offset of $54\farcs2$ corresponding to a physical projected distance of 9.2\,kpc. Although the transient was not detected by ZTF around peak light, we find the public observations combined with the later follow-up to be consistent with that of a Ca-rich gap transient. First, the initial detection and prior non-detection of the source reported by P. Wiggins suggest a fast rise to an absolute magnitude of $M \approx -16.0\,\rm{mag}$ (in clear filter; Figure \ref{fig:photometry}). The transient subsequently declined rapidly by $\approx 2$\,mags within $\approx 30\,\rm{d}$ after peak, confirming the faint peak luminosity and fast photometric evolution of the event. The peak light spectra together with the SEDM spectrum taken at $\approx 30\,\rm{d}$ are consistent with a Type Ib/c SN\footnote{Although not used as a defining characteristic of the class, the only Type Ib/c SNe reported thus far in the outskirts of early-type galaxies are found to be Ca-rich gap transients.} in the photospheric phase, which exhibited a fast transition to the nebular phase. We obtained a follow-up spectrum of the transient with LRIS on the Keck-I telescope on 2018 Dec 04, which exhibited strong [Ca II] emission and weak [O I] emission, confirming the classification of this source of a Ca-rich gap transient. 

\subsubsection{SN\,2018kjy / ZTF\,18acsodbf / PS\,18cfh}
ZTF\,18acsodbf (= SN\,2018kjy) was first detected in ZTF difference imaging pipeline on 2018-12-03.36 ($\rm{MJD}=58455.36$) at J2000 coordinates $\alpha = $ 06:47:17.96 and $\delta =$ +74:14:05.9, at a magnitude of $r = 19.56 \pm 0.14\,\rm{mag}$. The transient was saved as a candidate supernova on its second detection on 2018-12-04, and assigned for spectroscopic follow-up. The transient was detected by the Pan-STARRS1 survey \citep{Chambers2016} on 2018-12-17 and reported to TNS on 2018-12-22 \citep{SN2018kjy}, acquiring the IAU name AT\,2018kjy. With forced photometry on the ZTF difference images, we find three more lower significance detections up to $\approx 7\,\rm{d}$ before the first alert was issued. The last non-detection of the source was on 2018-11-19.48 ($\rm{MJD}= 58442.48$) to a 5$\sigma$ limit of $g \approx 20.45\,\rm{mag}$.

The transient was found in the outskirts of NGC 2256, an E-type galaxy at $z = 0.017$ (Figure \ref{fig:cutouts}), at a projected offset of $\approx 17\farcs2$, corresponding to a physical separation of $\approx 6.0$ kpc. We obtained a spectrum of ZTF\,18acsodbf with DBSP on 2018-12-14, which exhibited narrow P-Cygni features of He I, O I and Ca II, and a reddened continuum similar to the Ca-rich gap transient PTF\,12bho \citep{Lunnan2017}. Subsequent photometry from ZTF and follow-up with the P60 + SEDM confirmed a faint and fast evolving (rise time $\approx 17\,\rm{d}$) light curve peaking at an absolute magnitude of $M\approx-15.6\,\rm{mag}$ (Figure \ref{fig:photometry}). Subsequent spectra of the source taken with Keck/LRIS $+30\,\rm{d}$ and $+120\,\rm{d}$ from peak show a fast transition to the nebular phase dominated by [Ca II] emission, confirming its classification as a Ca-rich gap transient.

\subsubsection{SN\,2019hty / ZTF\,19aaznwze / ATLAS\,19nhp / PS\,19bhn}

ZTF\,19aaznwze (= SN\,2019hty) was first detected in the ZTF difference imaging pipeline on 2019-06-14.18 ($\rm{MJD}= 58648.18$) at J2000 coordinates $\alpha = $ 12:55:33.03 and $\delta = $ +32:12:21.7, at a magnitude of $g = 19.62 \pm 0.19\,\rm{mag}$. The transient passed machine learning thresholds on the GROWTH Marshal on 2019-06-20 and was saved for spectroscopic follow-up. The transient was detected by the ATLAS survey \citep{Tonry2018} on 2019-06-19 \citep{SN2019hty} and reported to TNS on the same date, and assigned the IAU name AT\,2019hty. Since the source was detected in unreleased public survey data from ZTF, we are unable to perform forced photometry on the images and report the last non-detection as in the ZTF alert packets on 2019-06-11.24 ($\rm{MJD}= 58645.24$) to a 5$\sigma$ limiting magnitude of $g = 19.82\,\rm{mag}$.

The transient was found in the outskirts of WISEA J125534.50+321221.5, an E-type galaxy at $z = 0.023$ (Figure \ref{fig:cutouts}), at a projected offset of $\approx 18\farcs7$, corresponding to a physical separation of $\approx 8.7$\,kpc. We obtained a spectrum of ZTF\,19aaznwze with SEDM and DBSP on 2019-07-01, which exhibited a Type Ib-like spectrum with a reddened continuum and a broad P-Cygni feature of the Ca NIR triplet. Photometric follow-up with SEDM and data from ZTF show a faint peak magnitude of $M \approx -16.1\,\rm{mag}$ and a rise time of $\approx 15\,\rm{d}$ (Figure \ref{fig:photometry}). We obtained an additional spectrum of the source with DSBP at $\approx 40\,\rm{d}$ from peak light, which showed broad emerging line of [Ca II] and weak [O I], confirming a fast transition to the nebular phase dominated by [Ca II] emission, and classifying this source as a Ca-rich gap transient.

\subsubsection{SN\,2019ofm / ZTF\,19abrdxbh / ATLAS\,19tjf}

ZTF\,19abrdxbh (= SN\,2019ofm) was first detected in the ZTF difference imaging pipeline on 2019-08-20.15 ($\rm{MJD}= 58715.15$) at J2000 coordinates $\alpha = $ 14:50:54.65 and $\delta = $ +27:34:57.6, at a magnitude of $g = 20.43 \pm 0.24\,\rm{mag}$. The transient met machine learning thresholds and was saved to the GROWTH Marshal on 2019-08-24. The ZTF detection was reported by the AMPEL \citep{Nordin2019} automatic stream to TNS on 2019-08-23 \citep{SN2019ofm}, acquiring the IAU name AT\,2019ofm. Since the source was detected in unreleased public survey data from ZTF, we are unable to perform forced photometry on the images and instead report the last non-detection as in the ZTF alert packets on 2019-08-17.24 ($\rm{MJD}= 58712.24$) to a 5$\sigma$ limiting magnitude of $r = 19.53\,\rm{mag}$.

The transient was found on top of the SB-type galaxy IC\,4514 at $z = 0.030$ (Figure \ref{fig:cutouts}), at a projected offset of $\approx 18\farcs2$, corresponding to a physical offset of $\approx 11.0$\,kpc. We obtained a spectrum of the source with DBSP on 2019-08-27, which exhibited clear features of a 1991bg-like Type Ia SN at the host redshift. Following the first detection, the source rose to a peak absolute magnitude of $M_r \approx -16.6\,\rm{mag}$, suggesting a sub-luminous Type Ia SN consistent with the spectroscopic classification. We obtained a follow-up spectrum of the source with LRIS at $\approx 175\,\rm{d}$ after peak light, which showed that the source had transitioned to the nebular phase exhibiting [Ca II] as the only detectable broad feature in the spectrum. Together with the non-detection of iron-group features typically seen in 1991bg-like objects, the strong [Ca II] feature suggested that SN\,2019ofm was similar to the Ca-rich gap transients PTF\,09dav and SN\,2016hnk, thus classifying the source as a Ca-rich gap transient.

\subsubsection{SN\,2019pxu / ZTF\,19abwtqsk / ATLAS\,19uvg / PS\,19fwq}

ZTF\,19abwtqsk (= SN\,2019pxu) was first detected in the ZTF difference imaging pipeline on 2019-09-04.50 ($\rm{MJD}= 58730.50$) at J2000 coordinates $\alpha = $ 05:10:12.61 and $\delta = $ $-$00:46:38.6, at a magnitude of $r = 20.09 \pm 0.20\,\rm{mag}$. The transient met machine learning thresholds and was saved to the GROWTH Marshal on 2019-09-22 on its third detection in the ZTF alert stream. The transient was detected by the ATLAS survey on 2019-09-10 \citep{SN2019pxu} and reported on the same date to TNS, acquiring the IAU name AT\,2019pxu. The field was not covered by the survey in $> 30\,\rm{d}$ before the first detection, and hence we are unable to determine a recent upper limit from the first detection. 

The transient was found in the outskirts of the spiral galaxy WISEA J051011.32-004702.5 at $z = 0.028$, at a projected offset of $30\farcs94$ (Figure \ref{fig:cutouts}), corresponding to a physical projected offset of $17.5$ kpc. We obtained a peak light spectrum of the source with P200 + DBSP on 2019-10-03, which exhibited a Type Ib-like spectrum and a reddened continuum similar to the Ca-rich gap transient PTF\,12bho. Subsequent photometry from ZTF and P60 + SEDM indicated a rise time of $\approx 17\,\rm{d}$ and a faint peak absolute magnitude of $M = -16.4\,\rm{mag}$. We obtained a follow-up spectrum with LRIS on 2019-10-27, which showed a faint continuum with emerging broad emission lines of [Ca II] and Ca II, confirming a fast nebular transition and the classfication of the source as a Ca-rich gap transient.

\subsection{Photometry}
\label{sec:photometry}

We obtained $gri$ photometry of transients from data taken with the P48 ZTF camera \citep{Bellm2019a}, that were processed with the ZTF data processing system \citep{Masci2019}. Light curves were extracted using forced point spread function (PSF) photometry \citep{Masci2019} at the location of the transient in the difference images, where the location was determined from the median position of the source reported in all alerts of the transient in the ZTF transient detection pipeline. We report detections of the transients in the forced photometry for epochs where the signal to noise ratio is higher than $3\sigma$, while $5\sigma$ upper limits are reported for other epochs. We include data acquired in the public as well as the higher cadence internal collaboration and Caltech surveys. In cases where the transient was covered by an internal survey and has more than one visit per night, we performed an inverse variance weighted binning of the flux measurements in bins of $1.5\,\rm{d}$ to improve the signal to noise ratio of the measurements. We perform the same binning for reporting upper limits, where we use inverse variance weighted flux uncertainty to report the $5\sigma$ upper limit for that epoch. At the time of writing, data acquired after the second data release of the ZTF public survey (i.e. data taken after 2019-06-30) is not available for forced photometry, and hence we use the ZTF light curves and upper limits as reported in the alert packets \citep{Masci2019}.

\begin{figure*}[!ht]
\centering
\includegraphics[width=0.9613\textwidth]{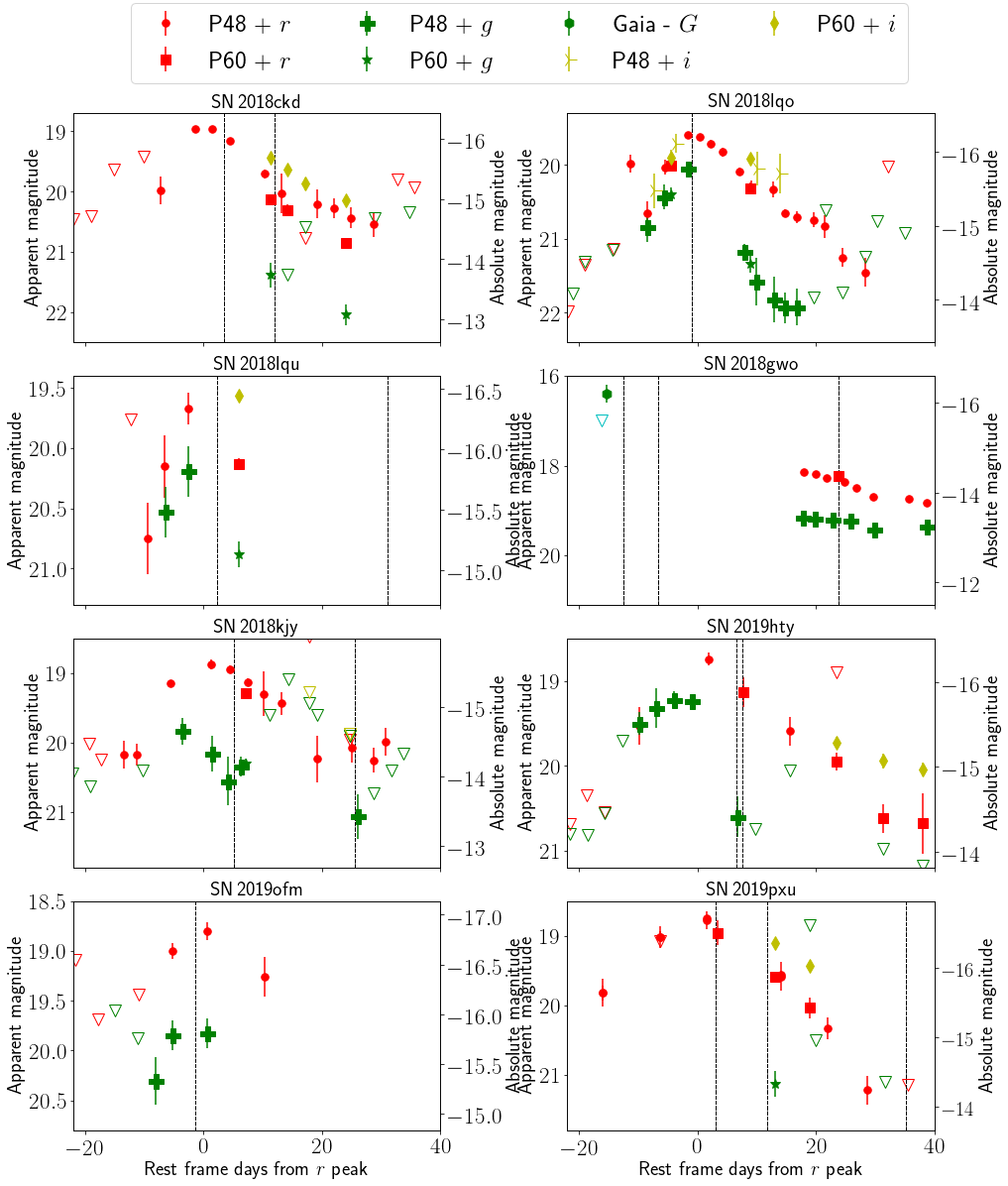}
\caption{Photometric evolution near peak light of the Ca-rich gap transients presented in this sample, with time presented with respect to the $r$-band light curve of the individual sources (corrected for Galactic extinction). Filled symbols denote detections from forced photometry on the ZTF difference images (see figure legend), while hollow inverted triangles denote 5$\sigma$ upper limits at the position of the transient. Red points denote $r$-band photometry, green points denote $g$-band photometry, yellow points denote $i$-band photometry and cyan points denote photometry in the clear filter.  The vertical dashed lines denote epochs of spectroscopy. For SN\,2018gwo, we also show public TNS photometry from Gaia (as in legend) and \citet{SN2018gwo} (in cyan).}
\label{fig:photometry}
\end{figure*}

We obtained additional multi-color photometry near peak light with the SEDM rainbow camera on the Palomar 60-inch telescope, and the data were processed using the SEDM image reduction pipeline. Image subtraction against archival references from SDSS and PS1 were performed, and difference magnitudes were obtained using the pipeline described in \citet{Fremling2016}. We show the photometric evolution of these transients near peak light in Figure \ref{fig:photometry}, while the data are presented in Table \ref{tab:photometry}. We correct the photometry for foreground Galactic extinction using the maps in \citet{Schlafly2011} and the extinction law of \citet{Cardelli1989}, assuming $R_V = 3.1$.

\subsection{Late-time imaging}

\begin{figure*}[!ht]
    \centering
    \includegraphics[width=\textwidth]{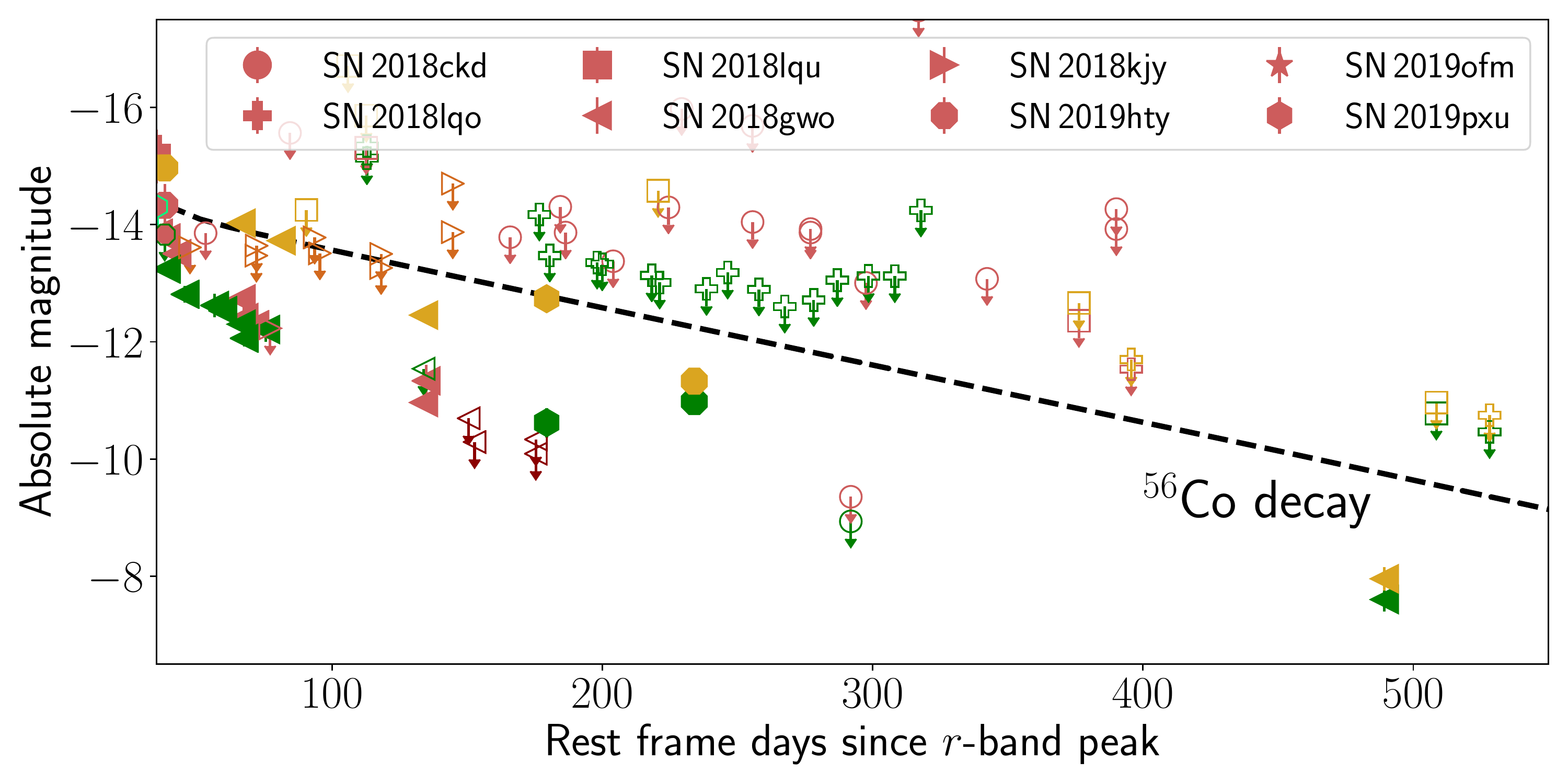}
    \caption{Late-time photometric evolution of the Calcium rich gap transients in the CLU sample, as function of rest-frame time from $r$-band peak. Each object is assigned a separate symbol for late-time detections as indicated in the legend. The colors denote individual filters, with brown for $r$-band, light green for $g$-band and golden for $i$-band. Upper limits are indicated as hollow symbols with arrows in the respective filter colors. The black dashed line shows the decline rate expected from radioactive decay of $^{56}$Co using an Arnett model with a $^{56}$Ni mass of 0.015 \Msun, under the assumption of complete trapping of $\gamma$-rays.}
    \label{fig:latephoto}
\end{figure*}{}

We obtained additional late-time photometry of the transients with the Wafer Scale Imager for Prime (WASP) on the Palomar 200-inch telescope, which were reduced with the pipeline described in \citet{De2019b}. We used LRIS on the Keck-I telescope for late-time imaging of some transients reported in this paper, and the data were reduced using the automated \texttt{lpipe} pipeline \citep{lpipe}. Image subtraction was not necessary for the late-time imaging for most sources since the majority were far from their host galaxies, and we report aperture photometry measurements (accounting for aperture corrections) for these sources, calibrated against the PS1 \citep{Chambers2016} catalog. For sources where the photometry was deemed to be contaminated by host galaxy light from visual inspection (which was found to be true only for SN\,2018kjy), we performed image subtraction using reference images from WASP obtained $> 1$ year after the peak of the transient light curve. The image subtraction was performed by first aligning the science image to the reference image by aligning the two images to the same system calibrated against Gaia DR2 \citep{Gaia2018}. The alignment was performed by first extracting a source catalog for both the science and the reference image using \texttt{SExtractor} \citep{Bertin2006} followed by astrometric alignment using the \texttt{scamp} code with Gaia DR2 as the reference catalog. 

The images were then resampled to the same output grid using \texttt{Swarp} \citep{Bertin2002} and flux scaled to a common zero-point. The image subtraction was performed using the \texttt{ZOGY} code \citep{Zackay2016} using an input PSF model for the science and reference image using \texttt{PSFEx} \citep{Bertin2011}. Forced PSF photometry was performed on the generated difference image to estimate the flux and flux uncertainty at the transient position, including an additional Monte Carlo simulation of the PSF-fit flux variance across the difference images to account for uncorrected correlated noise in the difference image output.  However, we caution that in several cases, the latest images from LRIS show evidence of diffuse sources near the transient which we are unable to subtract due to the absence of a template (ideally acquired several hundred days after the latest observation), and thus report the host-contaminated aperture photometry fluxes only. As in the case of the ZTF photometry, we report detections of sources where the signal to noise ratio is higher than 3$\sigma$, and $5\sigma$ upper limits otherwise. The late-time photometric evolution for the sample is shown in Figure \ref{fig:latephoto}. The nearby environments of the Ca-rich gap transients as observed in the late-time imaging are shown in Figure \ref{fig:lateimage}.

\subsection{Spectroscopy}

\begin{figure*}[]
    \centering
    \includegraphics[width=0.46\textwidth]{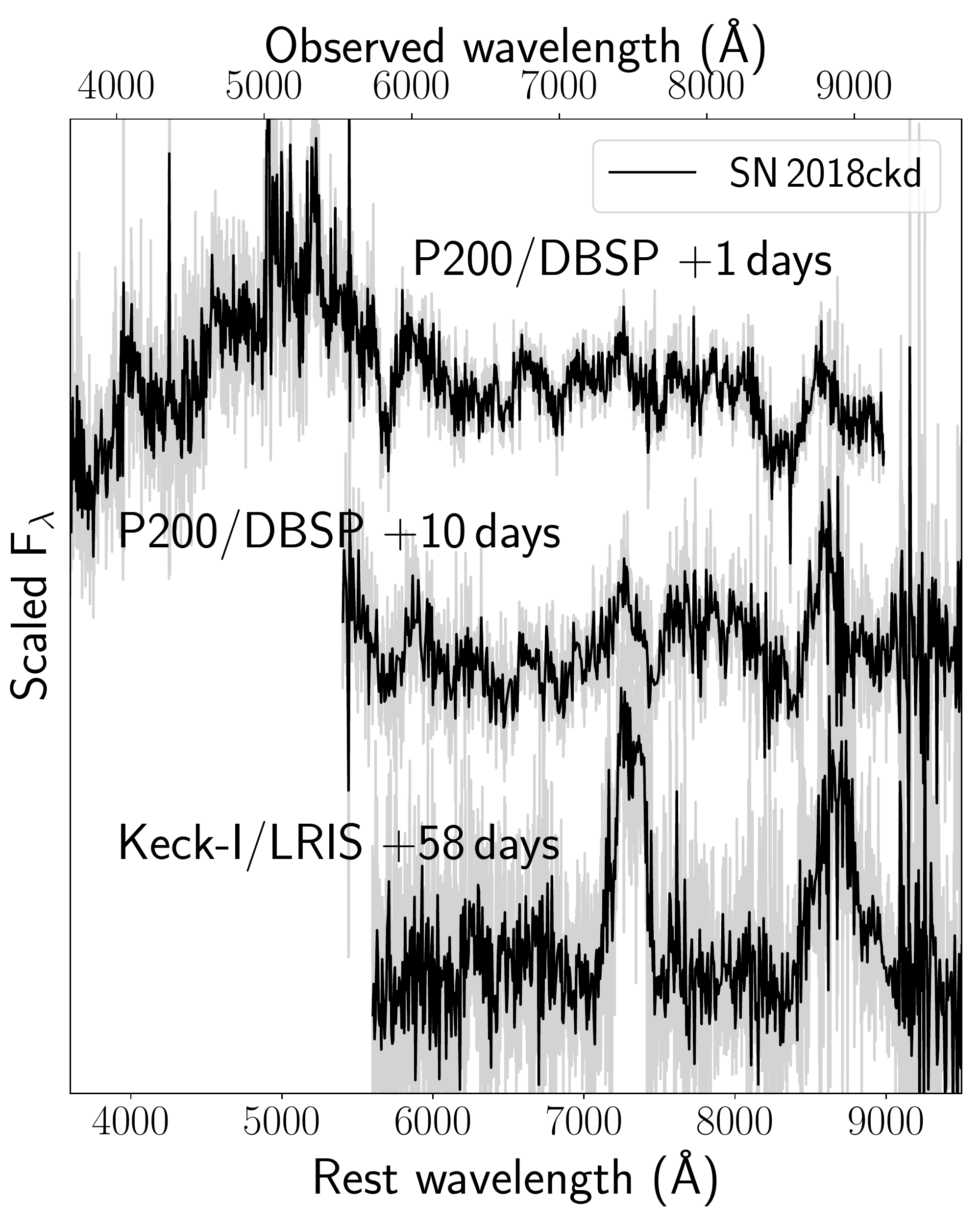}
    \includegraphics[width=0.46\textwidth]{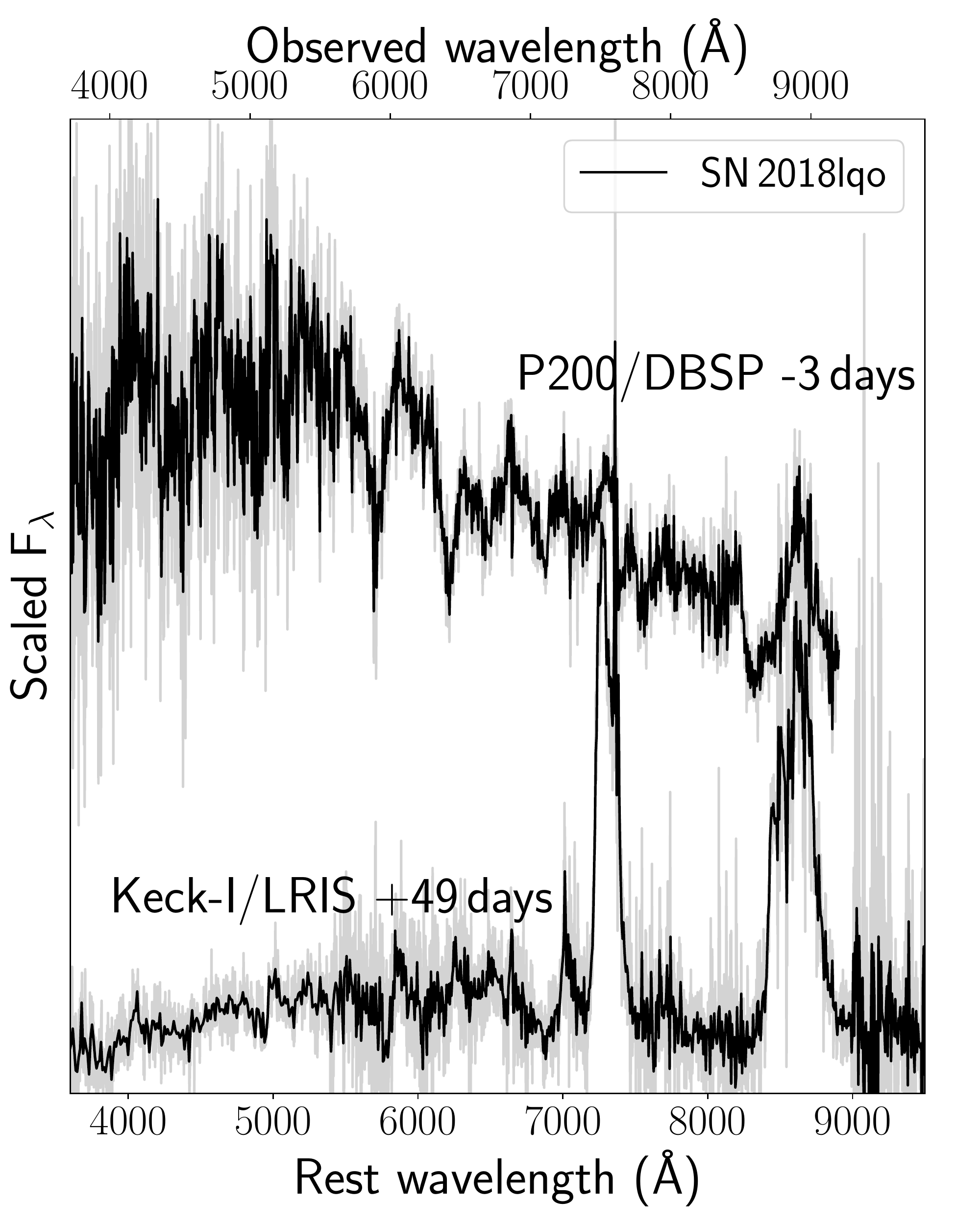}
    \includegraphics[width=0.46\textwidth]{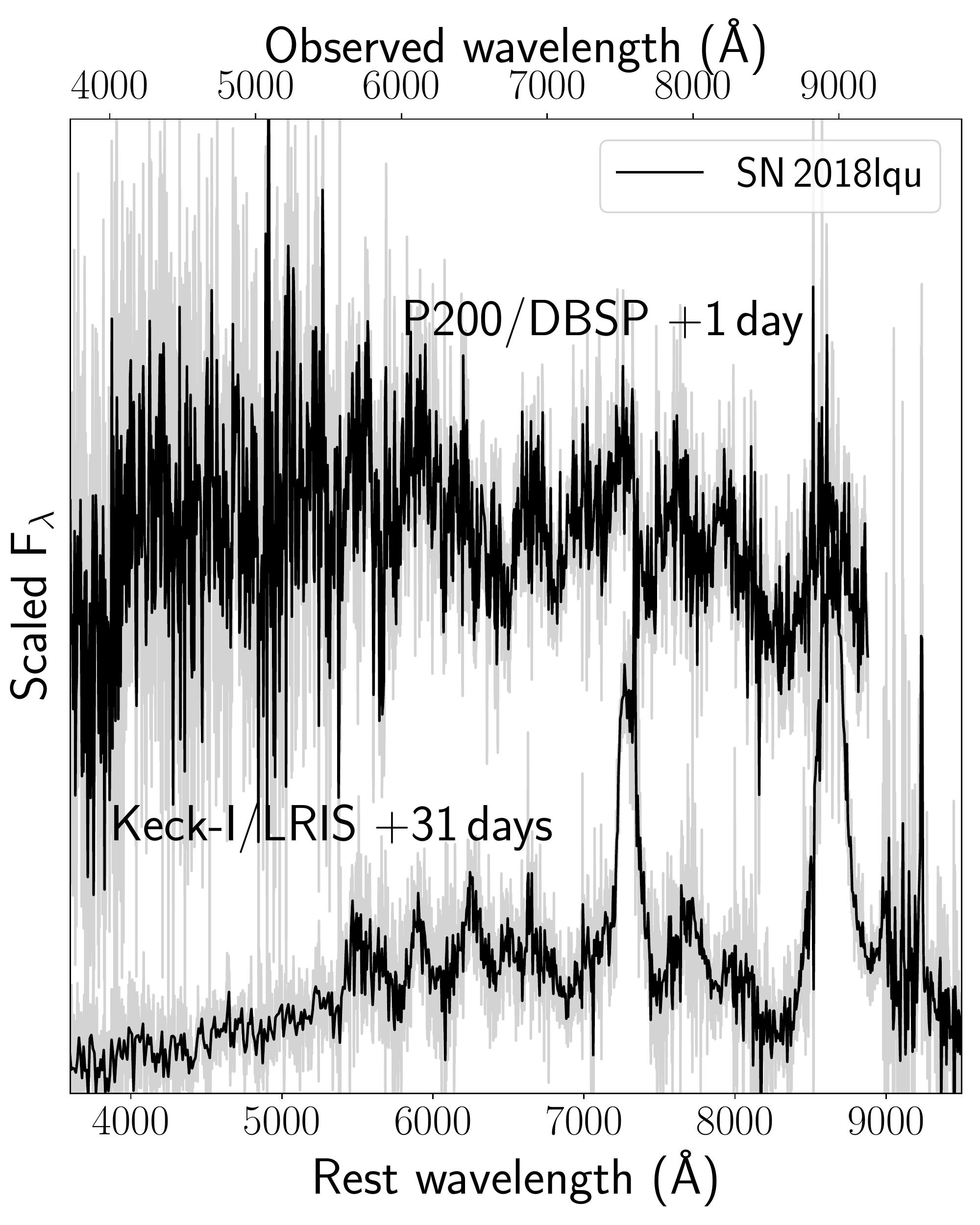}
    \includegraphics[width=0.46\textwidth]{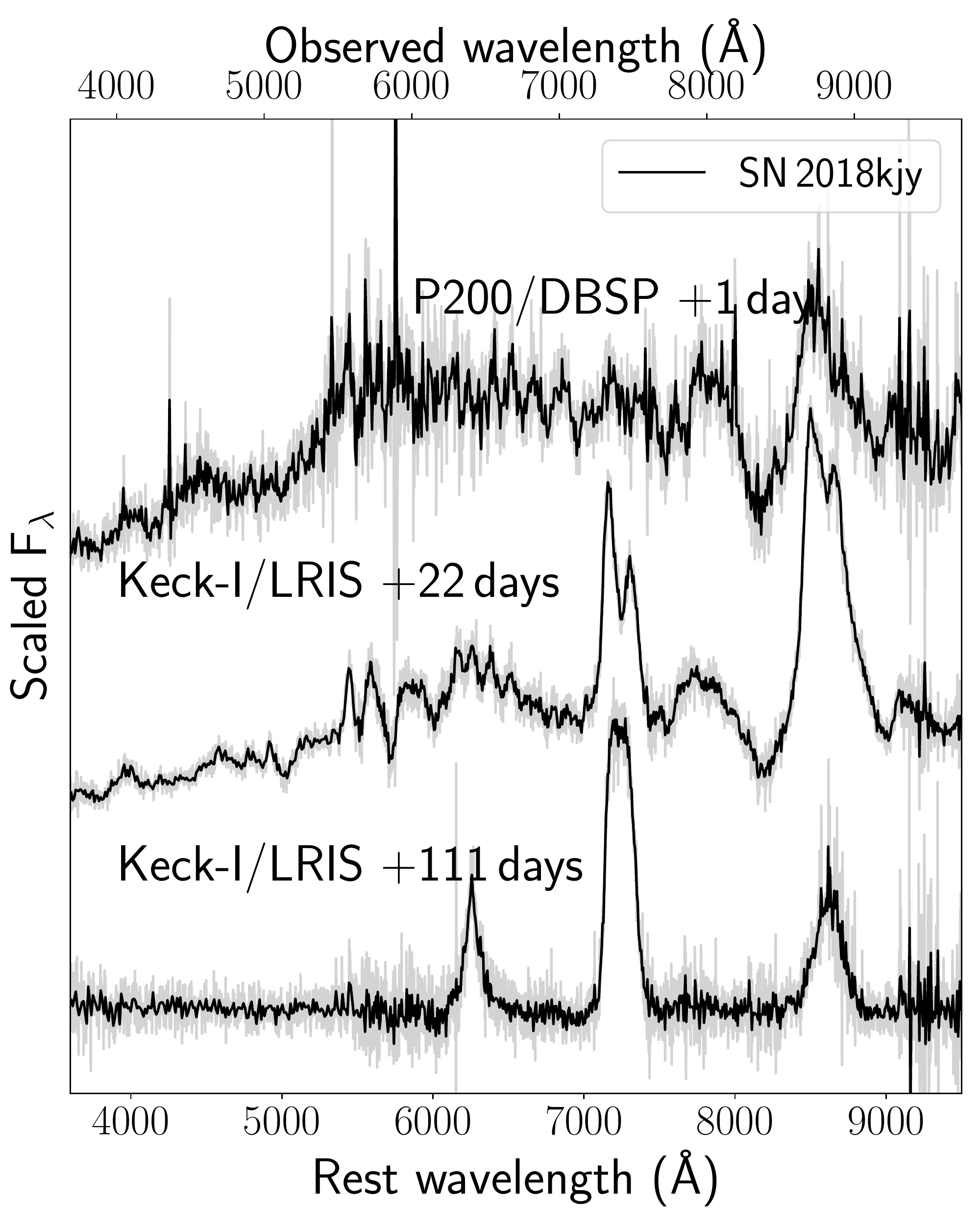}
    \caption{Spectroscopic sequence of the sample of Ca-rich gap transients presented in this paper. In each panel, the object name is indicated in the legend and the phase of the spectrum is denoted next to each spectrum with respect to the peak of the $r$-band light curve. The gray lines show the unbinned spectra while black lines show the same spectra binned to improve the signal-to-noise ratio.}
    \label{fig:spztf1}
\end{figure*}

\begin{figure*}[]
\ContinuedFloat
    \centering
    \includegraphics[width=0.46\textwidth]{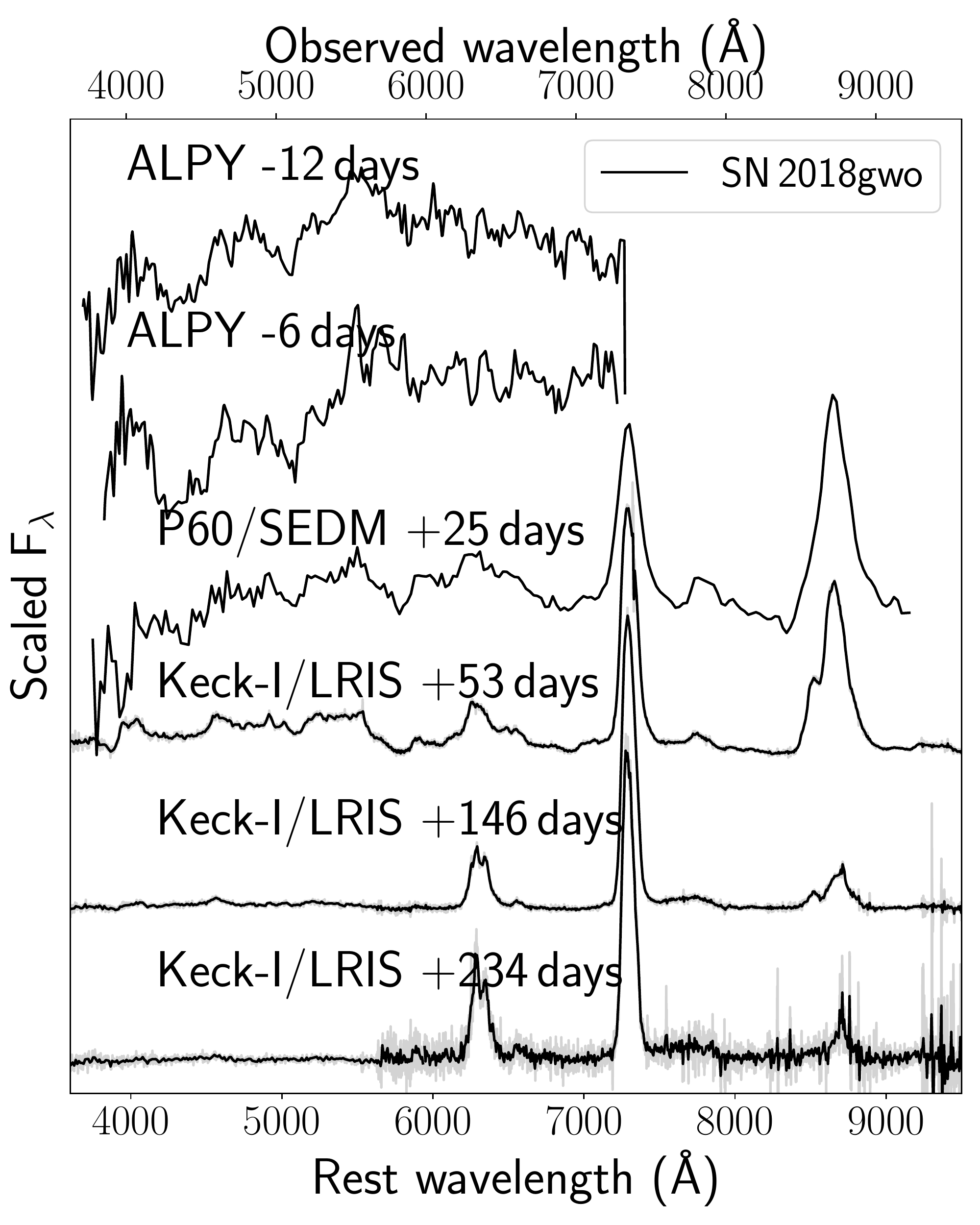}
    \includegraphics[width=0.46\textwidth]{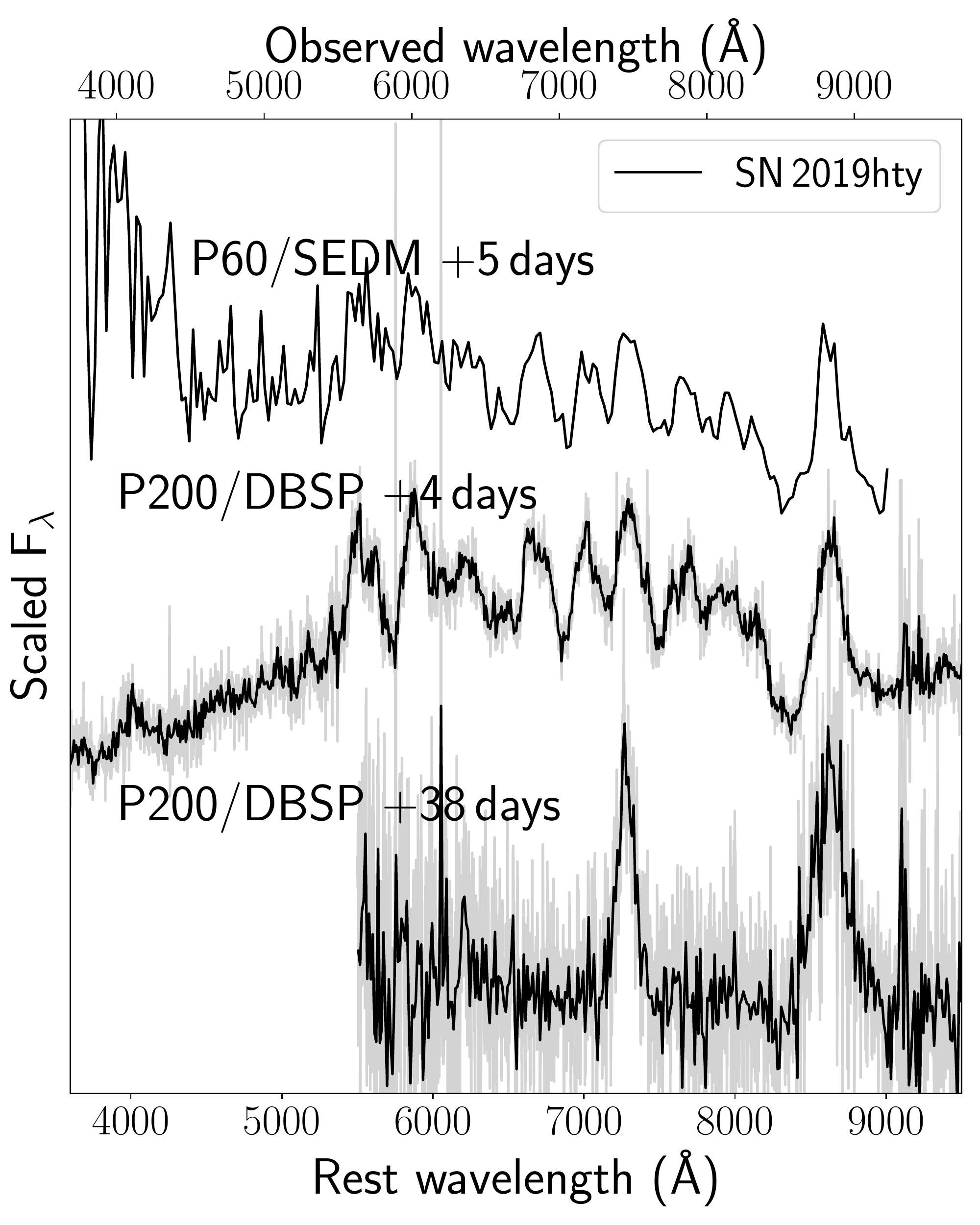}
    \includegraphics[width=0.46\textwidth]{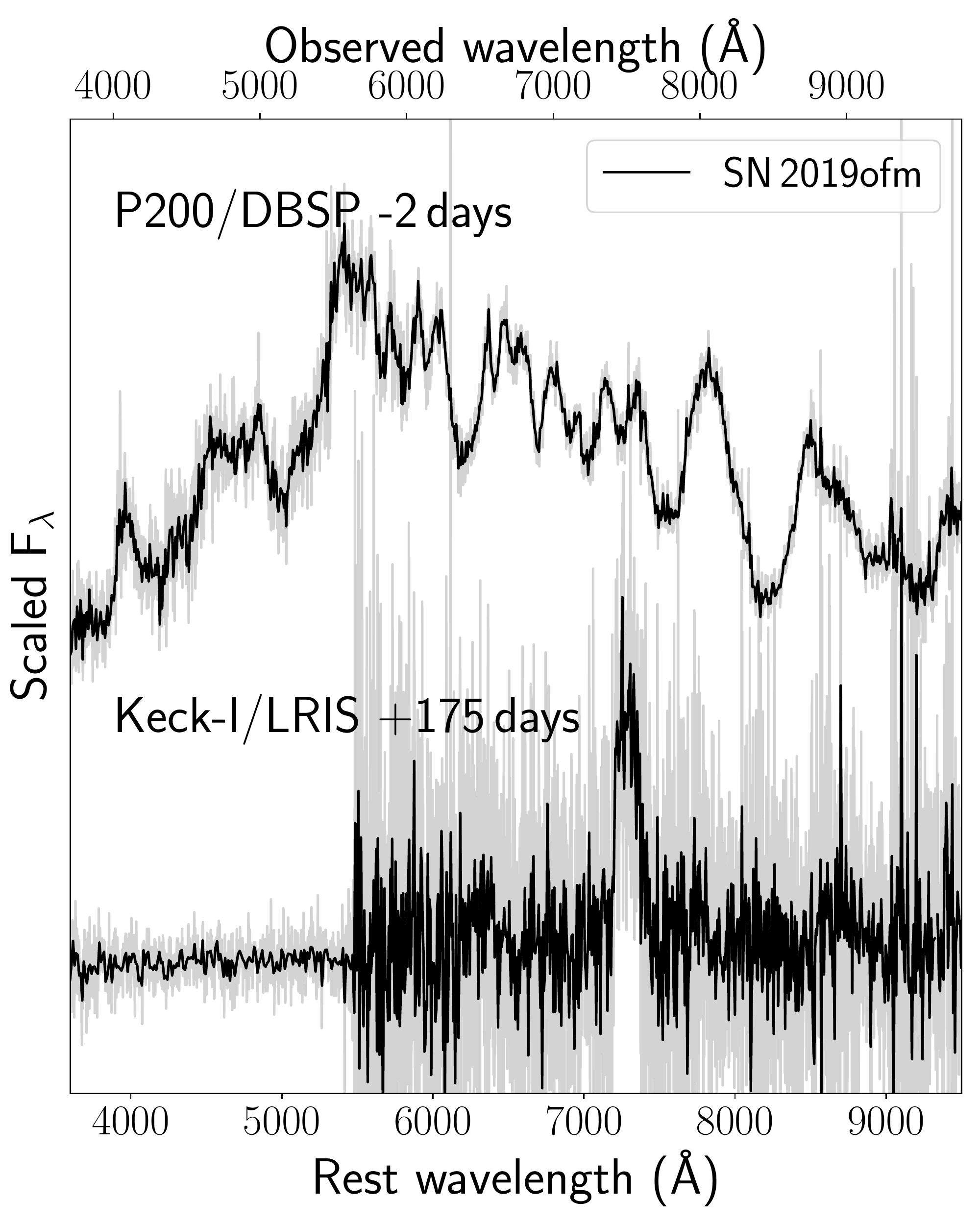}
    \includegraphics[width=0.46\textwidth]{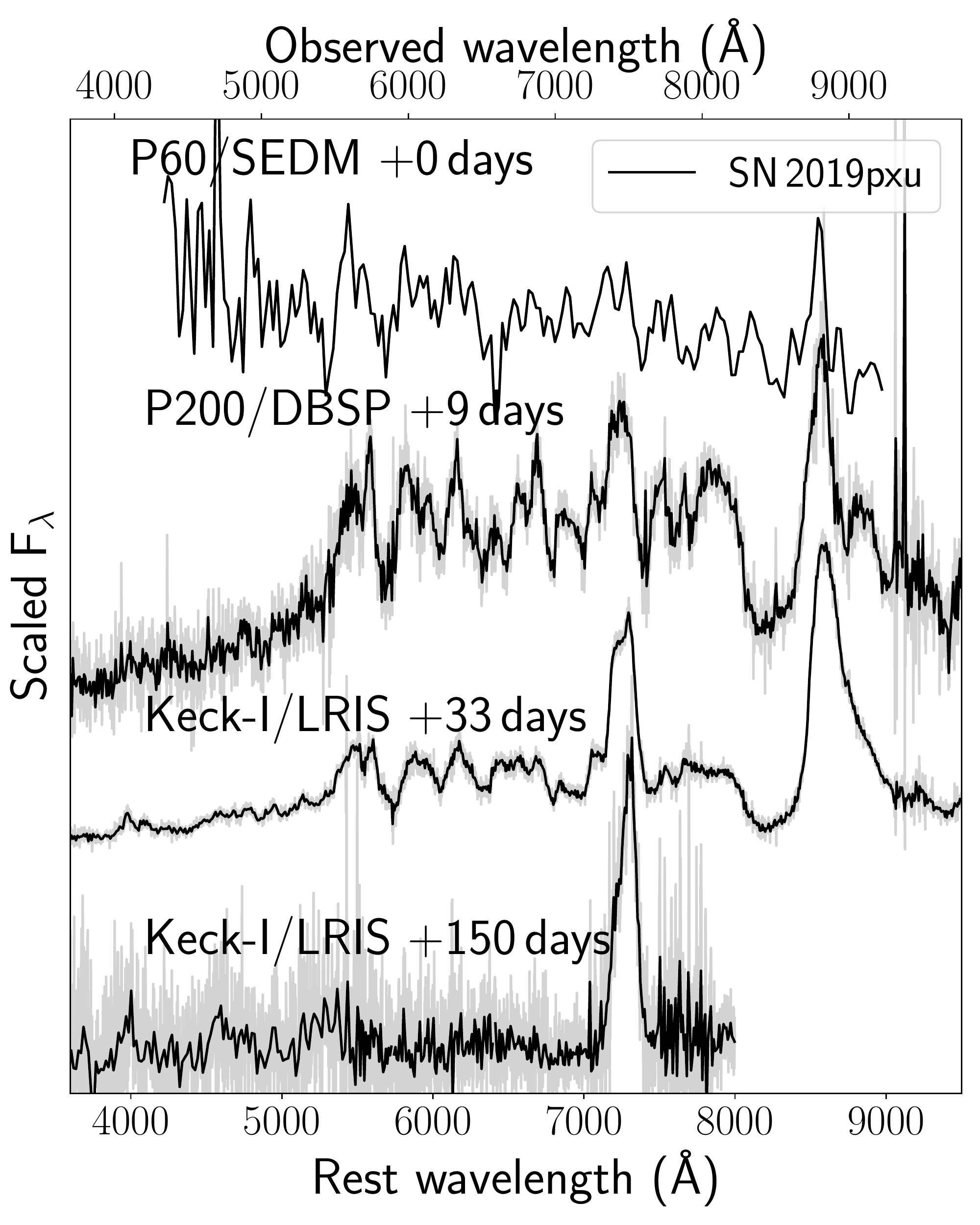}
    \caption{Continued}
\end{figure*}

Spectroscopic follow-up of transients near peak light was obtained as a part of the regular classification effort of the CLU experiment. Typically one spectrum was obtained near peak light for initial spectroscopic classification and a sequence of nebular phase spectra were obtained starting at least $\approx 30\,\rm{d}$ after peak. The SED Machine spectrograph was used for spectroscopy for only two of these sources (SN\,2018gwo and SN\,2019hty) since they were typically too faint ($> 19$\,mag) for SEDM spectroscopy. The SEDM data were reduced using the \texttt{pysedm} \citep{Rigault2019} automatic pipeline. Peak light spectroscopy for the rest of the sample was obtained using the DBSP, and the data were reduced using the \texttt{pyraf-dbsp} pipeline \citep{Bellm2016}. We obtained two epochs of spectroscopy of SN\,2018gwo with the APLY200 spectrograph at Three Hills Observatory. The spectra were reduced using ISIS software\footnote{by C. Buil; \url{http://www.astrosurf.com/buil/isis-software.html}}. The spectrum images were bias and dark subtracted, flat field corrected using a tungsten halogen lamp, corrected for geometric distortions and the sky background subtracted before extracting  the spectrum profile. The spectrum was wavelength calibrated using a Ne/Ar reference lamp and  calibrated in relative flux using as a reference, a hot star (HD123299) from the MILES library of spectra\footnote{\url{http://miles.iac.es/}} measured the same night at similar airmass. We obtained follow-up spectroscopy for some sources using the Alhambra Faint Object Spectrograph and Camera (ALFOSC) on the Nordic Optical Telescope (NOT). The NOT data were reduced using the {\sc foscgui} pipeline\footnote{\url{http://graspa.oapd.inaf.it/foscgui.html}}.

Late-time nebular spectroscopy was obtained using LRIS on the Keck-I telescope starting from $\approx 30\,\rm{d}$ after peak light. For some sources, we obtained up to four epochs of nebular phase spectra using LRIS. The data were reduced using the automated \texttt{lpipe} pipeline. We present the complete list of spectroscopic observations in Table \ref{tab:spectra}, while the spectra are presented in Figure \ref{fig:spztf1}. In addition, we use publicly available spectra from TNS for SN\,2018gwo and some events in the control sample (Appendix \ref{sec:control}). Spectra for the literature sample of events were obtained from WISEReP \citep{Yaron2012} and attributed to the original source where relevant. All data presented in this paper will be publicly released on WISEReP and as an electronic supplement upon publication.

\section{Analysis of the combined sample}
\label{sec:analysis}

Here, we present a combined analysis of the spectroscopic and photometric properties of the sample of Ca-rich gap transients presented in this paper with those in the literature sample of events that satisfy our selection criteria. We begin with a qualitative analysis of the spectroscopic properties, in particular, to highlight the existence of a continuum of spectroscopic characteristics in the full sample of events. We outline the procedures used for a quantitative analysis of the full sample of events, and present quantitative results on the photometric and spectroscopic properties of the sample to highlight trends across the continuum of spectroscopic properties. We use these results to discuss implications for the progenitor channels in Section \ref{sec:discussion}.

\subsection{Photospheric phase spectra}
\label{sec:peakspecclass}
We perform a combined analysis of the spectroscopic properties of the sample of Ca-rich gap transients presented in this paper with previous events in Figures \ref{fig:peakspec_comparison}, \ref{fig:nebspec_comparison} and \ref{fig:IaClass}.
Since we aim to characterize the peak light spectral diversity in this class, we discuss objects where a medium resolution spectrum was available within $\approx 10\,\rm{d}$ of peak light -- 16 of the total sample of 18 objects\footnote{SN\,2007ke and PTF\,11bij are the only objects that do not have a peak light spectrum} have photospheric phase spectra acquired near the peak of the light curve. The photospheric phase spectra of this sample are diverse, and most notably separate into SN\,Ib/c-like (absence of a strong Si II line, with a continuum of He I line strengths) and SN\,Ia-like (with strong Si II lines) objects. This distinction is a natural parallel to the traditional classification scheme invoked for the broader population of hydrogen poor supernovae \citep{Filippenko1997, Gal-Yam2017}. We thus proceed by defining two spectroscopic classes within the sample of Ca-rich gap transients based on their similarity to either SNe\,Ib/c or SNe\,Ia near peak light -- \textbf{and refer to them as Ca-Ib/c and Ca-Ia objects}. Within the photometric selection criteria defined in this experiment, the relative occurrence rate of the Ca-Ib/c to Ca-Ia objects are 6:1, although a true rate estimate would require incorporating the luminosity functions of the two classes (see Section \ref{sec:rates}).

\subsubsection{The Ca-Ib/c class}

\begin{figure*}[]
\centering
\includegraphics[width=0.75\textwidth]{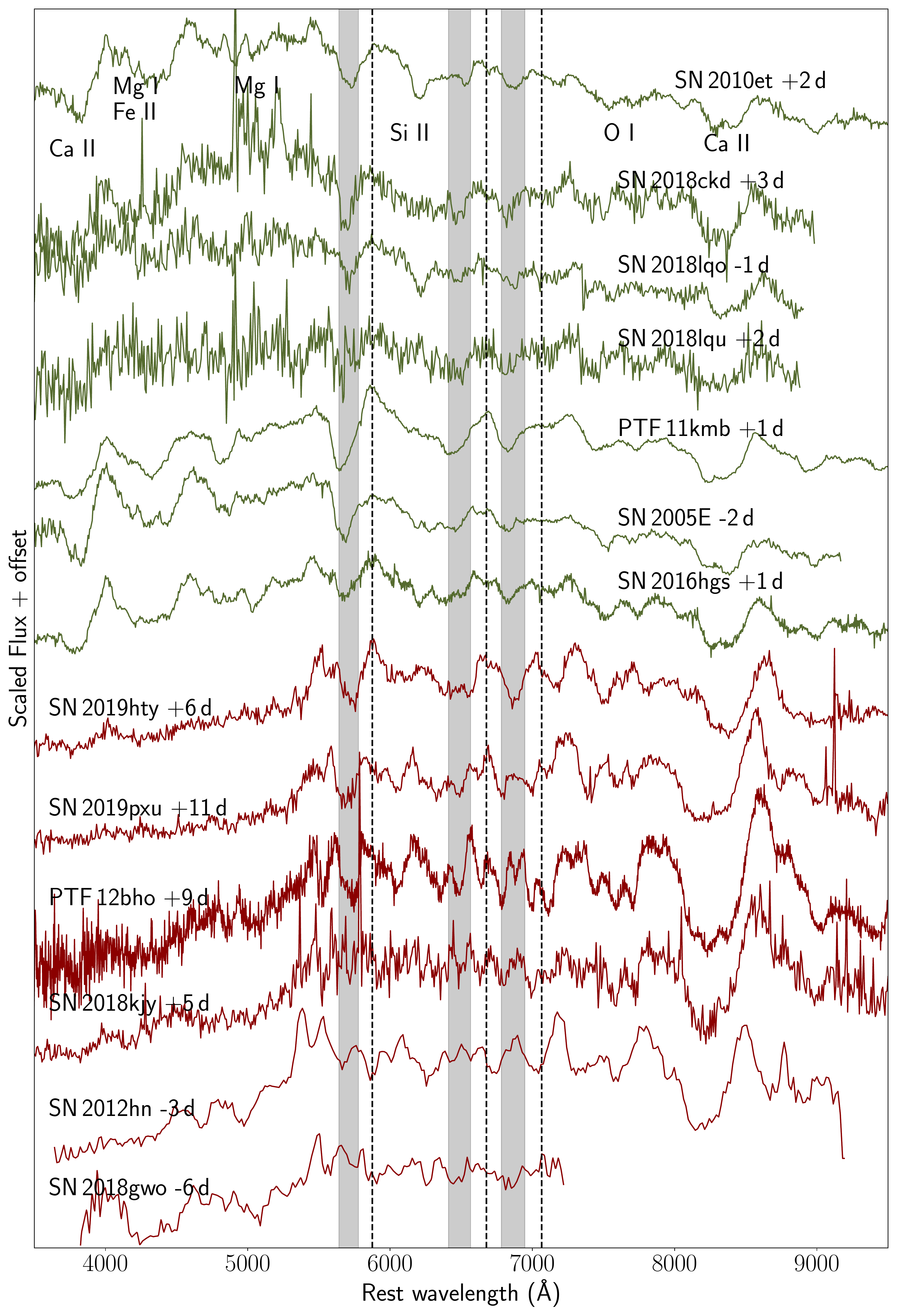}
\caption{Photospheric phase spectra of Ca-rich gap transients that exhibit SN\,Ib/c-like features (termed Ca-Ib/c objects) in the ZTF sample combined with the same for events in the literature. The transient name and phase of the spectrum is indicated next to each spectrum. The spectrum color separates the two primary spectral types in the sample based on the shape of the continuum -- the events plotted in green have peak spectra characterized by flat or green continua, while spectra in red show events that exhibit strong suppression of flux at bluer wavelengths thus exhibiting red continua. The solid dashed lines show the rest frame wavelengths of three optical He I lines, while the shaded bars show the expected P-Cygni absorption minima for velocities ranging from 5000 -- 12000 km s$^{-1}$.}
\label{fig:peakspec_comparison}
\end{figure*}

In Figure \ref{fig:peakspec_comparison}, we plot the photospheric phase spectra of the Ca-Ib/c objects. Prominent spectral lines detected in the photospheric phase are marked, along with three optical lines of He I and their P-Cygni absorption regions due to the known similarity of these objects to Type Ib SNe at peak \citep{Perets2010}. The ZTF sample of events is dominated by Type Ib-like spectra near peak light (exhibiting He I $\lambda5876$, $\lambda6678$ and $\lambda 7065$) in the photospheric phase spectra albeit with a range of line strengths and velocities. The He I $\lambda6678$ line is usually contaminated by the nearby Si II $\lambda\lambda 6347, 6371$ lines \citep{Sullivan2011, Kasliwal2012a, De2018b}. Other common features in the peak light spectra include P-Cygni features of O I $\lambda7774$ and Ca\,II H\&K and the NIR triplet. SN\,2012hn is the only object that does not show any evidence for He I in its peak spectrum \citep{Valenti2014}. We also do not conclusively identify He I in the peak light spectrum of SN\,2018gwo. We thus tentatively classify SN\,2018gwo as a Ca-Ic although the low SNR and resolution of the peak spectrum precludes a definite classification.

\begin{figure}
    \centering
    \includegraphics[width=0.49\textwidth]{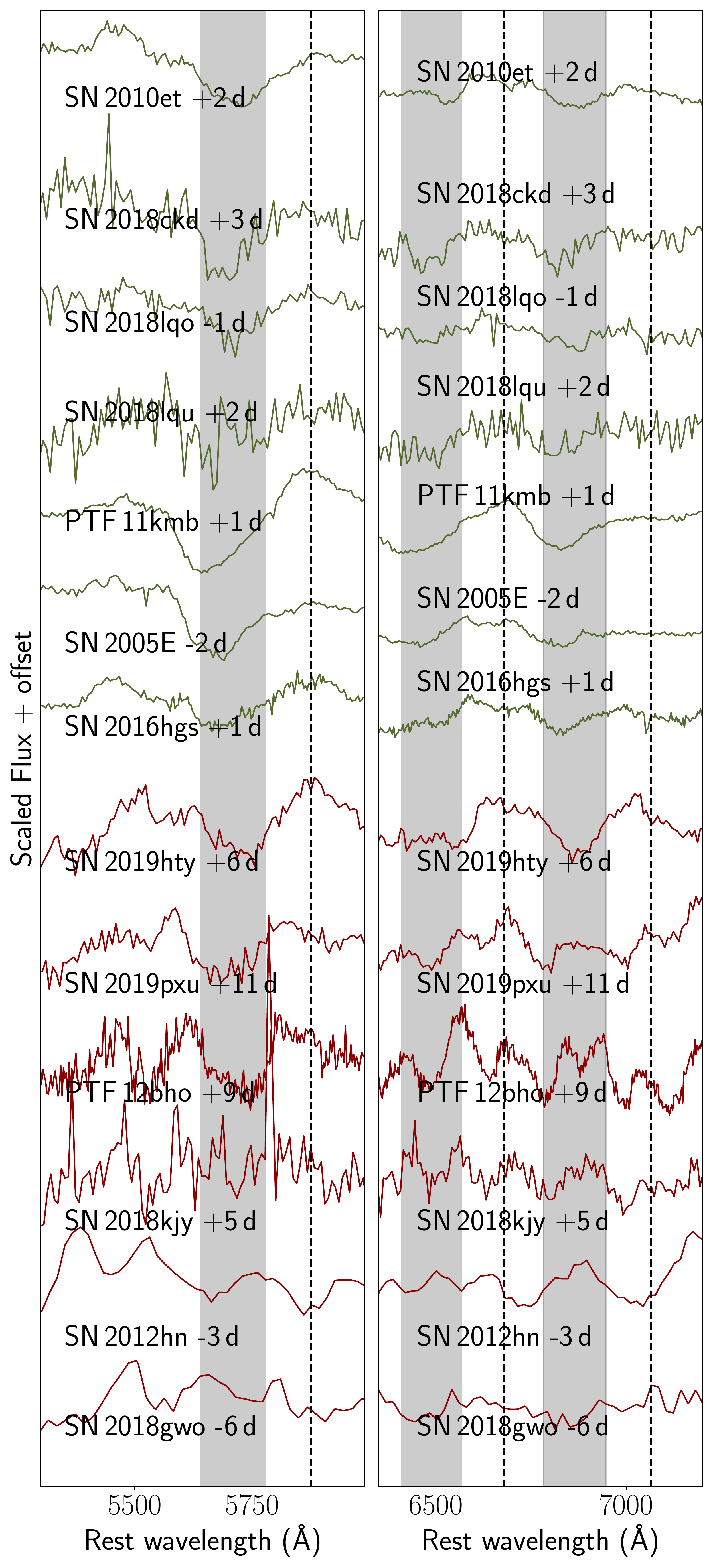}
    \caption{Zoomed-in plots of the photospheric phase spectra of the Ca-Ib/c objects around the expected positions of He I transitions. The left panel shows the region around the He I $\lambda5876$ line and the right panel shows the region including the He I $\lambda 6678$ and $\lambda 7065$ lines. The color coding of the spectra is the same as in Figure \ref{fig:peakspec_comparison}. The black dashed lines show the rest frame positions of the He I lines, while the grey shaded regions show the absorption region for the a velocity range of $5000 - 12000$ km s$^{-1}$.}
    \label{fig:helines}
\end{figure}{}

Upon closer inspection, the set of peak light spectra shown in Figure \ref{fig:peakspec_comparison} demarcates into two groups of events -- one with events characterized by flat continua across the entire spectral range and one with events characterized by strong suppression of flux at bluer wavelengths and red continua. We indicate these two classes of events with different colors (dark green and red) in Figure \ref{fig:peakspec_comparison}, and throughout the rest of this manuscript. The spectra of events in the first class are relatively homogeneous, and show clear evidence of strong He I at normal photospheric velocities ($\approx 8000 - 11000$ km s$^{-1}$; see Section \ref{sec:specfit}). Notably, these objects exhibit strong continuum in the blue side of the spectrum (below $5500$\,\AA) and clear absorption features of Ca II, Mg I and Fe II superimposed on the blue side continuum. On the other hand, events in the latter group show strong suppression of the continuum flux in the blue side of the spectrum (below $5500$\,\AA) producing a spectrum with redder colors. These exhibit a diverse range of line velocities, ranging from events with normal photospheric velocities (SN\,2019pxu) to peculiar low velocity ($\approx 4000 - 6000$ km s$^{-1}$) events such as PTF\,12bho and SN\,2018kjy. Absorption features of metals blue-wards of 5500\,\AA\ (Ca II, Mg I and Fe II), are only weakly detected due to the strong suppression of flux in this region. Notably, SN\,2012hn and SN\,2018gwo in this group do not show evidence of He I (and hence would be Ca-Ic objects nominally), while SN\,2018kjy exhibits only weak signatures of low velocity He I in its spectrum. Given the small number of events, it is unclear whether there is a continuum of events between these two types of objects. We proceed by referring to the two classes of objects as objects with green and red continua respectively, and use the same color scheme as in Figure \ref{fig:peakspec_comparison}. 

\begin{figure*}[!ht]
    \centering
    \includegraphics[width=0.8\textwidth]{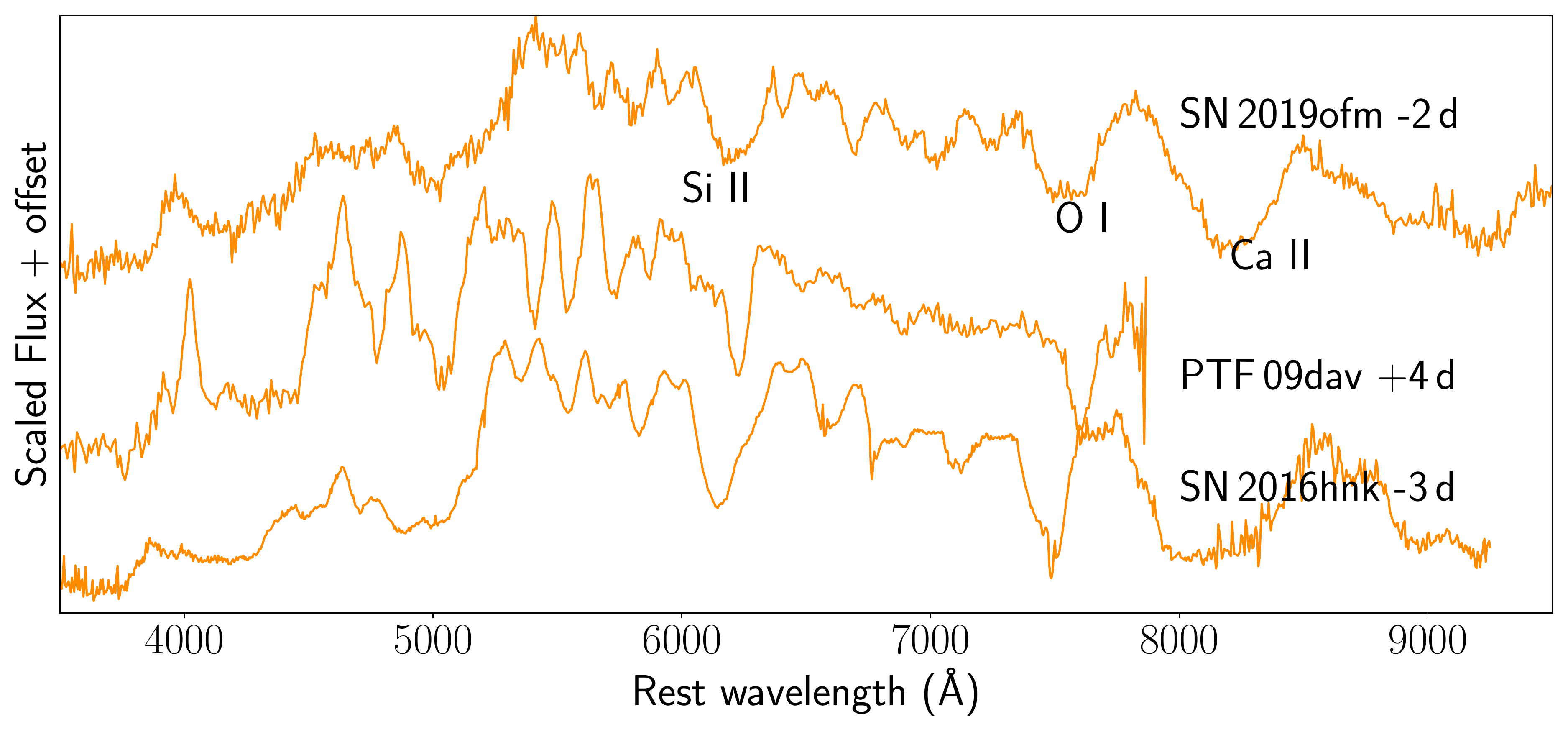}
    \includegraphics[width=0.8\textwidth]{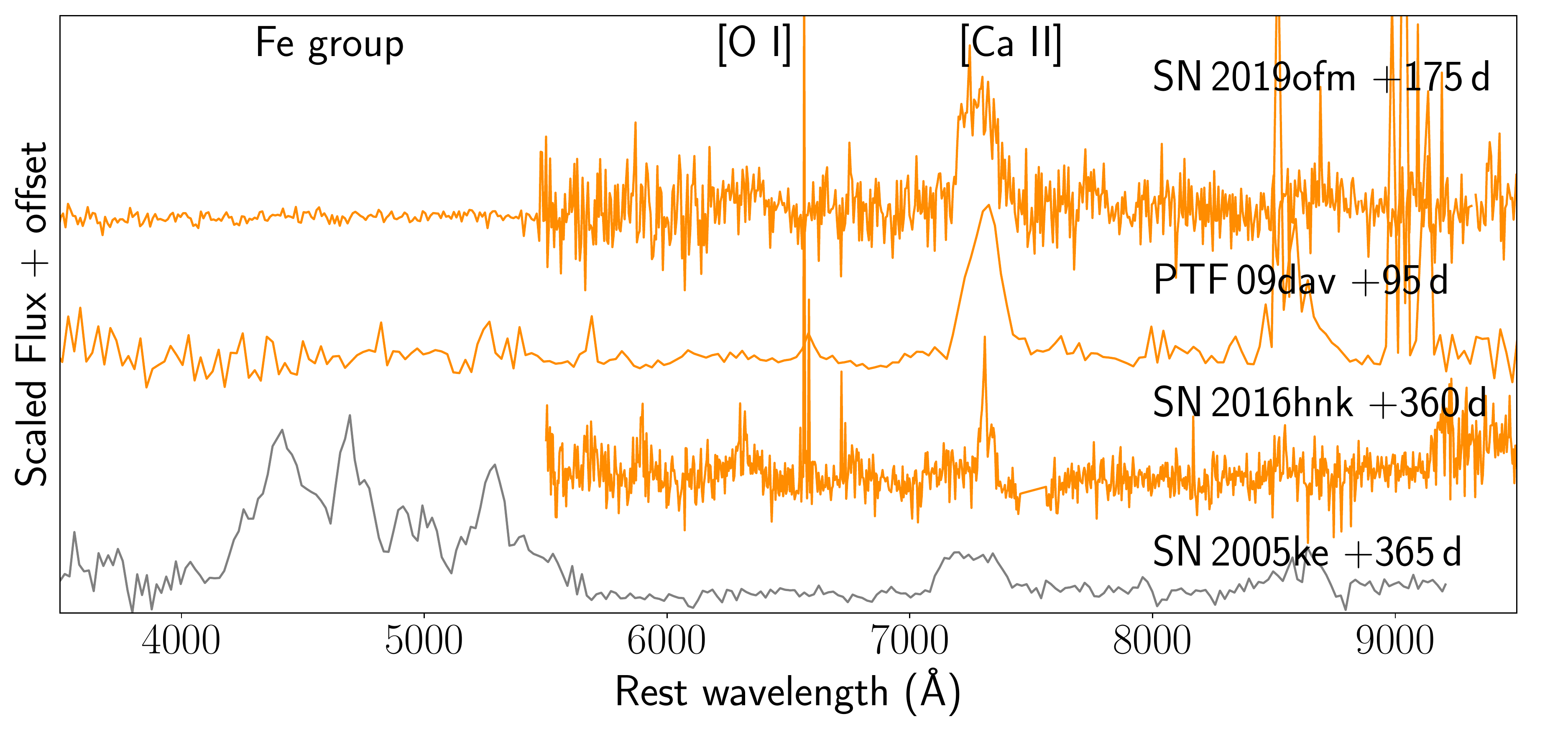}
    \caption{Photospheric (top panel) and nebular (bottom panel) phase spectra of Ca-rich gap transients that exhibit SN\,Ia-like features (termed Ca-Ia objects) in the ZTF sample combined with the same for events in the literature. The transient name and phase of the spectrum is indicated next to each spectrum. The prominent photospheric lines of Si II, O I and Ca II are marked in the peak light spectra plot while the nebular lines of [O I], [Ca II] and Fe group elements are marked in the lower panel. In the lower panel, we also show a nebular phase spectrum of the SN\,1991bg-like event SN\,2005ke to highlight the differences between 1991bg-like objects and the Ca-rich objects in terms of the absence of Fe-group features in the blue part of the spectrum.}
    \label{fig:IaClass}
\end{figure*}{}

There is considerable diversity in the presence and strength of He lines in the peak light spectra of the Ca-Ib/c objects. The identification of He is a crucial aspect for understanding the progenitors of these explosions since the presence of He in the ejecta is indicative of a He-rich progenitor system. However, He lines visible in the optical region are non-thermally excited \citep{Dessart2012, Hachinger2012} and hence their absence does not necessarily preclude the presence of He in the ejecta.  The detection of He lines is dependent on the amount of $^{56}$Ni mixing in the ejecta to be able to excite the He I transitions non-thermally and thus these lines also constrain on the mixing in the ejecta. While most of the Ca-Ib/c objects exhibit prominent and unambiguous He I lines in the optical, similar to the prototype event SN\,2005E, these lines are difficult to unambiguously identify in peculiar events. In Figure \ref{fig:helines}, we show zoomed in regions of the peak light spectra of the sample around the optical He I lines at 5876\,\AA, 6678\,\AA\, and 7065\,\AA. The identification of He is complicated by the contamination of the He I $5876$\,\AA line with the nearby Na I line seen in SNe\,Ic, thus requiring the detection of multiple He I lines at similar velocities to conclusively confirm the presence of He. This necessitates a careful examination of the features around the other optical He I lines at 6678\,\AA\, and 7065\,\AA. We note that all the green Ca-Ib/c events exhibit unambiguous evidence of He I at similar velocities at all the optical transitions. 

However, the family of events with red continua exhibit much more diverse properties around the He I transitions, which include peculiar events like  SN\,2012hn, PTF\,12bho and SN\,2018kjy. Only SN\,2019hty exhibits unambiguous P-Cygni absorption in all the He I lines and thus He can be confirmed. In the progression from SN\,2019pxu to SN\,2012hn, we see a gradual change in the strength and absorption depth of the He lines. Specifically, we note the appearance of an emission feature at the expected absorption position of the He I $\lambda7065$ line that gets stronger from SN\,2019pxu to SN\,2012hn. This emission feature was attributed to C II in the spectral modeling of SN\,2012hn, although it could also be associated with Al II \citep{Kasliwal2010, Sullivan2011, De2018b}. In the same sequence of objects the He I $\lambda7065$ line gets progressively weaker until it is not detected at all in SN\,2012hn. The same trend is also detected in the He I $\lambda 6678$ line although the identification of He I $\lambda 6678$ in SN\,2012hn is complicated by the presence of the nearby Si II line. SN\,2012hn does not show any evidence of He either in its optical or near-infrared spectra \citep{Valenti2014}. We thus find evidence of a continuum of He line strengths in these events, which range from events with strong He lines to very weak or absent He lines.

\subsubsection{The Ca-Ia class}

Figure \ref{fig:IaClass} shows a comparison of the photospheric and nebular phase spectra of the three Ca-Ia objects in the sample, indicated by orange markers throughout this manuscript. These objects exhibit typical features of 1991bg-like objects defined by the strong Ti II trough in the blue side of the spectrum. In addition, all of these objects exhibit mild to strong line blanketing features short-wards of 5000\,\AA\, in the blue side of the spectrum. Such features are typically indicative of the outer ejecta being rich in Fe-group material that efficiently absorbs the blue flux \citep{Nugent1997, Polin2019a}. We note that SN\,2019ofm exhibits several similarities to the peak light spectrum of SN\,2016hnk (although with lower line velocities), while PTF\,09dav exhibits some different features and line strengths, some of which were attributed to rare elements like Sc\,II and Sr\,II in \citet{Sullivan2011}. Although the peak light spectra are broadly similar, such exotic elements were not required in the spectral modeling of SN\,2016hnk in \citet{Galbany2019} and \citet{Jacobson-Galan2019}, and hence is unlikely for SN\,2019ofm given the spectral similarity between the two objects. 

\begin{figure}[!ht]
    \centering
    \includegraphics[width=\columnwidth]{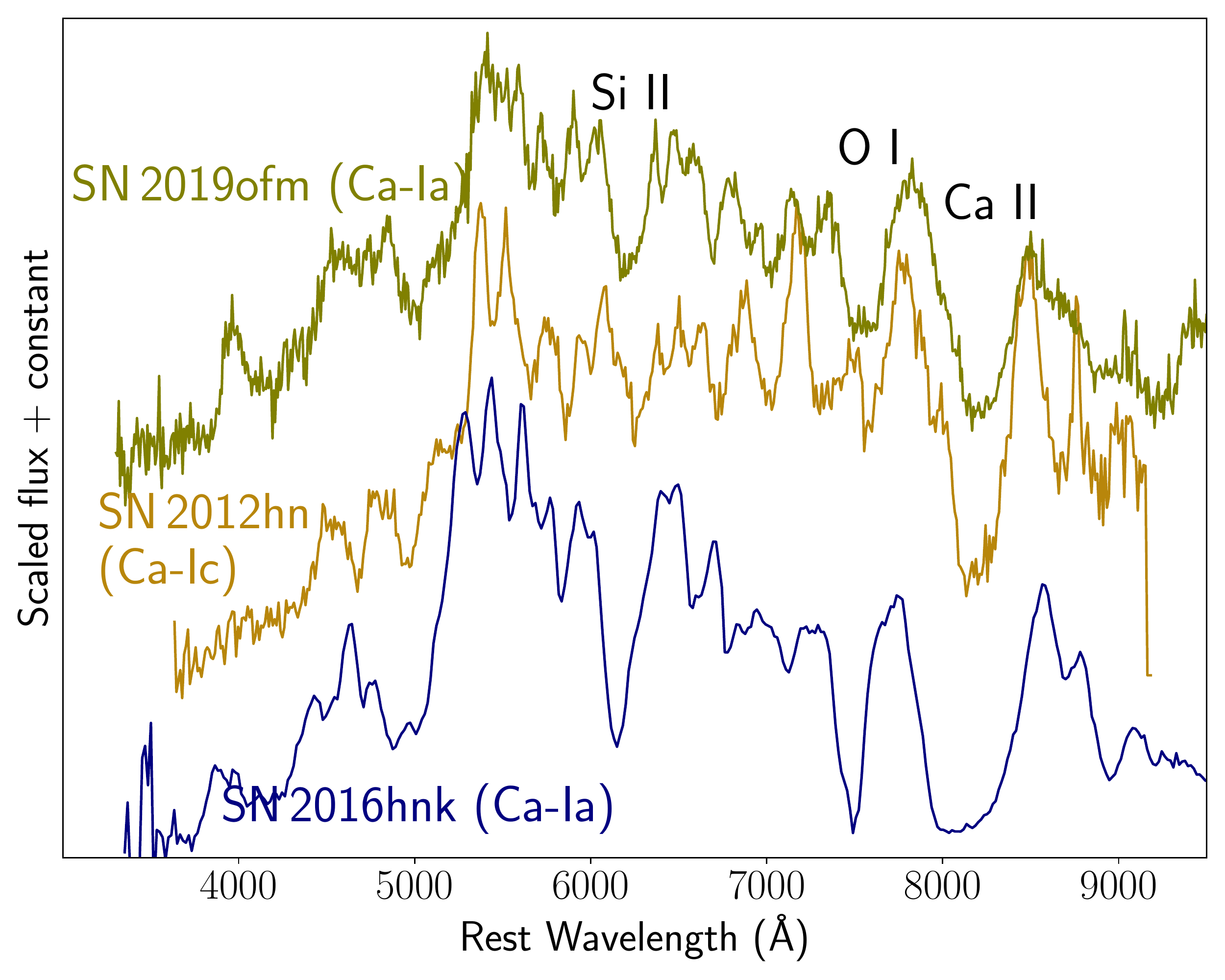}
    \caption{Comparison of the peak light optical spectra of He poor Ca-Ib/c event SN\,2012hn and the two Ca-Ia events SN\,2016hnk and SN\,2019ofm. Prominent spectral features of Si II, O I and Ca II are marked.}
    \label{fig:Ibc_Ia_compare}
\end{figure}{}

Given the lack of the prominent He lines in both the Ca-Ic objects and Ca-Ia objects, it is instructive to compare the peak light spectra of the two classes. In Figure \ref{fig:Ibc_Ia_compare}, we plot the peak light spectrum of SN\,2012hn -- the only unambiguous Ca-Ic object in the sample along with the peak light spectra of two Ca-Ia objects SN\,2019ofm and SN\,2016hnk. It is worth noting the striking resemblance between the spectra of SN\,2012hn and SN\,2019ofm, barring the weaker strength of the Si II line in SN\,2012hn (which leads to its Ca-Ic classification). Specifically, we find that although the velocities are different in the three objects, they show similar features over the entire optical spectrum. The only discrepancies are in the bluer part of the spectrum where the Ca-Ia objects show features from Fe group elements (near $\approx 4000$\,\AA). SN\,2016hnk exhibits a very strong Si\,II line similar to normal / sub-luminous SNe\,Ia \citep{Gal-Yam2017}, while SN\,2019ofm exhibits a weaker Si\,II line but with all the characteristic SN\,Ia features, and SN\,2012hn exhibits the weakest Si\,II line and nearly the same spectral features as SN\,2019ofm. 

\citet{Sun2017} demonstrated that Type I SN subtypes (Ia/Ib/Ic) occupy different loci on the line depth diagram of Si\,II $\lambda 6150$\,\AA\, and O\,I $\lambda 7774$\,\AA\, measured in peak-brightness spectra (see their Figure 9). In order to quantitatively investigate the strking similarities between the Ca-Ia and Ca-Ic objects, we performed these measurements in a manner similar to that of \citet{Sun2017}. SN\,2016hnk exhibits a $\lambda 6150$ line depth of $\approx 0.6$ and a depth ratio of $\lambda 6150 / \lambda 7774 \approx 0.75$, similar to 91bg-like SNe\,Ia in the \citet{Sun2017} sample. However, we find that SN\,2019ofm exhibits a $\lambda 6150$ line depth of $\approx 0.35$ and a depth ratio of $\lambda 6150 / \lambda 7774 \approx 0.83$ which is exactly at the SN\,Ia-SN\,Ic classification boundary suggested in that work. Similarly, the peak spectrum of SN\,2012hn exhibits a $\lambda 6150$ depth of $\approx 0.3$ and a depth ratio of $\lambda 6150 / \lambda 7774 \approx 1.0$, which falls exactly on the classification boundary for SNe\,Ib/c in that sample. In particular, we note that SN\,2019ofm and SN\,2012hn occupy an empty phase space in the classification diagram of \citet{Sun2017} -- with SN\,2019ofm being a transitional Ia-Ic object and SN\,2012hn being a transitional Ib-Ic object. We thus conclude that there may be a continuum of events leading from Ca-Ia to Ca-Ic to Ca-Ib objects based on their peak light photospheric spectral properties. We discuss this sequence together with the photometric and nebular phase properties in Section \ref{sec:discussion}.

\subsubsection{Photospheric velocities}
\label{sec:specfit}

\begin{figure*}[!ht]
    \centering
    \includegraphics[width = \textwidth]{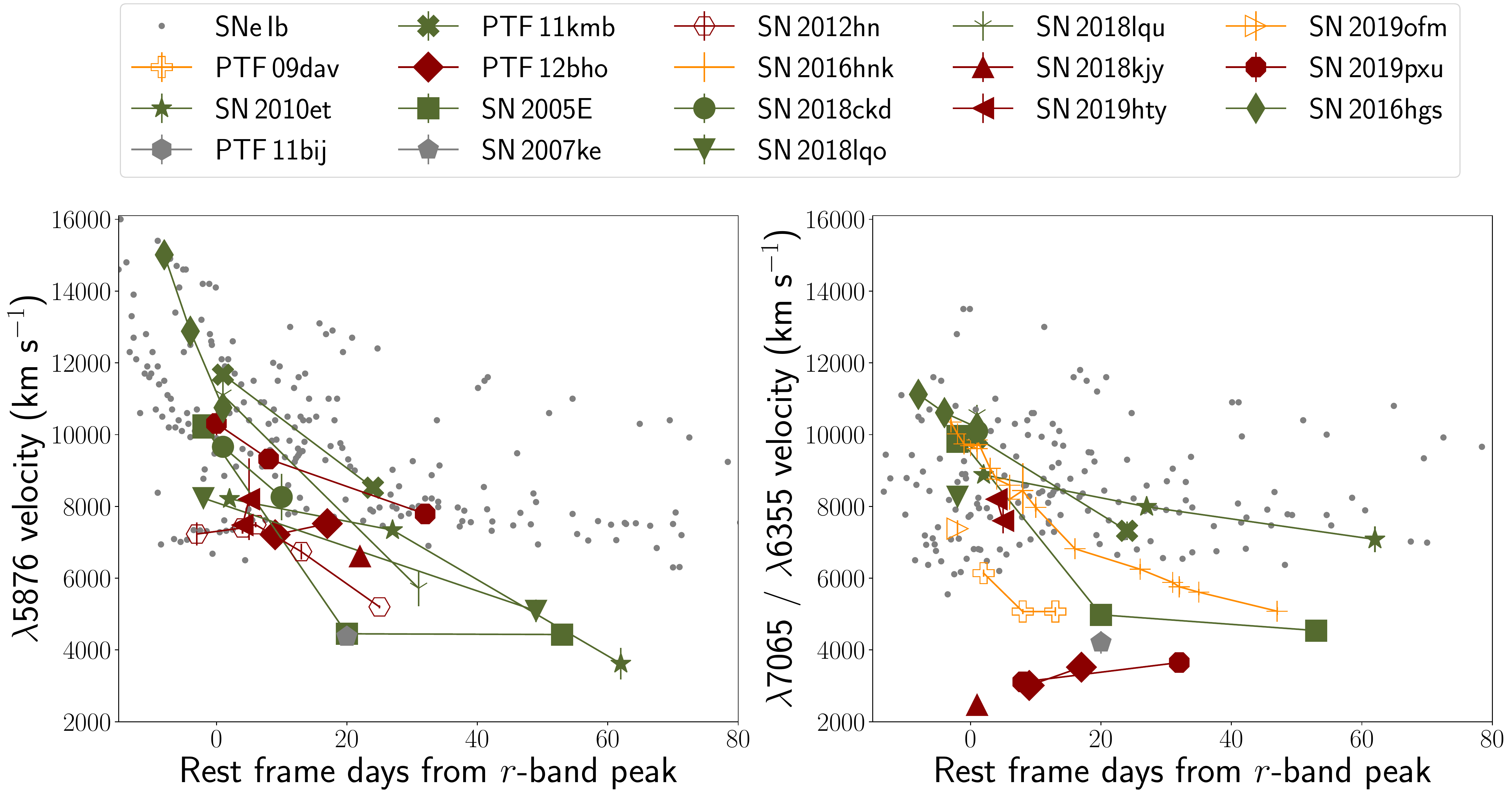}
    \caption{Evolution of photospheric line velocity as a function of phase from $r$-band peak for the combined sample of Calcium rich gap transients discussed in this paper. Points joined by solid lines represent the velocity evolution for the same object. We also plot the velocity evolution of the respective lines observed in normal SNe\,Ib in the sample of \citet{Liu2016}, as gray dots in the background. Individual events are shown by markers as indicated in the legend, with their marker colors indicating whether they belong to the Ca-Ib/c class with green (shown in dark green) or red (shown in red) continua, or to the Ca-Ia class (shown in orange). The left panel shows the velocity evolution of the prominent He I $\lambda5876$ line in cases where it is unambiguously identified (with solid symbols), or the likely nearby Na I feature (for SN\,2012hn in hollow symbols). The right panel shows the velocity evolution of the He I $\lambda7065$ feature for the Ca-Ib/c events and the Si II $\lambda6355$ feature for the Ca-Ia events.}
    \label{fig:photolineevolution}
\end{figure*}

For a quantitative comparison of the spectral features, we performed fits of the most prominent spectroscopic features of the combined sample of events, and list the derived parameters in Table \ref{tab:specfits}. The aim of this exercise is to elucidate the photospheric velocity evolution of the most prominent spectral lines in these transients as they hold clues to the density structure of the ejecta and internal emission powering mechanism \citep{Piro2014,Sell2015}. These velocity estimates are also necessary for quantitative estimates of the explosion kinetics \citep{Arnett1982, Arnett1985}. We follow procedures similar to those used in \citet{Liu2016} and \citet{Fremling2018} for normal stripped envelope SNe. He I is the most common spectral feature near peak light in the sample of Ca-Ib/c events, and hence we estimate the velocity of the He I $\lambda5876$ line by fitting a low order polynomial to the flux around the line. We estimate the uncertainties in the velocity by Monte Carlo sampling of the flux in the relevant wavelength region. We estimate the flux uncertainty by subtracting the smooth polynomial fit from the spectral data and compute the noise RMS as the standard deviation of the flux from the smoothed spectrum. We then add a Gaussian distribution of noise to the spectrum using the wavelength dependent flux RMS as the standard deviation, and compute the spectral fit parameters for the several realizations of the input spectrum. While O I $\lambda7774$ is also detected in most of the peak light spectra, we do not fit the absorption in this line since it appears to be uniformly contaminated by another nearby absorption feature (likely Mg II; e.g. \citealt{Valenti2014, De2018b}). We perform the same fitting for the He I $\lambda7065$ line in the spectra where it is detected in the Ca-Ib/c events, while the same is computed for the Si II $\lambda6355$ line in the case of the Ca-Ia events. 

The velocity estimates from fitting the He I $\lambda5876$ and $\lambda7065$ are largely consistent in the Ca-Ib/c events. However, the velocities and detection of He in the red events is complicated by several factors. While SN\,2019hty shows unambiguous presence of both the He I lines we measure, the $\lambda 7065$ line absorption in the other objects is contaminated by emission from a nearby blueward feature (see Figure \ref{fig:helines}), which could be associated with C II \citep{Valenti2014} or Al II \citep{Kasliwal2012a, De2018b}. Thus, the He I velocities are discrepant between the two lines for these objects. However, we report all these measurements for completeness. SN\,2012hn does not show clear signs of He I in its peak light spectra, and hence we compute the velocity for the nearby 5800\,\AA\, feature (suggested to be due to Na I or Cr II in \citealt{Valenti2014}) assuming that the peak of the feature near maximum light corresponds to the rest wavelength of the line. The low signal-to-noise ratio and resolution of the spectrum of SN\,2018gwo does not allow us to conclusively identify He I, and hence we do not measure the corresponding velocities for these objects. SN\,2018kjy exhibits a peculiar peak light spectrum with a large number of low-velocity ($\sim 2500$ km s$^{-1}$) lines. While He I $\lambda7065$ is identifiable in the peak light spectrum, we caution that due to the large number of low velocity lines, we only tentatively identify He I $\lambda5876$ in the peak light spectrum.

In Figure \ref{fig:photolineevolution}, we plot the evolution of the prominent photospheric phase He I $\lambda5876$ and $\lambda7065$ (for the Ca-Ib/c events), and Si II $\lambda 6355$ line (for the Ca-Ia events) velocities as a function of phase from $r$-band peak. For comparison, we also plot the evolution of the He I line velocities observed in the sample normal SNe\,Ib presented in \citet{Liu2016}. The He~I $\lambda7065$ velocity in most of the red events is much lower than in the events with green continua. As stated in Section \ref{sec:specfit}, this discrepancy is due to blending of a nearby emission feature blue-wards of $\lambda7065$ that contaminates the velocity measurement, thus making the velocity evolution uncertain. SN\,2018kjy stands out as a peculiar low velocity event as evident from the large number of narrow lines visible in its peak light spectrum (see Figure \ref{fig:spztf1}), exhibiting several similarities to the low velocity spectrum observed in PTF\,12bho. SN\,2012hn does not exhibit signatures of He I in its peak light spectrum and hence we show the velocity evolution of the nearby 5800\,\AA\, feature, likely associated with Na I. It is worth noting the diversity of Si II $\lambda 6355$ velocities in the Ca-Ia events -- both PTF\,09dav and SN\,2019ofm exhibit lower (by $\approx 3000$ km s$^{-1}$) velocities than SN\,2016hnk at peak ($\approx 11000$ km s$^{-1}$). The combined sample of photospheric phase velocities shows a consistent trend of decreasing velocities with time, consistent with a receding photosphere (in mass coordinates) in the SN ejecta. Comparing to the sample of normal Type Ib events in the comparison sample, we find that while the photospheric velocities are similar near peak light, the Ca-rich gap transients exhibit a much faster drop to low photospheric velocites around $\approx 20\,\rm{d}$ after peak light. A faster drop in photospheric phase velocities suggests that the photospheric line forming regions recede into the inner and slower layers of the ejecta faster than normal SNe\,Ib, consistent with the lower ejecta masses and faster transition to the optically thin nebular phase observed in these events. 

\subsection{Nebular phase spectra}

\begin{figure*}[]
\centering
\includegraphics[width=0.75\textwidth]{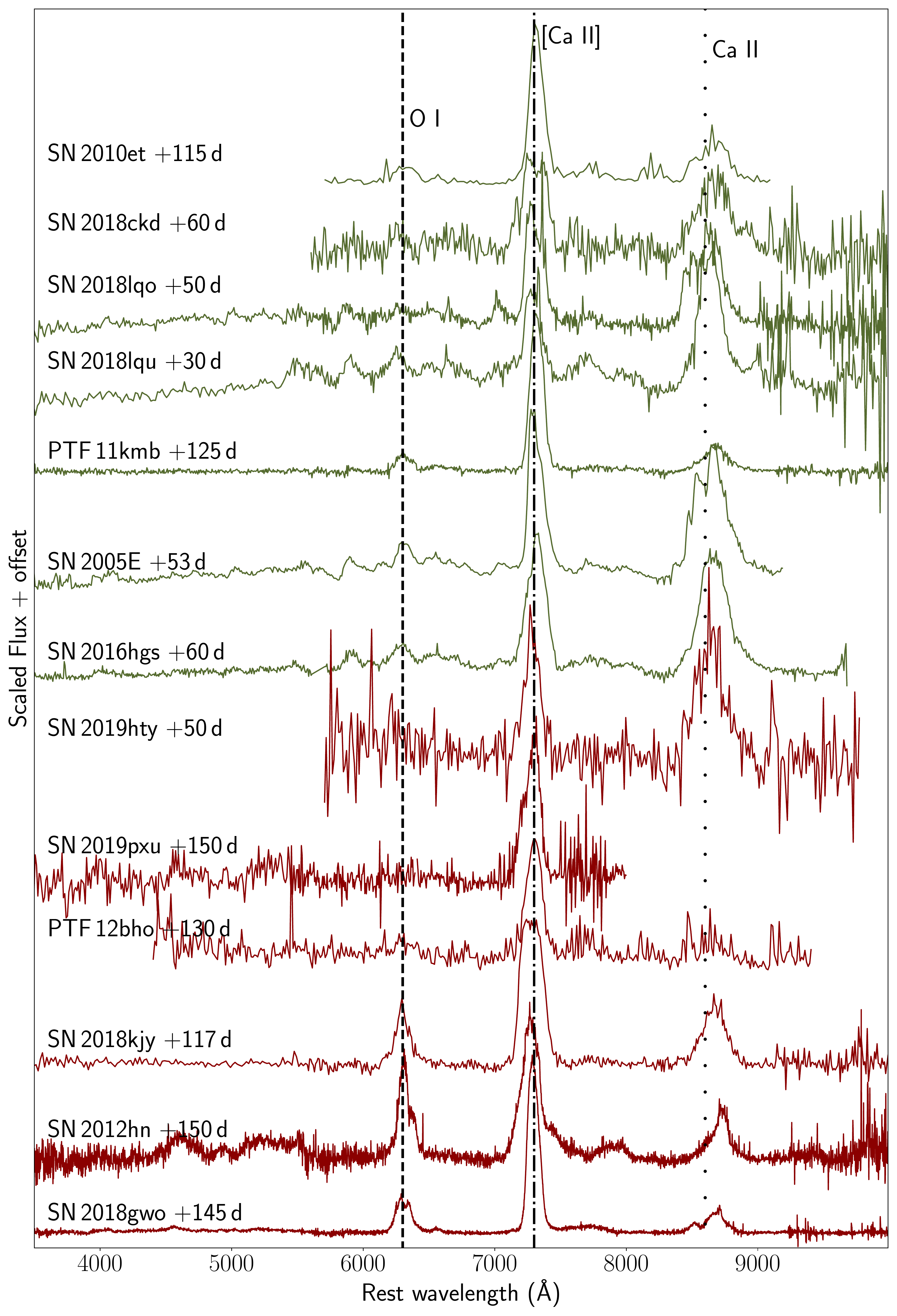}
\caption{Nebular phase spectra of the Ca-Ib/c events in the ZTF sample combined with the same for events in the literature. The transient name and phase of the spectrum is indicated next to each spectrum. Color coding of the spectra are the same as in Figure \ref{fig:peakspec_comparison}. The prominent nebular lines of [O I], [Ca II] and Ca II are marked.}
\label{fig:nebspec_comparison}
\end{figure*}

\subsubsection{The Ca-Ib/c class}

In Figure \ref{fig:nebspec_comparison}, we show a comparison of the nebular phase spectra of the Ca-Ib/c class of objects. Despite the diversity in the photospheric phase colors and velocities, the nebular phase spectra are relatively homogeneous, and dominated by strong [Ca II] emission and weak [O I] emission (if [O I] is detected at all).  The low velocity events SN\,2018kjy and SN\,2019pxu exhibit double-peaked lines near the [Ca II] doublet early in the nebular phase, similar to that observed in PTF\,12bho \citep{Lunnan2017}, although later spectra exhibit a single unresolved [Ca II] feature.  Since SN\,2018gwo was a relatively nearby event at 30 Mpc, our nebular phase spectral sequence extends out to $\approx 235\,\rm{d}$, showing that [Ca II] emission continues to dominate the spectrum from the earliest to these very late phases and hence the Ca-rich classification is independent of the exact phase of the nebular spectrum. We note that the events with red continua SN\,2012hn, SN\,2018kjy and SN\,2018gwo exhibit relatively stronger [O I] lines relative to [Ca II], when compared with other objects observed at similar phases. SN\,2012hn and SN\,2018gwo are also notable for exhibiting clear signatures of Fe group elements around $4000 - 6000$\,\AA\, similar to late-time spectra of sub-luminous SNe\,Ia. 

\subsubsection{The Ca-Ia class}

Figure \ref{fig:IaClass} shows the nebular phase spectra for the Ca-Ia objects, which are dominated by [Ca II] lines. It is important to note the difference between the Ca-Ia objects and the class of 1991bg-like objects in the nebular phase -- while 1991bg-like objects exhibit strong features of Fe-group elements in the blue part of the spectrum in the nebular phase, the Ca-Ia objects show nearly no signatures of such features in the blue-side spectra at similar phases. We demonstrate this by plotting a nebular phase spectrum of the 1991bg-like object SN\,2005ke \citep{Silverman2012} in Figure \ref{fig:IaClass}. SN\,2005ke also exhibits a strong emission feature at 7290\,\AA\, near [Ca II]; however this feature could be associated with [Fe II] and [Ni II] emission given the strong Fe group elements observed in the blue side of the spectrum (\citealt{Floers2020}; see also \citealt{Polin2019b} who suggest that this feature is due to [Ca II]).  Unlike the Ca-Ib/c objects, the Ca-Ia objects exhibit very weak or no [O I] emission in the nebular phase. The nebular phase spectrum of SN\,2016hnk \citep{Galbany2019} exhibits a narrow double-peaked feature at the [Ca II] line, and is noticably narrower than other objects \footnote{However, the late-time spectrum of SN\,2016hnk reported by \citealt{Jacobson-Galan2019} does not show a clear double peak}. However this could be due to the very late phase of the spectrum; unavailable for the other, fainter objects in this class.

\subsubsection{[Ca II] / [O I] ratio}

In the nebular phase spectra of the combined sample of events, we fit a Gaussian to the [Ca II] and [O I] emission features to estimate their flux ratio. We compute the fluxes by fitting a single Gaussian (see e.g. \citealt{Jerkstrand2017}) to the respective line emission features.  We do not compute the absolute flux in these lines as the spectro-photometric calibration is not available for several spectra in the literature\footnote{Absolute calibration would also require contemporaneous photometry, which is largely unavailable}. Instead, we compute the [Ca II]/[O I] ratio for each spectrum. For several late-time spectra [O I] is barely or not detected. In such cases, we compute an upper limit on the [O I] flux by using the RMS of the flux around the expected position of the line center to compute a $1\sigma$ upper limit on the [O I] flux assuming that [Ca II] and [O I] lines have the same velocity width (this provides a lower limit on the [Ca II]/[O I] ratio). We estimate the uncertainties in the line ratios by computing the standard deviations in these quantities from 1000 Monte Carlo realizations of the spectra, where the samples are created by adding flux uncertainties in the same way as with the photospheric phase spectra. Table \ref{tab:spectra} lists the best fit parameters for the sample of events in this paper as well as all the published spectra in the literature. 

\begin{figure}
     \centering
     \includegraphics[width=\columnwidth]{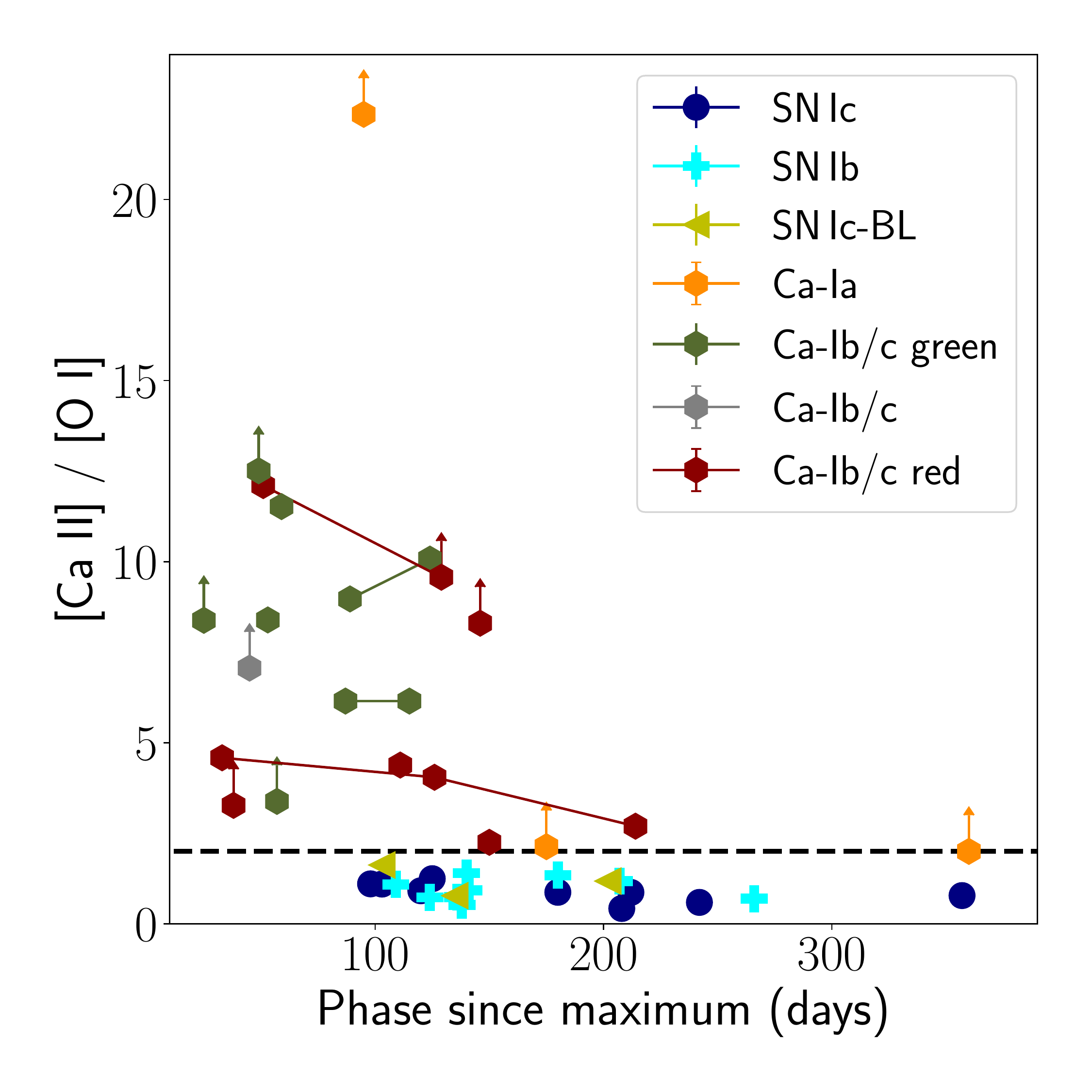}
     \caption{[Ca II]/[O I] ratio for the Ca-rich gap transients and stripped envelope SNe in the ZTF volume limited experiment. The black horizontal dashed line shows the [Ca II]/[O I] ratio defined as the threshold in this paper for classification as a Ca-rich transient. Symbols with upward arrows indicate lower limits on the [Ca II]/[O I] ratio where the [O I] feature is not detected with statistical significance. The Ca-rich events are indicated by hexagons with colors that reflect their peak light spectroscopic appearance as discussed in the text. }
     \label{fig:caoratio}
\end{figure}{}

We plot the evolution of the [Ca II]/[O I] ratio in these sources compared to a sample of nebular phase spectra of other types of stripped envelope SNe in ZTF in Figure \ref{fig:caoratio}. These spectra were obtained either as a part of confirmation spectra for the candidate Ca-rich gap transients identified from photometry or as a part of nebular phase follow-up of a volume-limited sample of stripped envelope SNe (C. Fremling et al., in prep.). The Ca-rich gap transients occupy a unique phase space in this plot with high [Ca II]/[O I] at all phases in their evolution. We also show the threshold of [Ca II]/[O I] = 2 used to select the sample of Ca-rich gap transients in this paper, which clearly separates out the Ca-rich events from the normal events, which primarily occupy the phase space of [Ca II]/[O I] $\lesssim 1$ at all phases.  We note that several of the red Ca-Ib/c events exhibit relatively small [Ca II] / [O I] at all phases, where the [Ca II]/[O I] values are smaller by a factor of $\approx 2$ when compared to the green Ca-Ib/c events. This trend is consistent with the qualitative analysis in Section \ref{sec:analysis} where we noted the stronger [O I] features in the red events. The [Ca II]/[O I] estimates result in lower limits for all the Ca-Ia events, suggesting that [O I] is not detected in the nebular phase spectra of these objects.

\subsection{Photometric evolution}

\begin{figure*}[!ht]
\centering
\includegraphics[width=\textwidth]{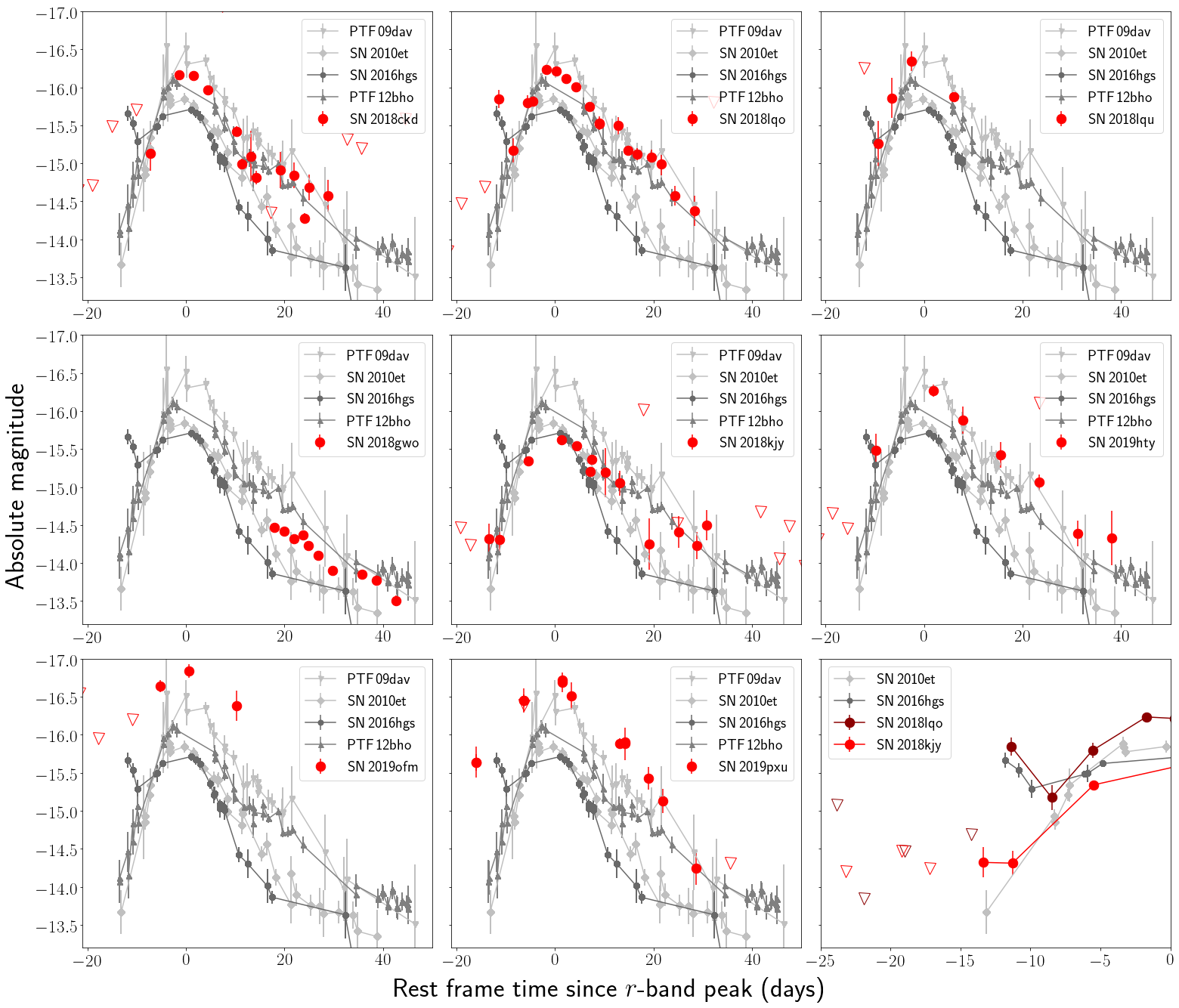}
\caption{Comparison of the $r$-band evolution of this sample of Ca-rich gap transients to some previously confirmed events -- PTF\,09dav \citep{Sullivan2006}, SN\,2010et \citep{Kasliwal2012a}, PTF\,12bho \citep{Lunnan2017} and SN\,2016hgs \citep{De2018b}. In each panel, the $r$-band photometry for the ZTF sample are presented as red points while the archival sources are plotted in shades of grey. The photometric evolution is shown as a function of rest-frame time from the best estimate of the $r$-band peak (except in the case of SN\,2018gwo; see text). Inverted triangles denote 5$\sigma$ upper limits. \textbf{Bottom right panel:} Comparison of the early time bumps seen in the light curves of some of the transients in the ZTF Ca-rich sample to that of the literature events SN\,2016hgs (which exhibited an early time bump; \citealt{De2018b}) and SN\,2010et (which exhibited a monotonic rise; \citealt{Kasliwal2012a}).}
\label{fig:phot_comp}
\end{figure*}

In Figure \ref{fig:phot_comp}, we compare the $r$-band light curves of the ZTF sample of Ca-rich gap transients near peak light, to four characteristic light curves of the literature Ca-rich events PTF\,09dav \citep{Sullivan2011}, PTF\,10iuv \citep{Kasliwal2012a}, PTF\,12bho \citep{Lunnan2017} and SN\,2016hgs \citep{De2018b}. The comparison sample was chosen to encompass the diversity of photospheric phase spectral properties reported in the literature sample. The $r$-band light curves of the full ZTF sample are largely similar to the comparison sample, with the exception of SN\,2019ofm and SN\,2019pxu, which exhibit more luminous and broader light curves than the comparison objects. SN\,2018lqo exhibits a prominent early time `bump' of $\approx 1$\,mag compared to the light curve of SN\,2010et. Excess emission is also marginally detected in the light curve of SN\,2018kjy, although at much lower significance. 

In the bottom right panel of Figure \ref{fig:phot_comp}, we show a zoomed inset of the early $r$-band light curves of these two objects together with that of SN\,2016hgs, which was previously reported as a peculiar Ca-rich gap transient with a prominent double-peaked light curve. For comparison, we plot the well-sampled light curve of SN\,2010et, which shows a purely monotonic rise in its early light curve, as is found for all the other events in the full sample. The early excess emission in the light curve of SN\,2018lqo is similar in luminosity and time scale to the early emission in SN\,2016hgs. The early excess in SN\,2018kjy is detected but of lower significance, and hence we do not discuss it further here. SN\,2018gwo has no ZTF coverage near peak light, although the peak light photometry published on TNS and late time decay tail are consistent with the literature sample of events if we assume that the source was first detected $\approx 7\,\rm{d}$ before peak light. We caution, however, that we are unable to measure the time of peak or any other light curve parameter for this object due to the absence of photometry around peak.

\subsubsection{Light curve parameters}
\label{sec:lcfits}
 We fit the light curve of each transient (in every filter available) around peak light (within $20\,\rm{d}$ of peak) with a low order polynominal (order 3 to 4) and derive parameters describing the light curve peak and timescale. We perform the same fitting for all multi-color photometry data available for the literature sample of Ca-rich gap transients.  We use the functional fits to determine the times of peak in each filter, peak apparent and absolute magnitudes ($m_p$ and $M_p$ respectively), and the characteristic rise time and decline time. We define the rise and decay time ($t_{r, 1/2}$ and $t_{f, 1/2}$) of the light curve as the time it takes to rise or decline to half flux from peak light. In addition, we compute the characteristic decay of the light curve (in magnitudes) in $7\,\rm{d}$ from peak light, denoted by $\Delta m_7$. 
 
 We estimate uncertainties on these quantities by Monte Carlo sampling of the derived parameters from 1000 realizations of each light curve using the photometric uncertainties of each point in the light curve. We restrict our fitting to photometric bands that have at least one data point before peak light, since it is not possible to estimate the peak magnitude without a corresponding detection before peak. In addition, for sources that do not have photometry sampling the relevant time period of the rise or fall of the light curve, we do not compute the respective $\Delta m_7$, rise or fall times. For SN\,2018gwo, photometry was not available around peak light, and hence we use the reported photometry near peak on the TNS \citep{SN2018gwo} as a lower limit on the peak magnitude. In computing the peak apparent and absolute magnitude, we also correct the photometry of the literature events for Galactic extinction using the maps of \citet{Schlafly2011} and the extinction law of \citet{Cardelli1989} with $R_V = 3.1$. We do not correct for any additional host galaxy extinction due to the absence of Na I D absorption in their spectra and the remote locations of these events. The only exception is SN\,2012hn, for which we assume a host extinction of $E(B-V) = 0.2$\,mag which was estimated from Na I D absorption in its spectra \citep{Valenti2014}. The best-fit parameters from the light curve fitting are given in Table \ref{tab:lcfits}.
 
\subsubsection{Color evolution}
\begin{figure*}
    \centering
    \includegraphics[width=\textwidth]{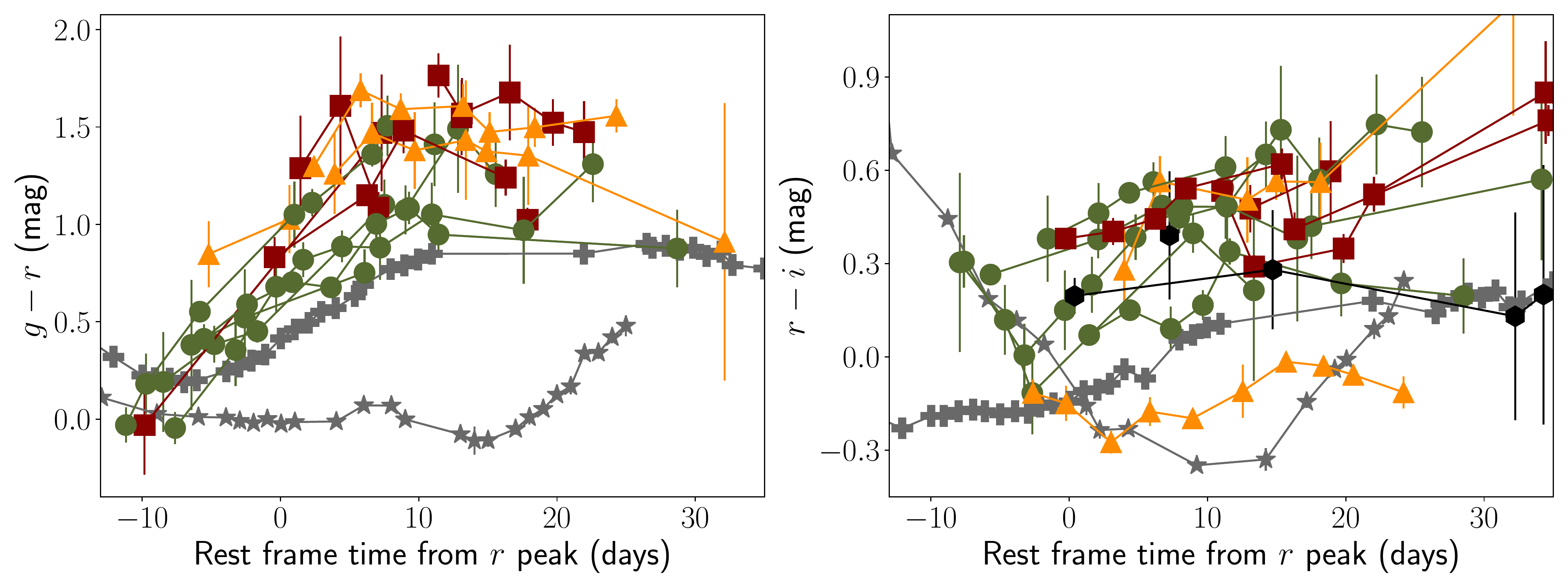}
    \caption{Comparison of the $g-r$ and $r-i$ color curves of the known sample of Calcium rich gap transients, color coded by their spectroscopic membership of the green Ca-Ib/c (green circles), red Ca-Ib/c (red squares) or the Ca-Ia (orange triangles) class. For comparison, we also show the color evolution curves of the Type Ia SN\,2011fe \citep{Nugent2011} as dark gray stars and the Type Ib iPTF\,13bvn \citep{Cao2013} as dark gray crosses, which are systematically bluer in $g-r$ than the Ca-rich gap transients at similar epochs. Events in the combined sample without a peak light spectrum are shown in black circles.}
    \label{fig:colorsample}
\end{figure*}

Figure \ref{fig:colorsample} shows the $g-r$ and $r-i$ color evolution of the complete sample of Ca-rich gap transients discussed in this paper. For comparison to other types of SNe, we also show the well-sampled color curves of the nearby SN Ia 2011fe \citep{Nugent2011} and the SN\,Ib iPTF\,13bvn \citep{Cao2013}. Ca-rich gap transients redden rapidly in $g-r$ color compared to SNe Ia, although the color evolution has a similar trend compared to the SN\,Ib iPTF13bvn. However, all the Ca-rich gap transients are redder than iPTF\,13bvn in $g-r$ at similar epochs. The same trend is also seen in the $r-i$ color evolution.  The $g-r$ evolution of both the Ca-Ib/c objects with red continua and the Ca-Ia objects are systematically redder than the Ca-Ib/c objects with green continua, consistent with the suppressed blue flux in the spectra of the the former objects. We note that the complete sample of Ca-rich gap transients occupies a narrow distribution around $g-r \approx 0.7\,\rm{mag}$ near peak light, which we later use to simulate their light curves for estimation of volumetric rates from the ZTF survey (Section \ref{sec:rates}).

\subsubsection{Luminosity, width and color relationship}

\begin{figure*}
    \centering
    \includegraphics[width=\textwidth]{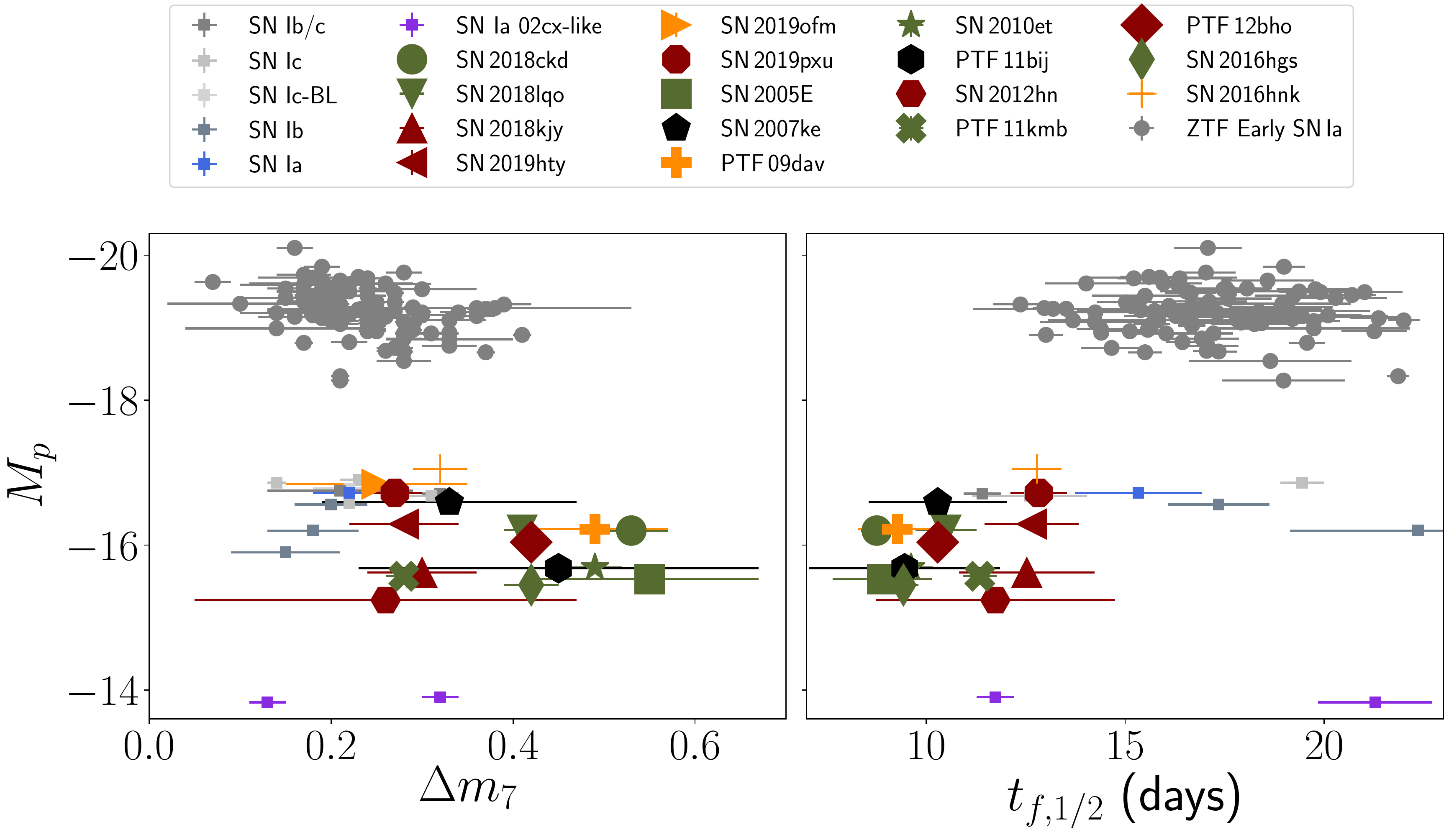}
    \caption{Luminosity - width phase space for the $r$-band light curves of Calcium rich gap transients (symbols are shown with the object names in the legend). The individual events are colored by their spectral type at peak light. Green Ca-Ib/c events are indicated by green and those with reddened continua are indicated by red. Ca-Ia events are indicated in orange and marker symbols for all events are indicated in the legend. We are unable to constrain the peak light spectroscopic properties of SN\,2007ke and PTF\,11bij, and hence show these objects in black. For comparison, we plot the same phase space of timescales for the ZTF 2018 early SN Ia sample in \citet{Yao2019} in gray dots, together with the same parameters for objects in the control sample (in squares; indicated in the legend by their spectroscopic type at peak).}
    \label{fig:lumwidth}
\end{figure*}{}

In Figure \ref{fig:lumwidth}, we plot the peak $r$-band magnitude of the transients as a function of the decline in $r$-band in $7\,\rm{d}$ from peak ($\Delta m_7$) and time taken to fall to half the maximum flux ($t_{f, 1/2}$). While some objects have well sampled light curves on the rise to estimate the time taken to rise from half-maximum in flux to maximum, the majority of literature events do not have well constrained pre-peak light curves and hence we only plot the fall time from peak. We choose $r$-band as it is the most commonly available filter for the combined photometric sample and allows us to perform a homogeneous analysis on the largest number of objects. For comparison of this phase space to the general trend followed by thermonuclear SNe, we plot the same parameters for SNe\,Ia. We use the sample of ZTF SN\,Ia light curves published in \citet{Yao2019}, and compute the same quantities using the fitting techniques mentioned above with the $r$ band light curves. The distribution of these SNe\,Ia shows the expected luminosity - width relationship (the Phillips relation; \citealt{Phillips1993}) with more luminous events being systematically slower evolving\footnote{Note that the canonical relationship for SNe Ia is defined using the magnitude decline in $15\,\rm{d}$ after peak in the B band versus peak absolute magnitude $M_B$. However, we choose to conduct this analysis on $\Delta m_7$ against peak absolute magnitude in the $r$ band $M_r$ since the faint and fast declining Calcium rich gap transients usually lack photometry extending beyond $\approx 15\,\rm{d}$ from maximum}. In order to investigate the presence of a luminosity-width relationship in the full sample of Ca-rich objects, we compute a Spearman correlation coefficient between the two pairs of plotted paramaters and find no significant evidence of correlation. The corresponding correlation coefficients are 0.25 between $M_p$ and $\Delta m_7$ (p-value of $0.35$) and $-0.36$ between $M_p$ and $t_{f, 1/2}$ (p value of $0.18$), suggesting no statistically significant evidence of a correlation between these parameters. 

\begin{figure}
    \centering
    \includegraphics[width=\columnwidth]{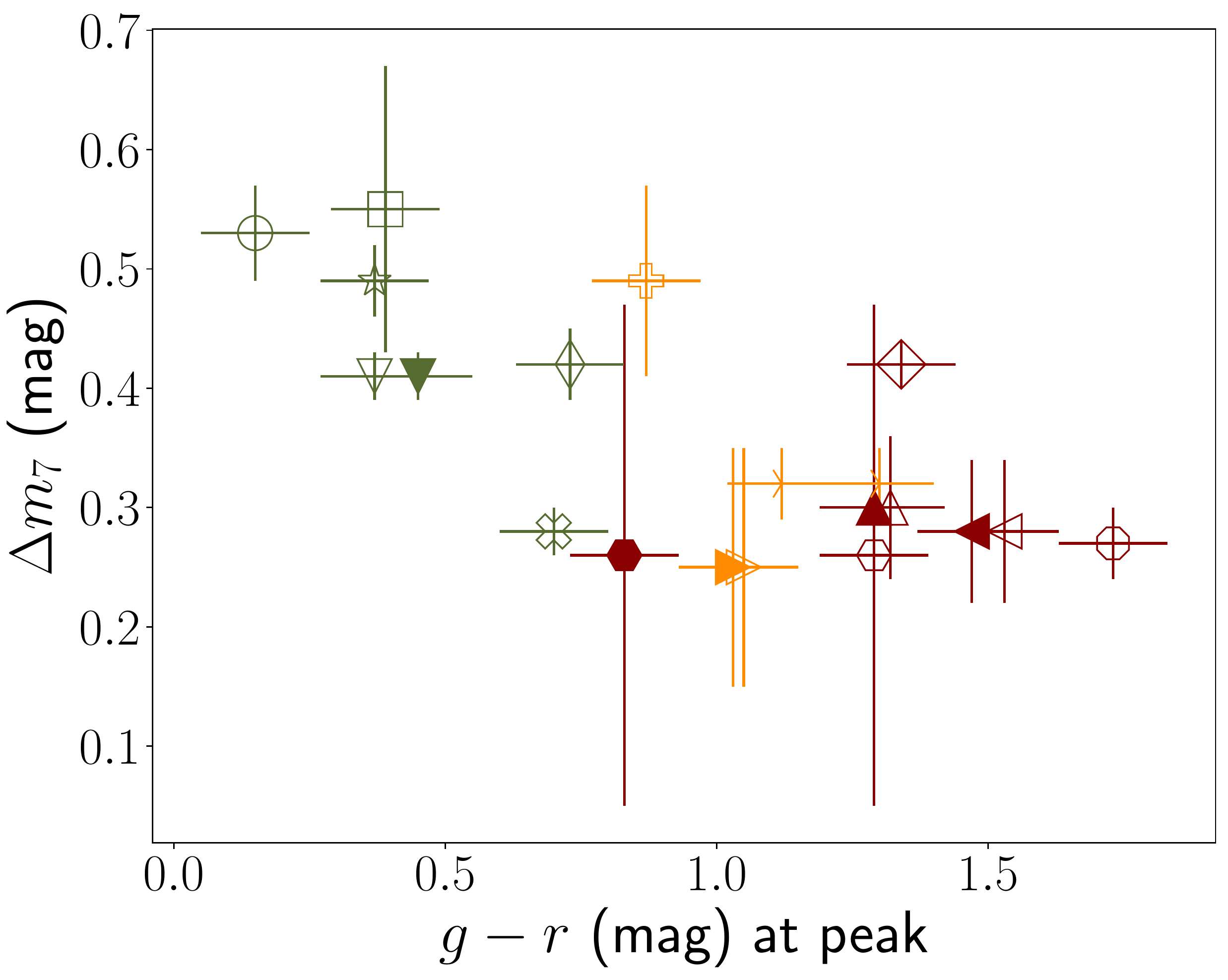}
    \caption{Dependence of the rate of light curve evolution post-peak in $r$-band ($\Delta m_7$) on the transient $g-r$ color at peak. The symbol color coding and markers are the same as those used in Figure \ref{fig:lumwidth}. Hollow symbols indicate colors computed from spectro-photometry on observed spectra within $\approx 10\,\rm{d}$ of peak light, while solid symbols indicate colors derived from peak light photometry where available. The colors have been corrected for foreground Galactic extinction.}
    \label{fig:colorlc}
\end{figure}

We now examine possible differences in the photometric properties between the Ca-Ib/c and the Ca-Ia objects. Due to the absence of peak light spectroscopy for SN\,2007ke and PTF\,11bij, we are unable to ascertain the nature of the blue continuum at peak light; however, their early nebular phase spectra show lines characteristic of SNe\,Ib/c suggesting their membership in the class of Ca-Ib/c objects. The peak luminosity distributions of the green and red events among the Ca-Ib/c objects are consistent with each other; however, the Ca-Ia objects are typically more luminous. The total number of events remain small to draw a conclusion on the statistical significance of the differences. Specifically, we note that Ca-Ib/c events with red continua exhibit systematically slower light curves than the Ca-Ib/c events with green continua, as shown by their smaller $\Delta m_7$ and larger $t_{f,1/2}$ values. A 2-sample Kolmogorov-Smirnov (KS) test suggests that the null hypothesis probability that the two sets of values are drawn from the same underlying population is $< 5$\% for $t_{f,1/2}$ and $< 18$\% for $\Delta m_7$. While only marginally significant due to the small number of events, we further examine this trend in Figure \ref{fig:colorlc}.

We show the dependence of $\Delta m_7$ as a function of the $g-r$ color of the transient near peak light in Figure \ref{fig:colorlc}. The $g-r$ photometric colors at peak light are not available for several events in the combined sample, and hence we use both the spectro-photometric colors derived from peak light spectroscopy as well as photometric colors where available. Since the photometric colors are not always available at the same phase as the spectroscopy, there are differences between the derived photometric and spectro-photometric colors. The Ca-Ib/c events separate into two classes of events based on their $g-r$ colors, while Ca-Ia events exhibit colors intermediate between the two classes but redder than the green Ca-Ib/c events. Specifically, we note that Ca-Ib/c events with bluer $g-r$ colors at peak exhibit larger $\Delta m_7$ (faster photometric evolution) at peak, while the red Ca-Ib/c events are slower evolving, consistent with the trend observed in Figure \ref{fig:lumwidth}. However, we caution against drawing conclusions about any correlations between these two parameters as the photometric and spectro-photometric colors were not available at the same phase in all cases. We find that the green and red events in the Ca-Ib/c sample are separated at $g-r \approx 1\,\rm{mag}$ at peak light; the Ca-Ia objects also exhibit redder colors of $g-r > 1\,\rm{mag}$ at peak, consistent with their line blanketed spectra.

\subsubsection{Dependence on spectroscopic properties}

\begin{figure*}
    \centering
    \includegraphics[width=0.9\textwidth]{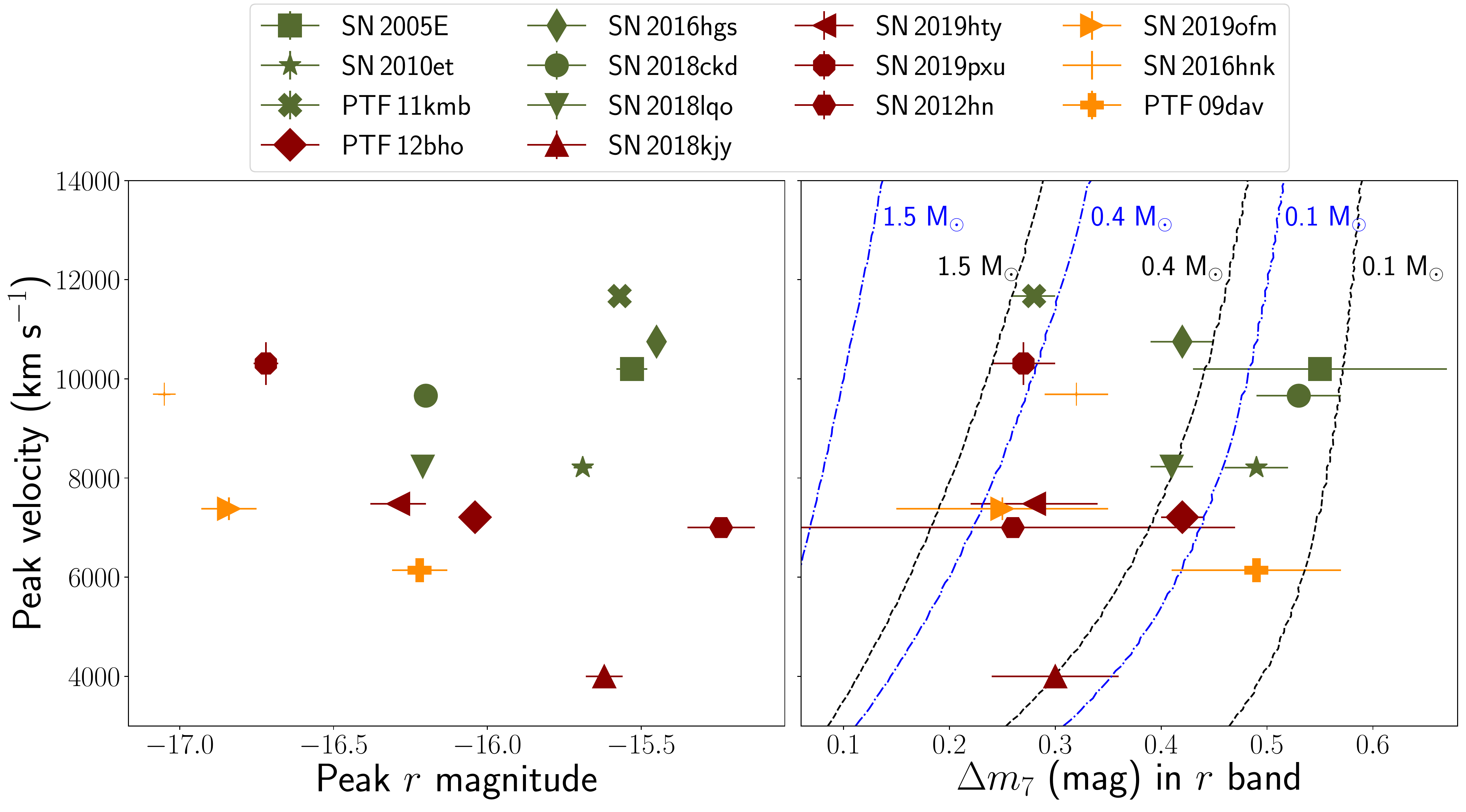}
    \caption{Photospheric phase line velocities as a function of the light curve peak luminosity (in $r$ band) and timescale of evolution (characterized by $\Delta m_7$) for the sample of Ca-rich gap transients analyzed in this paper. We use the velocity of the He I $\lambda5876$ feature from the spectrum taken closest to the estimated time of peak light in $r$ band (if available within $10\,\rm{d}$ of peak) for the Ca-Ib/c objects, and the Si II $\lambda6355$ velocity for the Ca-Ia objects. The left panel shows the peak velocity as a function of the peak $r$-band magnitude and the right panel shows the peak velocity as a function of $\Delta m_7$. The right panel also shows contours of constant ejecta mass computed using the analytical model of \citet{Arnett1982} to guide the eye on the range of ejecta masses found in the sample. We show the ejecta mass contours for two different optical opacities of $\kappa = 0.07$ cm$^2$ g$^{-1}$ (black dashed lines) and $\kappa = 0.2$ cm$^2$ g$^{-1}$ (blue dot-dashed lines). }
    \label{fig:speclc}
\end{figure*}

In Figure \ref{fig:speclc}, we plot the photopsheric velocity at peak light against the peak $r$-band magnitude of the sources to examine the dependence of the photospheric phase velocity on the peak luminosity of each event. In the case of SN\,2018kjy, we are unable to measure the He line velocities directly due to the large number of narrow lines, and hence we estimate typical line velocities in the spectra from the P-Cygni absorption velocities of $\approx 4000$ km s$^{-1}$. We do not find a clear dependence of the photospheric phase velocity on the peak luminosity of the events, but note that events with green continua have higher photospheric phase velocities compared to the red events.

The right panel of Figure \ref{fig:speclc} shows the dependence of the peak light photospheric velocity on the rate of decay of the light curve, $\Delta m_7$. As per the formalism for radioactively powered light curves laid down in \citet{Arnett1982}, the peak light photospheric velocity and light curve evolution near peak are indicators of the ejecta mass in the explosion. We thus also plot lines of constant ejecta mass in the right panel to guide the eye to the range of ejecta masses in the sample. We construct these lines by creating analytic light curves using the formalism of \citet{Arnett1982} assuming two constant opacities of $\kappa = 0.07$ cm$^2$ g$^{-1}$ (as relevant for Type Ib/c SNe; \citealt{Cano2013, Taddia2018}) and $\kappa = 0.2$ cm$^2$ g$^{-1}$ (relevant for completely ionized hydrogen-free material). We caution however, that the Arnett diffusion model has several assumptions which may not be satisfied in these explosions (see \citealt{Khatami2019} for a review). We do not find a dependence of the peak photospheric velocity on the light curve evolution near peak. It is important to note that the inferred ejecta masses can vary significantly depending on the assumed opacity. The majority of Ca-Ib/c events with green continua lie on contours of lower ejecta masses (between $0.1$ and $0.4$ M$_{\odot}$), while the red Ca-Ib/c and Ca-Ia events lie near larger ejecta masses (up to $\approx 1$ M$_{\odot}$ for $\kappa = 0.07$ cm$^2$ g$^{-1}$, but $< 0.5$ \Msun for $\kappa = 0.2$ cm$^2$ g$^{-1}$). However, we stress that redder events also likely have higher effective optical opacity in the ejecta than their green counterparts, as evident from the strong suppression of flux in the blue, which would suggest increased bound-bound opacity from Fe group material. As such, this effect would decrease the ejecta masses inferred from assuming a constant opacity across all events.

\subsubsection{Late-time photometric evolution}

In Figure \ref{fig:latephoto}, we show the late-time ($> 40\,\rm{d}$ after $r$-band peak) photometric evolution of the sample of Ca-rich gap transients presented in this paper together with published photometry of events in the literature. While most of the late-time photometry presented here was obtained using targeted follow-up observations using the P60 +SEDM, P200 + WASP and Keck-I + LRIS, we also stacked several epochs (over 3 - 7 days) of the high cadence ZTF observations to place limits on the flux at late times. Owing to its small distance, SN\,2018gwo has good photometric follow-up from ZTF up to $\approx 100\,\rm{d}$ from the estimated peak time. We also show a numerically computed Arnett model \citep{Arnett1982} for the decline rate expected from a radioactive powered light curve with ejecta mass of $0.5$ \Msun and $^{56}$Ni mass of 0.015 \Msun. The model parameters were chosen based on the typical values found in previous studies \citep{Perets2010, Kasliwal2012a, Valenti2014, De2018b}. Compared to the expected $^{56}$Ni decay tail shown in Figure \ref{fig:latephoto}, the luminosity at late times is much fainter than the prediction from the Arnett model, while the decay slope is also steeper for these events. The characteristics are consistent with the fast rising light curves of these events, which suggest low ejecta masses and incomplete $\gamma$-ray trapping at late times. Figure \ref{fig:latephoto} also shows the last deep photometric limits obtained using P200 and Keck for this sample of events, extending out to $\approx 1.5$ years after peak light. Although SN\,2018gwo is detected with Keck (owing to its small distance) out to $\approx 500\,\rm{d}$ after peak light, the very late-time follow-up photometry for all the objects could be potentially contaminated by underlying host systems. We discuss the presence of potential underlying host stellar systems from late-time imaging in Section \ref{sec:environments}.

\section{Locations and Host environments}
\label{sec:environments}

Here, we examine the environments and host properties of the Ca-rich gap transients analyzed in this sample. The host environments of the literature sample of transients have been noted in several works previously for their preference of old environments located far away from their host galaxies,  with no evidence for parent stellar populations at the location of the transients. \citep{Perets2010, Perets2011, Kasliwal2012a, Yuan2013, Lyman2014, Mulchaey2014, Lyman2016b, Lunnan2017, De2018b, Shen2019}. Additionally, \citet{Yuan2013} find that the offset distribution of a subset of events in the literature sample was inconsistent with the stellar mass profiles of their host galaxies, while \citet{Yuan2013} and \citet{Shen2019} show that their offset distributions are consistent with globular clusters or old metal poor stellar populations. \citet{Frohmaier2018} show that the preference for large host offsets in the PTF sample cannot be explained by the reduced recovery efficiency on top of bright galaxies. However, all of the works were based on heterogeneous samples of events gathered from different surveys with diverse selection effects that are difficult to quantify. As the first unbiased systematic experiment to classify a large sample of Ca-rich gap transients, we perform a systematic analysis of their locations and host environments.

\subsection{Host galaxy morphology}

We begin with analyzing the host galaxy morphologies of these events. Our selection criteria for these events did not include any restrictions on the host galaxy type. Six out of the eight events in this sample were found in S0 / E early-type host galaxies (four out of the eight were in E type galaxies), and all of the events were found at projected galactocentric offsets $> 6$\,kpc. Two events (SN\,2019ofm and SN\,2019pxu) were found in late-type galaxies, although at relatively large host offsets (11 and 17.5\,kpc respectively, corresponding to host-normalized offsets of 2.1 and 4.5 $R_e$\footnote{$R_e$ is defined as the half-light radius of the apparent host galaxy}). Notably, both the Ca-Ia objects SN\,2016hnk and SN\,2019ofm were found in late-type star forming galaxies, while PTF\,09dav was found to be hostless in late-time imaging and the nearest galaxy with a known redshift was a star forming spiral $\approx 40$\,kpc from the transient location. However, we caution that PTF\,09dav was close to several faint extended sources (see Figure 3 in \citealt{Kasliwal2012a}) which could be nearby dwarf galaxies; it also showed H$\alpha$ emission in its latest nebular phase spectrum, although it could be associated with circumstellar photo-ionized gas around the SN \citep{Kasliwal2012a}. 

Nevertheless, the preference of these transients for old host environments with large offsets is striking compared to other types of transients in the local universe. The preference for early type galaxies in this sample is reminiscent of that observed for 91bg-like SNe\,Ia \citep{Perets2010, Howell2001, Neill2009, Taubenberger2017}, albeit the preference of Ca-rich gap transients for early-type galaxies is more extreme than for these events. A total of 19 91bg-like SNe\,Ia were classified in the CLU experiment in the span of time considered in this paper, out of which 10 events were found in early type galaxies while the rest were in late-type galaxies. The near equal distribution in early and late type galaxies for 91bg-like SNe\,Ia is consistent with the stellar mass distribution between early-type and late-type galaxies in the local universe \citep{Kochanek2001, Bell2003}. 

\subsection{Locations and offset distribution}

\begin{figure*}[!ht]
\centering
\includegraphics[width=0.49\textwidth]{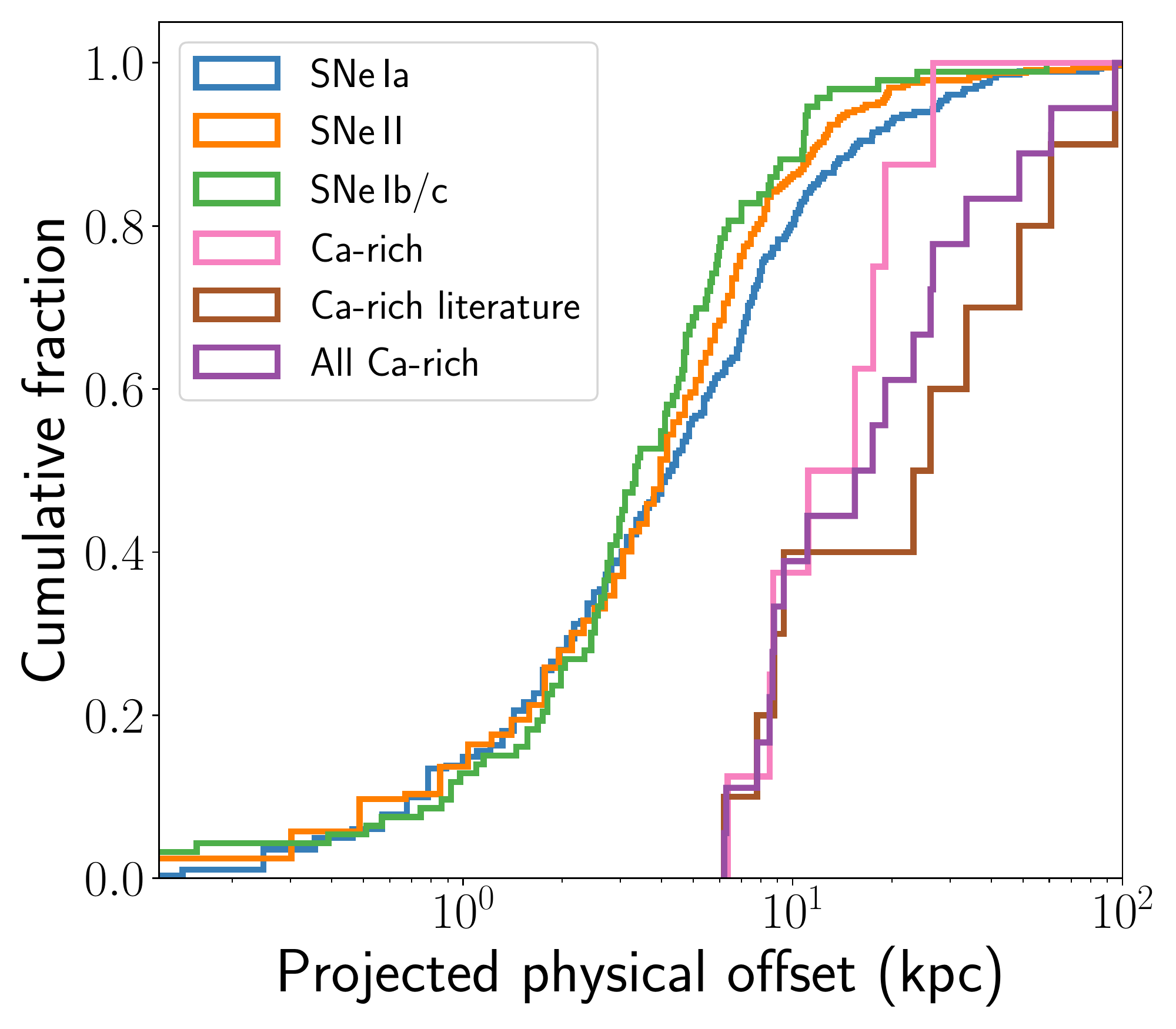}
\includegraphics[width=0.49\textwidth]{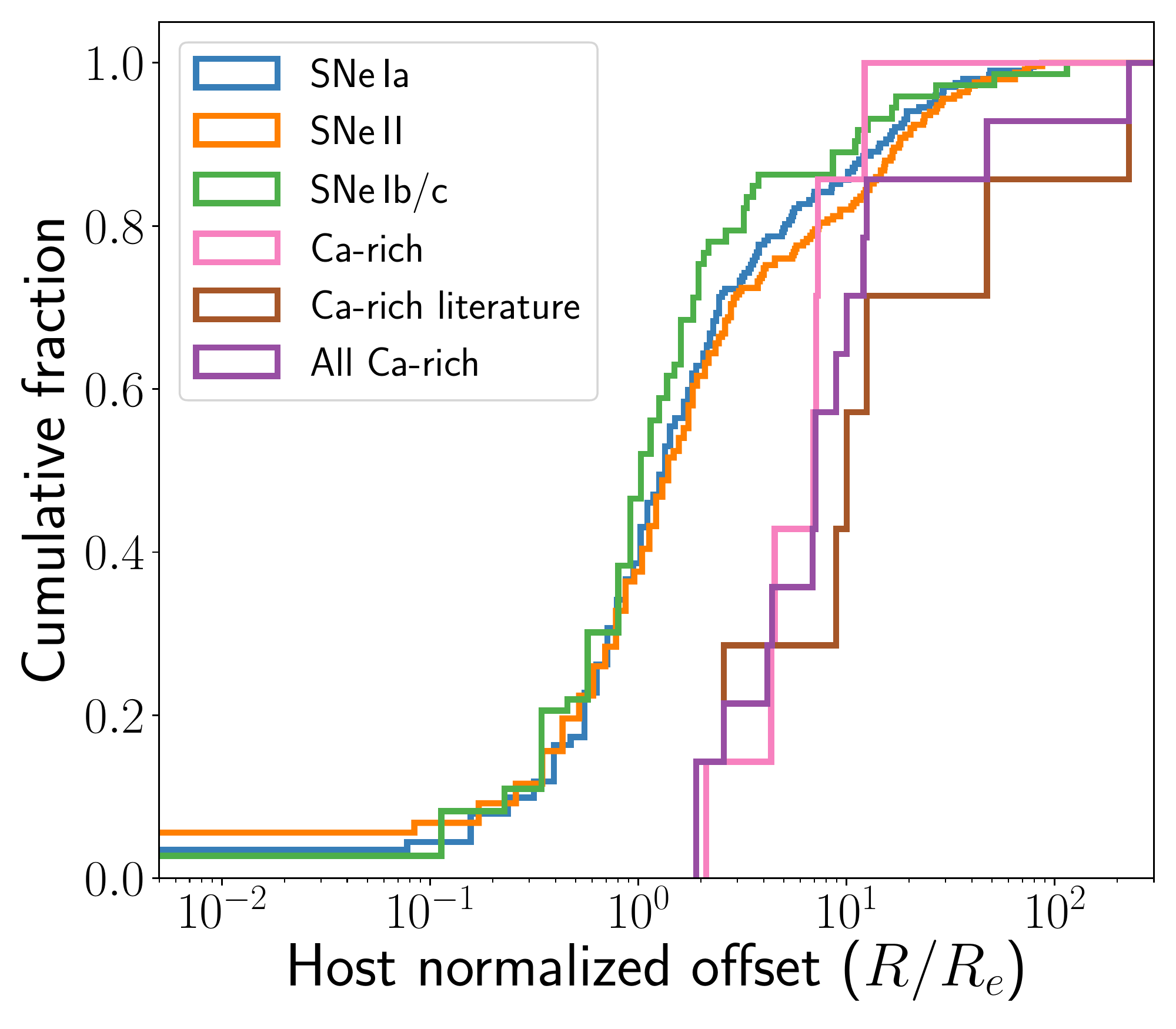}
\caption{Projected offset distribution of all transients in the CLU experiment. The left panel shows the distribution in units of physical projected distance (kpc), while the right panel shows the distribution in host offset normalized by the half-light Petrosian radius of the host galaxy. The half-light Petrosian radii were derived from the Sloan Digitial Sky Survey (SDSS; \citealt{Abolfathi2017}) catalog, and hence are limited to transients occurring in the SDSS footprint.}
\label{fig:offsets}
\end{figure*}

In Figure \ref{fig:offsets}, we compare the host offset distribution of the ZTF Ca-rich gap transients (both in physical projected distance and host-normalized distance) to the other types of SNe in the volume limited experiment -- SNe Ia, SNe II and SNe Ib/c. The host offsets for each event in the sample are computed from the host galaxy in the CLU catalog, which was confirmed to be at the same redshift as the SN. As shown in Figure \ref{fig:offsets}, SNe Ib/c show systematically smaller physical offsets than SNe II, while SNe II show systematically smaller offsets than SNe Ia. However, the Ca-rich gap transients exhibit a significantly skewed distribution of larger offsets (both in terms of physical offsets and host normalized offsets) than any of the object types in the comparison sample. Using a two sample KS test, we can rule out the possibility that the entire population of SNe\,Ia and Ca-rich gap transients in the ZTF sample originate from the same underlying population at 99.9\% confidence. For comparison, we also show the host offset distribution of the literature sample of Ca-rich gap transients and the total combined sample of Ca-rich gap transients, whose distribution appears to extend out to larger galactocentric offsets. A KS test between the ZTF and literature sample of offsets does not indicate a statistically significant difference between the two distributions (p value of 0.28); regardless, the limitation of the CLU experiment to finding transients within 100\arcsec of their host galaxies prevents us from finding objects with very large host offsets. The consistent offset distributions justify the use of the full sample of offset distributions to estimate the incompleteness of the CLU sample (Section \ref{sec:rates}).

\begin{figure}
    \centering
    \includegraphics[width=\columnwidth]{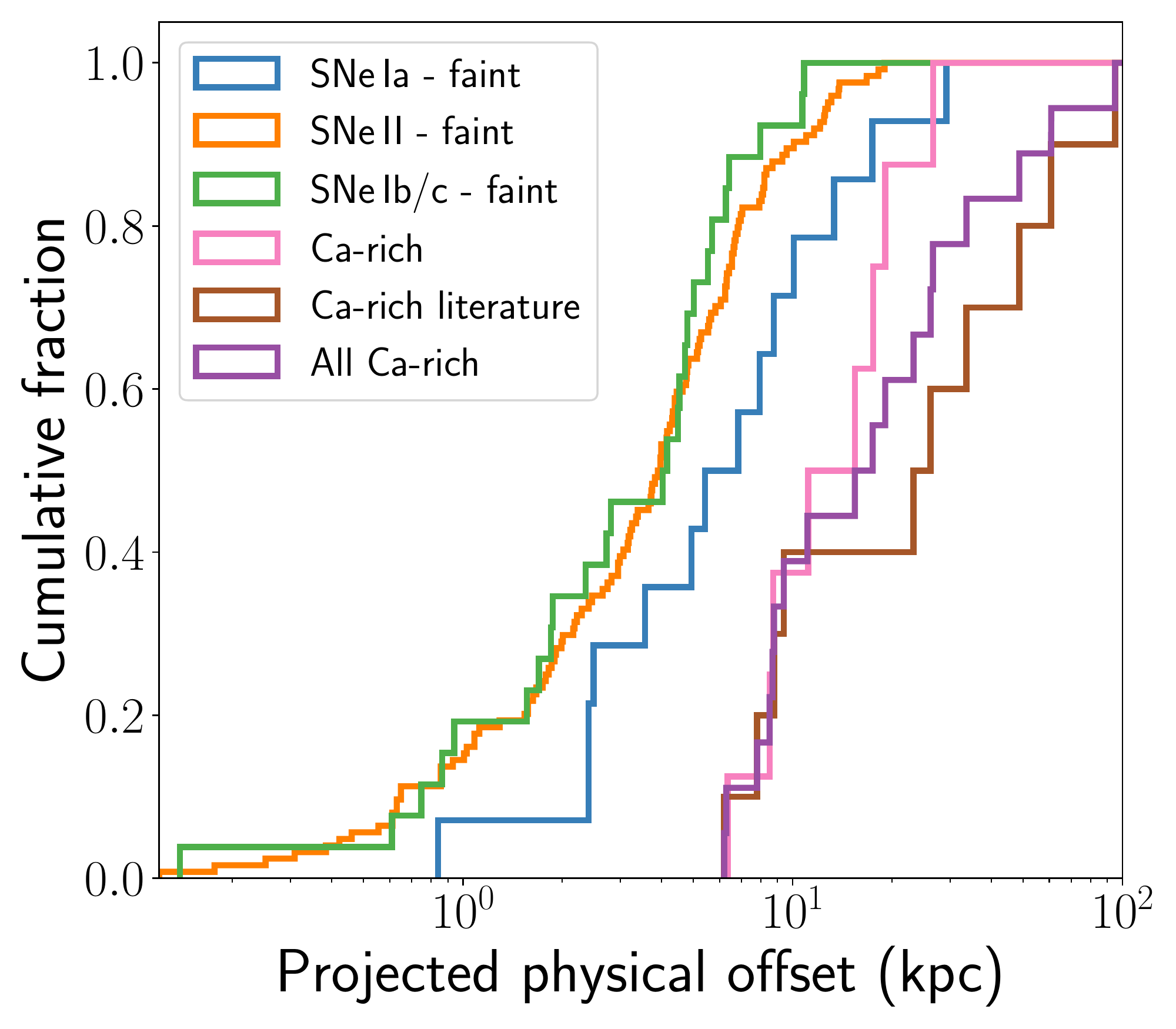}
    \caption{Projected offset distribution of all faint transients that were detected at $M>-17$ in the CLU experiment. }
    \label{fig:offsets_faint}
\end{figure}

Several previous works have suggested that the lack of Ca-rich gap transients at small host offsets may be due to their faint light curves, which would make them difficult to detect on top of high surface brightness regions on galaxies \citep{Foley2015}. \citet{Frohmaier2017} presented the recovery efficiency for the PTF pipeline as a function of the source magnitude and local surface brightness, demonstrating that the recovery efficiency is indeed lower in regions of high surface brightness. Yet, \citet{Frohmaier2018} showed that the preference of the small PTF sample of Ca-rich gap transients for large host offsets cannot be explained by the reduced recovery efficiency on bright galaxy backgrounds. While the recovery efficiency for the ZTF pipeline is currently not available, we can empirically examine if the offset locations for the faint Ca-rich gap transients in the ZTF sample can be primarily explained by poor recovery efficiency on the cores of galaxies. In Figure \ref{fig:offsets_faint}, we show the projected offset distribution of all transients fainter than $M = -17\,\rm{mag}$ in the CLU experiment. The offset distribution of the SNe\,II, SNe\,Ib/c and SNe\,Ia in this low luminosity sample extend from the smallest offsets at $< 1$\,kpc to $\approx 30$\,kpc. Notably the Ca-rich gap transients continue to stand out with large host offsets of at least $5$\,kpc and extending out to $\approx 40$\,kpc.  The skewed offset distribution of the Ca-rich gap transients even in this sample of low-luminosity transients suggests that low recovery efficiency of faint transients cannot completely explain the remote locations of these events.

\begin{figure}
    \centering
    \includegraphics[width=\columnwidth]{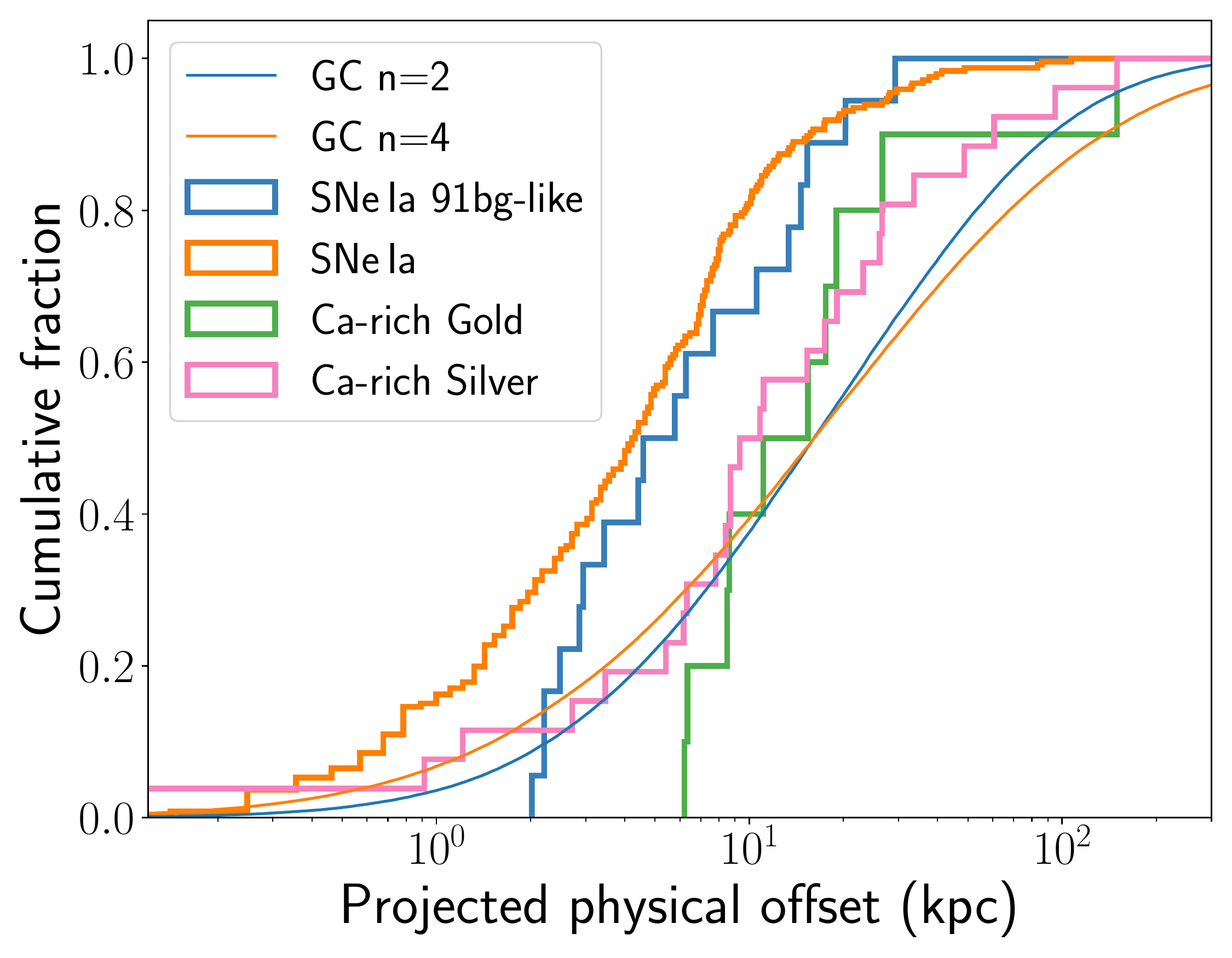}
    \caption{Offset distribution comparison of the `gold' and `silver' sample of Calcium rich gap transients discussed in this paper to that of all SNe Ia and 91bg-like SNe Ia in the volume limited experiment. Our gold sample includes the literature sample of events (which includes the gold sample discussed in \citealt{Shen2019} in addition to the Ca-Ia objects discussed in this work) and the ZTF sample, while the silver sample includes the silver sample discussed in \citet{Shen2019} in addition to the ZTF sample of events. We also show the simulated distribution of globular clusters for two Sersic indices presented in \citet{Shen2019}.}
    \label{fig:gcoffset}
\end{figure}{}

We further compare the environments of 91bg-like SNe\,Ia to that of the Ca-rich gap transients. We plot the projected offset distributions of all SNe\,Ia, 91bg-like SNe\,Ia and Ca-rich gap transients in Figure \ref{fig:gcoffset}. We also show the simulated radial distribution of globular clusters from \citet{Shen2019} for two difference Sersic indices. The literature sample analyzed in this paper includes the Ca-Ia events from the literature, in addition to the `gold' sample in \citet{Shen2019}. Thus, in plotting the radial offset distribution of the sample of Ca-rich events, we define our new `gold' sample by adding the ZTF sample of events to the literature sample (including all the Ca-Ia objects, amounting to a total of 18 events). We also define a new `silver' sample by adding the ZTF sample of events to the silver sample in \citet{Shen2019}, which included Ca-rich events without photometric constraints at peak light (amounting to a total of 24 events). 91bg-like SNe\,Ia show systematically smaller offsets than the Ca-rich gap transients in this sample, and systematically larger offsets compared to the full population of SNe\,Ia. A two sample KS test between the offset distribution of SN\,91bg-like SNe\,Ia and the gold and silver samples of Ca-rich gap transients produce null hypothesis probabilities of being drawn from the same underlying population at $<2.5$\% and $<10$\% respectively. We conclude that the discrepancies between the environments of 91bg-like and Ca-rich events suggest that stellar mass alone does not dictate the rates of Ca-rich gap transients. Long delay times and / or low metallicities in these remote environments of early type galaxies have thus been suggested to play an important role \citep{Perets2010, Yuan2013, Meng2015, Shen2019}. 

In addition to being skewed towards larger offsets compared to the SNe\,Ia, the Ca-rich gold and silver samples exhibit offset distributions different than that of the globular clusters. Specifically, the gold sample distribution is skewed towards larger offsets than the globular clusters. However, the silver sample offset distribution is more consistent with the globular cluster distributions, and consistent with the result in \citet{Shen2019}, although there are discrepancies at relatively large offsets. Notably, all of the Ca-rich gap transients found in this experiment were in `rich' environments with $\gtrsim 10$ galaxies clustered near the nominal host, while \citet{Shen2019} find that $17$\% of globular cluster hosts are in rich clusters (see Section \ref{sec:clusters}). 

\begin{figure*}[!ht]
    \centering
    \includegraphics[width=\textwidth]{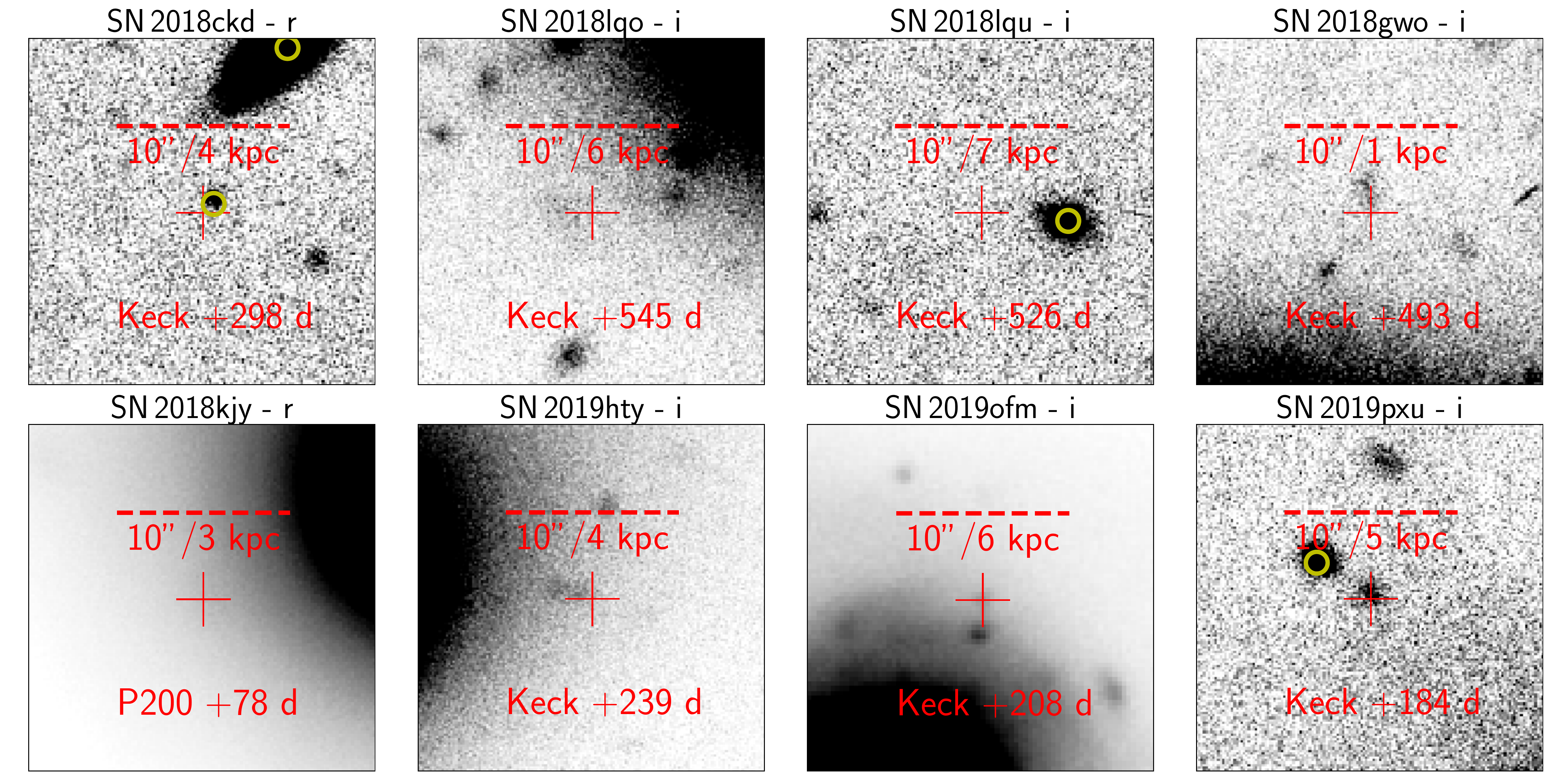}
    \caption{Cutouts of the locations of the Ca-rich gap transients presented in this paper from late-time ground based imaging of the transient locations. North is up and East is left in each cutout. The source name, filter and phase of observation, instrument used and physical scale of the image at the redshift of the transient is shown in each panel. The crosses show the location of the transient. Yellow circles mark locations of potential host systems also detected in pre-explosion archival imaging (see text).}
    \label{fig:lateimage}
\end{figure*}{}

We examine the deep late-time images of the locations of these transients to investigate if there is any evidence for underlying stellar populations at the locations of these transients. We show these images in Figure \ref{fig:lateimage}. SN\,2018ckd shows clear evidence for a relatively bright point-like source (marked with a yellow circle) offset by $<1$\,kpc from the transient location. This source is also detected in archival imaging of the field in the Dark Energy Legacy Survey DR8 \citep{Dey2019} at a magnitude of $r \approx 23.5\,\rm{mag}$ and $g-r \approx 0.3\,\rm{mag}$, corresponding to an absolute magnitude of $\approx -11.5\,\rm{mag}$ if at the redshift of the transient. However, the source is unlikely to be a globular cluster as it is more luminous than nearly the entire known luminosity function of globular clusters \citep{Harris1996}. We oriented the slit to include the extended galaxy $\approx 10$\arcsec to the north of SN\,2018ckd (marked with a yellow circle) during nebular phase spectroscopy, and find it to be an unrelated background galaxy at $z=0.1$, consistent with the photometric redshift of the object in SDSS.

An extended source is detected near the location of SN\,2018lqu (marked with a yellow circle), although its redshift is unknown. If at the redshift of the transient, its magntiude of $r \approx 21.4$ would imply an absolute magnitude of $M_r \approx -14.5\,\rm{mag}$ similar to a dwarf galaxy. SN\,2018lqo, SN\,2018gwo and SN\,2019hty show evidence of faint and extended sources underneath their locations, which likely contaminate our photometry measurements during very late-time imaging (Figure \ref{fig:lateimage}).  However, we caution that the density of unrelated background sources at the depths of the late-time images ($\approx 25$\,mag) is high \citep{Hogg1997}. Using the methodology of \citet{Bloom2002}, we find that the chance coincidence probability of an unrelated $\approx 25$\,mag galaxy within a 5\arcsec\, radius of the transient is $\sim 50$\%, while the same for a 10\arcsec\, radius is $\sim 95$\%.  Thus, the association of these sources to the transients can only be determined from deep spectroscopy in the future. SN\,2018kjy is notable for the smallest host offset ($\approx 6$\,kpc) in the ZTF sample, and is located within the halo of its host galaxy. SN\,2019ofm is found to be on top of its spiral host galaxy, while SN\,2019pxu is at a large offset from its spiral host and has a point source (marked with a yellow circle) within 5\,\arcsec of its location ($M_r \approx -12.1\,\rm{mag}$ if at the redshift of the transient). The latest Keck images of SN\,2019ofm and SN\,2019pxu still show the transient clearly. Future deep imaging for these events will help disentangle the potential host contamination in our latest photometry measurements, as well as allow inspection of potential underlying hosts.

\subsection{Group and cluster environments}
\label{sec:clusters}

\begin{figure*}[]
    \centering
    \includegraphics[width = \textwidth]{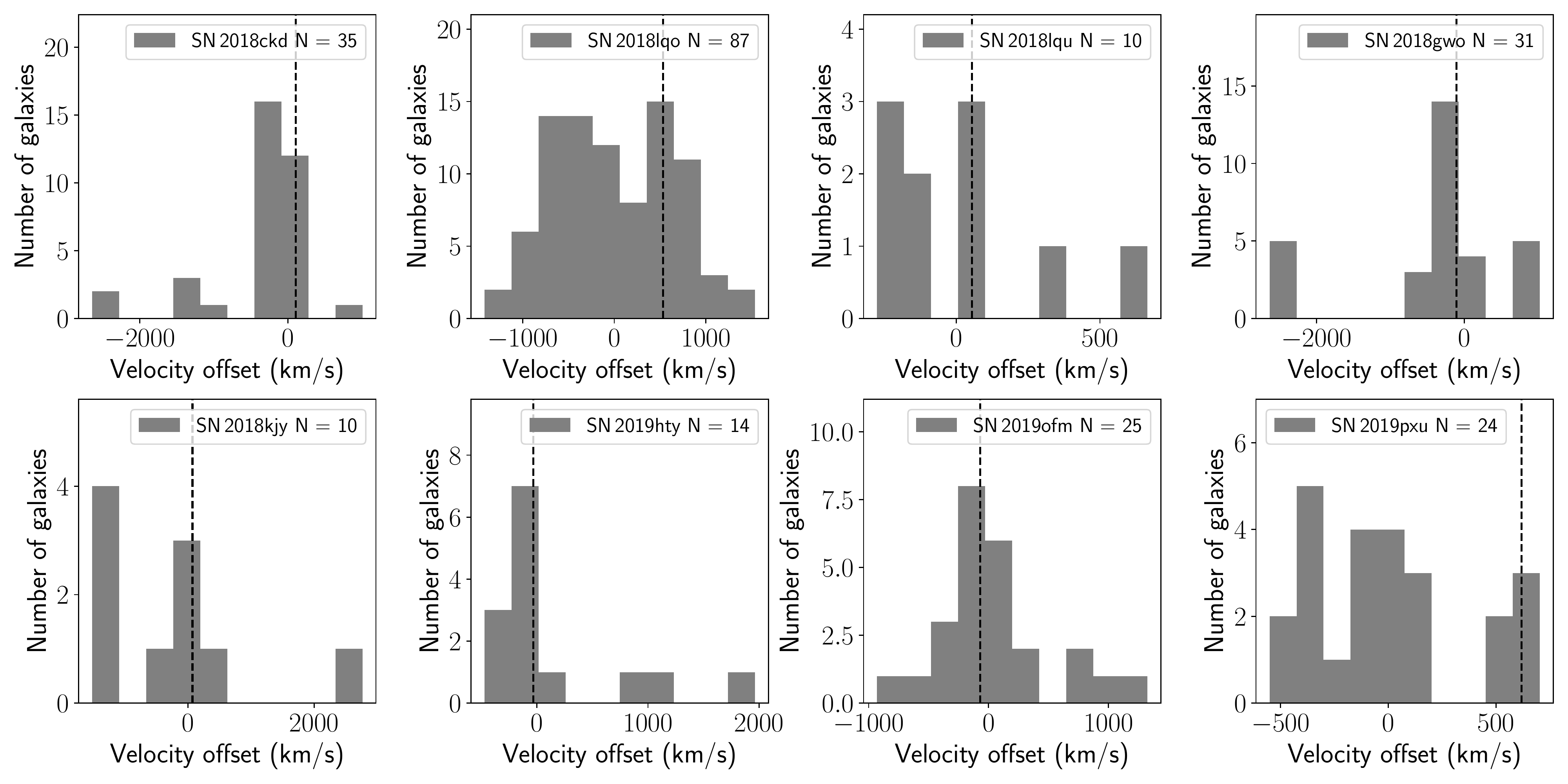}
    \caption{Velocity distributions of galaxies in the environments of the ZTF Ca-rich gap transient sample, that have previously known spectroscopic redshifts in the CLU catalog. In each distribution, we define zero velocity as the median of the redshift distribution of all the galaxies in the projected vicinity of the transient, and show the velocity of the assumed (nearest) host galaxy of the transient with a dashed line. The transient name and number of galaxies in each histogram is indicated in the legend.}
    \label{fig:hostcluster}
\end{figure*}

Given the large host offsets of the Ca-rich gap transient sample, we analyze the environments of the assumed host galaxies to check if they are part of a larger group or cluster that may explain the remote locations of these transients. \citet{Mulchaey2014}, \citet{Lunnan2017} and \citet{De2018b} performed a similar analysis on the literature sample of Ca-rich gap transients, and demonstrated that these objects preferred host galaxies in groups and clusters. For each transient, we construct the sample of nearby galaxies by selecting galaxies from the CLU catalog (with a previously known spectroscopic redshift) within a projected radius of 1 Mpc and a recession velocity difference of $\pm 3000$ km s$^{-1}$ from the location of the transient.

We emphasize that since galaxy catalogs are highly incomplete at the redshifts of these transients \citep{Kulkarni2018, Fremling2019}, these distributions are only lower limits to the true number of galaxies in the environments of these transients. We show the velocity histograms of the identified galaxies in Figure \ref{fig:hostcluster}. As shown, the environments of the Ca-rich gap transient host galaxies are largely dominated by groups and clusters with at least 10 known objects within the selection criteria defined above. SN\,2018lqo is in the densest environment with 87 known nearby galaxies. We thus conclude that all of the transients in the ZTF sample are in group or cluster environments, consistent with that reported for the literature sample presented in \citet{Lunnan2017} and \citet{De2018b}. 

\section{Volumetric rates in the local universe}
\label{sec:rates}

In this section, we estimate the volumetric rates of Ca-rich gap transients using the ZTF CLU experiment. As a large scale systematic and controlled experiment, the volume-limited SN classification effort provides the first direct way to estimate the volumetric rates of this class  within $200$\,Mpc due to its high spectroscopic completeness ($\approx 90$\%) down to the experiment limiting magnitude of 20\,mag. 

\subsection{Demographics from the volume limited experiment}

Since the volume-limited experiment has high spectroscopic classification completeness, a straightforward way to estimate the volumetric rates of Ca-rich gap transients relative to other SNe is by comparing the number of events. We perform this analysis by restricting the sample of transients in the CLU experiment to within the volume where Ca-rich gap transients are detectable. Based on the luminosity function of known events discussed in Section \ref{sec:lumfunc}, we find that the average Ca-rich gap transient (peaking at $M = -16\,\rm{mag}$) is detectable out to 150\,Mpc for a flux limit of $r = 20.0\,\rm{mag}$, which is the target limiting magnitude of the CLU experiment. Within the experiment duration mentioned above, a total of eight Ca-rich gap transients were detected along with 133 SNe Ia that were classified in the same volume. So, we can place a lower limit on the rates of Ca-rich gap transients of 6\% of the volumetric rates of SNe\,Ia, or $1.5 \times 10^{-6}$ Mpc$^{-3}$ yr$^{-1}$ (assuming the SN Ia rate to be $2.5 \times 10^{-5}$ Mpc$^{-3}$ yr$^{-1}$; \citealt{Frohmaier2019}). Note that this estimate is likely an underestimate of the true rate of Ca-rich gap transients since SNe\,Ia will be bright in the 150\,Mpc volume (peaking at $\approx 17$\,mag) andso detectable during the survey for at least $\approx 100\,\rm{d}$ around peak, while the average Ca-rich gap transient in this volume is visible for $\approx 10 - 20\,\rm{d}$. In addition, our selection criteria requires the detection of each source on the rise to peak, which further limits the sample of fast evolving Ca-rich sample relative to the slow rising SNe\,Ia. A true estimate of the volumetric rate of Ca-rich gap transients thus requires a simulation of the cadence of the ZTF survey together with the characteristic luminosity evolution of a Ca-rich gap transient to be able to estimate the number of sources that pass our light curve selection criteria as a function of the input volumetric rate.

\subsection{Luminosity function}
\label{sec:lumfunc}
\begin{figure}
    \centering
    \includegraphics[width = \columnwidth]{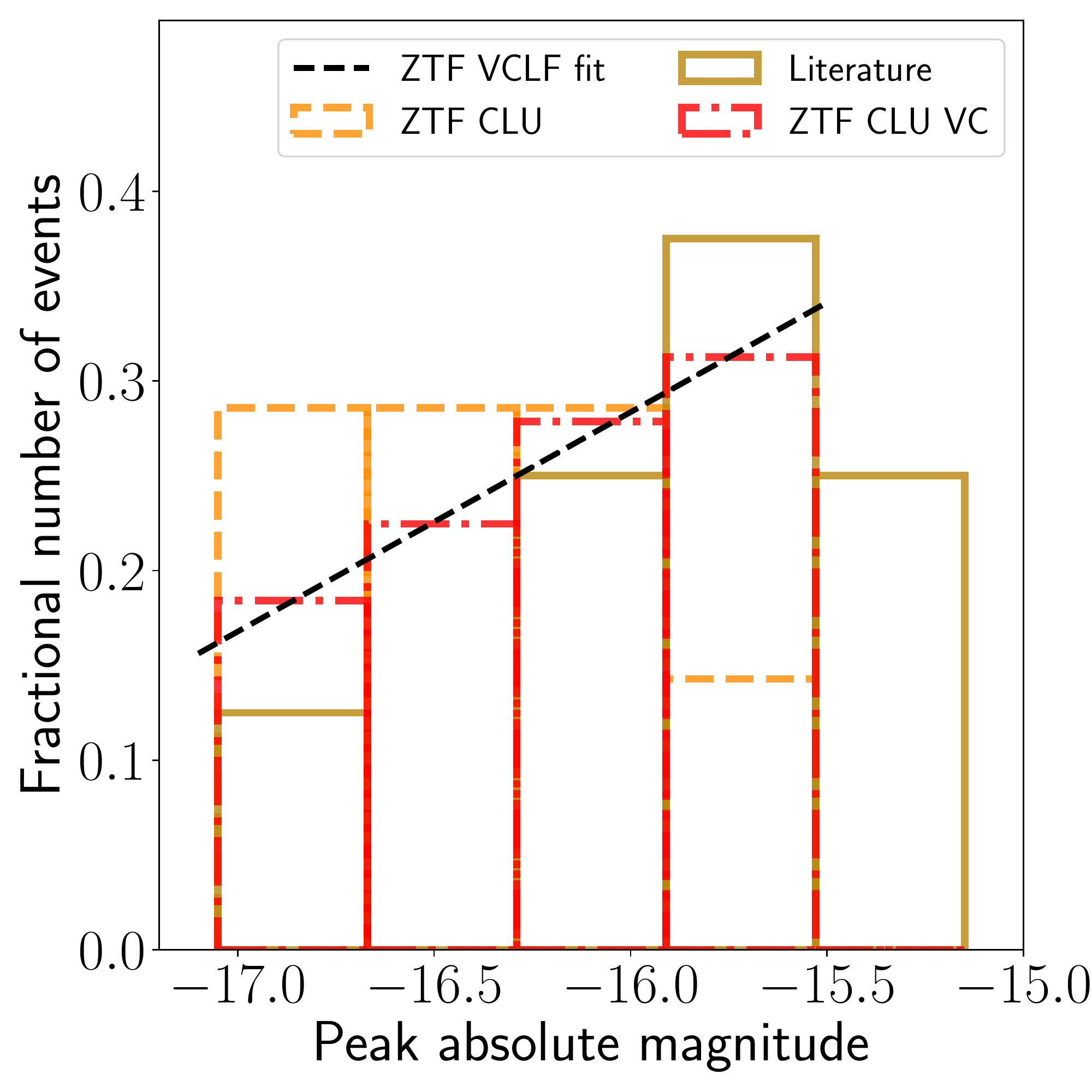}
    \caption{Peak $r$-band absolute magnitude distribution of the literature sample of Ca-rich gap transients (gold solid) and the ZTF sample (orange dashed). The dot-dashed distribution in red shows the volume corrected luminosity function (VCLF) of the ZTF sample, where a $\frac{1}{V_{max}}$ weighting was applied, with $V_{max}$ being the maximum volume out to which the transient would be detected by the volume limited experiment. The black dashed line shows a linear fit to the volume-corrected luminosity function.}
    \label{fig:lumfunc}
\end{figure}{}

We aim to derive the luminosity function of Ca-rich gap transients using the controlled sample of events from the CLU experiment. Due to the small number of events, we do not separate the spectroscopic sub-types discussed in this work while estimating luminosity functions and volumetric rates. In Figure \ref{fig:lumfunc}, we show the observed histograms of the peak $r$-band magnitudes of the literature sample of Ca-rich objects and the ZTF sample of objects. The majority of the literature events exhibit peak magnitudes between $M = -16.5\,\rm{mag}$ and $M = -15.5\,\rm{mag}$, while the ZTF sample shows a near uniform distribution between $M = -17\,\rm{mag}$ and $M = -15.5\,\rm{mag}$. However, it is difficult to quantify the selection biases when combining the literature sample of events with the ZTF sample due to the diverse selection criteria of the surveys that detected the literature objects. In order to derive an unbiased luminosity function, we focus specifically on the ZTF sample of events. Despite being classified as part of a CLU experiment, most of the objects in the ZTF sample are not luminous enough to be detectable across the entire experiment volume (200 Mpc) for a limiting magnitude of $r = 20\,\rm{mag}$, and thus a volume correction needs to be applied to recover the true luminosity function. 

We show a volume-corrected luminosity function (VCLF) histogram of the ZTF Ca-rich sample in Figure \ref{fig:lumfunc}. For each object, we apply a volume-correction of $\frac{1}{V_{max}}$ where $V_{max}$ is the maximum volume out to which object would be detectable for a limiting magnitude of 20. For sources that would be detectable at $> 200$\,Mpc, we set the relevant volume to 200\,Mpc given the volume-limited nature of the experiment. The VCLF shows evidence of an increasing number of events down to the faintest event, and a simple linear polynomial describes the VCLF well. It is difficult to constrain the luminosity function below the faintest observed events, since it is unclear if there is a population of fainter events which would be detectable only within a very small volume ($< 100$\,Mpc). Hence, for the purpose of the simulations, we restrict the observed function between $M = -17\,\rm{mag}$ (the upper luminosity limit for our sample) and $M = -15.3\,\rm{mag}$ (faintest observed event SN\,2012hn). We caution that the ZTF CLU experiment is not sensitive to transients brighter than $M = -17\,\rm{mag}$ at peak due to the sample selection criteria, if such a population exists. Our rate estimates are thus limited to events peaking at $-15.3 > M > -17$.

\subsection{Template light curve for a Ca-rich gap transient}

\begin{figure*}[!ht]
\centering
\includegraphics[width=0.49\textwidth]{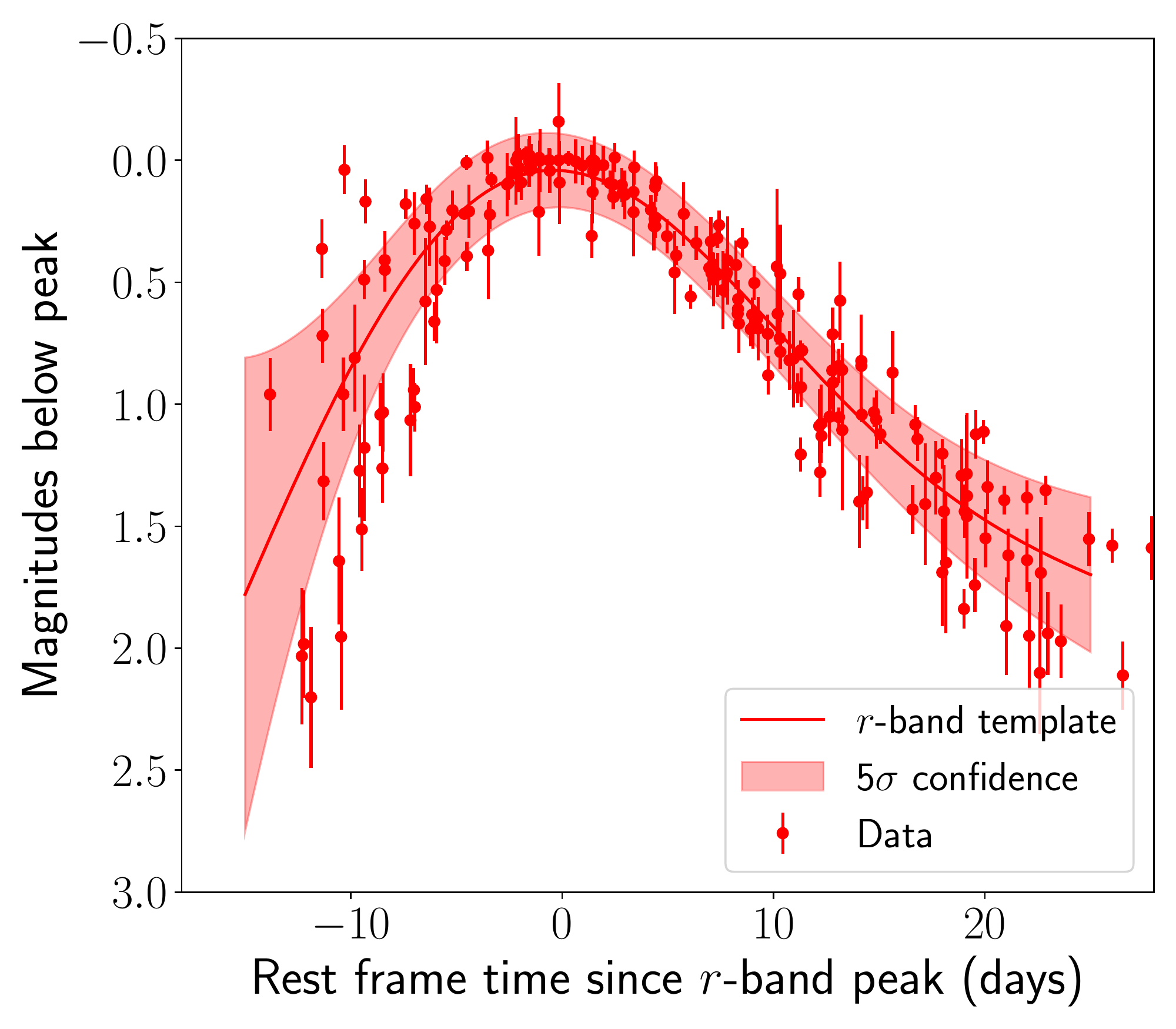}
\includegraphics[width=0.49\textwidth]{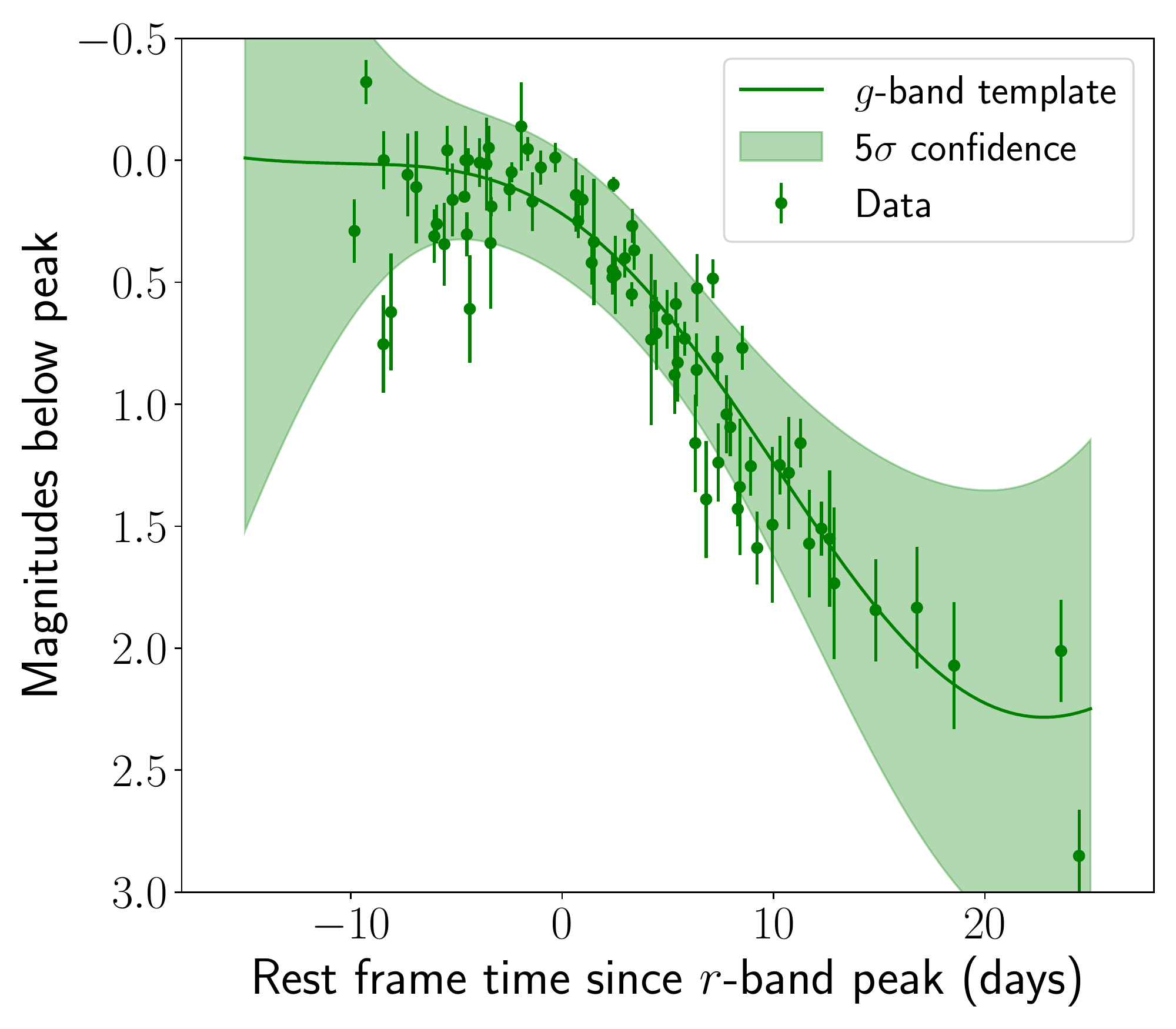}
\caption{$r$-band and $g$-band light curve templates for Ca-rich gap transients, normalized to peak magnitude. The points with error bars show the observed light curves with 1$\sigma$ error bars in the respective filters while the solid lines indicate the best-fit light curve from Gaussian process fitting. The shaded regions indicate the uncertainty intervals derived from the Gaussian process fitting, corresponding to 5$\sigma$ confidence regions.}
\label{fig:templates}
\end{figure*}

In order to provide a more robust estimate of the volumetric rates of Ca-rich gap transients, accounting for their low-luminosity light curves and the ZTF survey cadence, we start by constructing a template light curve of a Calcium rich gap transient in the $r$ and $g$ bands (which have the most photometric coverage) using the data available for the events in the combined sample. Since we are interested in the photometric evolution timescale of each event, we first normalize each light curve by its peak magnitude measured in the respective filter. Time is measured with respect to the best-fit $r$-band peak time. We then fit a Gaussian process model with a constant kernel to the normalized light curves in each filter to construct a normalized light curve template in the $r$ and $g$ filters.  We perform the fit in the phase space of magnitude versus time ranging from $-15\,\rm{d}$ before $r$-band peak to $+25\,\rm{d}$ after $r$-band peak, where there is photometric coverage for more than one object in both the filters. This produces the average peak-normalized template light curve and its uncertainty as a function of phase from $r$-band maximum. We do not include photometry upper limits in the fit. In Figure \ref{fig:templates}, we show the peak normalized data and best fit templates together with their uncertainties for the two filters. In particular, we note that the sample of peak normalized light curves in both filters are fairly homogeneous around peak light (even though there is a dispersion in the peak magnitudes), suggesting that a single light curve template normalized to peak magnitude can well capture the shape of the light curve. 

\subsection{Simulations of the ZTF survey}

\begin{figure*}[]
    \centering
    \includegraphics[width=\columnwidth]{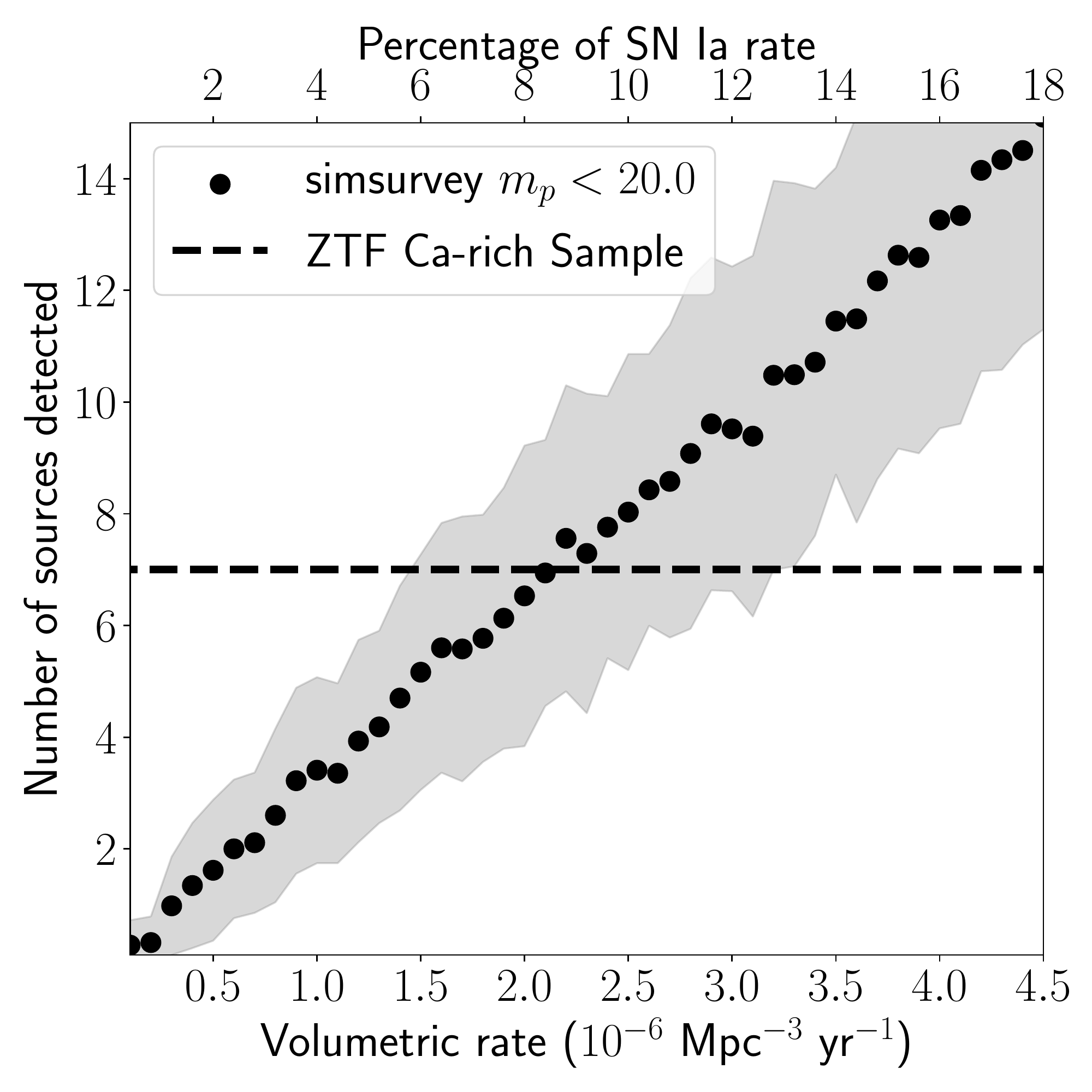}
    \includegraphics[width=\columnwidth]{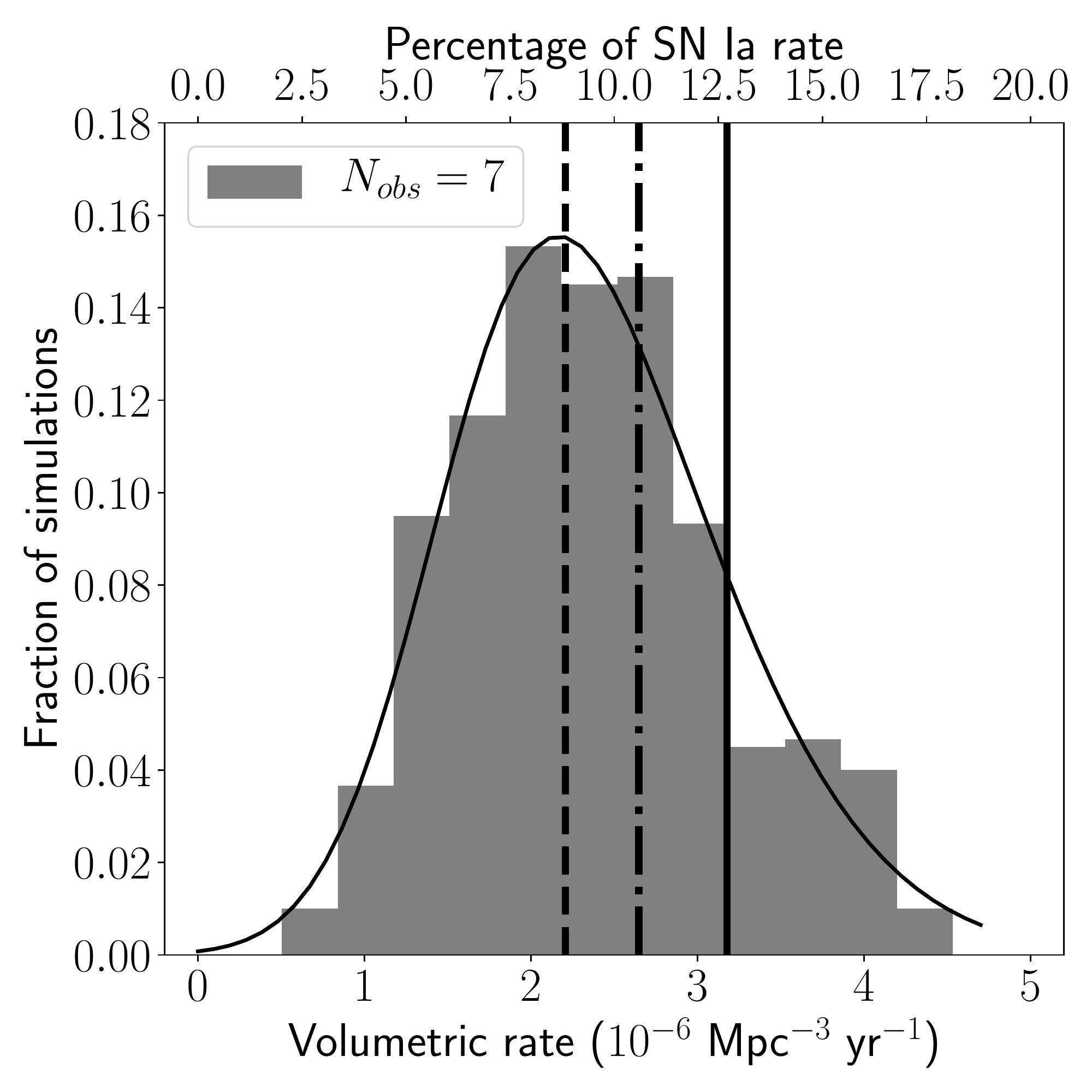}
    \caption{Estimate of the volumetric rate of Ca-rich gap transients with simulations of the ZTF survey using the \texttt{simsurvey} code. (Left panel) We show the number of sources passing our selection criteria for the CLU experiment as a function of the input volumetric rate (see text). The points and error bars are the mean and standard deviation of the number of transients recovered as a function of the volumetric rate, while the dashed black line shows the observed number of sources in the experiment. (Right panel) The fraction of simulations producing the observed number of transients as a function of the input volumetric rate. The dependence is fit with a skewed Gaussian distribution shown by the solid line, which we use to derive the best estimate of the volumetric rate and its confidence interval (see text). The dashed line shows the mean of the distribution, the dot-dashed line shows the rate after correcting for galaxy catalog incompleteness and the solid line shows the rate estimate after accounting for the transients missed beyond 100\arcsec from their host galaxies.}
    \label{fig:rate}
\end{figure*}{}
Using the derived luminosity function and light curve templates for the class of Ca-rich gap transients, we estimate their volumetric rates in the local universe using the \texttt{simsurvey} code \citep{Feindt2019}. \texttt{simsurvey} is capable of simulating transient light curves as would be observed by ZTF for a given input SN template (provided using the \texttt{sncosmo} package from \citealt{Barbary2016}), and an input survey tiling pattern and duration (termed as a survey plan). We use the best-fit $r$ and $g$ band templates to construct a \texttt{TimeSeriesSource} model in the \texttt{sncosmo} package to simulate the spectral evolution of a Ca-rich gap transient between $15\,\rm{d}$ before and $25\,\rm{d}$ after $r$-band peak. We then use the actual ZTF observing history between 2018 June 01 and 2019 September 30 in any of the public or collaboration surveys as the input survey plan. Since the ZTF reference images were created shortly before the start of the survey and extended well into the time period discussed here for some fields, we only consider pointings that were acquired at least $60\,\rm{d}$ after the end of reference creation to avoid contamination of the reference images by transient light.

We then simulate ZTF light curves of Ca-rich gap transients for a range of input volumetric rates, performing 100 simulations of the ZTF observing plan for each input rate. Based on the observed volume-corrected luminosity function, we fix the peak absolute magnitude distribution of the injected transients to be a linear function between $M = -15.3\,\rm{mag}$ and $M = -17\,\rm{mag}$ in $r$-band. We assume a color of $g-r = 0.7\,\rm{mag}$ at $r$-band peak (based on the observed color evolution of the sample). Transients are injected out to a redshift of $z = 0.05$. In order to select transient candidates that would have passed the selection criteria defined in this experiment, we perform quality cuts on the simulated light curves as follows:
\begin{enumerate}
    \item At least 2 detections of the source are required above signal to noise ratio (SNR) of 5 in either $g_{\text{ZTF}}$ or $r_{\text{ZTF}}$ filters.
    \item The peak detected magnitude of the transient should be $m < 20$ in either $g_{\text{ZTF}}$ or $r_{\text{ZTF}}$ filters.
    \item The transient is observed before peak such that there is at least one detection with SNR $>5$ before the peak of the light curve in either $g_{\text{ZTF}}$ or $r_{\text{ZTF}}$ filters.
\end{enumerate}
Applying this selection criteria to the sample presented in this paper, all but SN\,2018gwo, which was recovered after peak light, satisfy the criteria. We use the average number of transients qualifying these cuts as the best estimate of the number of detected transients for each input volumetric rate, while the standard deviation of the number detected in the simulations is taken as the uncertainty. We show the expected number of detected transients as a function of the input volumetric rate in Figure \ref{fig:rate}. Figure \ref{fig:rate} also shows the fraction of simulations that produce the observed number of transients as a function of the input volumetric rate. 

Note that \texttt{simsurvey} is designed to inject simulated transients over the entire sky for a given input volumetric rate, while the CLU experiment is restricted to transients coincident within 100\,\arcsec of galaxies with known spectroscopic redshifts. Hence we denote the \texttt{simsurvey} derived rate as $r_{Ca, u, o}$ with the $u$ subscript indicating uncorrected for the galaxy catalog completeness, and $o$ indicating uncorrected for offset distribution. The distribution of the fraction of simulations (Figure \ref{fig:rate}) is well described by a skewed Gaussian function and we fit the fraction of simulations with this functional form to derive the best estimate of the volumtric rate and its 68\% confidence interval. We find a volumetric rate of 
\begin{equation}
    r_{Ca, u, o} = (2.21^{+1.01}_{-0.67}) \times 10^{-6} \text{Mpc}^{-3} \text{yr}^{-1}
\end{equation}{}
This rate corresponds to $\approx 9^{+4}_{-3}$\% of the SN Ia rate in the local universe ($\approx 2.5 \times 10^{-5}$ Mpc$^{-3}$ yr$^{-1}$; \citealt{Frohmaier2019}).

In order to estimate the effect of incompleteness of galaxy catalogs in our estimate of the volumetric rate, we use the estimated redshift completeness factor (RCF) from the ZTF BTS \citep{Fremling2019} as a function of the WISE $W_1$ (3.36\,$\mu$m) magnitude and redshift of the host galaxies. Taking the observed distribution of redshift $z$ and $W_1$ absolute AB magnitude $M_{W1}$ (as obtained from the \texttt{Tractor} catalogs described in \citealt{Lang2016}) of the ZTF sample of Ca-rich gap transients, we weight each event by $\frac{1}{RCF(M_{W1}, z)}$ for its host galaxy, and sum up over the sample of seven events relevant for the simulation. With this exercise, we find that the incompleteness of galaxy catalogs leads to an underestimate of the Ca-rich gap transient rate by $\approx 20$\%. Next, analyzing the full sample of 18 events, we find that 3 out of the 18 objects exhibited offsets larger than 100\arcsec of their host galaxies, although we caution that it is hard to quantify the systematic biases associated with the literature events. Accounting for this effect would increase the inferred rate by another $\approx 20$\%.
Adjusting for these incompleteness, we derive a rate of 
\begin{equation}
    r_{Ca} = (3.19^{+1.45}_{-0.96}) \times 10^{-6} \text{Mpc}^{-3} \text{yr}^{-1}
\end{equation}{}
which is $13^{+6}_{-4}$\% of the volume-averaged SN Ia rate in the local volume. Owing to the predominance of early type hosts in this sample, we also compare this rate against the SN\,Ia rate in early type galaxies. \citet{Li2011} report a luminosity-function averaged SN\,Ia rate in early type galaxies of $\approx 0.05$ per 100 yr per $10^{10}$ L$_{\odot, K}$. Using the local K-band luminosity density in early type galaxies \citep{Kochanek2001}, we find the corresponding volumetric rate of SNe\,Ia in early type galaxies to be $\approx 1.1 \times 10^{-5}$ Mpc$^{-3}$ yr$^{-1}$. Thus, the inferred rate of Ca-rich gap transients is $\approx 30$\% of the volume-averaged SN\,Ia rate in early type galaxies. We further note that the rate of SNe\,Ia in early type galaxies in cluster environments \citep{Mannucci2008} is $\approx 40$\% lower  than the volume-averaged rate per unit mass in \citet{Li2011}. This suggests that the rate of Ca-rich gap transients in early type galaxies in clusters (which is true for nearly all events in our sample) is nearly $\approx 50$\% of the SN\,Ia rate in these environments.

However, we caution that this rate estimate is still strictly a lower limit as we did not include the detection efficiency of the ZTF image subtraction pipeline as a function of transient magnitude and underlying surface brightness. An accurate estimate of the true rate would require us to assign a probabilistic detection likelihood for each simulated detection, using measured detection efficiencies of the ZTF pipeline, which are currently not available. We thus proceed taking the derived rate as the first measured lower limit to the volumetric rate of Ca-rich gap transients from a large systematic volume-limited experiment. The derived rate is consistent with those estimated by \citet{Perets2010} from LOSS and \citet{Kasliwal2012a} from PTF. \citealt{Frohmaier2018} presented the first quantitative analysis of the volumetric rates of Ca-rich gap transients using three events reported by PTF, incorporating the detection efficiency of the PTF pipeline as a function of transient magnitude and background surface brightness. They infer a rate that could be larger than in this work, of $\approx 30 - 90$\% of the SN\,Ia rate in the local universe, although their estimate has large error bars owing to the small number of three events in the PTF sample.

\section{Discussion}
\label{sec:discussion}

In this paper, we have presented observations and analysis of eight new Calcium rich gap transients classified as a part of the ZTF CLU experiment, nearly doubling the known sample of events in the literature. In Section \ref{sec:analysis} we demonstrated that the ZTF sample shares several similarities to ten events reported in the literature, while also increasing the diversity of several observed properties among the combined sample.  Utilizing the systematic selection criteria of the ZTF sample, we presented an analysis of the host environment properties (Section \ref{sec:environments}) and volumetric rates (Section \ref{sec:rates}) of the class. In this section, we gather all of these findings to discuss their constraints and implications on the progenitor channels for this class.

\subsection{Spectroscopic sub-classes and correlations}
\label{sec:snibc_cont}
\begin{figure*}
    \centering
    \includegraphics[width=0.8\textwidth]{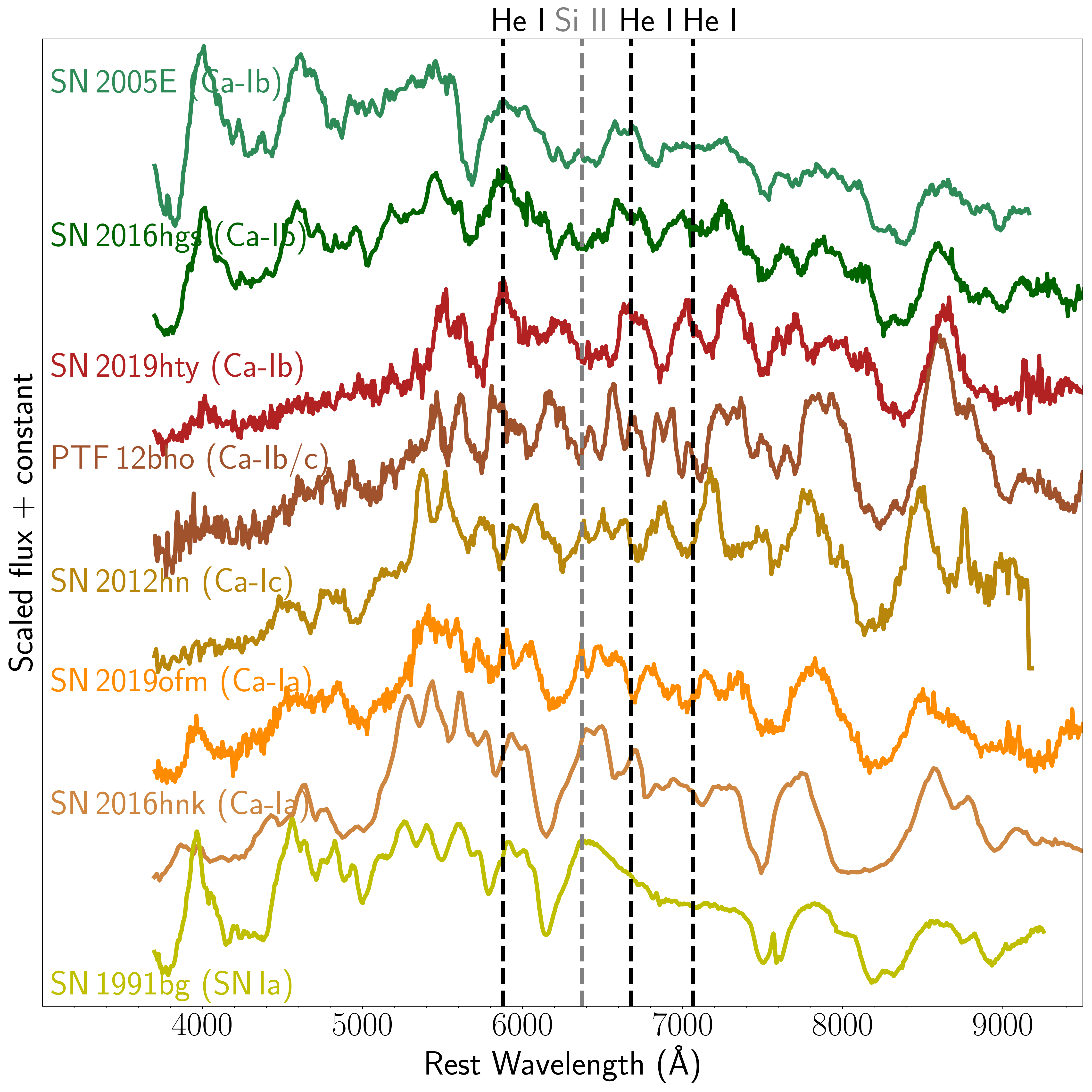}
    \caption{A continuum of spectroscopic and photometric properties in the sample of Ca-rich gap transients. For comparison, we also show a peak light spectrum of SN\,1991bg, which shows striking similarities to the Ca-Ia objects. From bottom to top, we find a sequence of events that appear similar to SNe\,Ia-91bg at peak (Ca-Ia events) with strong Si\,II lines, to SNe\,Ib/c at peak (Ca-Ib/c events) with weak He lines and line-blanketed continua, to SNe\,Ib at peak (Ca-Ib events) with strong optical He I lines and no line blanketing. The colors reflect the photometric colors of the transient at peak. In general, going up the same sequence from bottom to top, the light curves become faster evolving while the peak $g-r$ colors become bluer. The black dashed lines show the rest wavelength of the optical He I lines and the gray dashed line shows the position strong Si\,II line at $\approx 6360$\,\AA. Note the increasing depth and higher velocity of the Ca II NIR triplet going from the Ca-Ib to Ca-Ia events. }
    \label{fig:ca_sequence}
\end{figure*}

In Section \ref{sec:analysis}, we noted the existence of two classes of Ca-rich gap transients distinguished by their spectroscopic appearance at peak -- the Ca-Ib/c events and the Ca-Ia events (see Table \ref{tab:caclass}). We further found a possible continuum of peak light spectral characteristics within the Ca-Ib/c class, wherein the events evolve from green continua with strong P-Cygni features in the blue to events with featureless reddened continua at short ($\lesssim 5500$\,\AA) wavelengths. While the Ca-Ia events are distinguished by their strong Si\,II features at peak light, we demonstrated a continuum in the peak-light spectral features going from the Ca-Ia to Ca-Ic to Ca-Ib events. We briefly summarize them here:
\begin{enumerate}
    \item Ca-Ib/c objects with green continua show strong optical He I features, while events with redder continua show a continuum of strong to weak to no He lines in their peak spectra (Figure \ref{fig:helines}). 
    \item He line velocities in red Ca-Ib/c events ($\approx 7000$ km s$^{-1}$) are lower than those observed in the green Ca-Ib/c events, which show high photospheric velocities  of $\approx 10000$ km s$^{-1}$).
    \item Red Ca-Ib/c events exhibit slower evolving light curves (as quantified by $t_{f, 1/2}$) compared to the green events, with a null hypothesis probability (that the two classes are drawn from the same underlying population of $t_{f, 1/2}$) of $< 5$\% (Figure \ref{fig:lumwidth}). This dependence is corroborated by the $g-r$ color dependence of the photometric evolution ($\Delta m_7$) shown in Figure \ref{fig:colorlc}.
    \item Red Ca-Ib/c events typically exhibit smaller [Ca II]/[O I] compared to green Ca-Ib/c events (by a factor of $\approx 2$) at similar phases when the transient spectrum becomes optically thin (Figure \ref{fig:caoratio}). 
    \item Ca-Ia events exhibit more luminous light curves (peak absolute magnitude $M \lesssim -16.5$) than the Ca-Ib/c class, and have similar red spectroscopic colors / line blanketing and slow photometric evolution as the red Ca-Ib/c events (Figure \ref{fig:lumwidth}).
    \item Although the strong Si\,II line distinguishes the Ca-Ia events from the Ca-Ib/c events, there are striking similarities between the Ca-Ia and Ca-Ib/c events with redder continua, and we find evidence of a continuum of Si\,II line strengths going from Ca-Ia to Ca-Ib/c events (Figure \ref{fig:Ibc_Ia_compare}). 
\end{enumerate}

\begin{table}[!ht]
    \centering
    \footnotesize
    \begin{tabular}{c|ccc}
    \hline
    \hline
       Observable & Ca-Ia & Ca-Ib/c red & Ca-Ib/c green\\
       \hline
       Si\,II? & strong & strong to weak & weak \\
       He\,I? & No & weak to strong & strong  \\
       V ($10^3$ km s$^{-1}$) & 6 -- 10 (8) & 4 -- 10 (7) & 8 -- 12 (10) \\
       Blanketed? & Yes & Yes & No \\
       $M_p$ (mag) & -16.2 -- -17 & -15.3 -- -16.7 & -15.5 -- -16.2 \\
       $\Delta m_7$ (mag) & 0.4 & 0.3 & 0.5 \\
       $g - r$ (mag) & 1.0 & 1.5 & 0.4\\\relax
       [Ca II]/[O I] & No [O I] & 2.5 - 10 (4) & 7 - 13 (10) \\
         \hline
    \end{tabular}
    \caption{Primary observational differences between the Ca-Ia and Ca-Ib/c objects, also highlighting differences between the green and red Ca-Ib/c objects. The velocity row is indicated with $V$, and shows the range of observed peak velocities together with the typical value. The Peak mag and [Ca II]/[O I] row shows the range of peak absolute magnitudes in $r$-band and [Ca II]/[O I] respectively.}
    \label{tab:caclass}
\end{table}

It is worth noting that even most Ca-Ib/c events exhibit weak but identifiable Si II lines blended with the nearby $\lambda6678$ He line (Section \ref{sec:analysis}). We thus find evidence of a sequence in the Ca-rich gap transients going from Ca-Ia to red Ca-Ib/c to green Ca-Ib objects, potentially suggesting a continuum of underlying explosion conditions and progenitor systems. We show the corresponding spectroscopic sequence in Figure \ref{fig:ca_sequence}, where we plot the peak light spectra of a few representative members of the Ca-Ia and Ca-Ib/c class. In addition, we also show a peak light spectrum of SN\,1991bg \citep{Filippenko1992b}, which shows several similarities to the Ca-Ia objects. Thus, the underlying cause for this continuum of spectroscopic appearance could provide a clue to the nature of these explosions. We tabulate and quantify the key observational differences between the different spectroscopic classes in Table \ref{tab:caclass}. We now further this idea of a continuum of underlying progenitor systems as an explanation for the diversity observed in Ca-rich gap transients by probing physical explanations for this continuum.

\subsection{Constraints on ejecta composition and mixing}

Suppression of flux and features at short wavelengths can be explained as an outcome of line blanketing along the line of sight from Fe group material in the outer ejecta (e.g. \citealt{Woosley1994, Nugent1997}). Such features are often seen in peculiar thermonuclear SNe \citep{De2019a}, suggestive of Fe-group rich layers in the outer ejecta. The presence of such material is expected to produce redder colors in the transient \citep{Nugent1997,Kromer2010,Polin2019a} while increasing the effective opacity of the ejecta due to increase in bound-bound opacity from Fe group material. The increased opacity can also produce slower evolving light curves if we assume that the underlying ejecta mass distribution is the same (see Figure \ref{fig:speclc}; however, this is not expected to be dominant if the increased opacity is confined only to the outer layers of the ejecta). 

The variety in the strengths and velocities of the optical He lines provides yet another clue to the underlying ejecta composition. Due to the high ionization temperature of He, the optical He lines may or may not be excited depending on the ejecta temperature and non-thermal excitation from radioactive material \citep{Lucy1991}. \citet{Dessart2012} show that non-thermal excitation from radioactive decay of $^{56}$Ni is crucial for exciting optical He lines seen in the spectra of Type Ib/c SNe, thus suggesting that the SN\,Ib/c classification may be related to the amount of $^{56}$Ni mixing into the He layer instead of the actual He content of the ejecta (but see also \citealt{Hachinger2012}). While highly mixed radioactive material in the He layer would produce strong optical He lines, they would also redden the transient color due to the presence of Fe group material in the outer layers. However, we do not observe this trend in the Ca-rich sample -- events with redder continua exhibit weaker (or no) He lines, suggesting a reduction in the He content as the amount of Fe group elements increase in the outer ejecta. In addition, unlike the massive cores in core-collapse SNe, the low optical depth for $\gamma$-rays in these low ejecta mass explosions (see \citealt{De2018a} for a discussion) makes it difficult to hide non-thermal excitation of He.

While a continuum of Fe-group mass fraction in the outer ejecta appears to explain several of the observables, it does not explain the reason for such a continuum. This continuum could be associated with either a fundamental transition in the composition of the underlying ejecta, or be due to viewing angle effects. If the outer Fe group material is produced from He burning in the outer ejecta, one could explain these observations as a continuum of He burning efficiencies in the outer ejecta, wherein the outer ejecta become richer in Fe group elements (which cause the line blanketing) as the He burning is more complete, thus leading to weaker He lines (see \citealt{Townsley2012} for recent work on partial burning of He shells). 

Regardless, it is important to note that the change in the peak light continuum properties also appears to affect the appearance of the transient in the late-time nebular phase (its [Ca II]/[O I] ratio) when the ejecta are optically thin and viewing angle effects should be minimal. Both [Ca II] and [O I] are effective coolants of SN ejecta at late-times \citep{Fransson1989} being powered by $^{56}$Co decay in the case of normal core-collapse SNe. Together, these observations suggest that the cooling in the inner ejecta becomes progressively dominated by [O I] as the outer ejecta become poorer in He content / richer in Fe group content. Although not discussed thus far, Figure \ref{fig:ca_sequence} also shows a continuum in the depth and velocity of the prominent Ca\,II NIR triplet, wherein the absorption becomes deeper and moves to higher velocity over this sequence. Since Ca is a known He burning product \citep{Townsley2012}, this evolution in the Ca II NIR triplet may be associated with the He burning sequence discussed here.

For completeness, we can rule out dust extinction as a possible cause for this evolution -- while dust reddening can suppress red continua, it cannot explain the lack of blue side SN ejecta features seen in the red Ca-Ib/c events. In addition, the remote locations of all of these events and the lack of detectable Na I D absorption in the spectra argue against host galaxy dust extinction.

\subsection{Implications for the explosion mechanism}
\label{sec:progenitors}

\subsubsection{Models in the literature}
\citet{Shen2019} summarizes the circumstantial evidence used to rule out several progenitor channels in the literature sample based on their environments, hosts and volumetric rates. Owing to the striking similarities of the host demographics and environments of the ZTF sample and the literature events, our controlled experiment provides corroborating evidence for the preference of these transients for old environments. As in the literature sample, we find core-collapse SNe from massive stars as unlikely progenitor channels due to the prevalence of early type hosts and large offset locations with no signs of nearby star formation. On the other hand, the high inferred volumetric rates (lower limit of $\approx 15$\% of the SN\,Ia rate) rule out progenitor channels with low expected volumetric rates such as He WD - neutron star mergers, where the field rate is $\sim 100\times$ lower (\citealt{Toonen2018}; see also \citealt{Shen2019} for arguments against the viability of the high volumetric rates of these systems in globular clusters). 

We thus consider explosive burning of He shells on white dwarfs as the strongest candidates for the cause of these events, and proceed by discussing the implications of our findings on the possible explosion conditions. Assuming that Ca-rich gap transients arise from He shell explosions on white dwarfs, we aim to constrain variations in the underlying progenitor configurations and / or the burning mechanisms using the observed continuum of spectroscopic and photometric properties. 

In the He shell detonation scenario, a shell of accreted He on the WD surface (accreted from a He-rich companion) can undergo dynamical burning for large shell masses, and detonate explosively to produce a thermonuclear transient \citep{Iben1989, Bildsten2007, Shen2009, Woosley2011, Sim2012}. \citet{Shen2010} presented calculations of the optical signatures of these events termed as `.Ia' supernovae, and \citet{Perets2010} suggested that the prototype Ca-rich transient SN\,2005E was a result of such a detonation. However, the photometric evolution for the low ejecta mass ($\lesssim 0.2$\,\Msun) models presented in \citet{Shen2010} were substantially faster than SN\,2005E, leading \citet{Perets2010} to suggest that more massive shells could explain the slower light curve evolution. 

\citet{Waldman2011} carried out explosive nucleosynthesis calculations of the shell detonation scenario with a 0.2 \Msun shell on a 0.45 \Msun CO WD, and demonstrated nucleosynthesis of a large amount of intermediate mass elements together with unburned He in the ejecta. Including non-thermal excitation effects, \citet{Dessart2015a} showed that these events exhibit low-luminosity light curves, He spectroscopic signatures at peak light and [Ca II] emission in the nebular phase, and are thus consistent with Ca-rich gap transients. Yet, the relatively slow light curve evolution of most of the literature events required even more massive shells ($\gtrsim 0.2$ \Msun) than in these calculations if the underlying core is not detonated. In particular, \citet{Dessart2015a} showed that the ratio of [Ca II] to [O I] emission depends not only on their relative abundance, but also on where the $\gamma$-rays from late-time radioactive decay are being deposited, as these lines are primary coolants of the regions of the ejecta where they exist. They further showed that the ejecta continued to cool predominantly through [Ca II] emission even if [O I] was present owing to the higher efficiency of [Ca II] cooling, thus pointing out the importance of mixing of radioactive material as well as the Ca and O regions in determining [Ca II]/[O I] in the nebular phase \citep{Fransson1989}.

Modifications to this scenario involving the detonation of the underlying CO core and the conditions required thereof have also been explored in the literature, first in the context of double detonation models for Type Ia supernovae \citep{Nomoto1980, Nomoto1982a, Nomoto1982b, Woosley1986, Woosley1994, Livne1995}. These initial models invoked thick He shells ($\sim 0.1$\,\Msun) for this scenario and were largely ruled out due to the predicted red colors and strong line blanketing signatures found for these configurations (e.g. \citealt{Nugent1997}). Later studies found that thin He shells (as low as $0.01$\,\Msun) on relatively massive CO cores ($\gtrsim 0.8$ \Msun) can detonate the underlying white dwarf \citep{Bildsten2007, Fink2010, Shen2010, Shen2014b}, potentially producing luminous slow-evolving transients akin to normal and sub-luminous Type Ia SNe \citep{Kromer2010, Sim2010, Woosley2011, Polin2019a}. Specifically, the slower evolving light curves in these models may be consistent with the Ca-rich gap transients; however, since the luminosity and timescale of these light curves increase with the underlying total mass (owing to higher Ni production from higher density cores), Ca-rich gap transients are likely associated with explosions on lower mass WDs in this scenario. 

To this end, \citet{Sim2012} further extended these calculations to lowest mass CO WDs ($\approx 0.45$\,\Msun) with thick He shells ($0.2$\,\Msun) specifically to probe the parameter space for sub-luminous and fast evolving events like Ca-rich events. They find that secondary detonations are likely triggered for these shell masses (although there remain large uncertainties), and present a suite of simulations varying the extent and mechanism of the core detonation to demonstrate the corresponding effects on the nucleosynthetic signatures. They specifically note that their suite of models produce brighter transients (peak absolute magnitude $\lesssim -17$) than the prototype SN\,2005E, and thus reproducing the properties would require lower yields of radioactive material that could be possible in lower density He shells \citep{Shen2009, Woosley2011} or via significant pollution of the shell with C \citep{Kromer2010}.

\subsubsection{Ca-rich gap transients from He shell explosions}

\citet{Jacobson-Galan2019} suggested that the Ca-Ia object SN\,2016hnk was consistent with the detonation of a thin ($\approx 0.02$\,\Msun) He shell on a $\approx 0.8$\,\Msun WD. This interpretation was based on recent work by \citet{Polin2019b} showing that the ejecta in double detonation events could cool predominantly through [Ca II] lines in the nebular phase (instead of the Fe group lines as in other SNe\,Ia) for low total (WD core + He shell) masses ($\lesssim 0.9$\,\Msun), even if the Ca abundance in the ejecta is of the order of a few percent. This channel thus provides a promising scenario to explain the origins of the Ca-Ia objects, owing to their luminous and slow evolving light curves, strong line blanketing signatures with SN\,Ia-like spectra, and [Ca II] emission in the nebular phase. The observed diversity in the peak luminosities could then be associated with a range of white dwarf core masses.

Therefore, it is interesting to extend this mechanism to a continuum of He shell and CO core masses that may explain the diversity in the population of Ca-Ib/c events with double detonations. Unlike pure shell detonations that have had difficulty explaining the relatively slow evolving light curves of prototypical Ca-rich transients like SN\,2005E and SN\,2010et (see for example, \citealt{Waldman2011} and \citealt{Dessart2015a}), double detonations predict slower evolving light curves compared to pure shell detonations due to the higher ejecta mass involved \citep{Woosley2011, Sim2012}. Noting that SN\,2005E and SN\,2010et belong to the class of green Ca-Ib/c events that have faster evolving light curves, the problem with slow evolution is even worse for the red Ca-Ib/c events that exhibit slower light curves similar to the Ca-Ia events (Figure \ref{fig:lumwidth}), leading us to consider double detonations for the red Ca-Ib/c events. 

Owing to the explosive burning of He-rich material, a common spectroscopic prediction of the He shell double detonation scenario is a transient marked by strong line blanketing features of Fe group material and Ti\,II (at early times when the photopshere is in the He detonation material), in addition to the intermediate mass elements produced from the core burning \citep{Hoeflich1996,Nugent1997,Kromer2010,Woosley2011,Polin2019a}. Such signatures are also found in pure shell and double detonations on lower mass white dwarfs \citep{Dessart2015a, Sim2012}. While previous attempts to model the green Ca-Ib/c objects \citep{Dessart2015a, Sim2012} have had difficulty accounting for the relatively blue colors of these transients at peak, the red colors, line blanketing and slow evolution make the double detonation channel an attractive possibility for the red Ca-Ib/c events. The typically lower peak luminosity of the red Ca-Ib/c events would then require detonations on lower mass white dwarfs compared to the Ca-Ia events. 

However, despite having un-burned He in the ejecta \citep{Townsley2012, Sim2012, Moore2013, Polin2019a}, existing works on double detonations have not yet demonstrated whether He I lines can be reproduced as observed. On the other hand, Si\,II lines are the most conspicuous spectral features in these models as seen in the Ca-Ia objects. The red Ca-Ib/c events show weak but clear signatures of He, sometimes stronger than the Si\,II lines. We note that existing models have primarily explored this scenario with relatively higher mass white dwarfs ($\gtrsim 0.8$\,\Msun), and without including non-thermal radiation effects required to excite He lines (\citealt{Sim2012} explored lower mass WDs but without non-thermal effects). Given the continuum of strong to weak He lines observed in the red Ca-Ib/c population, it is possible that these explosions constitute a range of shell burning efficiencies, and correspondingly He content in the ejecta depending on the pressure at the base of the shell at the time of ignition \citep{Moore2013}. Ca-Ia events would then represent the extreme end of this population where the He is nearly completely burnt to iron group elements.

In the nebular phase, red Ca-Ib/c objects often show strong [O I] emission, unlike the Ca-Ia events. If both the red Ca-Ib/c events and the Ca-Ia events arise from double detonations but with different underlying white dwarf masses, this provides observational evidence that the core burning becomes less efficient as one moves towards smaller core masses. A possible explanation is if the core detonation transitions from converging shock detonations in high mass cores to edge lit detonations \citep{Nomoto1982b, Livne1990, Livne1991} at the base of the shell for lower mass white dwarfs. Owing to the lower density of the core in edge-lit detonations, \citet{Sim2012} show that these explosions produce incomplete burning of the core (thus lower Ni yields and fainter light curves) and larger amounts of O in the core, both consistent with the red Ca-Ib/c population. Alternatively, it is possible that a larger fraction of the Ni produced in the core detonation is convectively mixed into the O rich regions of the ejecta for lower mass white dwarf cores, thus producing stronger [O I] emission in the nebular phase. 

Ca-Ib/c objects with green continua pose several problems to the general double detonation picture above, especially with regards to their bluer colors at peak without line blanketing signatures. The lack of these signatures suggests a scenario where the outer ejecta are not significantly enriched with Fe group elements, while their strong He line signatures indicate low He burning efficiency. Taken together with their systematically faster evolving light curves, these signatures likely point to a low efficiency burning mechanism that ejects a small amount of material. Although pure He shell detonation scenarios also predict line blanketed spectra (due to abundance of Ti II near the photosphere; e.g. \citealt{Waldman2011, Dessart2015a}), lower He burning efficiency has been shown to be achievable if the pressure at the base of the shell is lower \citep{Perets2010, Moore2013}. If so, the higher photospheric velocities, relatively fast photometric evolution and weak [O I] emission in the nebular phase are consistent with properties of green Ca\,Ib/c events \citep{Waldman2011, Dessart2015a}. The relatively low peak luminosity of this group  (Table \ref{tab:caclass}) then translates to a lower amount of radioactive material synthesized in these shell-only explosions compared to double-detonations.

Alternatively, such low efficiency burning conditions may be achieved in pure shell deflagrations. The study of this mechanism is extremely limited at this time but \citet{Woosley2011} show that these explosions produce slowly evolving sub-luminous light curves, but with low photospheric velocities ($\approx 4000$ km s$^{-1}$). However, that study was performed with 1D simulations which are poorly suited for deflagrations and future work is required to determine if higher velocities could be achieved, making this a possible scenario for these events. We note that early excess emission has so far been clearly seen only in the green Ca-Ib/c objects SN\,2016hgs and SN\,2018lqo. If these explosions are indeed associated with He shell detonations or deflagrations, these observations suggest that the outer ejecta produced in the shell burning is enriched with short-lived isotopes (e.g. $^{48}$Cr, $^{52}$Fe and $^{56}$Ni with half-lives of $0.90\,\rm{d}$, $0.35\,\rm{d}$ and $6.07\,\rm{d}$ respectively), as has been found in previous simulations of the pure shell detonation scenario \citep{Shen2010}. Although the early bump in iPTF\,16hgs was potentially consistent with a core-collapse explosion from a giant He star (e.g. \citealt{Woosley2019}), the detection of this bump in SN\,2018lqo in an old elliptical galaxy strongly argues against a core-collapse interpretation.

\subsection{Implications for the progenitor stellar populations}

We have thus far discussed the implications of diversity in the photometric and spectroscopic properties of the Ca-rich gap transient population on the underlying explosion mechanism. Broadly, we find that the range of observed properties can be explained with scenarios involving explosive He burning on the surface of a CO WD. Specifically, the slowly evolving light curves of the Ca-Ia and red Ca-Ib/c objects are difficult to explain using scenarios involving pure shell detonations, and likely require relatively massive explosions that detonate the underlying core. On the other hand, the green Ca-Ib/c objects may be consistent with shell-only explosions with low He burning efficiency. We now extend this discussion to probe the implications of the environments of these transients on the inferred progenitors, i.e., white dwarfs accreting He from a companion. As noted in several previous works, this population is distinguished from other classes of transients by their striking preference for large offsets from their passive host galaxies in predominantly group and cluster environments. These two trends generally point to old stellar populations with long delay times, and argue against channels which have a significant rate at short delay times ($\lesssim 1$\,Gyr; \citealt{Perets2010, Meng2015}).  We note that several objects in our combined sample (iPTF\,16hgs, SN\,2016hnk and SN\,2019ofm) were found in star forming environments, suggesting that there is a small but likely non-zero contribution at short delay times as well. 

The pathways of stellar evolution to explosive burning of He shells have been explored in several previous works \citep{Nomoto1982a, Woosley1994, Bildsten2007, Shen2009, Brooks2015, Bauer2017}, with tight sdB + WD systems \citep{Geier2013, Kupfer2017} and AM-CVn binaries \citep{Nelemans2004} being possible well-observed evolutionary stages prior to the detonation. The observed formation rate of stable accreting AM-CVn systems in the Milky Way is $\sim 1.3 \times 10^{-4}$ yr$^{-1}$ \citep{Roelofs2007, Carter2013} or $\sim 2$\% of the SN\,Ia rate in Milky-way like galaxies \citep{Li2011}. \citet{Bildsten2007} use this to estimate the rate of He shell detonations to be $2-6$\% of the SN\,Ia rate in E/S0 galaxies assuming all AM-CVns detonate in a last flash. This rate is $\approx 3-7\times$ lower than our lower limit estimate for the volumetric rate of Ca-rich gap transients, and thus likely inconsistent. However, we cannot rule out the scenario where the green Ca-Ib events of the Ca-rich continuum discussed here originates in these explosions.

Similar detonation conditions can also be achieved in the case of mergers of He WDs with CO WDs (e.g. \citealt{Pakmor2013}), where a variety of detonation conditions could be achieved depending on how the He layer settles on the surface of the more massive WD \citep{Fryer2010, Dessart2015a}. The merger rate of He + CO WD binaries in the field for a Milky way-like galaxy is $\sim 3\times10^{-3}$ yr$^{-1}$ \citep{Brown2016} or $\approx 10$\% of the SN\,Ia rate, suggesting that most He + CO WD binaries end up merging to produce thermonuclear transients or stable long-lived remnants (e.g. RCrB stars; \citealt{Clayton1996, Schwab2019}). The high rate of these mergers within Milky-Way like galaxies is however in contradiction with the preference of large offset distributions (long delay times) of the Ca-rich gap transient population \citep{Shen2019}. The measured rate in the halo \citep{Brown2016} is only $\sim 10$\% of the rate within the galaxy ($\sim 1$\% of the SN\,Ia-rate) and hence inconsistent with our estimates for the Ca-rich population which reside in or outside their host halos. 

With population synthesis calculations, \citet{Meng2015} suggested that the relatively high rates and long delay times of the Ca-rich gap transient population could be consistent with double WD binaries with a CO WD $\lesssim 0.6$\,\Msun and He WD $\lesssim 0.25$\,\Msun, specifically highlighting the low mass as a key aspect of extending the delay time. This is consistent with our suggestion for low mass white dwarfs as the progenitor population, but the observed halo population of He-CO WD mergers remain too rare to explain the high rate \citep{Brown2016}. The measured rates are high enough such that there is one Ca-rich gap transient for every three SNe\,Ia in early type galaxies, suggesting that the progenitors must be nearly as common as the progenitors of SNe\,Ia. 

Together, we find that the observed merger rates of WDs in He-rich accreting systems is too low in the field in the outskirts of galaxies to explain the high rates of Ca-rich gap transients. However, the interaction rate may be enhanced in dense stellar systems like globular clusters (see e.g. \citealt{Pfahl2009}). Although such stellar systems are not detected exactly at the positions of known Ca-rich gap transients \citep{Lyman2014, Lyman2016b, Lunnan2017}, \citet{Shen2019} suggest these transients could be produced from being kicked out due to dynamical interactions inside dense stellar systems in the outskirts. \citet{Yuan2013} and \citet{Shen2019} argue that the locations and host environments are consistent with the globular cluster scenario; however, we continue to find that the number of Ca-rich gap transients at small offsets is too small compared to the globular cluster offset distribution in \citet{Shen2019}. Additionally, the preference for rich cluster environments is even more extreme compared to globular clusters and remains to be explained. 

Based on our deep late-time imaging observations of the transient locations, we find in 4 out of 8 cases evidence of faint and diffuse nearby or underlying stellar associations. We estimate the absolute magnitudes of these sources to be $\approx -9\,\rm{mag}$ to $-11\,\rm{mag}$ at the redshift of the transient. A similar diffuse system was also detected in late-time imaging of PTF\,11kmb \citep{Lunnan2017}. While we caution that the chance coincidence probability of background galaxies at this depth is high, we speculate about the implications of these being associated with the transient itself. Specifically, a population of such faint and ultra-diffuse galaxies have been detected in low surface brightness imaging surveys \citep{Abraham2014} in nearby clusters. These systems appear to exhibit an overabundance (by nearly an order of magnitude) of globular clusters when compared to the expected numbers for their luminosity \citep{vanDokkum2017, vanDokkum2016, vanDokkum2018}. Although work remains to be done to constrain the formation rate of double WD binaries in such dense stellar systems, it is an intriguing possibility that the offset locations of Ca-rich gap transients may be due to an overabundance of globular clusters in ultra-diffuse galaxies within nearby galaxy groups and clusters. While \citet{Shen2019} point out that a relatively large fraction of WDs in globular clusters could be required to explain the estimated rate in \citet{Frohmaier2018}, our lower rate estimates may provide a more viable solution to the problem. The locations of these transients in early type galaxies in predominantly rich cluster environments are indeed suggestive given that up to 30 -- 70\% of stellar light may be in intra-cluster light in these environments \citep{DSouza2014, Perets2014}.

\section{Summary}
\label{sec:summary}

\begin{figure*}[!ht]
    \centering
    \includegraphics[width=0.8\textwidth]{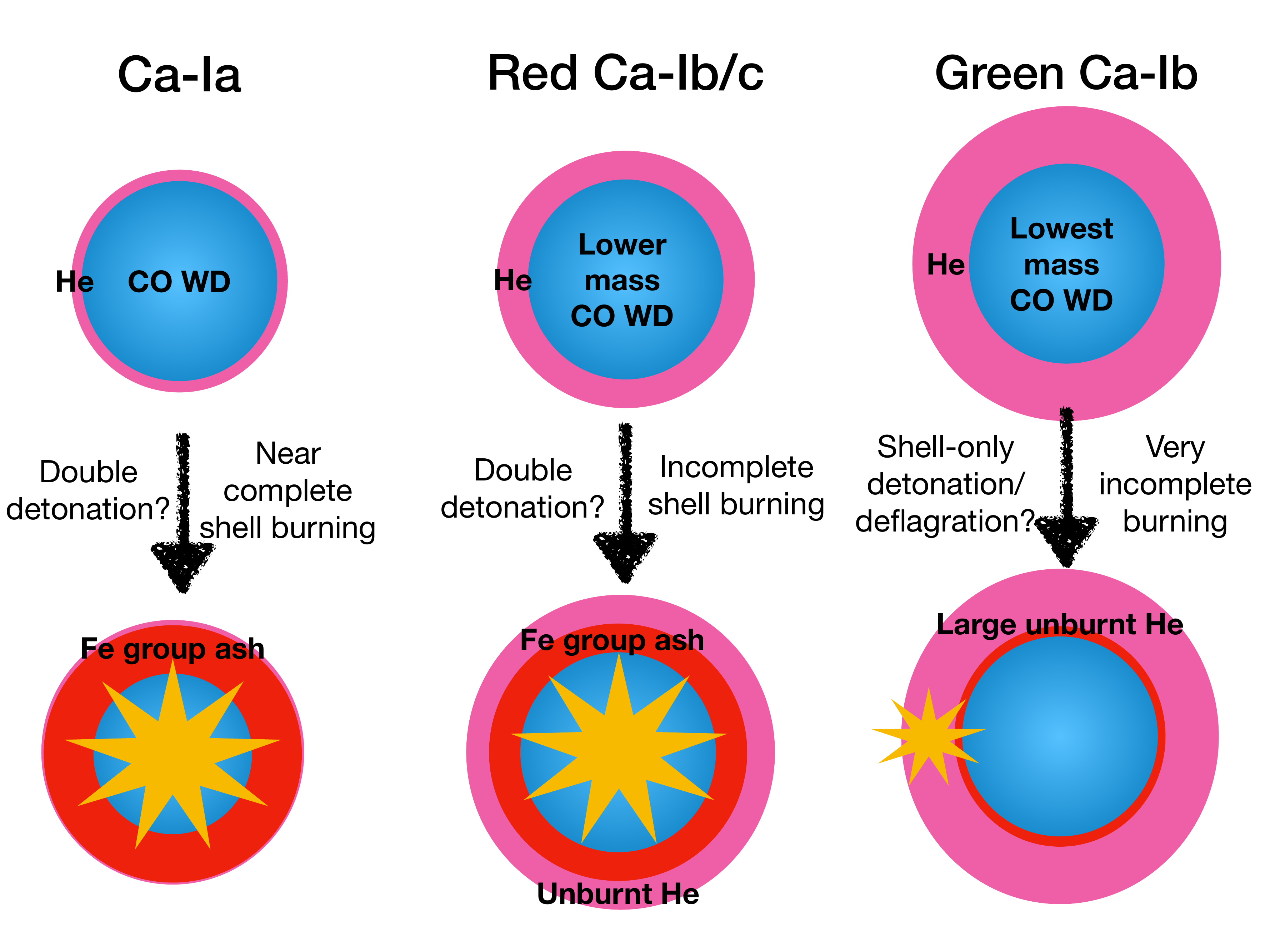}
    \caption{A cartoon schematic of a possible progenitor and explosion mechanism sequence in WD and He shell masses that may explain the diversity of properties of the population of Ca-rich gap transients. Double-detonation with thin He shells and relatively massive WDs produce Ca-Ia objects, while double detonations on lower mass WDs produce red Ca-Ib/c objects. Pure shell detonations or deflagrations could explain the incomplete burning and bluer spectra of green Ca-Ib objects.}
    \label{fig:progenitor}
\end{figure*}

In this paper, we have presented the design and completeness of the Census of the Local Universe experiment of the Zwicky Transient Facility, aimed at assembling a spectroscopically complete sample of transients in the local universe within 200\,Mpc. With a total sample of 754 spectroscopically classified SNe, we present the first systematic search for Ca-rich gap transients. Using simple and systematic selection criteria, we identified a sample of 22 low luminosity hydrogen poor supernovae as candidate Ca-rich gap transients from the first 16 months of operations, which were followed up with nebular phase spectroscopy. We report the detection of eight Ca-rich gap transients in this sample, which we combine with the literature sample of ten events that pass the same selection criteria. We perform the first systematic study of the spectroscopic and photometric properties of this sample and identify several trends and correlations that have implications for the underlying explosions and progenitor systems. Summarizing these correlations, we find:
\begin{itemize}
    \item Ca-rich gap transients broadly bifurcate into two classes based on their peak light spectroscopic similarity to SNe\,Ia (with strong Si\,II lines; Ca-Ia objects) and SNe\,Ib/c (without strong Si\,II lines; Ca-Ib/c objects). Ca-Ia objects do not show any He lines, red Ca-Ib/c objects show a continuum from weak to strong He lines (velocity $\approx 7000$ km\,s$^{-1}$) and green Ca-Ib/c objects always exhibit strong He lines at higher velocities of $\approx 10000$ km\,s$^{-1}$. 
    \item Ca-Ib/c objects further show a bimodal population in peak light spectral properties based on their spectroscopic behavior bluewards of $\approx 5500$\,\AA, wherein some events exhibit red continua and strong line blanketing of blue flux (Ca-Ib/c-red) while other events exhibit flat continua with clear P-Cygni absorption features of metals at short wavelengths (Ca-Ib/c-green). Ca-Ia objects always show line blanketed spectra at peak.
    \item Photometrically, red Ca-Ib/c events with line blanketing signatures exhibit redder colors at peak and slower evolving light curves ($\Delta m_7 \approx 0.3$\,mag) compared to green Ca-Ib/c events ($\Delta m_7 \approx 0.5$\,mag), where $\Delta m_7$ is the decay of the $r$-band light curve in 7\,days from peak light. Ca-Ia events also exhibit relatively slow photometric evolution compared to green Ca-Ib/c objects ($\Delta m_7 \approx 0.4$\,mag).
    \item Ca-Ia objects do not show [O I] in the nebular phase, red Ca-Ib/c objects show stronger [O I] lines in the nebular phase relative to [Ca II] and green Ca-Ib/c objects shower weaker [O I] lines (higher [Ca II]/[O I] flux ratio) relative to red Ca-Ib/c objects.
    \item We find tentative evidence of a spectroscopic continuum of properties going from Ca-Ia to Ca-Ic to Ca-Ib objects, where Si\,II lines get weaker, He\,I lines get stronger and line blanketing becomes less significant in the spectrum in moving along that sequence. Photometrically, the peak light colors get bluer while the light curve evolution becomes faster along this sequence.
\end{itemize}
We find that these spectroscopic properties and the corresponding trends can be broadly explained in scenarios involving the explosive burning of He shells on low mass white dwarfs ($\lesssim 0.8$\,\Msun). The slowly evolving light curves and line blanketed spectra of the Ca-Ia and red Ca-Ib/c events are consistent with scenarios involving the double detonation of He shells on low mass white dwarfs where the efficiency of He burning in the outer ejecta is high. On the other hand, the strong He lines, higher velocities and faster evolving light curves of the green Ca-Ib/c events suggest lower ejecta mass explosions likely involving He shell-only detonations with low He burning efficiency or even deflagrations. While the theoretical modeling of He shell explosions on low mass white dwarfs remains limited at this time, this data set together with the inferred correlations from the first systematic search for these explosions will be a useful benchmark for future modeling efforts.

The host environments of the sample are dominated by remote locations in the far outskirts of galaxies similar to what was found in previous studies, while the offset distribution are skewed towards large host offsets compared to globular clusters. The apparent host galaxies themselves are always found in rich group and cluster environments, and are likely an important clue to the progenitor channels. Using the systematic selection strategy of the experiment, together with the measured incompleteness of galaxy catalogs from the ZTF Bright Transient Survey, we infer the volumetric rates of these events to be at least $\approx 15\pm5$\% of the local SN\,Ia rate. While the explosion scenarios require white dwarfs with He-rich donors, the observed rates of these systems are apparently too low in the field to explain their high volumetric rates. However, we note that the observed environments of these events in the outskirts of early type galaxies in clusters are strikingly different from the Galactic disk. We speculate that the rich host environments, offset locations and faint nearby diffuse stellar associations in some events may point to a contribution from dynamical interactions within ultra-diffuse stellar associations in galaxy clusters, that are found to be extremely rich in globular clusters \citep{vanDokkum2016,vanDokkum2017}.

Our findings have broader implications for the population of thermonuclear supernovae in the local universe. While the double detonation scenario has been long proposed as a viable explosion triggering mechanism for Type Ia SNe, the strong line blanketing signatures produced by Fe group elements in the burning of the He shell have remained inconsistent for the broad population of normal Type Ia SNe \citep{Hoeflich1996, Nugent1997, Kromer2010, Woosley2011, Polin2019a}. However, recent discoveries of peculiar SNe such as SN\,2016jhr \citep{Jiang2017} and SN\,2018byg \citep{De2019a} show that thin and thick He shell double detonations may indeed be realized in nature but are intrinsically rare in the population. These transients were much more luminous at peak ($M \lesssim -18$) compared to the sample discussed here, consistent with the larger core masses ($\gtrsim 0.75$\,\Msun) inferred for those objects. The Ca-Ia objects are then lower luminosity analogs of this family with relatively lower mass white dwarfs, while red Ca-Ib/c objects could be manifestations of He shell double detonations on the lowest mass white dwarfs. While it remains unclear if double detonations are realized for the lowest mass white dwarfs, the green Ca-Ib/c objects could potentially be manifestations of `failed double-detonations' of the lowest mass white dwarfs in the universe that do not burn much of the progenitor He shell\footnote{See also \citealt{Kasliwal2010, Drout2011, Inserra2015} for promising candidates for He shell-only detonation transients.}. We summarize this possible progenitor sequence within Ca-rich gap transients in Figure \ref{fig:progenitor}.

Given the fast evolving light curves of Ca-rich gap transients, high cadence and wide-field time domain surveys will continue to be vital for unveiling large samples of these elusive events. With the combination of the flux-limited Bright Transient Survey (to estimate the redshift incompleteness of nearby galaxy catalogs; \citealt{Fremling2019}) and the volume-limited Census of the Local Universe experiment described in this work, we have demonstrated that focused experiments that spectroscopically classify a small fraction ($\lesssim 10$\%) of the total transient yield of a wide-field, all-sky survey like ZTF can yield statistically meaningful samples of transients that shed vital light on the underlying explosions and progenitor populations. As the current effort has been focused on a targeted search for these events in current (and incomplete) catalogs of nearby galaxies, the advent of heavily multiplexed spectrographs (e.g. \citealt{DESI2016, Kollmeier2017}) aiming to complete galaxy catalogs in the local universe will be vital for dedicated experiments to complete our census of transients in the local universe.

\section*{Acknowledgements}
We thank H. Perets, K. Shen, E. S. Phinney, J. Fuller, D. Kasen, B. Margalit, E. Ramirez-Ruiz, A. Filippenko, R. Fisher and S. Schulze for valuable discussions during this work. We thank M. Coughlin, S. Anand, A. Sagues Carracedo, L. Rauch and U. Feindt for several discussions on the use of \texttt{simsurvey}. 

This work was supported by the GROWTH (Global Relay of Observatories Watching Transients Happen) project funded by the National Science Foundation under PIRE Grant No 1545949. GROWTH is a collaborative project among California Institute of Technology (USA), University of Maryland College Park (USA), University of Wisconsin Milwaukee (USA), Texas Tech University (USA), San Diego State University (USA), University of Washington (USA), Los Alamos National Laboratory (USA), Tokyo Institute of Technology (Japan), National Central University (Taiwan), Indian Institute of Astrophysics (India), Indian Institute of Technology Bombay (India), Weizmann Institute of Science (Israel), The Oskar Klein Centre at Stockholm University (Sweden), Humboldt University (Germany), Liverpool John Moores University (UK) and University of Sydney (Australia). This research benefited from interactions at several ZTF Theory Network meetings, funded by the Gordon and Betty Moore Foundation through Grant GBMF5076. 

Based on observations obtained with the Samuel Oschin Telescope 48-inch and the 60-inch Telescope at the Palomar Observatory as part of the Zwicky Transient Facility project. ZTF is supported by the National Science Foundation under Grant No. AST-1440341 and a collaboration including Caltech, IPAC, the Weizmann Institute for Science, the Oskar Klein Center at Stockholm University, the University of Maryland, the University of Washington, Deutsches Elektronen-Synchrotron and Humboldt University, Los Alamos National Laboratories, the TANGO Consortium of Taiwan, the University of Wisconsin at Milwaukee, and Lawrence Berkeley National Laboratories. Operations are conducted by COO, IPAC, and UW. SED Machine is based upon work supported by the National Science Foundation under Grant No. 1106171. The ZTF forced-photometry service was funded under the Heising-Simons Foundation grant 12540303 (PI: Graham).  Some of the data presented herein were obtained at the W.M. Keck Observatory, which is operated as a scientific partnership among the California Institute of Technology, the University of California and the National Aeronautics and Space Administration. The Observatory was made possible by the generous financial support of the W.M. Keck Foundation. The authors wish to recognize and acknowledge the very significant cultural role and reverence that the summit of Mauna Kea has always had within the indigenous Hawaiian community. We are most fortunate to have the opportunity to conduct observations from this mountain. Based on observations made with the Nordic Optical Telescope (operated by the Nordic Optical Telescope Scientific Association at the Observatorio del Roque de los Muchachos, La Palma, Spain, of the Instituto de Astrofisica de Canarias.)

AGY’s research is supported by the EU via ERC grant No. 725161, the ISF GW excellence center, an IMOS space infrastructure grant and BSF/Transformative and GIF grants, as well as The Benoziyo Endowment Fund for the Advancement of Science, the Deloro Institute for Advanced Research in Space and Optics, The Veronika A. Rabl Physics Discretionary Fund, Paul and Tina Gardner, Yeda-Sela and the WIS-CIT joint research grant;  AGY is the recipient of the Helen and Martin Kimmel Award for Innovative Investigation. A.Y.Q.H. is supported by a National Science Foundation Graduate Research Fellowship under Grant No.\,DGE-1144469 and by the GROWTH project funded by the National Science Foundation under PIRE Grant No.\,1545949. R.L. is supported by a Marie Sk\l{}odowska-Curie Individual Fellowship within the Horizon 2020 European Union (EU) Framework Programme for Research and Innovation (H2020-MSCA-IF-2017-794467). Foscgui is a graphic user interface aimed at extracting SN spectroscopy and photometry obtained with FOSC-like instruments. It was developed by E. Cappellaro. A package description can be found at \url{http://sngroup.oapd.inaf.it/foscgui.html}.

\software{\texttt{astropy} \citep{Astropy2013}, \texttt{matplotlib} \citep{Hunter2007}, \texttt{scipy} \citep{Virtanen2019}, \texttt{pandas} \citep{McKinney2010}, \texttt{SExtractor} \citep{Bertin1996}, \texttt{scamp} \citep{Bertin2006}, \texttt{SWarp} \citep{Bertin2002}, \texttt{PSFEx} \citep{Bertin2011}, \texttt{pysedm} \citep{Rigault2019}, \texttt{pyraf-dbsp} \citep{Bellm2016}, \texttt{lpipe} \citep{lpipe}, \texttt{simsurvey} \citep{Feindt2019}}

\facilities{PO:1.2m (ZTF), PO:1.5m (SEDM), Hale (DBSP, WASP), NOT: ALFOSC, THO: ALPY200, Keck:I (LRIS).}

\clearpage

\bibliographystyle{aasjournal}
\bibliography{ZTF_CaRich}

\begin{thebibliography}{}
\expandafter\ifx\csname natexlab\endcsname\relax\def\natexlab#1{#1}\fi
\providecommand{\url}[1]{\href{#1}{#1}}

\bibitem[{{Abolfathi} {et~al.}(2017){Abolfathi}, {Aguado}, {Aguilar}, {Allende
  Prieto}, {Almeida}, {Tasnim Ananna}, {Anders}, {Anderson}, {Andrews},
  {Anguiano}, \& et~al.}]{Abolfathi2017}
{Abolfathi}, B., {Aguado}, D.~S., {Aguilar}, G., {et~al.} 2017, ArXiv e-prints,
  arXiv:1707.09322

\bibitem[{{Abraham} \& {van Dokkum}(2014)}]{Abraham2014}
{Abraham}, R.~G., \& {van Dokkum}, P.~G. 2014, \pasp, 126, 55

\bibitem[{{Arnett}(1982)}]{Arnett1982}
{Arnett}, W.~D. 1982, ApJ, 253, 785

\bibitem[{{Arnett} {et~al.}(1985){Arnett}, {Branch}, \& {Wheeler}}]{Arnett1985}
{Arnett}, W.~D., {Branch}, D., \& {Wheeler}, J.~C. 1985, \nat, 314, 337

\bibitem[{{Astropy Collaboration} {et~al.}(2013){Astropy Collaboration},
  {Robitaille}, {Tollerud}, {Greenfield}, {Droettboom}, {Bray}, {Aldcroft},
  {Davis}, {Ginsburg}, {Price-Whelan}, {Kerzendorf}, {Conley}, {Crighton},
  {Barbary}, {Muna}, {Ferguson}, {Grollier}, {Parikh}, {Nair}, {Unther},
  {Deil}, {Woillez}, {Conseil}, {Kramer}, {Turner}, {Singer}, {Fox}, {Weaver},
  {Zabalza}, {Edwards}, {Azalee Bostroem}, {Burke}, {Casey}, {Crawford},
  {Dencheva}, {Ely}, {Jenness}, {Labrie}, {Lim}, {Pierfederici}, {Pontzen},
  {Ptak}, {Refsdal}, {Servillat}, \& {Streicher}}]{Astropy2013}
{Astropy Collaboration}, {Robitaille}, T.~P., {Tollerud}, E.~J., {et~al.} 2013,
  \aap, 558, A33

\bibitem[{{Barbary} {et~al.}(2016){Barbary}, {Barclay}, {Biswas}, {Craig},
  {Feindt}, {Friesen}, {Goldstein}, {Jha}, {Rodney}, {Sofiatti}, {Thomas}, \&
  {Wood-Vasey}}]{Barbary2016}
{Barbary}, K., {Barclay}, T., {Biswas}, R., {et~al.} 2016, {SNCosmo: Python
  library for supernova cosmology}, , , ascl:1611.017

\bibitem[{{Bauer} {et~al.}(2017){Bauer}, {Schwab}, \& {Bildsten}}]{Bauer2017}
{Bauer}, E.~B., {Schwab}, J., \& {Bildsten}, L. 2017, \apj, 845, 97

\bibitem[{{Bell} {et~al.}(2003){Bell}, {McIntosh}, {Katz}, \&
  {Weinberg}}]{Bell2003}
{Bell}, E.~F., {McIntosh}, D.~H., {Katz}, N., \& {Weinberg}, M.~D. 2003, \apjs,
  149, 289

\bibitem[{{Bellm} \& {Sesar}(2016)}]{Bellm2016}
{Bellm}, E.~C., \& {Sesar}, B. 2016, {pyraf-dbsp: Reduction pipeline for the
  Palomar Double Beam Spectrograph}, , , ascl:1602.002

\bibitem[{{Bellm} {et~al.}(2019{\natexlab{a}}){Bellm}, {Kulkarni}, {Graham},
  {Dekany}, {Smith}, {Riddle}, {Masci}, {Helou}, {Prince}, {Adams},
  {Barbarino}, {Barlow}, {Bauer}, {Beck}, {Belicki}, {Biswas}, {Blagorodnova},
  {Bodewits}, {Bolin}, {Brinnel}, {Brooke}, {Bue}, {Bulla}, {Burruss}, {Cenko},
  {Chang}, {Connolly}, {Coughlin}, {Cromer}, {Cunningham}, {De}, {Delacroix},
  {Desai}, {Duev}, {Eadie}, {Farnham}, {Feeney}, {Feindt}, {Flynn},
  {Franckowiak}, {Frederick}, {Fremling}, {Gal-Yam}, {Gezari}, {Giomi},
  {Goldstein}, {Golkhou}, {Goobar}, {Groom}, {Hacopians}, {Hale}, {Henning},
  {Ho}, {Hover}, {Howell}, {Hung}, {Huppenkothen}, {Imel}, {Ip}, {Ivezi{\'c}},
  {Jackson}, {Jones}, {Juric}, {Kasliwal}, {Kaspi}, {Kaye}, {Kelley},
  {Kowalski}, {Kramer}, {Kupfer}, {Landry}, {Laher}, {Lee}, {Lin}, {Lin},
  {Lunnan}, {Giomi}, {Mahabal}, {Mao}, {Miller}, {Monkewitz}, {Murphy},
  {Ngeow}, {Nordin}, {Nugent}, {Ofek}, {Patterson}, {Penprase}, {Porter},
  {Rauch}, {Rebbapragada}, {Reiley}, {Rigault}, {Rodriguez}, {van Roestel},
  {Rusholme}, {van Santen}, {Schulze}, {Shupe}, {Singer}, {Soumagnac}, {Stein},
  {Surace}, {Sollerman}, {Szkody}, {Taddia}, {Terek}, {Van Sistine}, {van
  Velzen}, {Vestrand}, {Walters}, {Ward}, {Ye}, {Yu}, {Yan}, \&
  {Zolkower}}]{Bellm2019a}
{Bellm}, E.~C., {Kulkarni}, S.~R., {Graham}, M.~J., {et~al.}
  2019{\natexlab{a}}, \pasp, 131, 018002

\bibitem[{{Bellm} {et~al.}(2019{\natexlab{b}}){Bellm}, {Kulkarni}, {Barlow},
  {Feindt}, {Graham}, {Goobar}, {Kupfer}, {Ngeow}, {Nugent}, {Ofek}, {Prince},
  {Riddle}, {Walters}, \& {Ye}}]{Bellm2019b}
{Bellm}, E.~C., {Kulkarni}, S.~R., {Barlow}, T., {et~al.} 2019{\natexlab{b}},
  \pasp, 131, 068003

\bibitem[{{Bertin}(2006)}]{Bertin2006}
{Bertin}, E. 2006, in Astronomical Society of the Pacific Conference Series,
  Vol. 351, Astronomical Data Analysis Software and Systems XV, ed.
  C.~{Gabriel}, C.~{Arviset}, D.~{Ponz}, \& S.~{Enrique}, 112

\bibitem[{{Bertin}(2011)}]{Bertin2011}
{Bertin}, E. 2011, in Astronomical Society of the Pacific Conference Series,
  Vol. 442, Astronomical Data Analysis Software and Systems XX, ed. I.~N.
  {Evans}, A.~{Accomazzi}, D.~J. {Mink}, \& A.~H. {Rots}, 435

\bibitem[{{Bertin} \& {Arnouts}(1996)}]{Bertin1996}
{Bertin}, E., \& {Arnouts}, S. 1996, \aaps, 117, 393

\bibitem[{{Bertin} {et~al.}(2002){Bertin}, {Mellier}, {Radovich}, {Missonnier},
  {Didelon}, \& {Morin}}]{Bertin2002}
{Bertin}, E., {Mellier}, Y., {Radovich}, M., {et~al.} 2002, in Astronomical
  Society of the Pacific Conference Series, Vol. 281, Astronomical Data
  Analysis Software and Systems XI, ed. D.~A. {Bohlender}, D.~{Durand}, \&
  T.~H. {Handley}, 228

\bibitem[{{Bildsten} {et~al.}(2007){Bildsten}, {Shen}, {Weinberg}, \&
  {Nelemans}}]{Bildsten2007}
{Bildsten}, L., {Shen}, K.~J., {Weinberg}, N.~N., \& {Nelemans}, G. 2007, ApJL,
  662, L95

\bibitem[{{Blagorodnova} {et~al.}(2018){Blagorodnova}, {Neill}, {Walters},
  {Kulkarni}, {Fremling}, {Ben-Ami}, {Dekany}, {Fucik}, {Konidaris}, {Nash},
  {Ngeow}, {Ofek}, {O Sullivan}, {Quimby}, {Ritter}, \&
  {Vyhmeister}}]{Blagorodnova2018}
{Blagorodnova}, N., {Neill}, J.~D., {Walters}, R., {et~al.} 2018, \pasp, 130,
  035003

\bibitem[{{Blondin} \& {Tonry}(2007)}]{Blondin2007}
{Blondin}, S., \& {Tonry}, J.~L. 2007, \apj, 666, 1024

\bibitem[{{Bloom} {et~al.}(2002){Bloom}, {Kulkarni}, \&
  {Djorgovski}}]{Bloom2002}
{Bloom}, J.~S., {Kulkarni}, S.~R., \& {Djorgovski}, S.~G. 2002, \aj, 123, 1111

\bibitem[{{Brooks} {et~al.}(2015){Brooks}, {Bildsten}, {Marchant}, \&
  {Paxton}}]{Brooks2015}
{Brooks}, J., {Bildsten}, L., {Marchant}, P., \& {Paxton}, B. 2015, \apj, 807,
  74

\bibitem[{{Brown} {et~al.}(2016){Brown}, {Kilic}, {Kenyon}, \&
  {Gianninas}}]{Brown2016}
{Brown}, W.~R., {Kilic}, M., {Kenyon}, S.~J., \& {Gianninas}, A. 2016, \apj,
  824, 46

\bibitem[{{Burke} {et~al.}(2019){Burke}, {Arcavi}, {Hiramatsu}, {Howell},
  {McCully}, \& {Valenti}}]{Burke2019}
{Burke}, J., {Arcavi}, I., {Hiramatsu}, D., {et~al.} 2019, Transient Name
  Server Classification Report, 2019-1232, 1

\bibitem[{{Cano}(2013)}]{Cano2013}
{Cano}, Z. 2013, MNRAS, 434, 1098

\bibitem[{{Cao} {et~al.}(2013){Cao}, {Kasliwal}, {Arcavi}, {Horesh}, {Hancock},
  {Valenti}, {Cenko}, {Kulkarni}, {Gal-Yam}, {Gorbikov}, {Ofek}, {Sand},
  {Yaron}, {Graham}, {Silverman}, {Wheeler}, {Marion}, {Walker}, {Mazzali},
  {Howell}, {Li}, {Kong}, {Bloom}, {Nugent}, {Surace}, {Masci}, {Carpenter},
  {Degenaar}, \& {Gelino}}]{Cao2013}
{Cao}, Y., {Kasliwal}, M.~M., {Arcavi}, I., {et~al.} 2013, ApJL, 775, L7

\bibitem[{{Cardelli} {et~al.}(1989){Cardelli}, {Clayton}, \&
  {Mathis}}]{Cardelli1989}
{Cardelli}, J.~A., {Clayton}, G.~C., \& {Mathis}, J.~S. 1989, ApJ, 345, 245

\bibitem[{{Carter} {et~al.}(2013){Carter}, {Marsh}, {Steeghs}, {Groot},
  {Nelemans}, {Levitan}, {Rau}, {Copperwheat}, {Kupfer}, \&
  {Roelofs}}]{Carter2013}
{Carter}, P.~J., {Marsh}, T.~R., {Steeghs}, D., {et~al.} 2013, \mnras, 429,
  2143

\bibitem[{{Chambers} {et~al.}(2016){Chambers}, {Magnier}, {Metcalfe},
  {Flewelling}, {Huber}, {Waters}, {Denneau}, {Draper}, {Farrow}, {Finkbeiner},
  {Holmberg}, {Koppenhoefer}, {Price}, {Rest}, {Saglia}, {Schlafly}, {Smartt},
  {Sweeney}, {Wainscoat}, {Burgett}, {Chastel}, {Grav}, {Heasley}, {Hodapp},
  {Jedicke}, {Kaiser}, {Kudritzki}, {Luppino}, {Lupton}, {Monet}, {Morgan},
  {Onaka}, {Shiao}, {Stubbs}, {Tonry}, {White}, {Ba{\~n}ados}, {Bell},
  {Bender}, {Bernard}, {Boegner}, {Boffi}, {Botticella}, {Calamida},
  {Casertano}, {Chen}, {Chen}, {Cole}, {Deacon}, {Frenk}, {Fitzsimmons},
  {Gezari}, {Gibbs}, {Goessl}, {Goggia}, {Gourgue}, {Goldman}, {Grant},
  {Grebel}, {Hambly}, {Hasinger}, {Heavens}, {Heckman}, {Henderson}, {Henning},
  {Holman}, {Hopp}, {Ip}, {Isani}, {Jackson}, {Keyes}, {Koekemoer}, {Kotak},
  {Le}, {Liska}, {Long}, {Lucey}, {Liu}, {Martin}, {Masci}, {McLean}, {Mindel},
  {Misra}, {Morganson}, {Murphy}, {Obaika}, {Narayan}, {Nieto-Santisteban},
  {Norberg}, {Peacock}, {Pier}, {Postman}, {Primak}, {Rae}, {Rai}, {Riess},
  {Riffeser}, {Rix}, {R{\"o}ser}, {Russel}, {Rutz}, {Schilbach}, {Schultz},
  {Scolnic}, {Strolger}, {Szalay}, {Seitz}, {Small}, {Smith}, {Soderblom},
  {Taylor}, {Thomson}, {Taylor}, {Thakar}, {Thiel}, {Thilker}, {Unger},
  {Urata}, {Valenti}, {Wagner}, {Walder}, {Walter}, {Watters}, {Werner},
  {Wood-Vasey}, \& {Wyse}}]{Chambers2016}
{Chambers}, K.~C., {Magnier}, E.~A., {Metcalfe}, N., {et~al.} 2016, arXiv
  e-prints, arXiv:1612.05560

\bibitem[{{Chambers} {et~al.}(2018){Chambers}, {Huber}, {Flewelling},
  {Magnier}, {Schultz}, {Lowe}, {Bulger}, {Smartt}, {Smith}, {Tonry}, {Waters},
  {Wright}, \& {Young}}]{SN2018kjy}
{Chambers}, K.~C., {Huber}, M.~E., {Flewelling}, H., {et~al.} 2018, Transient
  Name Server Discovery Report, 2018-2068, 1

\bibitem[{{Chen} {et~al.}(2020){Chen}, {Dong}, {Stritzinger}, {Holmbo},
  {Strader}, {Kochanek}, {Peng}, {Benetti}, {Bersier}, {Brownsberger},
  {Buckley}, {Gromadzki}, {Moran}, {Pastorello}, {Aydi}, {Bose}, {Connor},
  {Boutsia}, {Mille}, {Elias-Rosa}, {French}, {Holoien}, {Mattila}, {Shappee},
  {Stark}, \& {Swihart}}]{Chen2020}
{Chen}, P., {Dong}, S., {Stritzinger}, M.~D., {et~al.} 2020, \apjl, 889, L6

\bibitem[{{Clayton}(1996)}]{Clayton1996}
{Clayton}, G.~C. 1996, \pasp, 108, 225

\bibitem[{{Cook} {et~al.}(2019){Cook}, {Kasliwal}, {Van Sistine}, {Kaplan},
  {Sutter}, {Kupfer}, {Shupe}, {Laher}, {Masci}, {Dale}, {Sesar}, {Brady},
  {Yan}, {Ofek}, {Reitze}, \& {Kulkarni}}]{Cook2019}
{Cook}, D.~O., {Kasliwal}, M.~M., {Van Sistine}, A., {et~al.} 2019, \apj, 880,
  7

\bibitem[{{Crts}(2018)}]{SN2018ckd}
{Crts}, N.~M.~W. 2018, Transient Name Server Discovery Report, 2018-804, 1

\bibitem[{{D{\'a}lya} {et~al.}(2018){D{\'a}lya}, {Galg{\'o}czi}, {Dobos},
  {Frei}, {Heng}, {Macas}, {Messenger}, {Raffai}, \& {de Souza}}]{Dalya2018}
{D{\'a}lya}, G., {Galg{\'o}czi}, G., {Dobos}, L., {et~al.} 2018, \mnras, 479,
  2374

\bibitem[{{De}(2019{\natexlab{a}})}]{SN2018lqo}
{De}, K. 2019{\natexlab{a}}, Transient Name Server Discovery Report, 2019-2208,
  1

\bibitem[{{De}(2019{\natexlab{b}})}]{SN2018lqu}
---. 2019{\natexlab{b}}, Transient Name Server Discovery Report, 2019-2307, 1

\bibitem[{{De} {et~al.}(2019{\natexlab{a}}){De}, {Tzanidakis}, {Kasliwal},
  {Fremling}, \& {Kulkarni}}]{De2019c}
{De}, K., {Tzanidakis}, A., {Kasliwal}, M.~M., {Fremling}, C., \& {Kulkarni},
  S.~R. 2019{\natexlab{a}}, The Astronomer's Telegram, 13262, 1

\bibitem[{{De} {et~al.}(2018{\natexlab{a}}){De}, {Kasliwal}, {Cantwell}, {Cao},
  {Cenko}, {Gal-Yam}, {Johansson}, {Kong}, {Kulkarni}, {Lunnan}, {Masci},
  {Matuszewski}, {Mooley}, {Neill}, {Nugent}, {Ofek}, {Perrott},
  {Rebbapragada}, {Rubin}, {O' Sullivan}, \& {Yaron}}]{De2018b}
{De}, K., {Kasliwal}, M.~M., {Cantwell}, T., {et~al.} 2018{\natexlab{a}}, \apj,
  866, 72

\bibitem[{{De} {et~al.}(2018{\natexlab{b}}){De}, {Kasliwal}, {Ofek}, {Moriya},
  {Burke}, {Cao}, {Cenko}, {Doran}, {Duggan}, {Fender}, {Fransson}, {Gal-Yam},
  {Horesh}, {Kulkarni}, {Laher}, {Lunnan}, {Manulis}, {Masci}, {Mazzali},
  {Nugent}, {Perley}, {Petrushevska}, {Piro}, {Rumsey}, {Sollerman},
  {Sullivan}, \& {Taddia}}]{De2018a}
{De}, K., {Kasliwal}, M.~M., {Ofek}, E.~O., {et~al.} 2018{\natexlab{b}},
  Science, 362, 201

\bibitem[{{De} {et~al.}(2019{\natexlab{b}}){De}, {Kasliwal}, {Polin}, {Nugent},
  {Bildsten}, {Adams}, {Bellm}, {Blagorodnova}, {Burdge}, {Cannella}, {Cenko},
  {Dekany}, {Feeney}, {Hale}, {Fremling}, {Graham}, {Ho}, {Jencson},
  {Kulkarni}, {Laher}, {Masci}, {Miller}, {Patterson}, {Rebbapragada},
  {Riddle}, {Shupe}, \& {Smith}}]{De2019a}
{De}, K., {Kasliwal}, M.~M., {Polin}, A., {et~al.} 2019{\natexlab{b}}, \apjl,
  873, L18

\bibitem[{{De} {et~al.}(2020){De}, {Hankins}, {Kasliwal}, {Moore}, {Ofek},
  {Adams}, {Ashley}, {Babul}, {Bagdasaryan}, {Burdge}, {Burnham}, {Dekany},
  {Declacroix}, {Galla}, {Greffe}, {Hale}, {Jencson}, {Lau}, {Mahabal},
  {McKenna}, {Sharma}, {Shopbell}, {Smith}, {Soon}, {Sokoloski}, {Soria}, \&
  {Travouillon}}]{De2019b}
{De}, K., {Hankins}, M.~J., {Kasliwal}, M.~M., {et~al.} 2020, \pasp, 132,
  025001

\bibitem[{{Dekany} {et~al.}(2016){Dekany}, {Smith}, {Belicki}, {Delacroix},
  {Duggan}, {Feeney}, {Hale}, {Kaye}, {Milburn}, {Murphy}, {Porter}, {Reiley},
  {Riddle}, {Rodriguez}, \& {Bellm}}]{Dekany2016}
{Dekany}, R., {Smith}, R.~M., {Belicki}, J., {et~al.} 2016, in \procspie, Vol.
  9908, Ground-based and Airborne Instrumentation for Astronomy VI, 99085M

\bibitem[{{DESI Collaboration} {et~al.}(2016){DESI Collaboration}, {Aghamousa},
  {Aguilar}, {Ahlen}, {Alam}, {Allen}, {Allende Prieto}, {Annis}, {Bailey},
  {Balland}, {Ballester}, {Baltay}, {Beaufore}, {Bebek}, {Beers}, {Bell},
  {Bernal}, {Besuner}, {Beutler}, {Blake}, {Bleuler}, {Blomqvist}, {Blum},
  {Bolton}, {Briceno}, {Brooks}, {Brownstein}, {Buckley-Geer}, {Burden},
  {Burtin}, {Busca}, {Cahn}, {Cai}, {Cardiel-Sas}, {Carlberg}, {Carton},
  {Casas}, {Castand er}, {Cervantes-Cota}, {Claybaugh}, {Close}, {Coker},
  {Cole}, {Comparat}, {Cooper}, {Cousinou}, {Crocce}, {Cuby}, {Cunningham},
  {Davis}, {Dawson}, {de la Macorra}, {De Vicente}, {Delubac}, {Derwent},
  {Dey}, {Dhungana}, {Ding}, {Doel}, {Duan}, {Ealet}, {Edelstein},
  {Eftekharzadeh}, {Eisenstein}, {Elliott}, {Escoffier}, {Evatt}, {Fagrelius},
  {Fan}, {Fanning}, {Farahi}, {Farihi}, {Favole}, {Feng}, {Fernandez},
  {Findlay}, {Finkbeiner}, {Fitzpatrick}, {Flaugher}, {Flender}, {Font-Ribera},
  {Forero-Romero}, {Fosalba}, {Frenk}, {Fumagalli}, {Gaensicke}, {Gallo},
  {Garcia-Bellido}, {Gaztanaga}, {Pietro Gentile Fusillo}, {Gerard},
  {Gershkovich}, {Giannantonio}, {Gillet}, {Gonzalez-de-Rivera},
  {Gonzalez-Perez}, {Gott}, {Graur}, {Gutierrez}, {Guy}, {Habib}, {Heetderks},
  {Heetderks}, {Heitmann}, {Hellwing}, {Herrera}, {Ho}, {Holland}, {Honscheid},
  {Huff}, {Hutchinson}, {Huterer}, {Hwang}, {Illa Laguna}, {Ishikawa},
  {Jacobs}, {Jeffrey}, {Jelinsky}, {Jennings}, {Jiang}, {Jimenez}, {Johnson},
  {Joyce}, {Jullo}, {Juneau}, {Kama}, {Karcher}, {Karkar}, {Kehoe}, {Kennamer},
  {Kent}, {Kilbinger}, {Kim}, {Kirkby}, {Kisner}, {Kitanidis}, {Kneib},
  {Koposov}, {Kovacs}, {Koyama}, {Kremin}, {Kron}, {Kronig}, {Kueter-Young},
  {Lacey}, {Lafever}, {Lahav}, {Lambert}, {Lampton}, {Land riau}, {Lang},
  {Lauer}, {Le Goff}, {Le Guillou}, {Le Van Suu}, {Lee}, {Lee}, {Leitner},
  {Lesser}, {Levi}, {L'Huillier}, {Li}, {Liang}, {Lin}, {Linder}, {Loebman},
  {Luki{\'c}}, {Ma}, {MacCrann}, {Magneville}, {Makarem}, {Manera}, {Manser},
  {Marshall}, {Martini}, {Massey}, {Matheson}, {McCauley}, {McDonald},
  {McGreer}, {Meisner}, {Metcalfe}, {Miller}, {Miquel}, {Moustakas}, {Myers},
  {Naik}, {Newman}, {Nichol}, {Nicola}, {Nicolati da Costa}, {Nie}, {Niz},
  {Norberg}, {Nord}, {Norman}, {Nugent}, {O'Brien}, {Oh}, {Olsen}, {Padilla},
  {Padmanabhan}, {Padmanabhan}, {Palanque-Delabrouille}, {Palmese},
  {Pappalardo}, {P{\^a}ris}, {Park}, {Patej}, {Peacock}, {Peiris}, {Peng},
  {Percival}, {Perruchot}, {Pieri}, {Pogge}, {Pollack}, {Poppett}, {Prada},
  {Prakash}, {Probst}, {Rabinowitz}, {Raichoor}, {Ree}, {Refregier}, {Regal},
  {Reid}, {Reil}, {Rezaie}, {Rockosi}, {Roe}, {Ronayette}, {Roodman}, {Ross},
  {Ross}, {Rossi}, {Rozo}, {Ruhlmann-Kleider}, {Rykoff}, {Sabiu}, {Samushia},
  {Sanchez}, {Sanchez}, {Schlegel}, {Schneider}, {Schubnell}, {Secroun},
  {Seljak}, {Seo}, {Serrano}, {Shafieloo}, {Shan}, {Sharples}, {Sholl},
  {Shourt}, {Silber}, {Silva}, {Sirk}, {Slosar}, {Smith}, {Smoot}, {Som},
  {Song}, {Sprayberry}, {Staten}, {Stefanik}, {Tarle}, {Sien Tie}, {Tinker},
  {Tojeiro}, {Valdes}, {Valenzuela}, {Valluri}, {Vargas-Magana}, {Verde},
  {Walker}, {Wang}, {Wang}, {Weaver}, {Weaverdyck}, {Wechsler}, {Weinberg},
  {White}, {Yang}, {Yeche}, {Zhang}, {Zhao}, {Zheng}, {Zhou}, {Zhou}, {Zhu},
  {Zou}, \& {Zu}}]{DESI2016}
{DESI Collaboration}, {Aghamousa}, A., {Aguilar}, J., {et~al.} 2016, arXiv
  e-prints, arXiv:1611.00036

\bibitem[{{Dessart} \& {Hillier}(2015)}]{Dessart2015a}
{Dessart}, L., \& {Hillier}, D.~J. 2015, \mnras, 447, 1370

\bibitem[{{Dessart} {et~al.}(2012){Dessart}, {Hillier}, {Li}, \&
  {Woosley}}]{Dessart2012}
{Dessart}, L., {Hillier}, D.~J., {Li}, C., \& {Woosley}, S. 2012, MNRAS, 424,
  2139

\bibitem[{{Dey} {et~al.}(2019){Dey}, {Schlegel}, {Lang}, {Blum}, {Burleigh},
  {Fan}, {Findlay}, {Finkbeiner}, {Herrera}, {Juneau}, {Landriau}, {Levi},
  {McGreer}, {Meisner}, {Myers}, {Moustakas}, {Nugent}, {Patej}, {Schlafly},
  {Walker}, {Valdes}, {Weaver}, {Y{\`e}che}, {Zou}, {Zhou}, {Abareshi},
  {Abbott}, {Abolfathi}, {Aguilera}, {Alam}, {Allen}, {Alvarez}, {Annis},
  {Ansarinejad}, {Aubert}, {Beechert}, {Bell}, {BenZvi}, {Beutler}, {Bielby},
  {Bolton}, {Brice{\~n}o}, {Buckley-Geer}, {Butler}, {Calamida}, {Carlberg},
  {Carter}, {Casas}, {Castander}, {Choi}, {Comparat}, {Cukanovaite}, {Delubac},
  {DeVries}, {Dey}, {Dhungana}, {Dickinson}, {Ding}, {Donaldson}, {Duan},
  {Duckworth}, {Eftekharzadeh}, {Eisenstein}, {Etourneau}, {Fagrelius},
  {Farihi}, {Fitzpatrick}, {Font-Ribera}, {Fulmer}, {G{\"a}nsicke},
  {Gaztanaga}, {George}, {Gerdes}, {Gontcho}, {Gorgoni}, {Green}, {Guy},
  {Harmer}, {Hernand ez}, {Honscheid}, {Huang}, {James}, {Jannuzi}, {Jiang},
  {Joyce}, {Karcher}, {Karkar}, {Kehoe}, {Kneib}, {Kueter-Young}, {Lan},
  {Lauer}, {Le Guillou}, {Le Van Suu}, {Lee}, {Lesser}, {Perreault Levasseur},
  {Li}, {Mann}, {Marshall}, {Mart{\'\i}nez-V{\'a}zquez}, {Martini}, {du Mas des
  Bourboux}, {McManus}, {Meier}, {M{\'e}nard}, {Metcalfe},
  {Mu{\~n}oz-Guti{\'e}rrez}, {Najita}, {Napier}, {Narayan}, {Newman}, {Nie},
  {Nord}, {Norman}, {Olsen}, {Paat}, {Palanque-Delabrouille}, {Peng},
  {Poppett}, {Poremba}, {Prakash}, {Rabinowitz}, {Raichoor}, {Rezaie},
  {Robertson}, {Roe}, {Ross}, {Ross}, {Rudnick}, {Safonova}, {Saha},
  {S{\'a}nchez}, {Savary}, {Schweiker}, {Scott}, {Seo}, {Shan}, {Silva},
  {Slepian}, {Soto}, {Sprayberry}, {Staten}, {Stillman}, {Stupak}, {Summers},
  {Sien Tie}, {Tirado}, {Vargas-Maga{\~n}a}, {Vivas}, {Wechsler}, {Williams},
  {Yang}, {Yang}, {Yapici}, {Zaritsky}, {Zenteno}, {Zhang}, {Zhang}, {Zhou}, \&
  {Zhou}}]{Dey2019}
{Dey}, A., {Schlegel}, D.~J., {Lang}, D., {et~al.} 2019, \aj, 157, 168

\bibitem[{{Dimitriadis} {et~al.}(2019){Dimitriadis}, {Siebert}, {Kilpatrick},
  {Rojas-Bravo}, {Foley}, \& {Rich}}]{Dimitriadis2019}
{Dimitriadis}, G., {Siebert}, M.~R., {Kilpatrick}, C.~D., {et~al.} 2019,
  Transient Name Server Classification Report, 2019-675, 1

\bibitem[{{Drake} {et~al.}(2009){Drake}, {Djorgovski}, {Mahabal}, {Beshore},
  {Larson}, {Graham}, {Williams}, {Christensen}, {Catelan}, {Boattini},
  {Gibbs}, {Hill}, \& {Kowalski}}]{Drake2009}
{Drake}, A.~J., {Djorgovski}, S.~G., {Mahabal}, A., {et~al.} 2009, \apj, 696,
  870

\bibitem[{{Drout} {et~al.}(2011){Drout}, {Soderberg}, {Gal-Yam}, {Cenko},
  {Fox}, {Leonard}, {Sand}, {Moon}, {Arcavi}, \& {Green}}]{Drout2011}
{Drout}, M.~R., {Soderberg}, A.~M., {Gal-Yam}, A., {et~al.} 2011, ApJ, 741, 97

\bibitem[{{D'Souza} {et~al.}(2014){D'Souza}, {Kauffman}, {Wang}, \&
  {Vegetti}}]{DSouza2014}
{D'Souza}, R., {Kauffman}, G., {Wang}, J., \& {Vegetti}, S. 2014, \mnras, 443,
  1433

\bibitem[{{Duev} {et~al.}(2019){Duev}, {Mahabal}, {Masci}, {Graham},
  {Rusholme}, {Walters}, {Karmarkar}, {Frederick}, {Kasliwal}, {Rebbapragada},
  \& {Ward}}]{Duev2019}
{Duev}, D.~A., {Mahabal}, A., {Masci}, F.~J., {et~al.} 2019, \mnras, 489, 3582

\bibitem[{{Feindt} {et~al.}(2019){Feindt}, {Nordin}, {Rigault}, {Brinnel},
  {Dhawan}, {Goobar}, \& {Kowalski}}]{Feindt2019}
{Feindt}, U., {Nordin}, J., {Rigault}, M., {et~al.} 2019, \jcap, 2019, 005

\bibitem[{{Filippenko}(1997)}]{Filippenko1997}
{Filippenko}, A.~V. 1997, \araa, 35, 309

\bibitem[{{Filippenko} {et~al.}(2003){Filippenko}, {Chornock}, {Swift},
  {Modjaz}, {Simcoe}, \& {Rauch}}]{Fillipenko2003}
{Filippenko}, A.~V., {Chornock}, R., {Swift}, B., {et~al.} 2003, \iaucirc, 8159

\bibitem[{{Filippenko} {et~al.}(1992){Filippenko}, {Richmond}, {Branch},
  {Gaskell}, {Herbst}, {Ford}, {Treffers}, {Matheson}, {Ho}, {Dey}, {Sargent},
  {Small}, \& {van Breugel}}]{Filippenko1992b}
{Filippenko}, A.~V., {Richmond}, M.~W., {Branch}, D., {et~al.} 1992, \aj, 104,
  1543

\bibitem[{{Fink} {et~al.}(2010){Fink}, {R{\"o}pke}, {Hillebrandt},
  {Seitenzahl}, {Sim}, \& {Kromer}}]{Fink2010}
{Fink}, M., {R{\"o}pke}, F.~K., {Hillebrandt}, W., {et~al.} 2010, \aap, 514,
  A53

\bibitem[{{Fl{\"o}rs} {et~al.}(2020){Fl{\"o}rs}, {Spyromilio}, {Taubenberger},
  {Blondin}, {Cartier}, {Leibundgut}, {Dessart}, {Dhawan}, \&
  {Hillebrandt}}]{Floers2020}
{Fl{\"o}rs}, A., {Spyromilio}, J., {Taubenberger}, S., {et~al.} 2020, \mnras,
  491, 2902

\bibitem[{{Foley}(2015)}]{Foley2015}
{Foley}, R.~J. 2015, \mnras, 452, 2463

\bibitem[{{Foley} {et~al.}(2016){Foley}, {Jha}, {Pan}, {Zheng}, {Bildsten},
  {Filippenko}, \& {Kasen}}]{Foley2016}
{Foley}, R.~J., {Jha}, S.~W., {Pan}, Y.-C., {et~al.} 2016, \mnras, 461, 433

\bibitem[{{Foley} {et~al.}(2009){Foley}, {Chornock}, {Filippenko},
  {Ganeshalingam}, {Kirshner}, {Li}, {Cenko}, {Challis}, {Friedman}, {Modjaz},
  {Silverman}, \& {Wood-Vasey}}]{Foley2009}
{Foley}, R.~J., {Chornock}, R., {Filippenko}, A.~V., {et~al.} 2009, \aj, 138,
  376

\bibitem[{{Fransson} \& {Chevalier}(1989)}]{Fransson1989}
{Fransson}, C., \& {Chevalier}, R.~A. 1989, \apj, 343, 323

\bibitem[{{Fremling} {et~al.}(2016){Fremling}, {Sollerman}, {Taddia}, {Ergon},
  {Fraser}, {Karamehmetoglu}, {Valenti}, {Jerkstrand}, {Arcavi}, {Bufano},
  {Elias Rosa}, {Filippenko}, {Fox}, {Gal-Yam}, {Howell}, {Kotak}, {Mazzali},
  {Milisavljevic}, {Nugent}, {Nyholm}, {Pian}, \& {Smartt}}]{Fremling2016}
{Fremling}, C., {Sollerman}, J., {Taddia}, F., {et~al.} 2016, A \& A, 593, A68

\bibitem[{{Fremling} {et~al.}(2018){Fremling}, {Sollerman}, {Kasliwal},
  {Kulkarni}, {Barbarino}, {Ergon}, {Karamehmetoglu}, {Taddia}, {Arcavi},
  {Cenko}, {Clubb}, {De Cia}, {Duggan}, {Filippenko}, {Gal-Yam}, {Graham},
  {Horesh}, {Hosseinzadeh}, {Howell}, {Kuesters}, {Lunnan}, {Matheson},
  {Nugent}, {Perley}, {Quimby}, \& {Saunders}}]{Fremling2018}
{Fremling}, C., {Sollerman}, J., {Kasliwal}, M.~M., {et~al.} 2018, \aap, 618,
  A37

\bibitem[{{Fremling} {et~al.}(2019){Fremling}, {Miller}, {Sharma}, {Dugas},
  {Perley}, {Taggart}, {Sollerman}, {Goobar}, {Graham}, {Neill}, {Nordin},
  {Rigault}, {Walters}, {Andreoni}, {Bagdasaryan}, {Belicki}, {Cannella},
  {Bellm}, {Cenko}, {De}, {Dekany}, {Frederick}, {Golkhou}, {Graham}, {Helou},
  {Ho}, {Kasliwal}, {Kupfer}, {Laher}, {Mahabal}, {Masci}, {Riddle},
  {Rusholme}, {Schulze}, {Shupe}, {Smith}, {Yan}, {Yao}, {Zhuang}, \&
  {Kulkarni}}]{Fremling2019}
{Fremling}, U.~C., {Miller}, A.~A., {Sharma}, Y., {et~al.} 2019, arXiv
  e-prints, arXiv:1910.12973

\bibitem[{{Frohmaier} {et~al.}(2018){Frohmaier}, {Sullivan}, {Maguire}, \&
  {Nugent}}]{Frohmaier2018}
{Frohmaier}, C., {Sullivan}, M., {Maguire}, K., \& {Nugent}, P. 2018, \apj,
  858, 50

\bibitem[{{Frohmaier} {et~al.}(2017){Frohmaier}, {Sullivan}, {Nugent},
  {Goldstein}, \& {DeRose}}]{Frohmaier2017}
{Frohmaier}, C., {Sullivan}, M., {Nugent}, P.~E., {Goldstein}, D.~A., \&
  {DeRose}, J. 2017, ApJS, 230, 4

\bibitem[{{Frohmaier} {et~al.}(2019){Frohmaier}, {Sullivan}, {Nugent}, {Smith},
  {Dimitriadis}, {Bloom}, {Cenko}, {Kasliwal}, {Kulkarni}, {Maguire}, {Ofek},
  {Poznanski}, \& {Quimby}}]{Frohmaier2019}
{Frohmaier}, C., {Sullivan}, M., {Nugent}, P.~E., {et~al.} 2019, \mnras, 486,
  2308

\bibitem[{{Fryer} {et~al.}(2010){Fryer}, {Ruiter}, {Belczynski}, {Brown},
  {Bufano}, {Diehl}, {Fontes}, {Frey}, {Holland }, {Hungerford}, {Immler},
  {Mazzali}, {Meakin}, {Milne}, {Raskin}, \& {Timmes}}]{Fryer2010}
{Fryer}, C.~L., {Ruiter}, A.~J., {Belczynski}, K., {et~al.} 2010, \apj, 725,
  296

\bibitem[{{Gaia Collaboration} {et~al.}(2018){Gaia Collaboration}, {Brown},
  {Vallenari}, {Prusti}, {de Bruijne}, {Babusiaux}, {Bailer-Jones}, {Biermann},
  {Evans}, {Eyer}, {Jansen}, {Jordi}, {Klioner}, {Lammers}, {Lindegren},
  {Luri}, {Mignard}, {Panem}, {Pourbaix}, {Randich}, {Sartoretti}, {Siddiqui},
  {Soubiran}, {van Leeuwen}, {Walton}, {Arenou}, {Bastian}, {Cropper},
  {Drimmel}, {Katz}, {Lattanzi}, {Bakker}, {Cacciari}, {Casta{\~n}eda},
  {Chaoul}, {Cheek}, {De Angeli}, {Fabricius}, {Guerra}, {Holl}, {Masana},
  {Messineo}, {Mowlavi}, {Nienartowicz}, {Panuzzo}, {Portell}, {Riello},
  {Seabroke}, {Tanga}, {Th{\'e}venin}, {Gracia-Abril}, {Comoretto},
  {Garcia-Reinaldos}, {Teyssier}, {Altmann}, {Andrae}, {Audard},
  {Bellas-Velidis}, {Benson}, {Berthier}, {Blomme}, {Burgess}, {Busso},
  {Carry}, {Cellino}, {Clementini}, {Clotet}, {Creevey}, {Davidson}, {De
  Ridder}, {Delchambre}, {Dell'Oro}, {Ducourant},
  {Fern{\'a}ndez-Hern{\'a}ndez}, {Fouesneau}, {Fr{\'e}mat}, {Galluccio},
  {Garc{\'\i}a-Torres}, {Gonz{\'a}lez-N{\'u}{\~n}ez}, {Gonz{\'a}lez-Vidal},
  {Gosset}, {Guy}, {Halbwachs}, {Hambly}, {Harrison}, {Hern{\'a}ndez},
  {Hestroffer}, {Hodgkin}, {Hutton}, {Jasniewicz}, {Jean-Antoine-Piccolo},
  {Jordan}, {Korn}, {Krone-Martins}, {Lanzafame}, {Lebzelter}, {L{\"o}ffler},
  {Manteiga}, {Marrese}, {Mart{\'\i}n-Fleitas}, {Moitinho}, {Mora}, {Muinonen},
  {Osinde}, {Pancino}, {Pauwels}, {Petit}, {Recio-Blanco}, {Richards},
  {Rimoldini}, {Robin}, {Sarro}, {Siopis}, {Smith}, {Sozzetti}, {S{\"u}veges},
  {Torra}, {van Reeven}, {Abbas}, {Abreu Aramburu}, {Accart}, {Aerts},
  {Altavilla}, {{\'A}lvarez}, {Alvarez}, {Alves}, {Anderson}, {Andrei},
  {Anglada Varela}, {Antiche}, {Antoja}, {Arcay}, {Astraatmadja}, {Bach},
  {Baker}, {Balaguer-N{\'u}{\~n}ez}, {Balm}, {Barache}, {Barata}, {Barbato},
  {Barblan}, {Barklem}, {Barrado}, {Barros}, {Barstow}, {Bartholom{\'e}
  Mu{\~n}oz}, {Bassilana}, {Becciani}, {Bellazzini}, {Berihuete}, {Bertone},
  {Bianchi}, {Bienaym{\'e}}, {Blanco-Cuaresma}, {Boch}, {Boeche}, {Bombrun},
  {Borrachero}, {Bossini}, {Bouquillon}, {Bourda}, {Bragaglia}, {Bramante},
  {Breddels}, {Bressan}, {Brouillet}, {Br{\"u}semeister}, {Brugaletta},
  {Bucciarelli}, {Burlacu}, {Busonero}, {Butkevich}, {Buzzi}, {Caffau},
  {Cancelliere}, {Cannizzaro}, {Cantat-Gaudin}, {Carballo}, {Carlucci},
  {Carrasco}, {Casamiquela}, {Castellani}, {Castro-Ginard}, {Charlot},
  {Chemin}, {Chiavassa}, {Cocozza}, {Costigan}, {Cowell}, {Crifo}, {Crosta},
  {Crowley}, {Cuypers}, {Dafonte}, {Damerdji}, {Dapergolas}, {David}, {David},
  {de Laverny}, {De Luise}, {De March}, {de Martino}, {de Souza}, {de Torres},
  {Debosscher}, {del Pozo}, {Delbo}, {Delgado}, {Delgado}, {Di Matteo},
  {Diakite}, {Diener}, {Distefano}, {Dolding}, {Drazinos}, {Dur{\'a}n},
  {Edvardsson}, {Enke}, {Eriksson}, {Esquej}, {Eynard Bontemps}, {Fabre},
  {Fabrizio}, {Faigler}, {Falc{\~a}o}, {Farr{\`a}s Casas}, {Federici},
  {Fedorets}, {Fernique}, {Figueras}, {Filippi}, {Findeisen}, {Fonti},
  {Fraile}, {Fraser}, {Fr{\'e}zouls}, {Gai}, {Galleti}, {Garabato},
  {Garc{\'\i}a-Sedano}, {Garofalo}, {Garralda}, {Gavel}, {Gavras}, {Gerssen},
  {Geyer}, {Giacobbe}, {Gilmore}, {Girona}, {Giuffrida}, {Glass}, {Gomes},
  {Granvik}, {Gueguen}, {Guerrier}, {Guiraud}, {Guti{\'e}rrez-S{\'a}nchez},
  {Haigron}, {Hatzidimitriou}, {Hauser}, {Haywood}, {Heiter}, {Helmi}, {Heu},
  {Hilger}, {Hobbs}, {Hofmann}, {Holland}, {Huckle}, {Hypki}, {Icardi},
  {Jan{\ss}en}, {Jevardat de Fombelle}, {Jonker}, {Juh{\'a}sz}, {Julbe},
  {Karampelas}, {Kewley}, {Klar}, {Kochoska}, {Kohley}, {Kolenberg},
  {Kontizas}, {Kontizas}, {Koposov}, {Kordopatis}, {Kostrzewa-Rutkowska},
  {Koubsky}, {Lambert}, {Lanza}, {Lasne}, {Lavigne}, {Le Fustec}, {Le
  Poncin-Lafitte}, {Lebreton}, {Leccia}, {Leclerc}, {Lecoeur-Taibi},
  {Lenhardt}, {Leroux}, {Liao}, {Licata}, {Lindstr{\o}m}, {Lister}, {Livanou},
  {Lobel}, {L{\'o}pez}, {Managau}, {Mann}, {Mantelet}, {Marchal}, {Marchant},
  {Marconi}, {Marinoni}, {Marschalk{\'o}}, {Marshall}, {Martino}, {Marton},
  {Mary}, {Massari}, {Matijevi{\v{c}}}, {Mazeh}, {McMillan}, {Messina},
  {Michalik}, {Millar}, {Molina}, {Molinaro}, {Moln{\'a}r}, {Montegriffo},
  {Mor}, {Morbidelli}, {Morel}, {Morris}, {Mulone}, {Muraveva}, {Musella},
  {Nelemans}, {Nicastro}, {Noval}, {O'Mullane}, {Ord{\'e}novic},
  {Ord{\'o}{\~n}ez-Blanco}, {Osborne}, {Pagani}, {Pagano}, {Pailler},
  {Palacin}, {Palaversa}, {Panahi}, {Pawlak}, {Piersimoni}, {Pineau}, {Plachy},
  {Plum}, {Poggio}, {Poujoulet}, {Pr{\v{s}}a}, {Pulone}, {Racero}, {Ragaini},
  {Rambaux}, {Ramos-Lerate}, {Regibo}, {Reyl{\'e}}, {Riclet}, {Ripepi}, {Riva},
  {Rivard}, {Rixon}, {Roegiers}, {Roelens}, {Romero-G{\'o}mez}, {Rowell},
  {Royer}, {Ruiz-Dern}, {Sadowski}, {Sagrist{\`a} Sell{\'e}s}, {Sahlmann},
  {Salgado}, {Salguero}, {Sanna}, {Santana-Ros}, {Sarasso}, {Savietto},
  {Schultheis}, {Sciacca}, {Segol}, {Segovia}, {S{\'e}gransan}, {Shih},
  {Siltala}, {Silva}, {Smart}, {Smith}, {Solano}, {Solitro}, {Sordo}, {Soria
  Nieto}, {Souchay}, {Spagna}, {Spoto}, {Stampa}, {Steele},
  {Steidelm{\"u}ller}, {Stephenson}, {Stoev}, {Suess}, {Surdej}, {Szabados},
  {Szegedi-Elek}, {Tapiador}, {Taris}, {Tauran}, {Taylor}, {Teixeira},
  {Terrett}, {Teyssand ier}, {Thuillot}, {Titarenko}, {Torra Clotet}, {Turon},
  {Ulla}, {Utrilla}, {Uzzi}, {Vaillant}, {Valentini}, {Valette}, {van Elteren},
  {Van Hemelryck}, {van Leeuwen}, {Vaschetto}, {Vecchiato}, {Veljanoski},
  {Viala}, {Vicente}, {Vogt}, {von Essen}, {Voss}, {Votruba}, {Voutsinas},
  {Walmsley}, {Weiler}, {Wertz}, {Wevers}, {Wyrzykowski}, {Yoldas},
  {{\v{Z}}erjal}, {Ziaeepour}, {Zorec}, {Zschocke}, {Zucker}, {Zurbach}, \&
  {Zwitter}}]{Gaia2018}
{Gaia Collaboration}, {Brown}, A.~G.~A., {Vallenari}, A., {et~al.} 2018, \aap,
  616, A1

\bibitem[{Gal-Yam(2017)}]{Gal-Yam2017}
Gal-Yam, A. 2017, Observational and Physical Classification of Supernovae
  (Cham: Springer International Publishing), 1--43.
\newblock \url{https://doi.org/10.1007/978-3-319-20794-0_35-1}

\bibitem[{{Gal-Yam} {et~al.}(2014){Gal-Yam}, {Arcavi}, {Ofek}, {Ben-Ami},
  {Cenko}, {Kasliwal}, {Cao}, {Yaron}, {Tal}, {Silverman}, {Horesh}, {De Cia},
  {Taddia}, {Sollerman}, {Perley}, {Vreeswijk}, {Kulkarni}, {Nugent},
  {Filippenko}, \& {Wheeler}}]{GalYam2014}
{Gal-Yam}, A., {Arcavi}, I., {Ofek}, E.~O., {et~al.} 2014, Nature, 509, 471

\bibitem[{{Galbany} {et~al.}(2019){Galbany}, {Ashall}, {H{\"o}flich},
  {Gonz{\'a}lez-Gait{\'a}n}, {Taubenberger}, {Stritzinger}, {Hsiao}, {Mazzali},
  {Baron}, {Blondin}, {Bose}, {Bulla}, {Burke}, {Burns}, {Cartier}, {Chen},
  {Della Valle}, {Diamond}, {Guti{\'e}rrez}, {Harmanen}, {Hiramatsu},
  {Holoien}, {Hosseinzadeh}, {Howell}, {Huang}, {Inserra}, {de Jaeger}, {Jha},
  {Kangas}, {Kromer}, {Lyman}, {Maguire}, {Marion}, {Milisavljevic},
  {Prentice}, {Razza}, {Reynolds}, {Sand}, {Shappee}, {Shekhar}, {Smartt},
  {Stassun}, {Sullivan}, {Valenti}, {Villanueva}, {Wang}, {Wheeler}, {Zhai}, \&
  {Zhang}}]{Galbany2019}
{Galbany}, L., {Ashall}, C., {H{\"o}flich}, P., {et~al.} 2019, \aap, 630, A76

\bibitem[{{Geier} {et~al.}(2013){Geier}, {Marsh}, {Wang}, {Dunlap}, {Barlow},
  {Schaffenroth}, {Chen}, {Irrgang}, {Maxted}, {Ziegerer}, {Kupfer},
  {Miszalski}, {Heber}, {Han}, {Shporer}, {Telting}, {G{\"a}nsicke},
  {{\O}stensen}, {O'Toole}, \& {Napiwotzki}}]{Geier2013}
{Geier}, S., {Marsh}, T.~R., {Wang}, B., {et~al.} 2013, \aap, 554, A54

\bibitem[{{Graham} {et~al.}(2019){Graham}, {Kulkarni}, {Bellm}, {Adams},
  {Barbarino}, {Blagorodnova}, {Bodewits}, {Bolin}, {Brady}, {Cenko}, {Chang},
  {Coughlin}, {De}, {Eadie}, {Farnham}, {Feindt}, {Franckowiak}, {Fremling},
  {Gezari}, {Ghosh}, {Goldstein}, {Golkhou}, {Goobar}, {Ho}, {Huppenkothen},
  {Ivezi{\'c}}, {Jones}, {Juric}, {Kaplan}, {Kasliwal}, {Kelley}, {Kupfer},
  {Lee}, {Lin}, {Lunnan}, {Mahabal}, {Miller}, {Ngeow}, {Nugent}, {Ofek},
  {Prince}, {Rauch}, {van Roestel}, {Schulze}, {Singer}, {Sollerman}, {Taddia},
  {Yan}, {Ye}, {Yu}, {Barlow}, {Bauer}, {Beck}, {Belicki}, {Biswas}, {Brinnel},
  {Brooke}, {Bue}, {Bulla}, {Burruss}, {Connolly}, {Cromer}, {Cunningham},
  {Dekany}, {Delacroix}, {Desai}, {Duev}, {Feeney}, {Flynn}, {Frederick},
  {Gal-Yam}, {Giomi}, {Groom}, {Hacopians}, {Hale}, {Helou}, {Henning},
  {Hover}, {Hillenbrand}, {Howell}, {Hung}, {Imel}, {Ip}, {Jackson}, {Kaspi},
  {Kaye}, {Kowalski}, {Kramer}, {Kuhn}, {Landry}, {Laher}, {Mao}, {Masci},
  {Monkewitz}, {Murphy}, {Nordin}, {Patterson}, {Penprase}, {Porter},
  {Rebbapragada}, {Reiley}, {Riddle}, {Rigault}, {Rodriguez}, {Rusholme}, {van
  Santen}, {Shupe}, {Smith}, {Soumagnac}, {Stein}, {Surace}, {Szkody}, {Terek},
  {Van Sistine}, {van Velzen}, {Vestrand}, {Walters}, {Ward}, {Zhang}, \&
  {Zolkower}}]{Graham2019}
{Graham}, M.~J., {Kulkarni}, S.~R., {Bellm}, E.~C., {et~al.} 2019, \pasp, 131,
  078001

\bibitem[{{Grzegorzek}(2019)}]{SN2019ehk}
{Grzegorzek}, J. 2019, Transient Name Server Discovery Report, 2019-666, 1

\bibitem[{{Guillochon} {et~al.}(2017){Guillochon}, {Parrent}, {Kelley}, \&
  {Margutti}}]{Guillochon2017}
{Guillochon}, J., {Parrent}, J., {Kelley}, L.~Z., \& {Margutti}, R. 2017, \apj,
  835, 64

\bibitem[{{Hachinger} {et~al.}(2012){Hachinger}, {Mazzali}, {Taubenberger},
  {Hillebrandt}, {Nomoto}, \& {Sauer}}]{Hachinger2012}
{Hachinger}, S., {Mazzali}, P.~A., {Taubenberger}, S., {et~al.} 2012, MNRAS,
  422, 70

\bibitem[{{Harris}(1996)}]{Harris1996}
{Harris}, W.~E. 1996, AJ, 112, 1487

\bibitem[{{Hinshaw} {et~al.}(2013){Hinshaw}, {Larson}, {Komatsu}, {Spergel},
  {Bennett}, {Dunkley}, {Nolta}, {Halpern}, {Hill}, {Odegard}, {Page}, {Smith},
  {Weiland}, {Gold}, {Jarosik}, {Kogut}, {Limon}, {Meyer}, {Tucker}, {Wollack},
  \& {Wright}}]{Hinshaw2013}
{Hinshaw}, G., {Larson}, D., {Komatsu}, E., {et~al.} 2013, \apjs, 208, 19

\bibitem[{{Hoeflich} \& {Khokhlov}(1996)}]{Hoeflich1996}
{Hoeflich}, P., \& {Khokhlov}, A. 1996, \apj, 457, 500

\bibitem[{{Hogg} {et~al.}(1997){Hogg}, {Pahre}, {McCarthy}, {Cohen},
  {Blandford}, {Smail}, \& {Soifer}}]{Hogg1997}
{Hogg}, D.~W., {Pahre}, M.~A., {McCarthy}, J.~K., {et~al.} 1997, \mnras, 288,
  404

\bibitem[{{Howell}(2001)}]{Howell2001}
{Howell}, D.~A. 2001, \apjl, 554, L193

\bibitem[{{Howell} {et~al.}(2005){Howell}, {Sullivan}, {Perrett}, {Bronder},
  {Hook}, {Astier}, {Aubourg}, {Balam}, {Basa}, {Carlberg}, {Fabbro},
  {Fouchez}, {Guy}, {Lafoux}, {Neill}, {Pain}, {Palanque-Delabrouille},
  {Pritchet}, {Regnault}, {Rich}, {Taillet}, {Knop}, {McMahon}, {Perlmutter},
  \& {Walton}}]{Howell2005}
{Howell}, D.~A., {Sullivan}, M., {Perrett}, K., {et~al.} 2005, \apj, 634, 1190

\bibitem[{{Hunter}(2007)}]{Hunter2007}
{Hunter}, J.~D. 2007, Computing in Science and Engineering, 9, 90

\bibitem[{{Iben} \& {Tutukov}(1989)}]{Iben1989}
{Iben}, Icko, J., \& {Tutukov}, A.~V. 1989, \apj, 342, 430

\bibitem[{{Inserra} {et~al.}(2015){Inserra}, {Sim}, {Wyrzykowski}, {Smartt},
  {Fraser}, {Nicholl}, {Shen}, {Jerkstrand}, {Gal-Yam}, {Howell}, {Maguire},
  {Mazzali}, {Valenti}, {Taubenberger}, {Benitez-Herrera}, {Bersier},
  {Blagorodnova}, {Campbell}, {Chen}, {Elias-Rosa}, {Hillebrandt},
  {Kostrzewa-Rutkowska}, {Koz{\l}owski}, {Kromer}, {Lyman}, {Polshaw},
  {R{\"o}pke}, {Ruiter}, {Smith}, {Spiro}, {Sullivan}, {Yaron}, {Young}, \&
  {Yuan}}]{Inserra2015}
{Inserra}, C., {Sim}, S.~A., {Wyrzykowski}, L., {et~al.} 2015, \apjl, 799, L2

\bibitem[{{Jacobson-Galan} {et~al.}(2019){Jacobson-Galan}, {Polin}, {Foley},
  {Dimitriadis}, {Kilpatrick}, {Margutti}, {Coulter}, {Jha}, {Jones},
  {Kirshner}, {Pan}, {Piro}, {Rest}, \& {Rojas-Bravo}}]{Jacobson-Galan2019}
{Jacobson-Galan}, W.~V., {Polin}, A., {Foley}, R.~J., {et~al.} 2019, arXiv
  e-prints, arXiv:1910.05436

\bibitem[{{Jerkstrand}(2017)}]{Jerkstrand2017}
{Jerkstrand}, A. 2017, {Spectra of Supernovae in the Nebular Phase}, ed. A.~W.
  {Alsabti} \& P.~{Murdin}, 795

\bibitem[{{Jiang} {et~al.}(2017){Jiang}, {Doi}, {Maeda}, {Shigeyama}, {Nomoto},
  {Yasuda}, {Jha}, {Tanaka}, {Morokuma}, {Tominaga}, {Ivezi{\'c}},
  {Ruiz-Lapuente}, {Stritzinger}, {Mazzali}, {Ashall}, {Mould}, {Baade},
  {Suzuki}, {Connolly}, {Patat}, {Wang}, {Yoachim}, {Jones}, {Furusawa}, \&
  {Miyazaki}}]{Jiang2017}
{Jiang}, J.-A., {Doi}, M., {Maeda}, K., {et~al.} 2017, \nat, 550, 80

\bibitem[{{Kasliwal} {et~al.}(2010){Kasliwal}, {Kulkarni}, {Gal-Yam}, {Yaron},
  {Quimby}, {Ofek}, {Nugent}, {Poznanski}, {Jacobsen}, {Sternberg}, {Arcavi},
  {Howell}, {Sullivan}, {Rich}, {Burke}, {Brimacombe}, {Milisavljevic},
  {Fesen}, {Bildsten}, {Shen}, {Cenko}, {Bloom}, {Hsiao}, {Law}, {Gehrels},
  {Immler}, {Dekany}, {Rahmer}, {Hale}, {Smith}, {Zolkower}, {Velur},
  {Walters}, {Henning}, {Bui}, \& {McKenna}}]{Kasliwal2010}
{Kasliwal}, M.~M., {Kulkarni}, S.~R., {Gal-Yam}, A., {et~al.} 2010, ApJL, 723,
  L98

\bibitem[{{Kasliwal} {et~al.}(2012){Kasliwal}, {Kulkarni}, {Gal-Yam}, {Nugent},
  {Sullivan}, {Bildsten}, {Yaron}, {Perets}, {Arcavi}, {Ben-Ami}, {Bhalerao},
  {Bloom}, {Cenko}, {Filippenko}, {Frail}, {Ganeshalingam}, {Horesh}, {Howell},
  {Law}, {Leonard}, {Li}, {Ofek}, {Polishook}, {Poznanski}, {Quimby},
  {Silverman}, {Sternberg}, \& {Xu}}]{Kasliwal2012a}
---. 2012, ApJ, 755, 161

\bibitem[{{Kasliwal} {et~al.}(2019){Kasliwal}, {Cannella}, {Bagdasaryan},
  {Hung}, {Feindt}, {Singer}, {Coughlin}, {Fremling}, {Walters}, {Duev},
  {Itoh}, \& {Quimby}}]{Kasliwal2019}
{Kasliwal}, M.~M., {Cannella}, C., {Bagdasaryan}, A., {et~al.} 2019, \pasp,
  131, 038003

\bibitem[{{Kawabata} {et~al.}(2010){Kawabata}, {Maeda}, {Nomoto},
  {Taubenberger}, {Tanaka}, {Deng}, {Pian}, {Hattori}, \&
  {Itagaki}}]{Kawabata2010}
{Kawabata}, K.~S., {Maeda}, K., {Nomoto}, K., {et~al.} 2010, \nat, 465, 326

\bibitem[{{Kawana} {et~al.}(2020){Kawana}, {Maeda}, {Yoshida}, \&
  {Tanikawa}}]{Kawana2020}
{Kawana}, K., {Maeda}, K., {Yoshida}, N., \& {Tanikawa}, A. 2020, \apjl, 890,
  L26

\bibitem[{{Khatami} \& {Kasen}(2019)}]{Khatami2019}
{Khatami}, D.~K., \& {Kasen}, D.~N. 2019, \apj, 878, 56

\bibitem[{{Kochanek} {et~al.}(2001){Kochanek}, {Pahre}, {Falco}, {Huchra},
  {Mader}, {Jarrett}, {Chester}, {Cutri}, \& {Schneider}}]{Kochanek2001}
{Kochanek}, C.~S., {Pahre}, M.~A., {Falco}, E.~E., {et~al.} 2001, \apj, 560,
  566

\bibitem[{{Kollmeier} {et~al.}(2017){Kollmeier}, {Zasowski}, {Rix}, {Johns},
  {Anderson}, {Drory}, {Johnson}, {Pogge}, {Bird}, {Blanc}, {Brownstein},
  {Crane}, {De Lee}, {Klaene}, {Kreckel}, {MacDonald}, {Merloni}, {Ness},
  {O'Brien}, {Sanchez-Gallego}, {Sayres}, {Shen}, {Thakar}, {Tkachenko},
  {Aerts}, {Blanton}, {Eisenstein}, {Holtzman}, {Maoz}, {Nandra}, {Rockosi},
  {Weinberg}, {Bovy}, {Casey}, {Chaname}, {Clerc}, {Conroy}, {Eracleous},
  {G{\"a}nsicke}, {Hekker}, {Horne}, {Kauffmann}, {McQuinn}, {Pellegrini},
  {Schinnerer}, {Schlafly}, {Schwope}, {Seibert}, {Teske}, \& {van
  Saders}}]{Kollmeier2017}
{Kollmeier}, J.~A., {Zasowski}, G., {Rix}, H.-W., {et~al.} 2017, arXiv
  e-prints, arXiv:1711.03234

\bibitem[{{Kromer} {et~al.}(2010){Kromer}, {Sim}, {Fink}, {R{\"o}pke},
  {Seitenzahl}, \& {Hillebrandt}}]{Kromer2010}
{Kromer}, M., {Sim}, S.~A., {Fink}, M., {et~al.} 2010, \apj, 719, 1067

\bibitem[{{Kulkarni} {et~al.}(2018){Kulkarni}, {Perley}, \&
  {Miller}}]{Kulkarni2018}
{Kulkarni}, S.~R., {Perley}, D.~A., \& {Miller}, A.~A. 2018, \apj, 860, 22

\bibitem[{{Kupfer} {et~al.}(2017){Kupfer}, {van Roestel}, {Brooks}, {Geier},
  {Marsh}, {Groot}, {Bloemen}, {Prince}, {Bellm}, {Heber}, {Bildsten},
  {Miller}, {Dyer}, {Dhillon}, {Green}, {Irawati}, {Laher}, {Littlefair},
  {Shupe}, {Steidel}, {Rattansoon}, \& {Pettini}}]{Kupfer2017}
{Kupfer}, T., {van Roestel}, J., {Brooks}, J., {et~al.} 2017, \apj, 835, 131

\bibitem[{{Lang} {et~al.}(2016){Lang}, {Hogg}, \& {Schlegel}}]{Lang2016}
{Lang}, D., {Hogg}, D.~W., \& {Schlegel}, D.~J. 2016, \aj, 151, 36

\bibitem[{{Leadbeater}(2018)}]{Leadbeater2018}
{Leadbeater}, R. 2018, Transient Name Server Classification Report, 2018-1486,
  1

\bibitem[{{Li} {et~al.}(2003){Li}, {Filippenko}, {Chornock}, {Berger},
  {Berlind}, {Calkins}, {Challis}, {Fassnacht}, {Jha}, {Kirshner}, {Matheson},
  {Sargent}, {Simcoe}, {Smith}, \& {Squires}}]{Li2003}
{Li}, W., {Filippenko}, A.~V., {Chornock}, R., {et~al.} 2003, \pasp, 115, 453

\bibitem[{{Li} {et~al.}(2011){Li}, {Leaman}, {Chornock}, {Filippenko},
  {Poznanski}, {Ganeshalingam}, {Wang}, {Modjaz}, {Jha}, {Foley}, \&
  {Smith}}]{Li2011}
{Li}, W., {Leaman}, J., {Chornock}, R., {et~al.} 2011, \mnras, 412, 1441

\bibitem[{{Liu} {et~al.}(2016){Liu}, {Modjaz}, {Bianco}, \& {Graur}}]{Liu2016}
{Liu}, Y.-Q., {Modjaz}, M., {Bianco}, F.~B., \& {Graur}, O. 2016, \apj, 827, 90

\bibitem[{{Livne} \& {Arnett}(1995)}]{Livne1995}
{Livne}, E., \& {Arnett}, D. 1995, \apj, 452, 62

\bibitem[{{Livne} \& {Glasner}(1990)}]{Livne1990}
{Livne}, E., \& {Glasner}, A.~S. 1990, \apj, 361, 244

\bibitem[{{Livne} \& {Glasner}(1991)}]{Livne1991}
---. 1991, \apj, 370, 272

\bibitem[{{Lucy}(1991)}]{Lucy1991}
{Lucy}, L.~B. 1991, \apj, 383, 308

\bibitem[{{Lunnan} {et~al.}(2017){Lunnan}, {Kasliwal}, {Cao}, {Hangard},
  {Yaron}, {Parrent}, {McCully}, {Gal-Yam}, {Mulchaey}, {Ben-Ami},
  {Filippenko}, {Fremling}, {Fruchter}, {Howell}, {Koda}, {Kupfer}, {Kulkarni},
  {Laher}, {Masci}, {Nugent}, {Ofek}, {Yagi}, \& {Yan}}]{Lunnan2017}
{Lunnan}, R., {Kasliwal}, M.~M., {Cao}, Y., {et~al.} 2017, ApJ, 836, 60

\bibitem[{{Lyman} {et~al.}(2014){Lyman}, {Levan}, {Church}, {Davies}, \&
  {Tanvir}}]{Lyman2014}
{Lyman}, J.~D., {Levan}, A.~J., {Church}, R.~P., {Davies}, M.~B., \& {Tanvir},
  N.~R. 2014, MNRAS, 444, 2157

\bibitem[{{Lyman} {et~al.}(2016){Lyman}, {Levan}, {James}, {Angus}, {Church},
  {Davies}, \& {Tanvir}}]{Lyman2016b}
{Lyman}, J.~D., {Levan}, A.~J., {James}, P.~A., {et~al.} 2016, \mnras, 458,
  1768

\bibitem[{{MacLeod} {et~al.}(2014){MacLeod}, {Goldstein}, {Ramirez-Ruiz},
  {Guillochon}, \& {Samsing}}]{Macleod2014}
{MacLeod}, M., {Goldstein}, J., {Ramirez-Ruiz}, E., {Guillochon}, J., \&
  {Samsing}, J. 2014, \apj, 794, 9

\bibitem[{{MacLeod} {et~al.}(2016){MacLeod}, {Guillochon}, {Ramirez-Ruiz},
  {Kasen}, \& {Rosswog}}]{Macleod2016}
{MacLeod}, M., {Guillochon}, J., {Ramirez-Ruiz}, E., {Kasen}, D., \& {Rosswog},
  S. 2016, ApJ, 819, 3

\bibitem[{{Mahabal} {et~al.}(2019){Mahabal}, {Rebbapragada}, {Walters},
  {Masci}, {Blagorodnova}, {van Roestel}, {Ye}, {Biswas}, {Burdge}, {Chang},
  {Duev}, {Golkhou}, {Miller}, {Nordin}, {Ward}, {Adams}, {Bellm}, {Branton},
  {Bue}, {Cannella}, {Connolly}, {Dekany}, {Feindt}, {Hung}, {Fortson},
  {Frederick}, {Fremling}, {Gezari}, {Graham}, {Groom}, {Kasliwal}, {Kulkarni},
  {Kupfer}, {Lin}, {Lintott}, {Lunnan}, {Parejko}, {Prince}, {Riddle},
  {Rusholme}, {Saunders}, {Sedaghat}, {Shupe}, {Singer}, {Soumagnac}, {Szkody},
  {Tachibana}, {Tirumala}, {van Velzen}, \& {Wright}}]{Mahabal2019}
{Mahabal}, A., {Rebbapragada}, U., {Walters}, R., {et~al.} 2019, \pasp, 131,
  038002

\bibitem[{{Mannucci} {et~al.}(2008){Mannucci}, {Maoz}, {Sharon}, {Botticella},
  {Della Valle}, {Gal-Yam}, \& {Panagia}}]{Mannucci2008}
{Mannucci}, F., {Maoz}, D., {Sharon}, K., {et~al.} 2008, \mnras, 383, 1121

\bibitem[{{Margalit} \& {Metzger}(2016)}]{Margalit2016}
{Margalit}, B., \& {Metzger}, B.~D. 2016, \mnras, 461, 1154

\bibitem[{{Marion} {et~al.}(2014){Marion}, {Vinko}, {Kirshner}, {Foley},
  {Berlind}, {Bieryla}, {Bloom}, {Calkins}, {Challis}, {Chevalier}, {Chornock},
  {Culliton}, {Curtis}, {Esquerdo}, {Everett}, {Falco}, {France}, {Fransson},
  {Friedman}, {Garnavich}, {Leibundgut}, {Meyer}, {Smith}, {Soderberg},
  {Sollerman}, {Starr}, {Szklenar}, {Takats}, \& {Wheeler}}]{Marion2014}
{Marion}, G.~H., {Vinko}, J., {Kirshner}, R.~P., {et~al.} 2014, \apj, 781, 69

\bibitem[{{Masci} {et~al.}(2019){Masci}, {Laher}, {Rusholme}, {Shupe}, {Groom},
  {Surace}, {Jackson}, {Monkewitz}, {Beck}, {Flynn}, {Terek}, {Landry},
  {Hacopians}, {Desai}, {Howell}, {Brooke}, {Imel}, {Wachter}, {Ye}, {Lin},
  {Cenko}, {Cunningham}, {Rebbapragada}, {Bue}, {Miller}, {Mahabal}, {Bellm},
  {Patterson}, {Juri{\'c}}, {Golkhou}, {Ofek}, {Walters}, {Graham}, {Kasliwal},
  {Dekany}, {Kupfer}, {Burdge}, {Cannella}, {Barlow}, {Van Sistine}, {Giomi},
  {Fremling}, {Blagorodnova}, {Levitan}, {Riddle}, {Smith}, {Helou}, {Prince},
  \& {Kulkarni}}]{Masci2019}
{Masci}, F.~J., {Laher}, R.~R., {Rusholme}, B., {et~al.} 2019, \pasp, 131,
  018003

\bibitem[{{McBrien} {et~al.}(2019){McBrien}, {Smartt}, {Chen}, {Inserra},
  {Gillanders}, {Sim}, {Jerkstrand}, {Rest}, {Valenti}, {Roy}, {Gromadzki},
  {Taubenberger}, {Fl{\"o}rs}, {Huber}, {Chambers}, {Gal-Yam}, {Young},
  {Nicholl}, {Kankare}, {Smith}, {Maguire}, {Mand el}, {Prentice},
  {Rodr{\'\i}guez}, {Pineda Garcia}, {Guti{\'e}rrez}, {Galbany}, {Barbarino},
  {Clark}, {Sollerman}, {Kulkarni}, {De}, {Buckley}, \& {Rau}}]{McBrien2019}
{McBrien}, O.~R., {Smartt}, S.~J., {Chen}, T.-W., {et~al.} 2019, \apjl, 885,
  L23

\bibitem[{McKinney(2010)}]{McKinney2010}
McKinney, W. 2010, in Proceedings of the 9th Python in Science Conference, ed.
  S.~van~der Walt \& J.~Millman, 51 -- 56

\bibitem[{{Meng} \& {Han}(2015)}]{Meng2015}
{Meng}, X., \& {Han}, Z. 2015, \aap, 573, A57

\bibitem[{{Mernier} {et~al.}(2016){Mernier}, {de Plaa}, {Pinto}, {Kaastra},
  {Kosec}, {Zhang}, {Mao}, {Werner}, {Pols}, \& {Vink}}]{Mernier2016}
{Mernier}, F., {de Plaa}, J., {Pinto}, C., {et~al.} 2016, \aap, 595, A126

\bibitem[{{Metzger}(2012)}]{Metzger2012}
{Metzger}, B.~D. 2012, MNRAS, 419, 827

\bibitem[{{Milisavljevic} {et~al.}(2017){Milisavljevic}, {Patnaude}, {Raymond},
  {Drout}, {Margutti}, {Kamble}, {Chornock}, {Guillochon}, {Sanders},
  {Parrent}, {Lovisari}, {Chilingarian}, {Challis}, {Kirshner}, {Penny},
  {Itagaki}, {Eldridge}, \& {Moriya}}]{Milisavljevic2017}
{Milisavljevic}, D., {Patnaude}, D.~J., {Raymond}, J.~C., {et~al.} 2017, ApJ,
  846, 50

\bibitem[{{Moore} {et~al.}(2013){Moore}, {Townsley}, \& {Bildsten}}]{Moore2013}
{Moore}, K., {Townsley}, D.~M., \& {Bildsten}, L. 2013, \apj, 776, 97

\bibitem[{{Moriya} {et~al.}(2017){Moriya}, {Mazzali}, {Tominaga}, {Hachinger},
  {Blinnikov}, {Tauris}, {Takahashi}, {Tanaka}, {Langer}, \&
  {Podsiadlowski}}]{Moriya2017}
{Moriya}, T.~J., {Mazzali}, P.~A., {Tominaga}, N., {et~al.} 2017, MNRAS, 466,
  2085

\bibitem[{{Mulchaey} {et~al.}(2014){Mulchaey}, {Kasliwal}, \&
  {Kollmeier}}]{Mulchaey2014}
{Mulchaey}, J.~S., {Kasliwal}, M.~M., \& {Kollmeier}, J.~A. 2014, \apjl, 780,
  L34

\bibitem[{{Neill} {et~al.}(2009){Neill}, {Sullivan}, {Howell}, {Conley},
  {Seibert}, {Martin}, {Barlow}, {Foster}, {Friedman}, {Morrissey}, {Neff},
  {Schiminovich}, {Wyder}, {Bianchi}, {Donas}, {Heckman}, {Lee}, {Madore},
  {Milliard}, {Rich}, \& {Szalay}}]{Neill2009}
{Neill}, J.~D., {Sullivan}, M., {Howell}, D.~A., {et~al.} 2009, \apj, 707, 1449

\bibitem[{{Nelemans} {et~al.}(2004){Nelemans}, {Yungelson}, \& {Portegies
  Zwart}}]{Nelemans2004}
{Nelemans}, G., {Yungelson}, L.~R., \& {Portegies Zwart}, S.~F. 2004, \mnras,
  349, 181

\bibitem[{{Nomoto}(1980)}]{Nomoto1980}
{Nomoto}, K. 1980, \ssr, 27, 563

\bibitem[{{Nomoto}(1982{\natexlab{a}})}]{Nomoto1982a}
---. 1982{\natexlab{a}}, \apj, 253, 798

\bibitem[{{Nomoto}(1982{\natexlab{b}})}]{Nomoto1982b}
---. 1982{\natexlab{b}}, \apj, 257, 780

\bibitem[{{Nordin} {et~al.}(2019{\natexlab{a}}){Nordin}, {Brinnel}, {Giomi},
  {Santen}, {Gal-Yam}, {Yaron}, \& {Schulze}}]{SN2019ofm}
{Nordin}, J., {Brinnel}, V., {Giomi}, M., {et~al.} 2019{\natexlab{a}},
  Transient Name Server Discovery Report, 2019-1594, 1

\bibitem[{{Nordin} {et~al.}(2019{\natexlab{b}}){Nordin}, {Brinnel}, {van
  Santen}, {Bulla}, {Feindt}, {Franckowiak}, {Fremling}, {Gal-Yam}, {Giomi},
  {Kowalski}, {Mahabal}, {Miranda}, {Rauch}, {Reusch}, {Rigault}, {Schulze},
  {Sollerman}, {Stein}, {Yaron}, {van Velzen}, \& {Ward}}]{Nordin2019}
{Nordin}, J., {Brinnel}, V., {van Santen}, J., {et~al.} 2019{\natexlab{b}},
  \aap, 631, A147

\bibitem[{{Nugent} {et~al.}(1997){Nugent}, {Baron}, {Branch}, {Fisher}, \&
  {Hauschildt}}]{Nugent1997}
{Nugent}, P., {Baron}, E., {Branch}, D., {Fisher}, A., \& {Hauschildt}, P.~H.
  1997, \apj, 485, 812

\bibitem[{{Nugent} {et~al.}(2011){Nugent}, {Sullivan}, {Cenko}, {Thomas},
  {Kasen}, {Howell}, {Bersier}, {Bloom}, {Kulkarni}, {Kandrashoff},
  {Filippenko}, {Silverman}, {Marcy}, {Howard}, {Isaacson}, {Maguire},
  {Suzuki}, {Tarlton}, {Pan}, {Bildsten}, {Fulton}, {Parrent}, {Sand},
  {Podsiadlowski}, {Bianco}, {Dilday}, {Graham}, {Lyman}, {James}, {Kasliwal},
  {Law}, {Quimby}, {Hook}, {Walker}, {Mazzali}, {Pian}, {Ofek}, {Gal-Yam}, \&
  {Poznanski}}]{Nugent2011}
{Nugent}, P.~E., {Sullivan}, M., {Cenko}, S.~B., {et~al.} 2011, \nat, 480, 344

\bibitem[{{Oke} \& {Gunn}(1982)}]{Oke1982}
{Oke}, J.~B., \& {Gunn}, J.~E. 1982, \pasp, 94, 586

\bibitem[{{Oke} {et~al.}(1995){Oke}, {Cohen}, {Carr}, {Cromer}, {Dingizian},
  {Harris}, {Labrecque}, {Lucinio}, {Schaal}, {Epps}, \& {Miller}}]{Oke1995}
{Oke}, J.~B., {Cohen}, J.~G., {Carr}, M., {et~al.} 1995, PASP, 107, 375

\bibitem[{{Pakmor} {et~al.}(2013){Pakmor}, {Kromer}, {Taubenberger}, \&
  {Springel}}]{Pakmor2013}
{Pakmor}, R., {Kromer}, M., {Taubenberger}, S., \& {Springel}, V. 2013, \apjl,
  770, L8

\bibitem[{{Patterson} {et~al.}(2019){Patterson}, {Bellm}, {Rusholme}, {Masci},
  {Juric}, {Krughoff}, {Golkhou}, {Graham}, {Kulkarni}, {Helou}, \& {Zwicky
  Transient Facility Collaboration}}]{Patterson2019}
{Patterson}, M.~T., {Bellm}, E.~C., {Rusholme}, B., {et~al.} 2019, \pasp, 131,
  018001

\bibitem[{{Perets}(2014)}]{Perets2014}
{Perets}, H.~B. 2014, arXiv e-prints, arXiv:1407.2254

\bibitem[{{Perets} {et~al.}(2011){Perets}, {Gal-yam}, {Crockett}, {Anderson},
  {James}, {Sullivan}, {Neill}, \& {Leonard}}]{Perets2011}
{Perets}, H.~B., {Gal-yam}, A., {Crockett}, R.~M., {et~al.} 2011, \apjl, 728,
  L36

\bibitem[{{Perets} {et~al.}(2010){Perets}, {Gal-Yam}, {Mazzali}, {Arnett},
  {Kagan}, {Filippenko}, {Li}, {Arcavi}, {Cenko}, {Fox}, {Leonard}, {Moon},
  {Sand}, {Soderberg}, {Anderson}, {James}, {Foley}, {Ganeshalingam}, {Ofek},
  {Bildsten}, {Nelemans}, {Shen}, {Weinberg}, {Metzger}, {Piro}, {Quataert},
  {Kiewe}, \& {Poznanski}}]{Perets2010}
{Perets}, H.~B., {Gal-Yam}, A., {Mazzali}, P.~A., {et~al.} 2010, \nat, 465, 322

\bibitem[{{Perley}(2019)}]{lpipe}
{Perley}, D.~A. 2019, \pasp, 131, 084503

\bibitem[{{Pfahl} {et~al.}(2009){Pfahl}, {Scannapieco}, \&
  {Bildsten}}]{Pfahl2009}
{Pfahl}, E., {Scannapieco}, E., \& {Bildsten}, L. 2009, \apjl, 695, L111

\bibitem[{{Phillips}(1993)}]{Phillips1993}
{Phillips}, M.~M. 1993, \apjl, 413, L105

\bibitem[{{Phillips} {et~al.}(2007){Phillips}, {Li}, {Frieman}, {Blinnikov},
  {DePoy}, {Prieto}, {Milne}, {Contreras}, {Folatelli}, {Morrell}, {Hamuy},
  {Suntzeff}, {Roth}, {Gonz{\'a}lez}, {Krzeminski}, {Filippenko}, {Freedman},
  {Chornock}, {Jha}, {Madore}, {Persson}, {Burns}, {Wyatt}, {Murphy}, {Foley},
  {Ganeshalingam}, {Serduke}, {Krisciunas}, {Bassett}, {Becker}, {Dilday},
  {Eastman}, {Garnavich}, {Holtzman}, {Kessler}, {Lampeitl}, {Marriner},
  {Frank}, {Marshall}, {Miknaitis}, {Sako}, {Schneider}, {van der Heyden}, \&
  {Yasuda}}]{Phillips2007}
{Phillips}, M.~M., {Li}, W., {Frieman}, J.~A., {et~al.} 2007, \pasp, 119, 360

\bibitem[{{Piro} \& {Nakar}(2014)}]{Piro2014}
{Piro}, A.~L., \& {Nakar}, E. 2014, \apj, 784, 85

\bibitem[{{Polin} {et~al.}(2019{\natexlab{a}}){Polin}, {Nugent}, \&
  {Kasen}}]{Polin2019a}
{Polin}, A., {Nugent}, P., \& {Kasen}, D. 2019{\natexlab{a}}, \apj, 873, 84

\bibitem[{{Polin} {et~al.}(2019{\natexlab{b}}){Polin}, {Nugent}, \&
  {Kasen}}]{Polin2019b}
---. 2019{\natexlab{b}}, arXiv e-prints, arXiv:1910.12434

\bibitem[{{Prentice} {et~al.}(2020){Prentice}, {Maguire}, {Fl{\"o}rs},
  {Taubenberger}, {Inserra}, {Frohmaier}, {Chen}, {Anderson}, {Ashall},
  {Clark}, {Fraser}, {Galbany}, {Gal-Yam}, {Gromadzki}, {Guti{\'e}rrez},
  {James}, {Jonker}, {Kankare}, {Leloudas}, {Magee}, {Mazzali}, {Nicholl},
  {Pursiainen}, {Skillen}, {Smartt}, {Smith}, {Vogl}, \&
  {Young}}]{Prentice2020}
{Prentice}, S.~J., {Maguire}, K., {Fl{\"o}rs}, A., {et~al.} 2020, \aap, 635,
  A186

\bibitem[{{Rigault} {et~al.}(2019){Rigault}, {Neill}, {Blagorodnova}, {Dugas},
  {Feeney}, {Walters}, {Brinnel}, {Copin}, {Fremling}, {Nordin}, \&
  {Sollerman}}]{Rigault2019}
{Rigault}, M., {Neill}, J.~D., {Blagorodnova}, N., {et~al.} 2019, \aap, 627,
  A115

\bibitem[{{Roelofs} {et~al.}(2007){Roelofs}, {Nelemans}, \&
  {Groot}}]{Roelofs2007}
{Roelofs}, G.~H.~A., {Nelemans}, G., \& {Groot}, P.~J. 2007, \mnras, 382, 685

\bibitem[{{Rosswog} {et~al.}(2008){Rosswog}, {Ramirez-Ruiz}, \&
  {Hix}}]{Rosswog2008}
{Rosswog}, S., {Ramirez-Ruiz}, E., \& {Hix}, W.~R. 2008, \apj, 679, 1385

\bibitem[{{Schlafly} \& {Finkbeiner}(2011)}]{Schlafly2011}
{Schlafly}, E.~F., \& {Finkbeiner}, D.~P. 2011, ApJ, 737, 103

\bibitem[{{Schwab}(2019)}]{Schwab2019}
{Schwab}, J. 2019, \apj, 885, 27

\bibitem[{{Sell} {et~al.}(2018){Sell}, {Arur}, {Maccarone}, {Kotak}, {Knigge},
  {Sand}, \& {Valenti}}]{Sell2018}
{Sell}, P.~H., {Arur}, K., {Maccarone}, T.~J., {et~al.} 2018, \mnras, 475, L111

\bibitem[{{Sell} {et~al.}(2015){Sell}, {Maccarone}, {Kotak}, {Knigge}, \&
  {Sand}}]{Sell2015}
{Sell}, P.~H., {Maccarone}, T.~J., {Kotak}, R., {Knigge}, C., \& {Sand}, D.~J.
  2015, MNRAS, 450, 4198

\bibitem[{{Shen} \& {Bildsten}(2009)}]{Shen2009}
{Shen}, K.~J., \& {Bildsten}, L. 2009, \apj, 699, 1365

\bibitem[{{Shen} {et~al.}(2010){Shen}, {Kasen}, {Weinberg}, {Bildsten}, \&
  {Scannapieco}}]{Shen2010}
{Shen}, K.~J., {Kasen}, D., {Weinberg}, N.~N., {Bildsten}, L., \&
  {Scannapieco}, E. 2010, ApJ, 715, 767

\bibitem[{{Shen} \& {Moore}(2014)}]{Shen2014b}
{Shen}, K.~J., \& {Moore}, K. 2014, \apj, 797, 46

\bibitem[{{Shen} {et~al.}(2019){Shen}, {Quataert}, \& {Pakmor}}]{Shen2019}
{Shen}, K.~J., {Quataert}, E., \& {Pakmor}, R. 2019, \apj, 887, 180

\bibitem[{{Silverman} {et~al.}(2009){Silverman}, {Mazzali}, {Chornock},
  {Filippenko}, {Clocchiatti}, {Phillips}, {Ganeshalingam}, \&
  {Foley}}]{Silverman2009}
{Silverman}, J.~M., {Mazzali}, P., {Chornock}, R., {et~al.} 2009, \pasp, 121,
  689

\bibitem[{{Silverman} {et~al.}(2012){Silverman}, {Foley}, {Filippenko},
  {Ganeshalingam}, {Barth}, {Chornock}, {Griffith}, {Kong}, {Lee}, {Leonard},
  {Matheson}, {Miller}, {Steele}, {Barris}, {Bloom}, {Cobb}, {Coil},
  {Desroches}, {Gates}, {Ho}, {Jha}, {Kandrashoff}, {Li}, {Mandel}, {Modjaz},
  {Moore}, {Mostardi}, {Papenkova}, {Park}, {Perley}, {Poznanski}, {Reuter},
  {Scala}, {Serduke}, {Shields}, {Swift}, {Tonry}, {Van Dyk}, {Wang}, \&
  {Wong}}]{Silverman2012}
{Silverman}, J.~M., {Foley}, R.~J., {Filippenko}, A.~V., {et~al.} 2012, \mnras,
  425, 1789

\bibitem[{{Sim} {et~al.}(2012){Sim}, {Fink}, {Kromer}, {R{\"o}pke}, {Ruiter},
  \& {Hillebrandt}}]{Sim2012}
{Sim}, S.~A., {Fink}, M., {Kromer}, M., {et~al.} 2012, \mnras, 420, 3003

\bibitem[{{Sim} {et~al.}(2010){Sim}, {R{\"o}pke}, {Hillebrandt}, {Kromer},
  {Pakmor}, {Fink}, {Ruiter}, \& {Seitenzahl}}]{Sim2010}
{Sim}, S.~A., {R{\"o}pke}, F.~K., {Hillebrandt}, W., {et~al.} 2010, \apjl, 714,
  L52

\bibitem[{{Srivastav} {et~al.}(2020){Srivastav}, {Smartt}, {Leloudas}, {Huber},
  {Chambers}, {Malesani}, {Hjorth}, {Gillanders}, {Schultz}, {Sim}, {Auchettl},
  {Fynbo}, {Gall}, {McBrien}, {Rest}, {Smith}, {Wojtak}, \&
  {Young}}]{Srivastav2020}
{Srivastav}, S., {Smartt}, S.~J., {Leloudas}, G., {et~al.} 2020, \apjl, 892,
  L24

\bibitem[{{Sullivan} {et~al.}(2006){Sullivan}, {Le Borgne}, {Pritchet},
  {Hodsman}, {Neill}, {Howell}, {Carlberg}, {Astier}, {Aubourg}, {Balam},
  {Basa}, {Conley}, {Fabbro}, {Fouchez}, {Guy}, {Hook}, {Pain},
  {Palanque-Delabrouille}, {Perrett}, {Regnault}, {Rich}, {Taillet}, {Baumont},
  {Bronder}, {Ellis}, {Filiol}, {Lusset}, {Perlmutter}, {Ripoche}, \&
  {Tao}}]{Sullivan2006}
{Sullivan}, M., {Le Borgne}, D., {Pritchet}, C.~J., {et~al.} 2006, \apj, 648,
  868

\bibitem[{{Sullivan} {et~al.}(2011){Sullivan}, {Kasliwal}, {Nugent}, {Howell},
  {Thomas}, {Ofek}, {Arcavi}, {Blake}, {Cooke}, {Gal-Yam}, {Hook}, {Mazzali},
  {Podsiadlowski}, {Quimby}, {Bildsten}, {Bloom}, {Cenko}, {Kulkarni}, {Law},
  \& {Poznanski}}]{Sullivan2011}
{Sullivan}, M., {Kasliwal}, M.~M., {Nugent}, P.~E., {et~al.} 2011, ApJ, 732,
  118

\bibitem[{{Sun} \& {Gal-Yam}(2017)}]{Sun2017}
{Sun}, F., \& {Gal-Yam}, A. 2017, arXiv e-prints, arXiv:1707.02543

\bibitem[{{Tachibana} \& {Miller}(2018)}]{Tachibana2018}
{Tachibana}, Y., \& {Miller}, A.~A. 2018, \pasp, 130, 128001

\bibitem[{{Taddia} {et~al.}(2018){Taddia}, {Stritzinger}, {Bersten}, {Baron},
  {Burns}, {Contreras}, {Holmbo}, {Hsiao}, {Morrell}, {Phillips}, {Sollerman},
  \& {Suntzeff}}]{Taddia2018}
{Taddia}, F., {Stritzinger}, M.~D., {Bersten}, M., {et~al.} 2018, \aap, 609,
  A136

\bibitem[{{Taubenberger}(2017)}]{Taubenberger2017}
{Taubenberger}, S. 2017, {The Extremes of Thermonuclear Supernovae}, ed. A.~W.
  {Alsabti} \& P.~{Murdin}, 317

\bibitem[{{Tauris} {et~al.}(2015){Tauris}, {Langer}, \&
  {Podsiadlowski}}]{Tauris2015}
{Tauris}, T.~M., {Langer}, N., \& {Podsiadlowski}, P. 2015, MNRAS, 451, 2123

\bibitem[{{Tomasella} {et~al.}(2020){Tomasella}, {Stritzinger}, {Benetti},
  {Elias-Rosa}, {Cappellaro}, {Kankare}, {Lundqvist}, {Magee}, {Maguire},
  {Pastorello}, {Prentice}, \& {Reguitti}}]{Tomasella2020}
{Tomasella}, L., {Stritzinger}, M., {Benetti}, S., {et~al.} 2020, arXiv
  e-prints, arXiv:2002.00393

\bibitem[{{Tonry} {et~al.}(2019{\natexlab{a}}){Tonry}, {Denneau}, {Heinze},
  {Weiland}, {Flewelling}, {Stalder}, {Rest}, {Stubbs}, {Smith}, {Smartt},
  {Young}, {Srivastav}, {McBrien}, {O'Neill}, {Clark}, {Fulton}, {Gillanders},
  {Chen}, \& {Wright}}]{SN2019hty}
{Tonry}, J., {Denneau}, L., {Heinze}, A., {et~al.} 2019{\natexlab{a}},
  Transient Name Server Discovery Report, 2019-1035, 1

\bibitem[{{Tonry} {et~al.}(2019{\natexlab{b}}){Tonry}, {Denneau}, {Heinze},
  {Weiland}, {Flewelling}, {Stalder}, {Rest}, {Stubbs}, {Smith}, {Smartt},
  {Young}, {Srivastav}, {McBrien}, {O'Neill}, {Clark}, {Fulton}, {Gillanders},
  {Dobson}, {Chen}, \& {Wright}}]{SN2019pxu}
---. 2019{\natexlab{b}}, Transient Name Server Discovery Report, 2019-1787, 1

\bibitem[{{Tonry} {et~al.}(2018){Tonry}, {Denneau}, {Heinze}, {Stalder},
  {Smith}, {Smartt}, {Stubbs}, {Weiland}, \& {Rest}}]{Tonry2018}
{Tonry}, J.~L., {Denneau}, L., {Heinze}, A.~N., {et~al.} 2018, \pasp, 130,
  064505

\bibitem[{{Toonen} {et~al.}(2018){Toonen}, {Perets}, {Igoshev}, {Michaely}, \&
  {Zenati}}]{Toonen2018}
{Toonen}, S., {Perets}, H.~B., {Igoshev}, A.~P., {Michaely}, E., \& {Zenati},
  Y. 2018, \aap, 619, A53

\bibitem[{{Townsley} {et~al.}(2012){Townsley}, {Moore}, \&
  {Bildsten}}]{Townsley2012}
{Townsley}, D.~M., {Moore}, K., \& {Bildsten}, L. 2012, \apj, 755, 4

\bibitem[{{Valenti} {et~al.}(2014){Valenti}, {Yuan}, {Taubenberger}, {Maguire},
  {Pastorello}, {Benetti}, {Smartt}, {Cappellaro}, {Howell}, {Bildsten},
  {Moore}, {Stritzinger}, {Anderson}, {Benitez-Herrera}, {Bufano},
  {Gonzalez-Gaitan}, {McCrum}, {Pignata}, {Fraser}, {Gal-Yam}, {Le Guillou},
  {Inserra}, {Reichart}, {Scalzo}, {Sullivan}, {Yaron}, \&
  {Young}}]{Valenti2014}
{Valenti}, S., {Yuan}, F., {Taubenberger}, S., {et~al.} 2014, MNRAS, 437, 1519

\bibitem[{{van Dokkum} {et~al.}(2016){van Dokkum}, {Abraham}, {Brodie},
  {Conroy}, {Danieli}, {Merritt}, {Mowla}, {Romanowsky}, \&
  {Zhang}}]{vanDokkum2016}
{van Dokkum}, P., {Abraham}, R., {Brodie}, J., {et~al.} 2016, \apjl, 828, L6

\bibitem[{{van Dokkum} {et~al.}(2017){van Dokkum}, {Abraham}, {Romanowsky},
  {Brodie}, {Conroy}, {Danieli}, {Lokhorst}, {Merritt}, {Mowla}, \&
  {Zhang}}]{vanDokkum2017}
{van Dokkum}, P., {Abraham}, R., {Romanowsky}, A.~J., {et~al.} 2017, \apjl,
  844, L11

\bibitem[{{van Dokkum} {et~al.}(2018){van Dokkum}, {Cohen}, {Danieli},
  {Kruijssen}, {Romanowsky}, {Merritt}, {Abraham}, {Brodie}, {Conroy},
  {Lokhorst}, {Mowla}, {O'Sullivan}, \& {Zhang}}]{vanDokkum2018}
{van Dokkum}, P., {Cohen}, Y., {Danieli}, S., {et~al.} 2018, \apjl, 856, L30

\bibitem[{{Virtanen} {et~al.}(2019){Virtanen}, {Gommers}, {Oliphant},
  {Haberland}, {Reddy}, {Cournapeau}, {Burovski}, {Peterson}, {Weckesser},
  {Bright}, {van der Walt}, {Brett}, {Wilson}, {Jarrod Millman}, {Mayorov},
  {Nelson}, {Jones}, {Kern}, {Larson}, {Carey}, {Polat}, {Feng}, {Moore}, {Vand
  erPlas}, {Laxalde}, {Perktold}, {Cimrman}, {Henriksen}, {Quintero}, {Harris},
  {Archibald}, {Ribeiro}, {Pedregosa}, {van Mulbregt}, \&
  {Contributors}}]{Virtanen2019}
{Virtanen}, P., {Gommers}, R., {Oliphant}, T.~E., {et~al.} 2019, arXiv
  e-prints, arXiv:1907.10121

\bibitem[{{Waldman} {et~al.}(2011){Waldman}, {Sauer}, {Livne}, {Perets},
  {Glasner}, {Mazzali}, {Truran}, \& {Gal-Yam}}]{Waldman2011}
{Waldman}, R., {Sauer}, D., {Livne}, E., {et~al.} 2011, \apj, 738, 21

\bibitem[{{Wiggins}(2018)}]{SN2018gwo}
{Wiggins}, P. 2018, Transient Name Server Discovery Report, 2018-1459, 1

\bibitem[{{Woosley}(2019)}]{Woosley2019}
{Woosley}, S.~E. 2019, \apj, 878, 49

\bibitem[{{Woosley} \& {Kasen}(2011)}]{Woosley2011}
{Woosley}, S.~E., \& {Kasen}, D. 2011, \apj, 734, 38

\bibitem[{{Woosley} {et~al.}(1986){Woosley}, {Taam}, \& {Weaver}}]{Woosley1986}
{Woosley}, S.~E., {Taam}, R.~E., \& {Weaver}, T.~A. 1986, \apj, 301, 601

\bibitem[{{Woosley} \& {Weaver}(1994)}]{Woosley1994}
{Woosley}, S.~E., \& {Weaver}, T.~A. 1994, \apj, 423, 371

\bibitem[{{Yao} {et~al.}(2019){Yao}, {Miller}, {Kulkarni}, {Bulla}, {Masci},
  {Goldstein}, {Goobar}, {Nugent}, {Dugas}, {Blagorodnova}, {Neill}, {Rigault},
  {Sollerman}, {Nordin}, {Bellm}, {Cenko}, {De}, {Dhawan}, {Feindt},
  {Fremling}, {Gatkine}, {Graham}, {Graham}, {Ho}, {Hung}, {Kasliwal},
  {Kupfer}, {Laher}, {Perley}, {Rusholme}, {Shupe}, {Soumagnac}, {Taggart},
  {Walters}, \& {Yan}}]{Yao2019}
{Yao}, Y., {Miller}, A.~A., {Kulkarni}, S.~R., {et~al.} 2019, \apj, 886, 152

\bibitem[{{Yaron} \& {Gal-Yam}(2012)}]{Yaron2012}
{Yaron}, O., \& {Gal-Yam}, A. 2012, PASP, 124, 668

\bibitem[{{Yuan} {et~al.}(2013){Yuan}, {Kobayashi}, {Schmidt}, {Podsiadlowski},
  {Sim}, \& {Scalzo}}]{Yuan2013}
{Yuan}, F., {Kobayashi}, C., {Schmidt}, B.~P., {et~al.} 2013, \mnras, 432, 1680

\bibitem[{{Zackay} {et~al.}(2016){Zackay}, {Ofek}, \& {Gal-Yam}}]{Zackay2016}
{Zackay}, B., {Ofek}, E.~O., \& {Gal-Yam}, A. 2016, \apj, 830, 27

\bibitem[{{Zenati} {et~al.}(2019){Zenati}, {Perets}, \& {Toonen}}]{Zenati2019}
{Zenati}, Y., {Perets}, H.~B., \& {Toonen}, S. 2019, \mnras, 486, 1805

\end{thebibliography}

\startlongtable
\begin{deluxetable*}{ccccccccc}
\tabletypesize{\footnotesize}
\tablecaption{Light curve fit parameters for the sample of Ca-rich gap transients presented in this paper, together with the fits for the literature sample of Ca-rich gap transients. $t_p$ denotes the time of peak in the respective filter, $m_p$ denotes the peak apparent magnitude (corrected for Galactic extinction using the maps of \citet{Schlafly2011}), $M_p$ denotes the peak absolute magnitude, $t_{r, 1/2}$ denotes the rise time to peak from half of the peak flux and $t_{f, 1/2}$ denotes the fall time from peak to half the peak flux. $\Delta m_7$ denotes the drop in magnitudes from time of peak to $7\,\rm{d}$ after peak. Photometry for some of the sources in the literature were obtained from the Open Supernova Catalog \citep{Guillochon2017}. The data were originally published in [1] \citet{Perets2010}, [2] \citet{Sullivan2011}, [3] \citet{Kasliwal2012a}, [4] \citet{Valenti2014}, [5] \citet{Lunnan2017}, [6] \citet{De2018b} and [7] \citet{Galbany2019}. This table will be available in its entirety in machine-readable form.}
\tablehead{
\colhead{Object} &
\colhead{Filter} &
\colhead{$t_p$}  &
\colhead{$m_p$} &
\colhead{ $M_p$} &
\colhead{$t_{r, 1/2}$}&
\colhead{$t_{f, 1/2}$} &
\colhead{$\Delta m_7$} &
\colhead{Ref.} \\
\colhead{} &
\colhead{} &
\colhead{MJD}  &
\colhead{} &
\colhead{} &
\colhead{(days)} &
\colhead{(days)} &
\colhead{(mag)} &
\colhead{}
}
\startdata
SN\,2018ckd & $r$ & $58277.67 \pm 0.30$ & $18.92 \pm 0.02$ & $-16.20$ & $6.17 \pm 0.29$ & $8.76 \pm 0.39$ & $0.53 \pm 0.04$ & This work \\
SN\,2018lqo & $r$ & $58351.92 \pm 0.23$ & $19.62 \pm 0.02$ & $-16.21$ & $7.35 \pm 0.26$ & $10.50 \pm 0.77$ & $0.41 \pm 0.02$ & This work \\
SN\,2018lqo & $g$ & $58350.03 \pm 0.38$ & $20.09 \pm 0.05$ & $-15.74$ & $5.81 \pm 0.30$ & $7.70 \pm 0.42$ & $0.64 \pm 0.06$ & This work \\
SN\,2018lqo & $i$ & $58352.75 \pm 2.51$ & $19.54 \pm 0.14$ & $-16.29$ & $7.73 \pm 1.73$ & ... & $0.30 \pm 0.11$ & This work \\
SN\,2018lqu & $r$ & $58370.83 \pm 2.24$ & $19.57 \pm 0.31$ & $-16.44$ & $7.40 \pm 1.35$ & ... & ... &This work \\
SN\,2018kjy & $r$ & $58460.93 \pm 0.39$ & $18.86 \pm 0.06$ & $-15.62$ & ... & $12.53 \pm 1.70$ & $0.30 \pm 0.06$ & This work \\
SN\,2018kjy & $g$ & $58457.83 \pm 1.42$ & $19.82 \pm 0.23$ & $-14.66$ & ... & ... & $0.58 \pm 0.34$ & This work \\
SN\,2019hty & $r$ & $58658.23 \pm 1.40$ & $18.72 \pm 0.09$ & $-16.29$ & $9.50 \pm 1.20$ & $12.65 \pm 1.18$ & $0.28 \pm 0.06$ & This work \\
SN\,2019hty & $g$ & $58655.40 \pm 1.76$ & $19.21 \pm 0.07$ & $-15.80$ & ... & $7.33 \pm 1.40$ & $0.68 \pm 0.27$ & This work \\
SN\,2019ofm & $r$ & $58724.45 \pm 2.06$ & $18.79 \pm 0.09$ & $-16.84$ & ... & ... & $0.25 \pm 0.10$ & This work \\
\enddata
\label{tab:lcfits}
\end{deluxetable*}

\startlongtable
\begin{deluxetable*}{lccccc}
\tabletypesize{\footnotesize}
\tablecaption{Spectral fit parameters for the sample of Ca-rich gap transients presented in this paper, together with the fits for the literature sample of Ca-rich gap transients. For each spectrum, we measure both the He\,I $\lambda5876$ and $\lambda7065$ velocity (if detected) for the Ca-Ib/c events, and only the Si\,II $\lambda6355$ velocity for the Ca-Ia events. We indicate the line measured in brackets next to the velocity measurements for each spectrum phase. Values denoted by * indicate epochs where the signal to noise ratio of the spectrum was not high enough in the region of interest to measure the specific parameter. The velocity for SN\,2012hn was measured using the feature near $5800$\,\AA at peak light since it does not exhibit He signatures. For phases where the spectrum exhibited nebular emission features, we report the measured [Ca II]/[O I] ratio or lower limits in case [O I] is not detected (see text). Archival spectra were obtained from the WISEReP repository \citep{Yaron2012}. The data were originally published in [1] \citet{Perets2010}, [2] \citet{Sullivan2011}, [3] \citet{Kasliwal2012a}, [4] \citet{Valenti2014}, [5] \citet{Lunnan2017}, [6] \citet{De2018b} and [7] \citet{Galbany2019}. This table will be available in its entirety in machine-readable form.}
\tablehead{
\colhead{Object}  & 
\colhead{Phase} & 
\colhead{$V_1$}&
\colhead{$V_2$} & 
\colhead{[Ca II] / [O I]} &
\colhead{Ref.}\\
\colhead{} &
\colhead{(days)} & 
\colhead{(km s$^{-1}$)} &
\colhead{(km s$^{-1}$)} & 
\colhead{} & \\
}
\startdata
    SN\,2018ckd & +1 & $9660 \pm 110$ ($\lambda5876$) & $10090 \pm 140$ ($\lambda7065$)  & ... &  This work\\
    SN\,2018ckd & +10 & $8260 \pm 650$ ($\lambda5876$) &  ... & ... &  This work\\
    SN\,2018ckd & +58 & ... & ... & $>3.38$ &   This work\\
    SN\,2018lqo & -2 & $8230 \pm 150$ ($\lambda5876$) & $8270 \pm 100$ ($\lambda7065$) & ... &  This work\\
    SN\,2018lqo & +49 & $5090 \pm 310$ ($\lambda5876$) & ...  & $>12.51$ &  This work\\
    SN\,2018lqu & +1 & $11100 \pm 410$ ($\lambda5876$) & $10550 \pm 280$ ($\lambda7065$) & ... &  This work\\
    SN\,2018lqu & +31 & $5720 \pm 500$ ($\lambda5876$) & ...* & $>8.38$ & This work \\
    SN\,2018gwo & +23 & $5150 \pm 570$ ($\lambda5876$) & ... & ... & This work\\
    SN\,2018gwo & +53 & $4780 \pm 100$ ($\lambda5876$) &  $6660 \pm 80$ ($\lambda7065$) & $5.16 \pm 0.08$ &  This work\\
    SN\,2018gwo  & +146 & ... & ... & $3.98 \pm 0.06$ &  This work\\
\enddata
\label{tab:specfits}
\end{deluxetable*}

\appendix
\section{Transients in the control sample}
\label{sec:control}

Here we discuss the photometric and spectroscopic properties of the transients that passed the selection criteria for follow-up but did not exhibit high [Ca II] / [O I] ratio in their nebular phase spectra. We summarize the photometric and spectroscopic properties of these transients in Table \ref{tab:summary}. The control sample consists of 4 SNe\,Ib, 5 SNe\,Ic, 1 SN\,Ic-BL, 2 SNe Ib/c and 3 SNe\,Ia. Figure \ref{fig:controllc1} shows forced photometry light curves of these transients, while Figure \ref{fig:control_spec_1} shows a collage of the spectroscopic data for each object. The complete log of the spectroscopic follow-up for these objects is presented in Table \ref{tab:spectra}, which will be released on WISeREP together with the photometry upon publication. We plot the original reduced spectra for the spectroscopy epochs near peak light for each object. For the nebular phase spectra, we show the original reduced spectra for events that did not have a large host contamination. For other nebular spectra, we attempted to fit a polynomial to the underlying host continuum to subtract the host features and show the subtracted spectrum to highlight the broad nebular emission features of [O I] and [Ca II]. In cases where the host background was not smooth and had features sharper than $\approx 1000$\,\AA\, (usually in the case of S0 / E type galaxies) such that the nebular emission features were not easily measurable, we attempted more careful host subtraction using \texttt{superfit} \citep{Howell2005}. In these cases, the spectra figures show the unsubtracted spectra, and the \texttt{superfit}-subtracted spectra are shown in Figure \ref{fig:hostsubneb}.

In addition to the objects discussed here, we note the case of the peculiar SN\,2019ehk in the galaxy M100. SN\,2019ehk was reported to the TNS by \citet{SN2019ehk} and an early spectrum was reported by \citet{Dimitriadis2019}, which exhibited a reddened featureless continuum with `flash'-ionized lines (see e.g. \citealt{GalYam2014}) of He~II $\lambda4686$ and H$\alpha$. We obtained follow-up spectra of the event near peak light with DBSP, which showed a reddened continuum with strong photospheric He absorption features and weaker H features, similar to Type IIb SNe. Specifically, the peak light spectra show signatures of H$\beta$ and H$\gamma$ absorption together with flat-bottomed feature near H$\alpha$ blended with the nearby He~I $\lambda6678$ line. The flat-bottomed H$\alpha$ feature is characteristic of several well-studied Type~IIb SNe like SN\,2001ig \citep{Silverman2009} and SN\,2011dh \citep{Marion2014}. Curiously, this object shows strong [Ca II] lines in our early nebular phase spectrum from SEDM and LRIS, similar to several Ca-rich transients in this sample. However, the clear presence of H in the early flash spectra and at peak light exclude it from our sample. Additionally, the deep Na~I D absorption detected in its spectrum suggests significant host reddening by $A_V \gtrsim 3$\,mag, making it intrinsically luminous ($M_p \lesssim -17$). This object may be similar to the Type~IIb iPTF\,15eqv which exhibited high [Ca~II]/[O~I] in late-time spectra \citep{Milisavljevic2017}, and we defer the discussion of this object to future work.

\subsection{Spectroscopic classification}
We summarize the detection, environments and properties of each transient in the control sample, and in particular, highlight how we exclude them from the group of Ca-rich gap transients. In the next section, we use this discussion to compare the properties of these transients to those in the Ca-rich sample.

\textit{SN\,2019ccm} was found on top of the spiral arm of a Sa-type galaxy at $z = 0.015$ and peaked at an absolute magnitude of $M_r \approx -16.4\,\rm{mag}$ (without correcting for its host extinction). A spectrum taken at peak light shows characteristic features of a SN\,Ib at peak on the red side (the blue side spectrum was not recorded due to a readout issue on the LRIS detector), as well as a deep Na I D feature at the host redshift, suggesting that the low luminosity is partly due to host extinction. A LRIS spectrum taken $\approx 180\,\rm{d}$ after peak shows [O I] and [Ca II] emission of nearly equal strengths, ruling out a Ca-rich classification. We note that the [Ca II]/[O I] ratio ($\approx 1.18$) is likely overestimated because the significant host extinction would only increase the observed ratio.

\textit{SN\,2019txl} was found on the arm of a spiral galaxy at $z = 0.034$ and peaked at an absolute magnitude of $M_r \approx -16.2\,\rm{mag}$. The peak light spectrum shows typical features of a SN\,Ib at peak, together with clear Na I D absorption at the host redshift, confirming it as a reddened normal Type Ib SN. A nebular phase spectrum taken $\approx +330\,\rm{d}$ from peak shows clear signatures of [O I] and [Ca II] emission with a [Ca II]/[O I] of $\approx 0.9$ (without correcting for host extinction), thus excluding it from the Ca-rich sample. 

\textit{SN\,2019txt} was found on the disk of a nearly edge-on disk galaxy at $z = 0.026$ and peaked at an absolute magnitude of $M_r \approx -15.9\,\rm{mag}$. The peak light spectrum is relatively noisy but still clearly shows features of a Type Ib SN at peak, as well as prominent Na I D absorption at the host redshift, consistent with a reddened Type Ib SN. The nebular phase spectrum at $\approx 220\,\rm{d}$ from peak light shows clear [O I] and [Ca II] emission with [Ca II]/[O I] of $\approx 1.3$ without host extinction correction, thus excluding it from the Ca-rich sample.

\textit{SN\,2019mjo} stands out as a peculiar SN\,Ib found in the outskirts of an elliptical galaxy at $z \approx 0.041$. Its peak light spectrum is reddened with strong He lines and it exhibits a very slow transition in spectroscopic properties. The source did not turn nebular even in our latest spectrum at $\approx +180\,\rm{d}$ from peak light, and hence does not satisfy our criterion for a fast nebular phase transition. We defer conclusions about the nature of this event to a forthcoming publication that will present the full dataset on this source (K. De et al. in prep.).

\textit{SN\,2018dbg} was found close to the nucleus of a grand spiral host galaxy at $z \approx 0.015$, peaking at an absolute magnitude of $M_r \approx -16.6\,\rm{mag}$. We were unable to secure a peak light spectrum of SN\,2018dbg, but secured a spectrum at $\approx 35\,\rm{d}$ after peak, where the spectrum is still dominated by strong photospheric phase lines of O, Ca II and possibly He I. Using \texttt{superfit} \citep{Howell2005} to subtract the underlying continuum, we find an excellent match to the spectrum of the Type Ib SN\,1990U at $\approx +41\,\rm{d}$ after peak, consistent with the photometric phase ($\approx + 30\,\rm{d}$). We thus classify it as a SN\,Ib/c. Since all Ca-rich gap transients start exhibiting strong nebular [Ca II] emission features at this phase, this object does not satisfy our early nebular phase transition criterion and is excluded from the sample.

\begin{figure*}[!ht]
    \centering
    \includegraphics[width=0.98\textwidth]{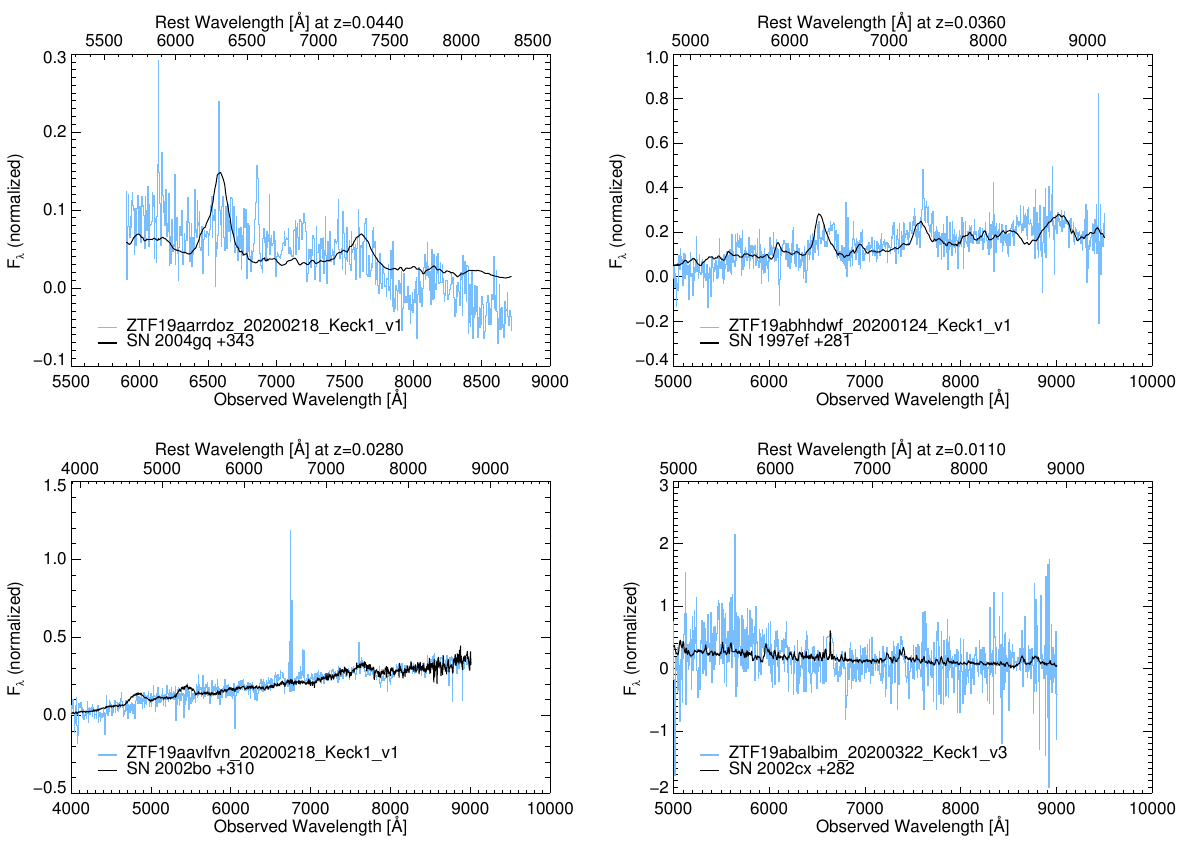}
    \caption{Host-subtracted late-time spectra of events in the sample that were heavily contaminated by host galaxy light and thus required host-subtraction using \texttt{superfit} \citep{Howell2005}. (Top Left) Late-time spectrum of SN\,2019txr with host features subtracted using \texttt{superfit}. The observed host-subtracted spectrum is shown in blue, while the black line shows the best match spectrum of the Type Ib SN\,2004gq at $297\,\rm{d}$ after peak. (Top right) Late-time spectrum of SN\,2019ouq plotted with the late-time spectrum of Type Ic SN\,1997ef at $\approx 270\,\rm{d}$ after peak light. (Bottom left) Host subtracted late-time spectrum of SN\,2019gau at $\approx 270\,\rm{d}$, compared against a late-time spectrum of the Type Ia SN\,2002bo. However, we do not find a convincing match as several features in the subtracted spectrum are not well matched, and present the fit here for completeness. (Bottom right) Host-subtracted late-time spectrum of SN\,2019ttf at $\approx 210\,\rm{d}$ compared to the late-time spectrum of SN\,2002cx at $\approx 280\,\rm{d}$.}
    \label{fig:hostsubneb}
\end{figure*}

\textit{SN\,2019txr} was found close to the nucleus of an irregular spiral galaxy at $z = 0.044$, and peaked at an absolute magnitude of $M_r \approx -16.7\,\rm{mag}$. The peak light spectrum is relatively noisy and we can only identify P-Cygni features of Ca II, O I and possibly He I; however, we classify it as a SN\,Ib/c due to the uncertain identification of He I. We cannot identify any Na I D absorption due to the noisy nature of the spectrum. We obtained a nebular phase spectrum at $\approx 270\,\rm{d}$ from peak, which we find to be dominated by host light. We visually identify a weak nebular emission peak around the [O I] transition, but the [Ca II] emission is not detected. We show a host-subtracted spectrum of the object matched with \texttt{superfit} with the Type Ib SN\,2004gq $\approx 300\,\rm{d}$ after peak. Although the features are very weak, the host-subtracted spectrum shows the existence of a broad emission feature around [O I] and possible [Ca II]. Given the host dominated spectrum, we are unable to measure a [Ca II]/[O I], but use the detection of [O I] and the weak detection of the nearby [Ca II] line to constrain the [Ca II]/[O I] ratio to $< 1$, excluding it from the Ca-rich sample.

\textit{SN\,2018fob} was found on the spiral arm of a disk galaxy at $z = 0.029$, and peaked at an absolute magnitude of $M_r \approx -16.9\,\rm{mag}$. The peak light spectra do not show any He signatures, and are consistent with a SN\,Ic. Na I D absorption is clearly detected at the host redshift, suggesting host extinction. The nebular phase spectrum at $\approx 210\,\rm{d}$ shows a strong [O I] line and a weak [Ca II] line with [Ca II]/[O I] $\approx 0.87$, thus excluding it from our sample.

\textit{SN\,2019yz} is the lowest redshift object in this sample, and is consistent with a reddened SN\,Ic in the disk of UGC\,09977 based on the prominent Na I D absorption in its peak light spectrum. The peak spectrum was obtained from TNS and was originally obtained by \citet{Burke2019}. The light curve peaks at an observed absolute magnitude of $M_r \approx -16.63\,\rm{mag}$. The nebular phase spectrum at $\approx +210\,\rm{d}$ shows strong [O I] emission with [Ca II]/[O I] $\approx 0.6$, excluding it from our sample of Ca-rich events.

\textit{SN\,2019abb} was found on top of an irregular blue galaxy at $z = 0.015$, and peaked at an absolute magnitude of $M_r \approx -16.6\,\rm{mag}$. The peak light spectrum shows characteristic features of a SN\,Ic with no obvious He signatures, as well as clear Na I D absorption at the host redshift suggesting significant host extinction. A spectrum taken $\approx 60\,\rm{d}$ after peak still shows photospheric phase features suggesting slow spectral evolution. The nebular phase spectrum obtained at $\approx +350\,\rm{d}$ is dominated by the underlying host, but clearly shows both the nebular [O I] and [Ca II] emission lines with [Ca II]/[O I] $\approx 0.8$, thus excluding the object from our sample.

\textit{SN\,2019ape} was detected on top of a yellow early type galaxy at $z = 0.020$, and peaked at an absolute magnitude of $M_r \approx -16.6\,\rm{mag}$. Although the galaxy morphology is early-type, the SDSS as well as the SN spectra show clear H$\alpha$ emission. The peak light spectrum is characteristic of a SN\,Ic with no He signatures. Na I D absorption is also detected in the peak light spectrum, confirming host reddening. The nebular phase spectrum taken at $\approx +280\,\rm{d}$ shows clear [O I] and [Ca II] emission lines with [Ca II]/[O I] $\approx 0.9$ thus excluding it from our sample. A complete analysis of this object will be presented in a forthcoming publication (I. Irani et al. in prep.).

\textit{SN\,2019ouq} was found in the disk of a nearly edge-on disk galaxy at $z = 0.036$, and peaked at an absolute magnitude of $M_r \approx -16.9\,\rm{mag}$. The peak light spectrum exhibits a highly reddened continuum but with clear broad P-Cygni features of O I and Ca II. Using \texttt{superfit} to subtract the host emission, we find that the peak spectrum is well matched to the Type Ic SN\,1994I about $7\,\rm{d}$ after peak. The same fit suggests that an extinction of $A_V \approx 1$\,mag is required to match the continuum. The nebular spectrum obtained at $\approx 170\,\rm{d}$ from maximum light is completely dominated by the underlying host galaxy continuum, making it difficult to identify the nebular [O I] and [Ca II] features directly. We thus used \texttt{superfit} to subtract the host features and show a host-subtracted spectrum in Figure \ref{fig:hostsubneb}. As shown, the nebular phase spectrum is consistent with a late-time spectrum of the Type Ic SN\,1997ef at $\approx 281\,\rm{d}$. The host-subtracted late-time spectrum also shows similarities to the late-time spectra of the Type Ic SNe SN\,2006aj and SN\,1994I. In particular, the host subtracted spectrum exhibits a stronger [O I] line compared to the [Ca II] line, constraining [Ca II]/[O I] $\lesssim 1$, thus excluding it from the Ca-rich sample.

\textit{SN\,2018kqr} was detected inside a blue irregular galaxy at $z = 0.045$, and peaked at an absolute magnitude of $M_r \approx -16.8\,\rm{mag}$. The peak light spectrum shows broad features consistent with a SN\,Ic-BL around peak light. Na I D absorption is not clearly detected at the host redshift. We obtained a follow-up spectrum $\approx 16\,\rm{d}$ after peak light where the source still exhibited strong photospheric features consistent with a SN\,Ic. However, we were unable to obtain a follow-up nebular spectrum for this object. Nevertheless, the slow photospheric phase evolution of this evolution is consistent with a normal SN\,Ic, and exclude it from the Ca-rich sample discussed in this work.

\textit{SN\,2019gau} was detected close to the core of a disk galaxy at $z = 0.028$, and peaked at an absolute magnitude of $M_r \approx -16.8\,\rm{mag}$. A low resolution SEDM spectrum taken near peak shows a strong Si II line and the Ca II NIR triplet in P-Cygni absorption, leading to the SN Ia classification of this event. We obtained a late-time spectrum of the source at $\approx 260\,\rm{d}$ from peak, and find it to be dominated by the host galaxy light. We attempted host subtraction using \texttt{superfit}, but were unable to find a good match to the host subtracted features. For completness, we present the best-fit host subtracted spectrum in Figire \ref{fig:hostsubneb} compared to the best match in \texttt{superfit} to the late time spectrum of SN\,2002bo. Unlike the Ca-Ia objects that show strong [Ca II] emission and no [O I] emission, this host-subtracted spectrum does not show any strong [Ca II] emission, leading us to exclude it from the Ca-rich sample.

\textit{SN\,2019gsc} was detected on a inside an irregular blue galaxy at $z = 0.011$, and peaked at a faint absolute magnitude of $M_r = -13.90\,\rm{mag}$. Its peak light spectrum exhibits low velocity Si\,II lines similar to 02cx-like SNe\,Ia \citep{Li2003}, while its faint peak luminosity makes it similar to the lowest luminosity member SN\,2008ha \citep{Foley2009}. Our spectrum taken at $\approx+30\,\rm{d}$ from peak shows that the spectrum is still photospheric and dominated by several low velocity P-Cygni features. Additional data on SN\,2019gsc has been published in \citealt{Srivastav2020} and \citealt{Tomasella2020}, which show that the source does not turn nebular even up to $\approx 60\,\rm{d}$ from peak light, consistent with our data and with what is typically observed in this class \citep{Foley2016}. The absence of an early nebular phase transition excludes it from the Ca-rich sample.

\textit{SN\,2019ttf} was found on top of an irregular star forming galaxy at $z = 0.011$, and peaked at a faint absolute magnitude of $M_r = -14.0\,\rm{mag}$. The spectrum taken at $\approx +10\,\rm{d}$ from peak shows low velocity lines similar to the 02cx-like SNe\,Ia SN\,2008ha \citep{Foley2009} and SN\,2005hk \citep{Phillips2007}. Its low peak luminosity is similar to SN\,2019gsc. We obtained a late-time spectrum of the object at $\approx 230\,\rm{d}$, which was dominated by the underlying host galaxy light. We show a host-subtracted spectrum using \texttt{superfit} in Figure \ref{fig:hostsubneb}. The subtracted spectrum is relatively noisy, but we note two detected features of Na I and [Ca II] (near 5800\AA\, and 7300\AA\, respectively) that are similar to SN\,2002cx at a similar phase. As in the case of 02cx-like objects that do not become completely nebular at late times, we find a weak Na I P-Cygni profile, and thus exclude this object from the Ca-rich sample.

\subsection{Candidate selection and false positives}
In Section \ref{sec:cand_select}, we presented our selection criterion for identifying candidate Ca-rich gap transients in order to prioritize nebular phase follow-up. We now evaluate the merits and disadvantages of our selection scheme in the context of understanding the broader population of Ca-rich transients. We start with a comparison of our selection criteria to that of \citet{Kasliwal2012a}. Unlike that work, we did not select candidates based on their photometric evolution or spectroscopic velocities. This choice makes us sensitive to events with broader light curves such as SN\,2019pxu, which appears to be a more luminous and slower evolving member of this class. 

However, we do select events based on their peak luminosity, with a cutoff at $M = -17\,\rm{mag}$. We thus caution that certain spectroscopic sub-types may be underrepresented in this sample, e.g. the Ca-Ia events appear to exhibit systematically higher peak luminosity than the Ca-Ib/c events. Indeed, the only Ca-Ia object in the sample SN\,2019ofm is also the faintest SN\,Ia in the CLU sample (barring the low velocity 08ha-like events), suggesting that Ca-Ia events discussed here could represent the tip of the Ca-rich sample in the broader 91bg-like SN\,Ia population. Resolving this issue would require a similar experiment targeting brighter SNe\,Ia, which will inevitably include other known populations of faint SNe\,Ia.

As in \citet{Kasliwal2012a}, we do not select events based on their spectroscopic type at peak light. However, the low spectroscopic velocities in SN\,2018kjy would not have passed the criteria in \citet{Kasliwal2012a} since they required `normal' photospheric phase velocities. Our follow-up campaign revealed that despite its peculiar low spectroscopic velocities at peak, the later evolution of SN\,2018kjy establishes its membership in the class of Ca-rich gap transients. In the context of low velocity events, it is important to highlight the contamination of 02cx-like SNe\,Ia. Two of the events in the control sample were spectroscopically similar to 02cx-like objects near peak light (similar to SN\,2008ha; \citealt{Foley2009}) with low spectroscopic velocities at peak (like SN\,2018kjy). Yet unlike SN\,2018kjy (and the rest of the Ca-rich class), their spectra do not turn nebular at late phases and exhibit a pseudo continuum of emission lines. 

Next, we discuss the host environments of these events, re-iterating that our selection criteria was agnostic to the host type and environment. In comparison to the SNe\,Ib/c in the control sample (which were primarily found in star-forming late-type galaxies), a striking characteristic of the sample of the Ca-Ib/c objects is their preference for early type galaxies and old environments. SN\,2019pxu is the only exception, but is still at a large projected offset of $\approx 18$\,kpc from its host, suggesting that spectroscopically classified SNe\,Ib/c in old environments could be used to select likely Ca-rich gap transients near peak light. Yet, SN\,2019ape serves as an important exception to this trend as a SN\,Ic in an early type galaxy, suggesting that this criterion also has its own false positives despite producing a relatively high success rate (six out of seven SNe\,Ib/c in early type galaxies in this sample turned out to be Ca-rich events). However, nearly all the low luminosity SNe\,Ia in the sample (including two 02cx-like events) are in late-type galaxies -- as such, the environment of the only Ca-Ia event in the sample SN\,2019ofm is not exceptional.

In terms of the photometric and spectroscopic properties of the transients, the sample presented in this work was selected using the smallest possible selection criteria to identify these faint transients in the local universe. The final confirmation of a Ca-rich gap transient, however, is derived from nebular phase spectroscopy at late times. As such, it is instructive to examine if the confirmed sample of Ca-rich gap transients can be differentiated based on the peak light spectra and photometric evolution in order to guide future searches for these events. In Figure \ref{fig:lumwidth}, we show the luminosity-width phase space of the control sample of transients compared to the Ca-rich gap transients analyzed in the work. We make this comparison in order to examine if Ca-rich gap transients can be identified from their peak luminosity and light curve evolution near peak. As shown, a striking trend is that the control sample of objects exhibit systematically slower evolving light curves (smaller $\Delta m_7$ and larger $t_{f,1/2}$) than nearly the entire sample of Ca-rich gap transients, suggesting typically larger diffusion time and correspondingly larger ejecta masses in the control sample. Yet, the Ca-rich and control objects occupy a common phase space near $\Delta m_7 \approx 0.3$ and $M_p$ between $-16$ and $-17$, noting that the fastest evolving objects in this sample are always Ca-rich. Thus, we conclude that while the fastest evolving Ca-rich gap transients $\Delta m_7 > 0.4$ can be identified with peak light photometry and spectroscopy, it is difficult to select a complete sample based on only these properties.

The broader light curves and low [Ca II]/[O I] of the Type Ib/c events in the control sample suggest that these events are likely consistent with being normal core-collapse SNe \citep{Gal-Yam2017} that are extinguished by foreground dust. Indeed, the detection of prominent Na I D absorption from the host galaxy in nearly all of these events suggest that host extinction likely plays an important role in making these events appear sub-luminous at peak similar to the Ca-rich events. The 02cx-like events \citep{Li2003} are consistent with being very low luminosity members of the SN\,Iax class similar to the lowest luminosity member SN\,2008ha \citep{Foley2009}. However, we did not have a high resolution spectrum of the only other Type Ia SN\,2019gau at peak to ascertain the role of host extinction. At the same time, we were unable to securely identify nebular emission features in the late time spectrum of this object, leaving this as an inconclusive low luminosity SN\,Ia in the sample.

We examine the spectroscopic properties of the control sample to identify potential spectroscopic clues at peak light for identifying Ca-rich gap transients. Ca-Ib exhibit strikingly similar spectra to normal SNe\,Ib in the sample; however, the line blanketed spectra of the red Ca-Ib objects are uncommon in the control sample. The only objects in the control sample that also exhibit suppressed blue continua are SN\,2019txl and SN\,2019mjo. The former object also exhibits a strong Na I D absorption, suggesting that host extinction likely suppresses the blue continuum in this otherwise normal core-collapse SN. SN\,2019mjo is a pecular SN\,Ib whose nature remains unclear, and we do not discuss its properties further here. Nevertheless, peculiar low velocity objects such as SN\,2018kjr and PTF\,12bho are not seen in the control sample, suggesting that the combination of low velocity and highly reddened continuum may be indicative of a Ca-rich object if its peak luminosity can be constrained. The Ca-Ia objects with strongly line blanketed spectra are unique in the broader population of SNe\,Ia that exhibit blue continua at peak (see discussion in \citealt{De2019a}), and thus strong line blanketing with prominent Si\,II lines could be used to identify likely Ca-Ia candidates at peak. Given that all the above selection procedures are unable to yield a complete sample of Ca-rich gap transients, we conclude in general that nebular phase spectroscopy of systematically selected samples of low luminosity transients in the local universe with 8 - 10 m class telescopes will continue to be important for our census of these elusive transients.

\begin{figure*}[]
    \centering
    \includegraphics[width=\textwidth]{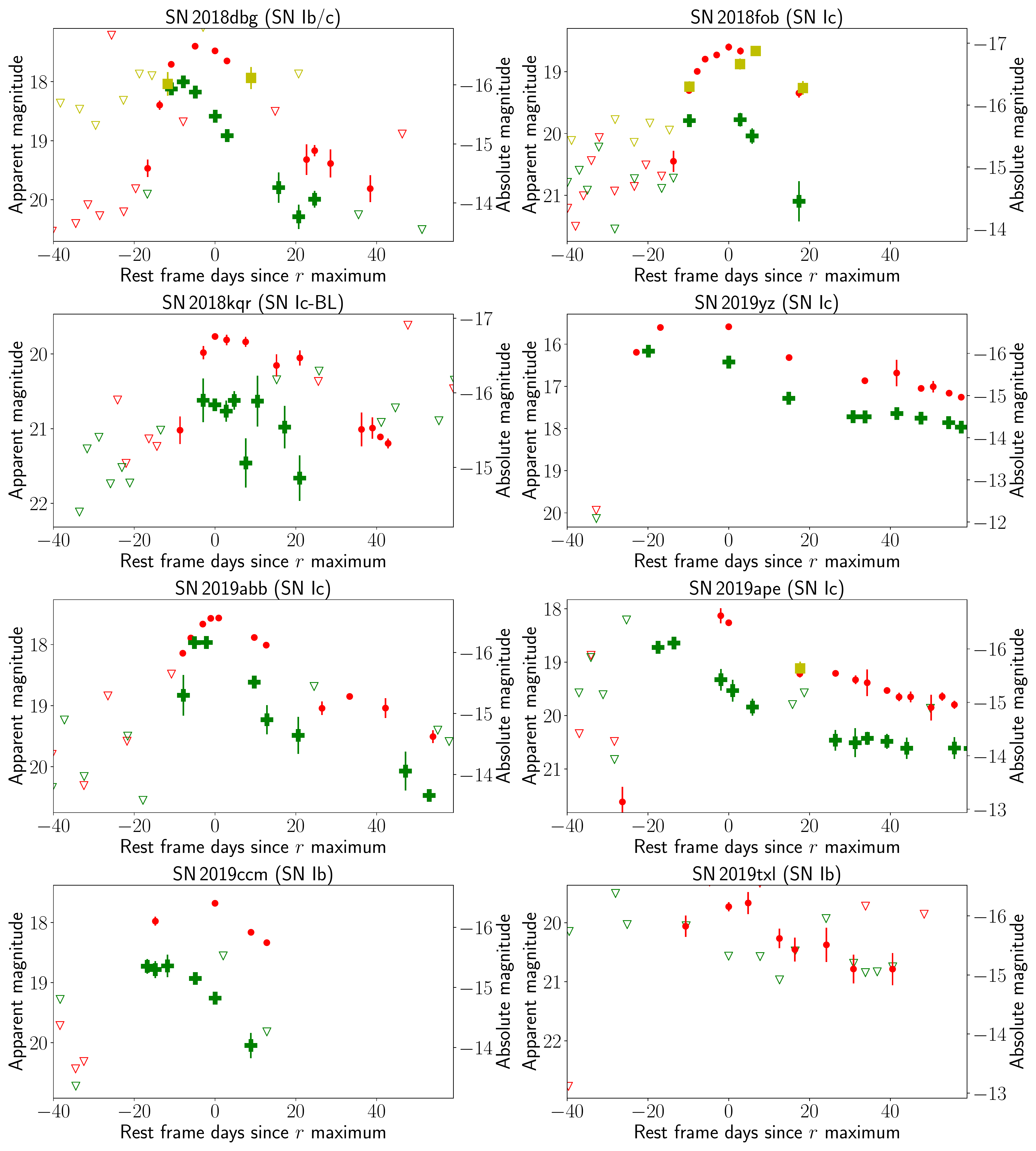}
    \caption{Forced photometry light curves of each object in the control sample that did not pass the [Ca II]/[O I] threshold defined in the sample. Each panel shows the photometric evolution near peak for the transient indicated in the figure title. We include photometry in $gri$ filters from ZTF and phase is defined with respect to time from $r$-band peak. Red circles denote $r$-band photometry, green plus symbols indicate $g$-band photometry and yellow squares indicate $i$-band photometry. Hollow inverted triangles denote 5$\sigma$ upper limits at the location of the transient.}
    \label{fig:controllc1}
\end{figure*}

\begin{figure*}[]
    \ContinuedFloat
    \centering
    \includegraphics[width=\textwidth]{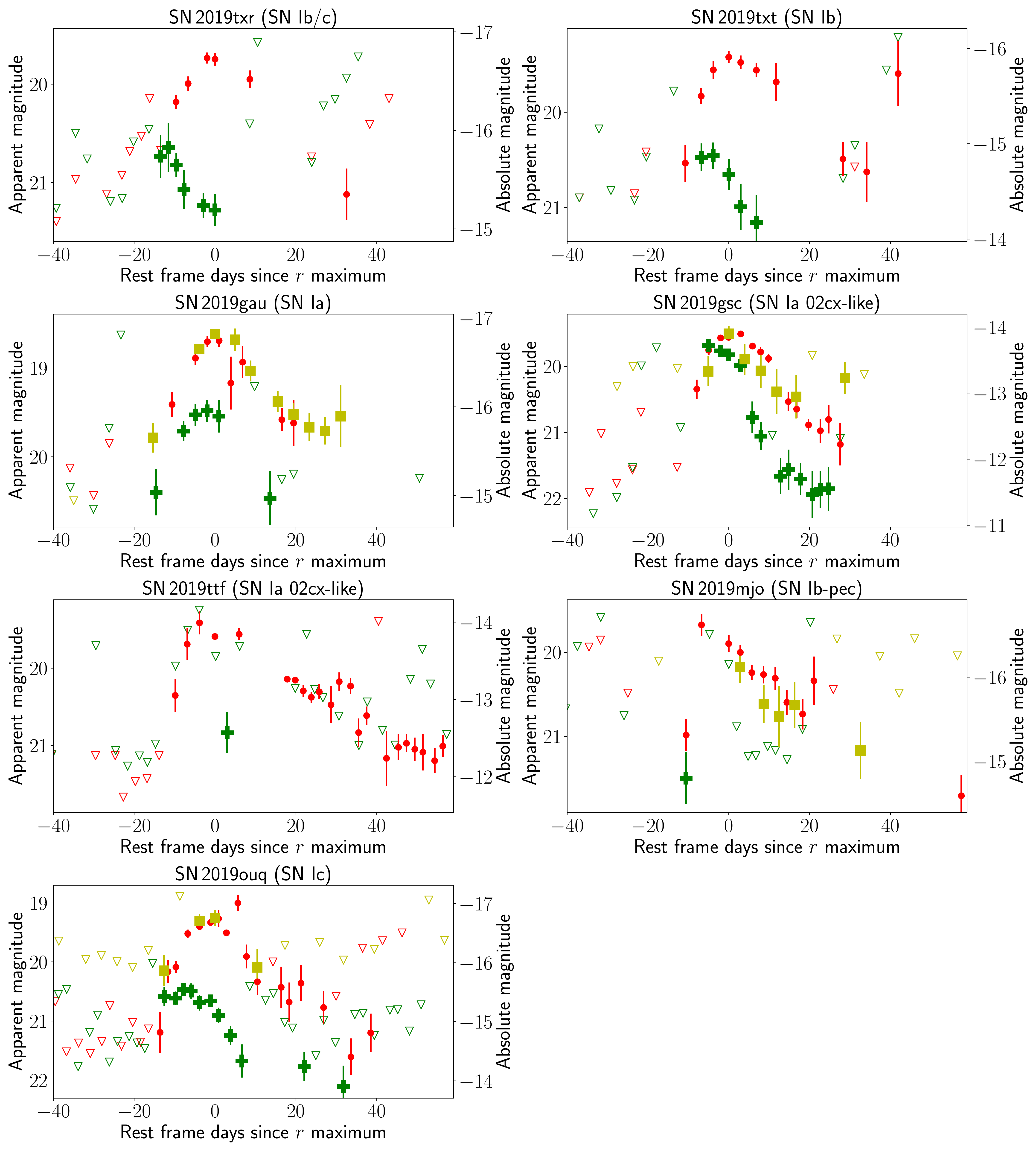}
    \caption{Continued}
\end{figure*}

\begin{figure*}[]
    \centering
    \includegraphics[width=0.45\textwidth]{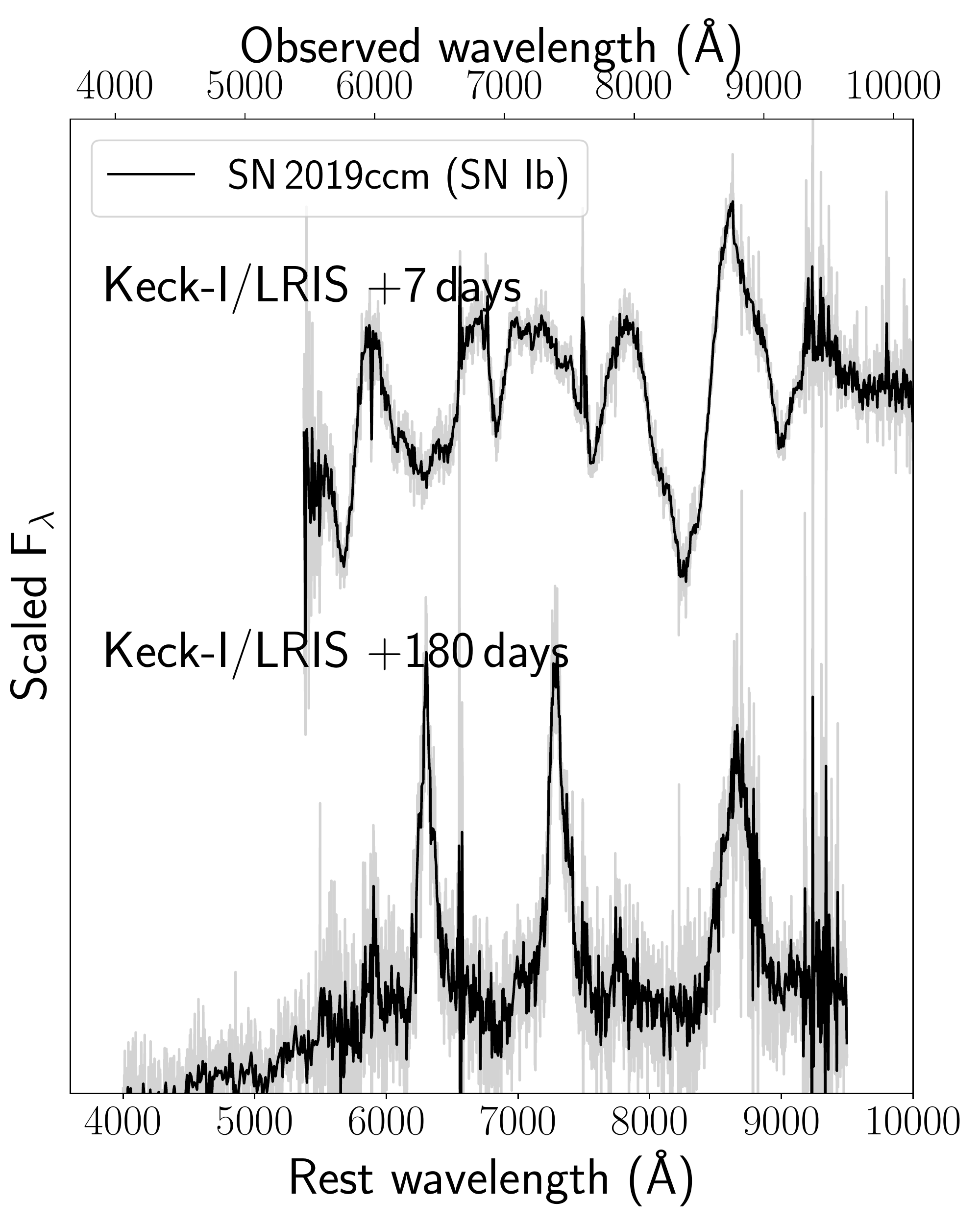}
    \includegraphics[width=0.45\textwidth]{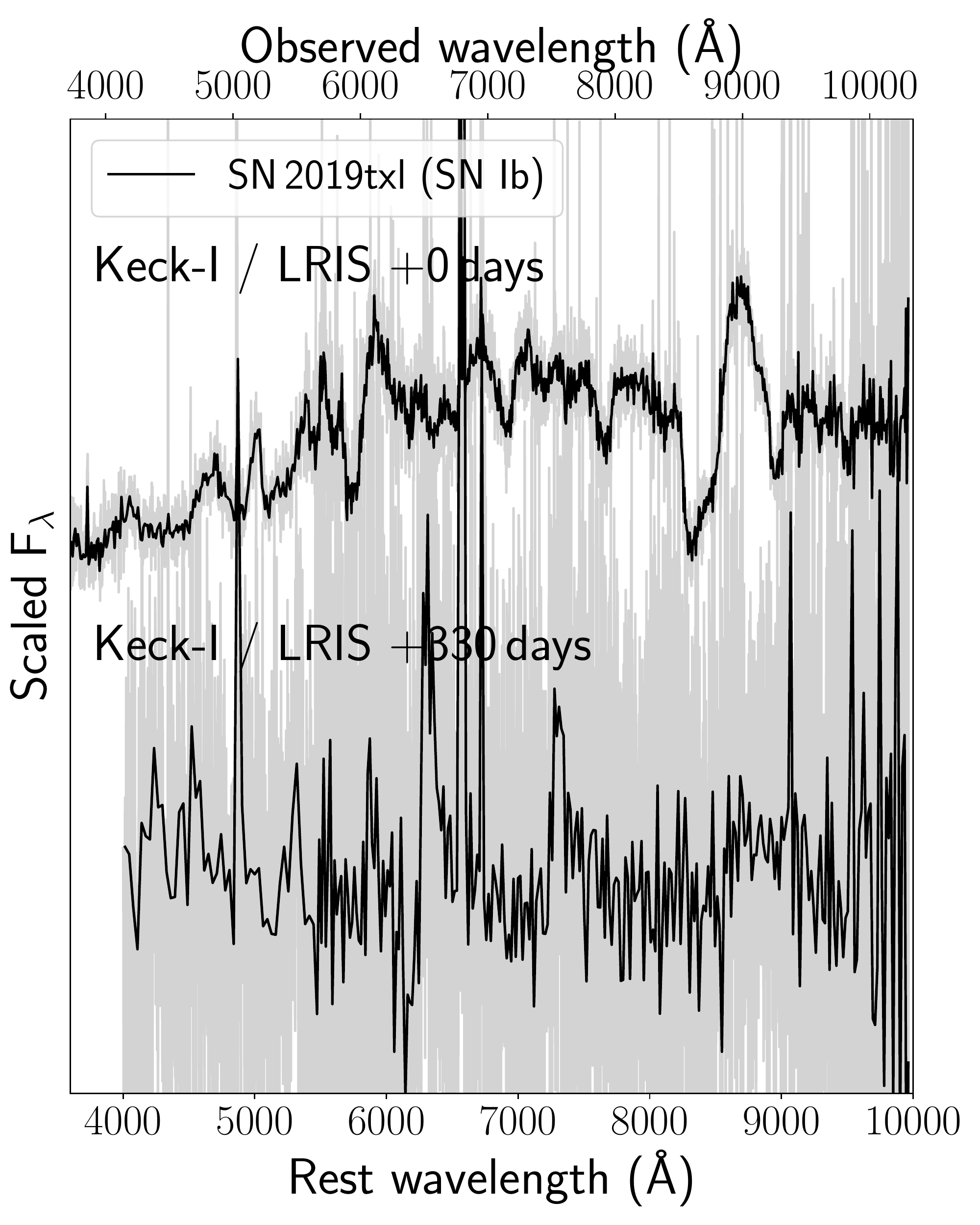}
    \includegraphics[width=0.45\textwidth]{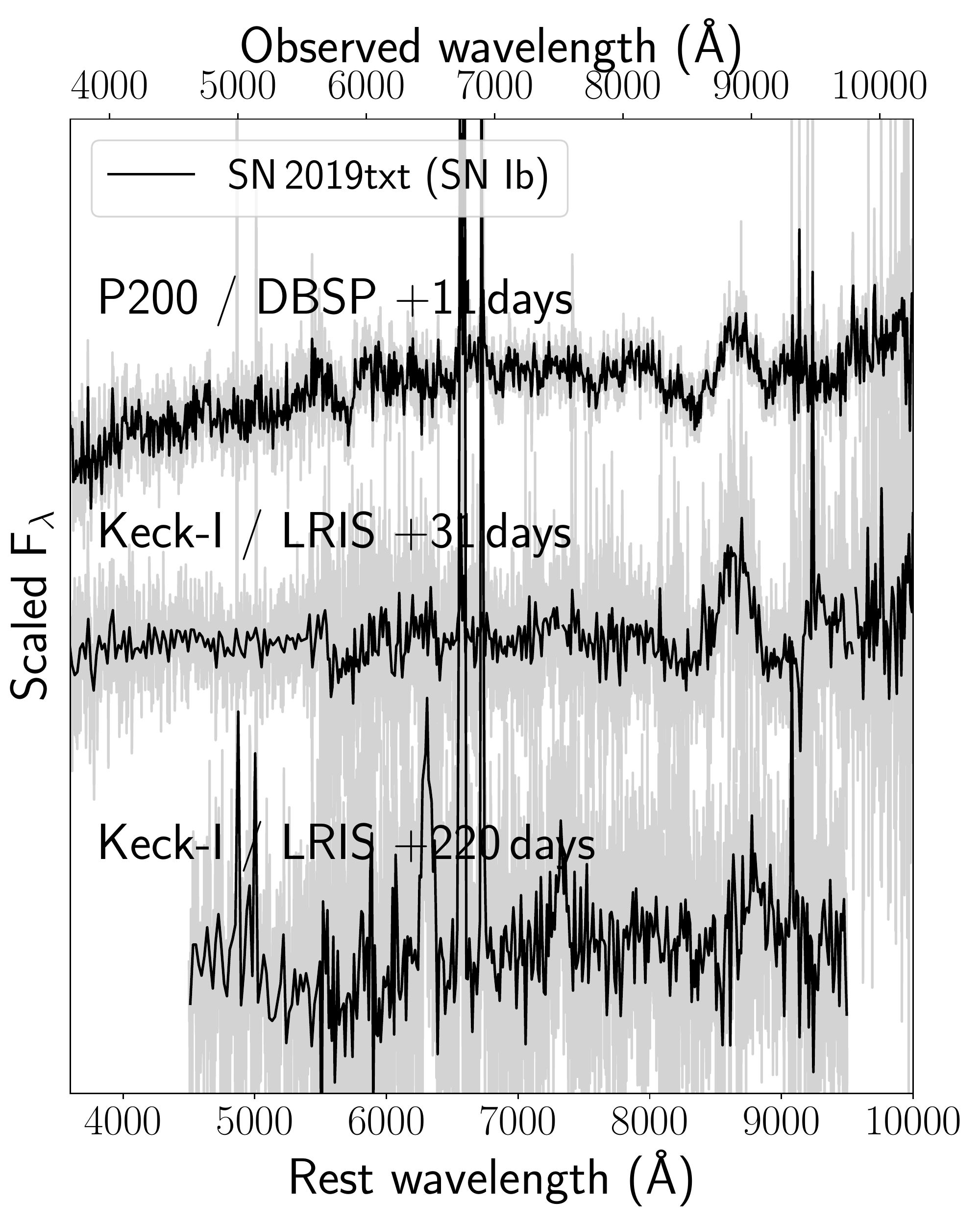}
     \includegraphics[width=0.45\textwidth]{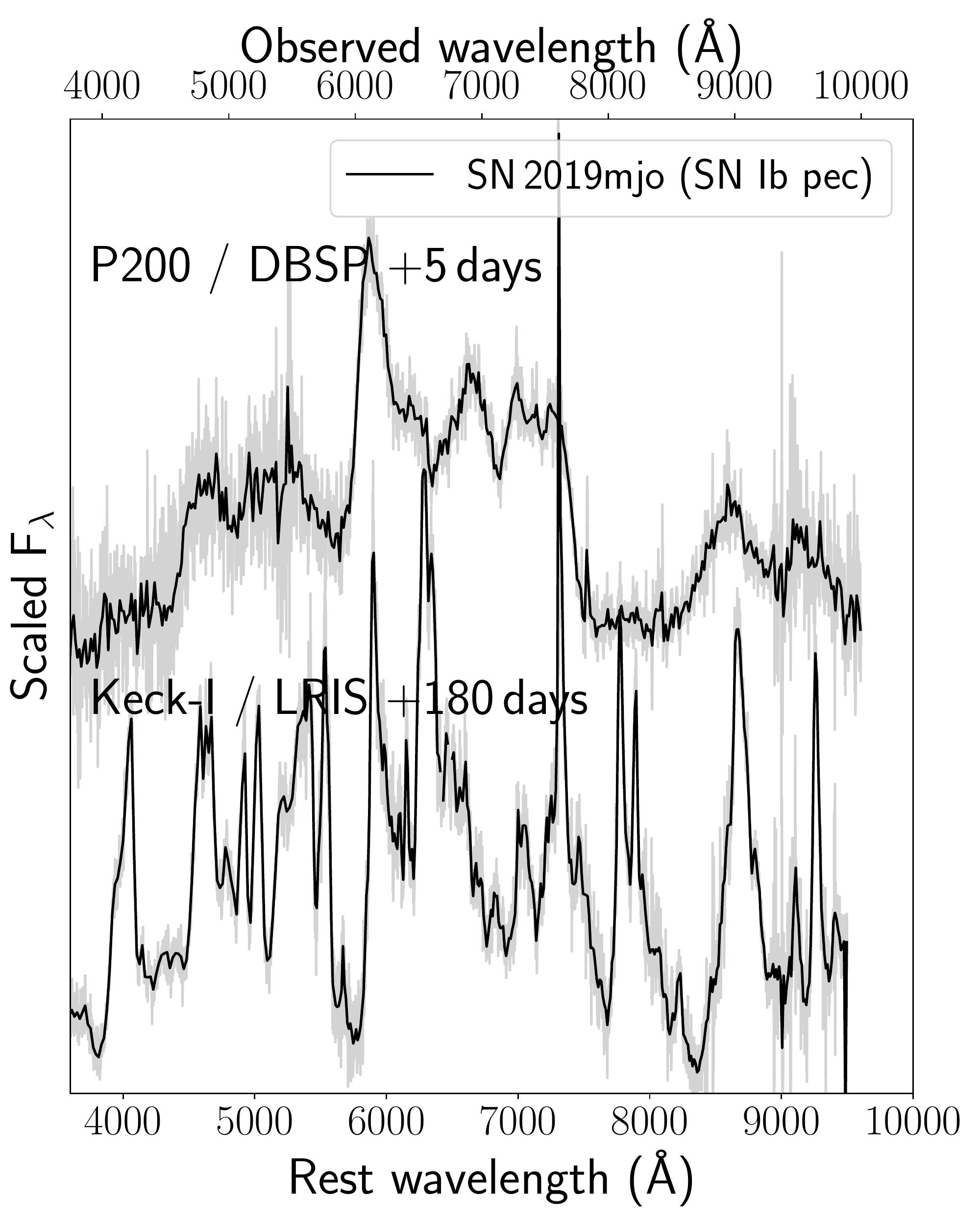}
        \caption{Photospheric and nebular phase spectrum of objects in the control sample that did not pass either the early nebular phase transition criterion or the nebular phase [Ca II]/[O I] threshold defined in the sample (see Appendix \ref{sec:control}). Each panel shows one object with its name and classification indicated in the legend. Gray lines show unbinned spectra while the black lines show spectra binned to improve the signal to noise ratio. The instrument used and phase of each spectrum is shown next to each spectrum.}
    \label{fig:control_spec_1}
\end{figure*}{}

\begin{figure*}[]
    \ContinuedFloat
    \centering
        \includegraphics[width=0.45\textwidth]{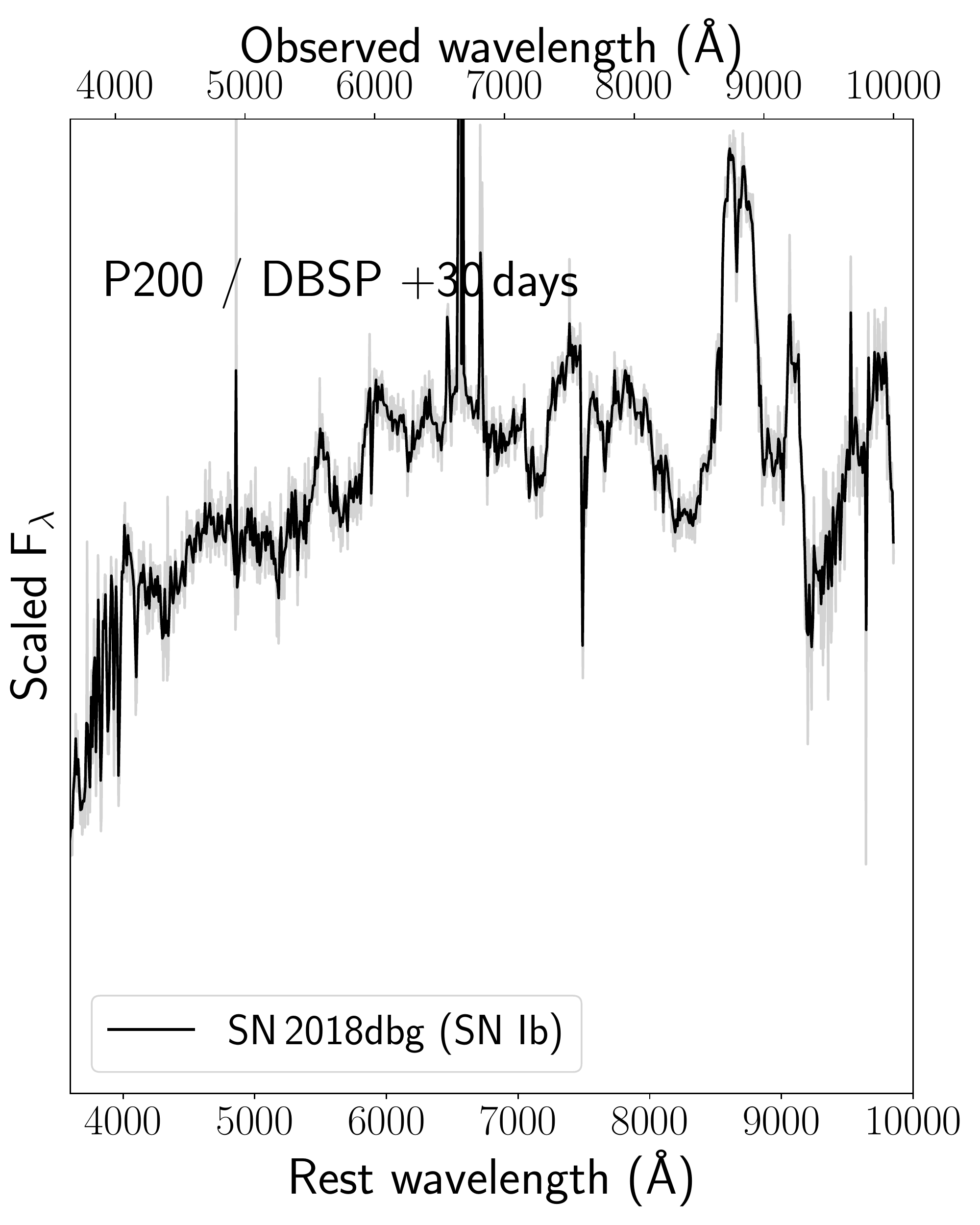}
    \includegraphics[width=0.45\textwidth]{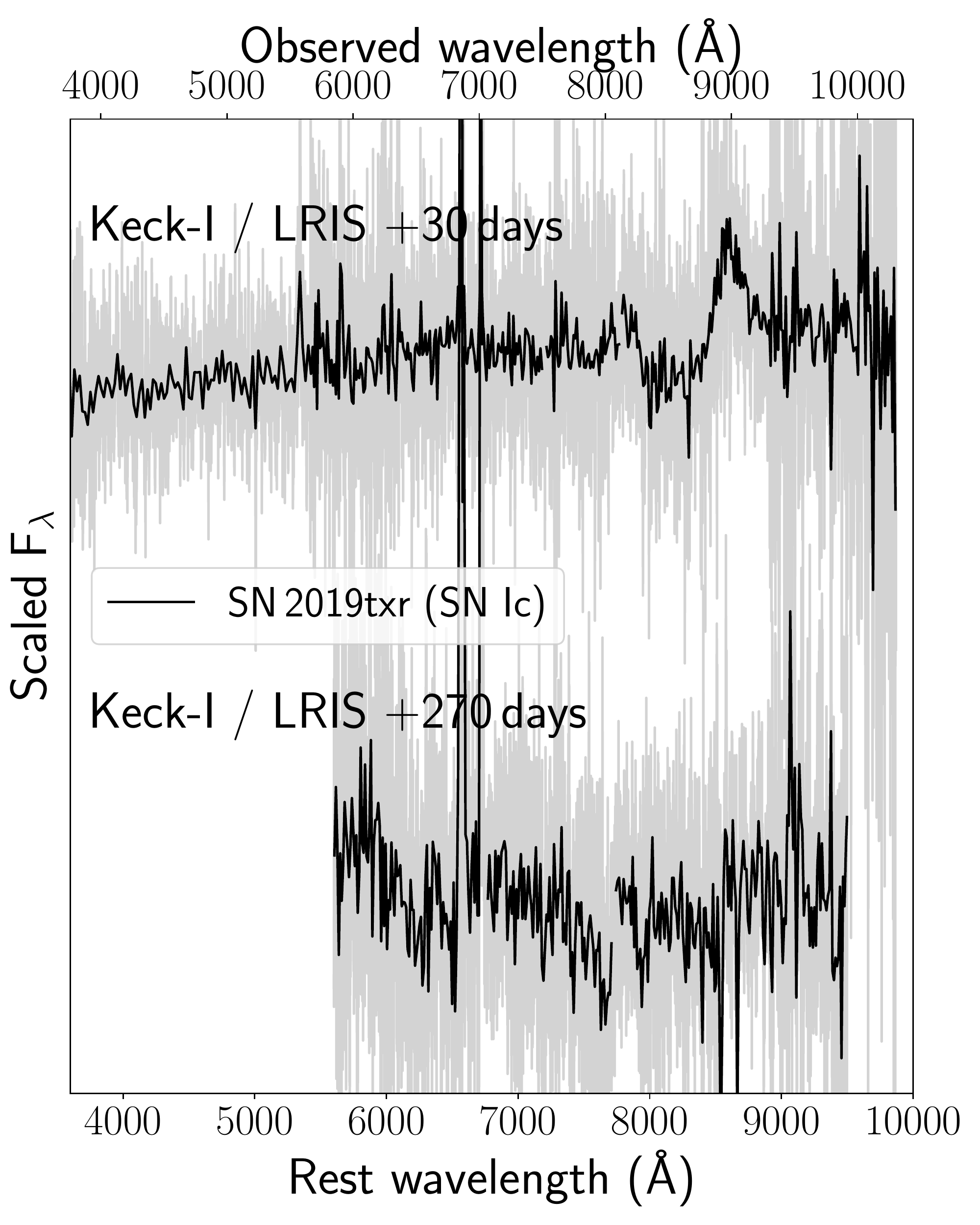}
    \includegraphics[width=0.45\textwidth]{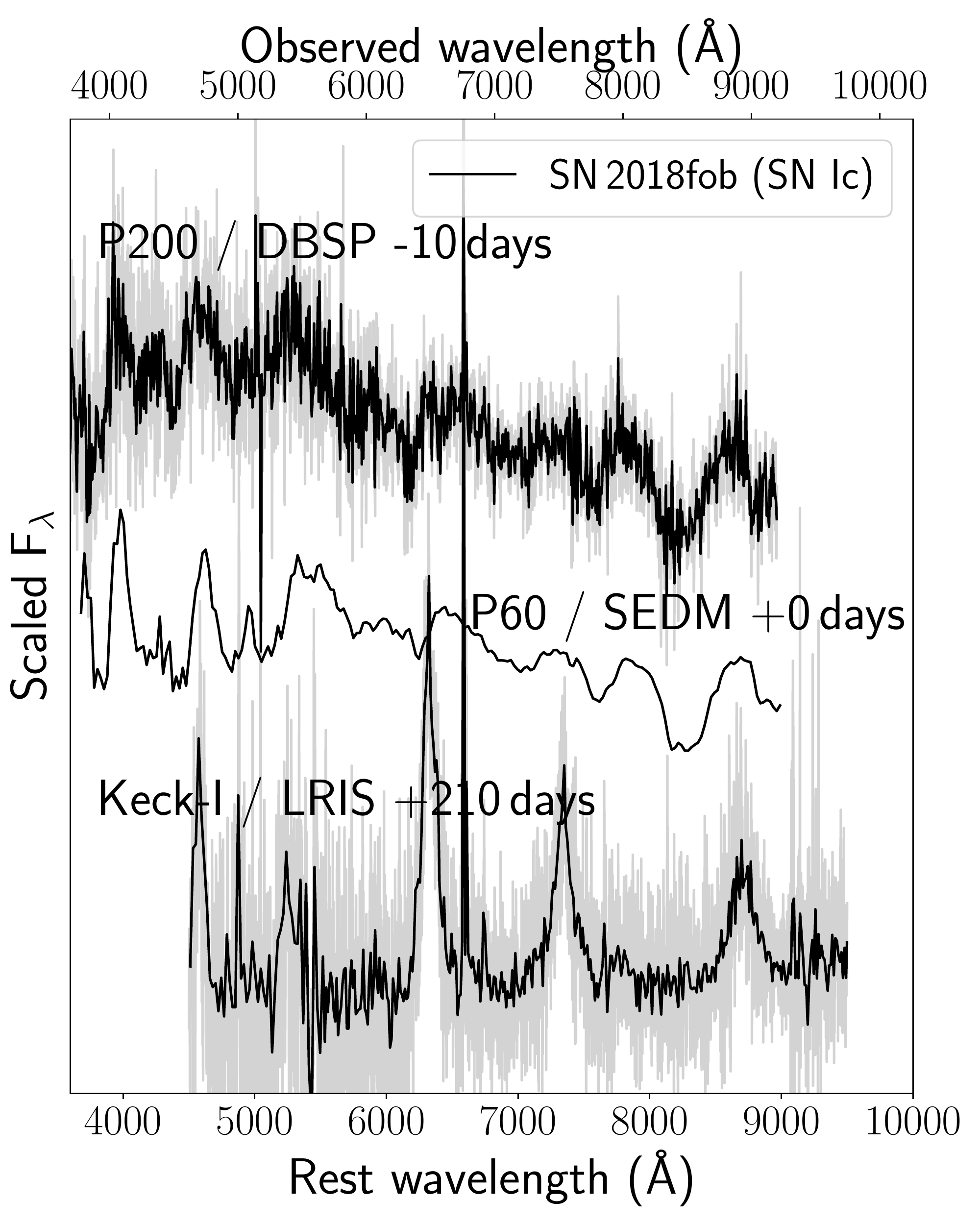}
    \includegraphics[width=0.45\textwidth]{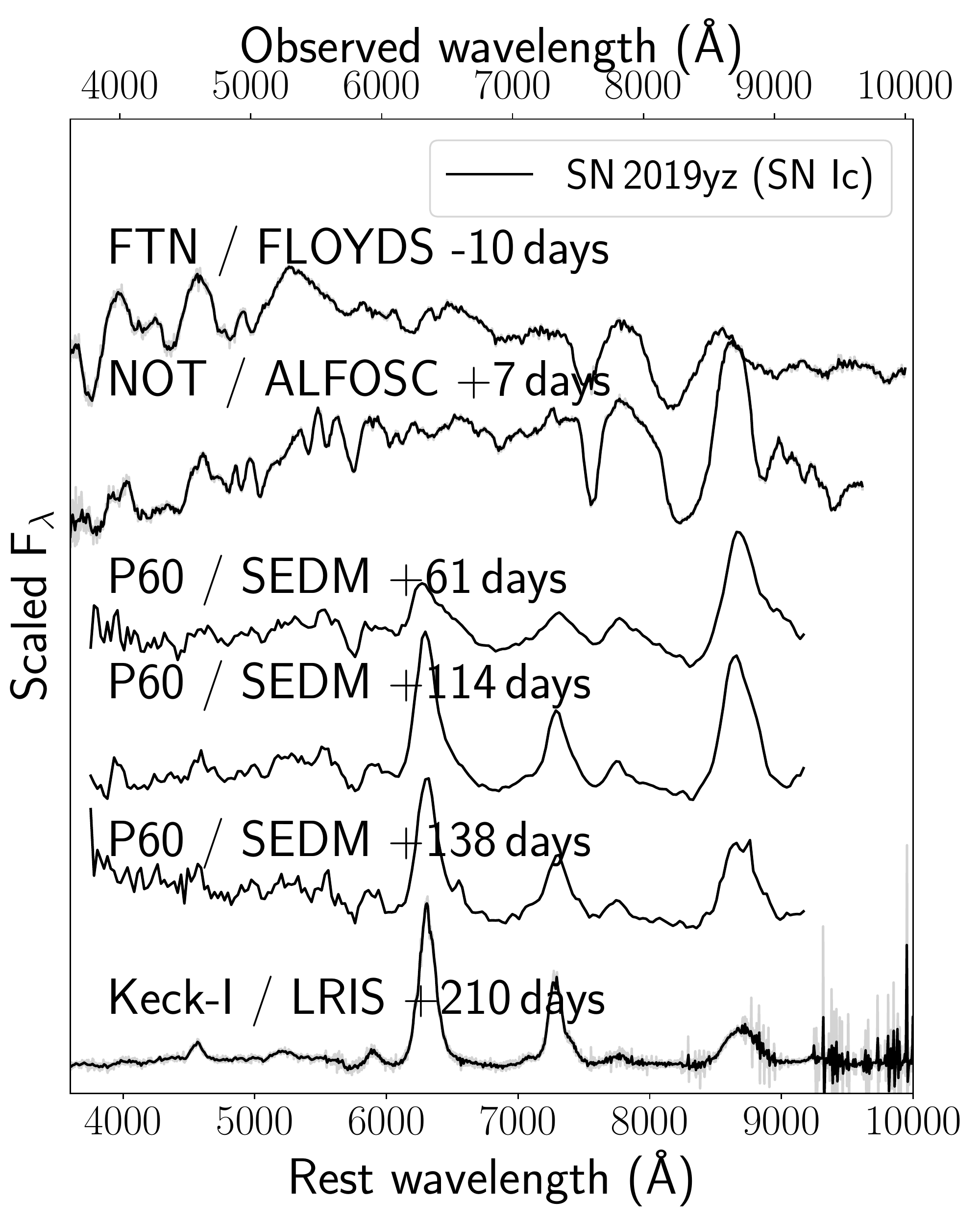}
    \caption{Continued}
\end{figure*}{}

\begin{figure*}[]
    \ContinuedFloat
    \centering
    \includegraphics[width=0.45\textwidth]{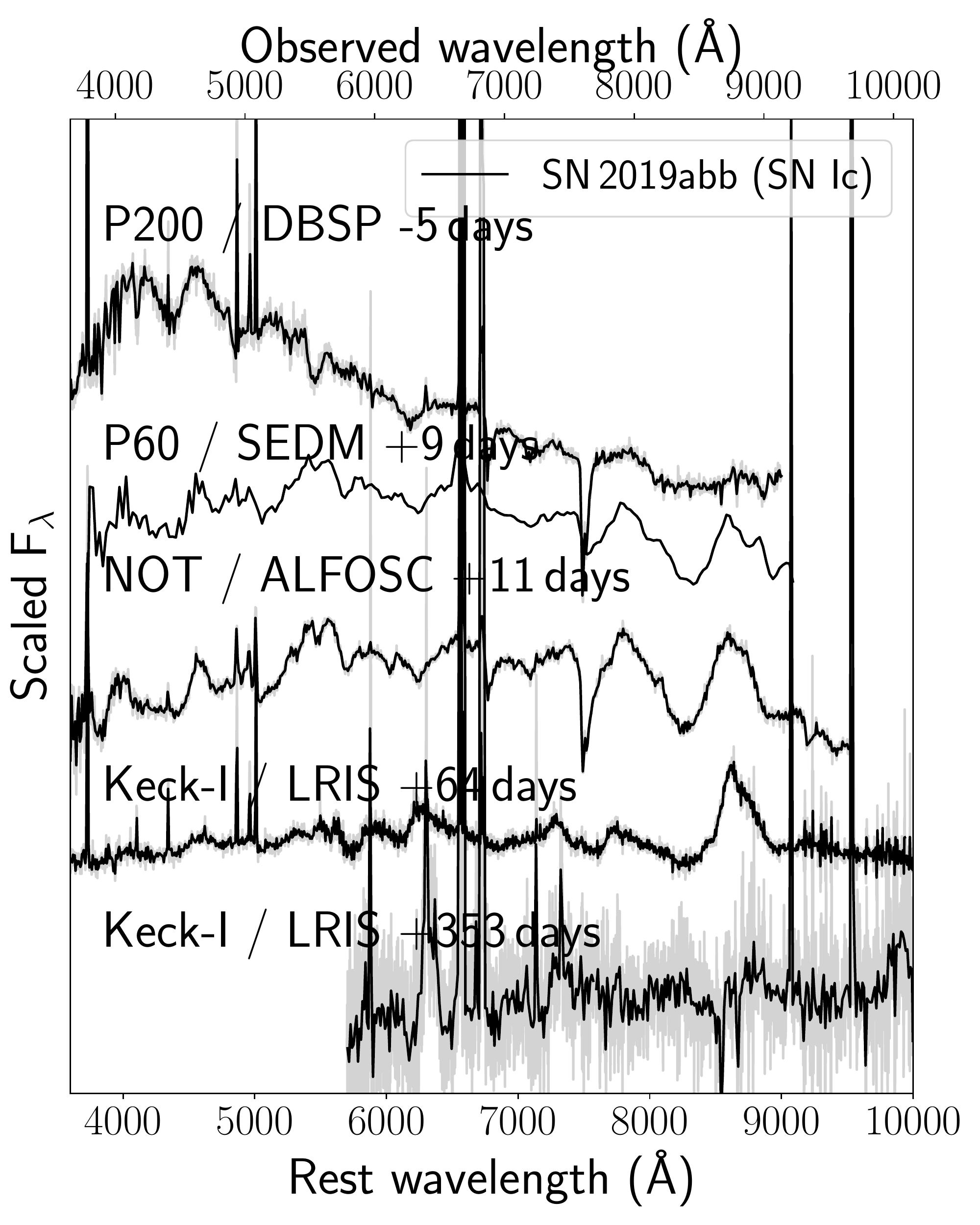}
    \includegraphics[width=0.45\textwidth]{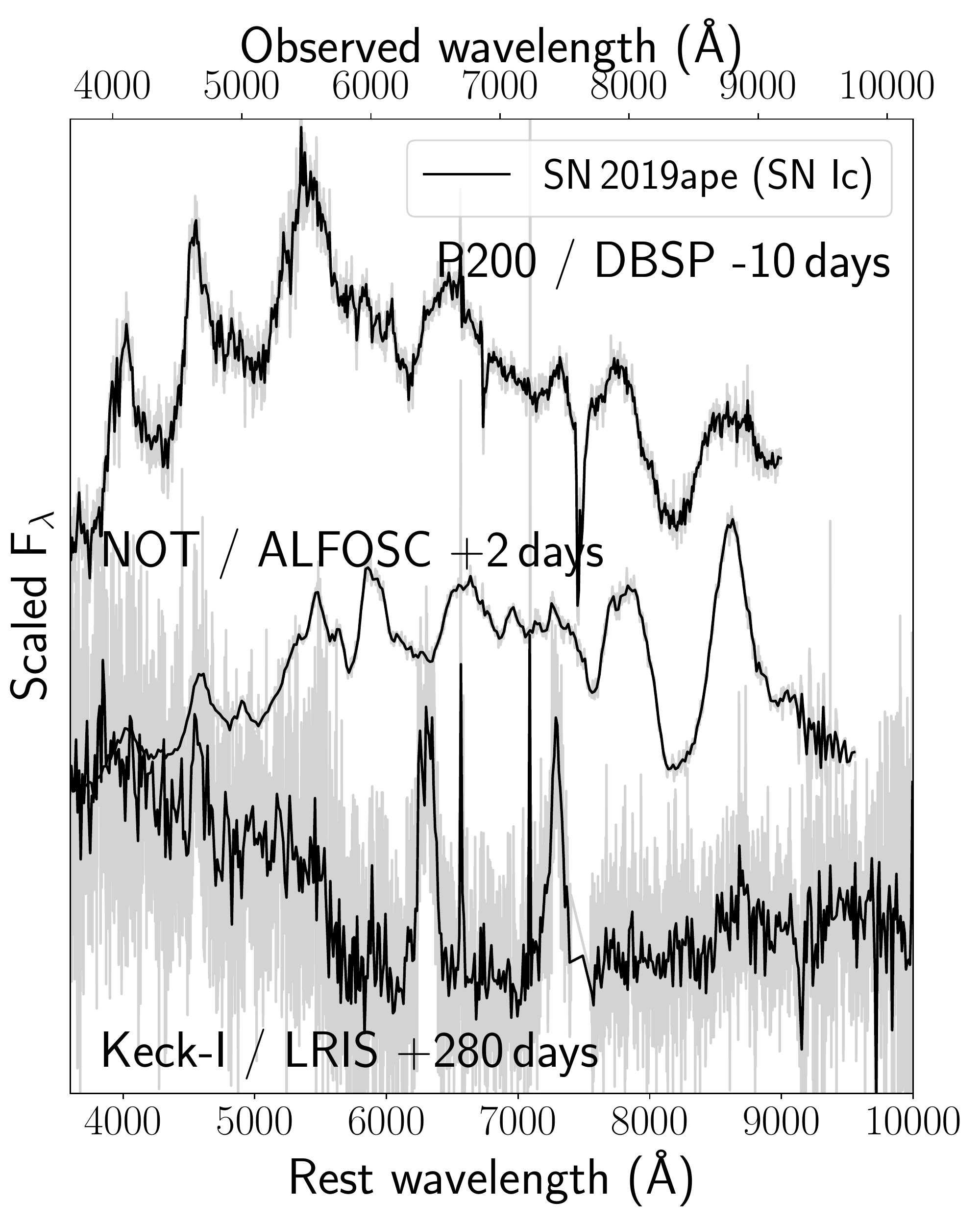}
    \includegraphics[width=0.45\textwidth]{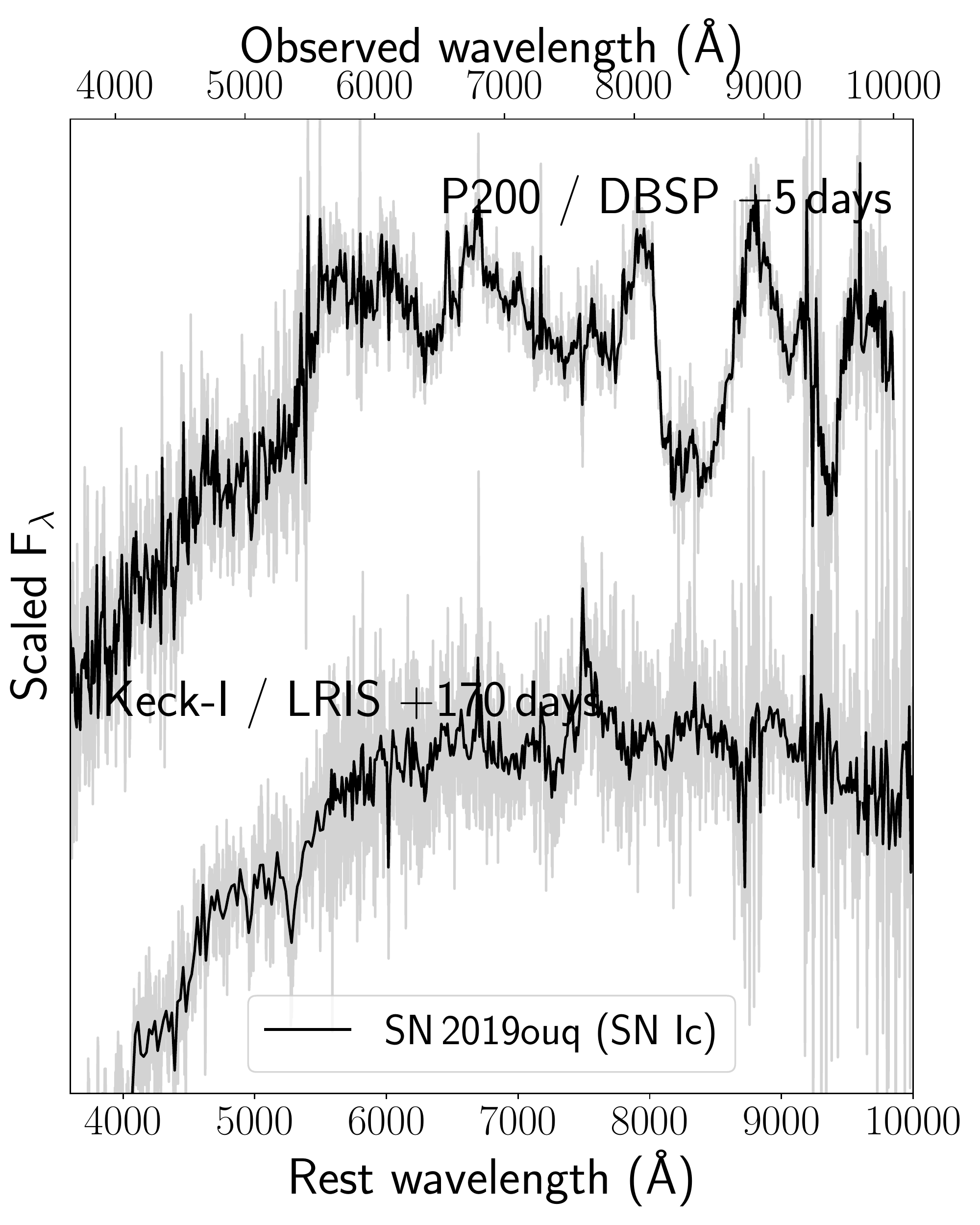}
    \includegraphics[width=0.45\textwidth]{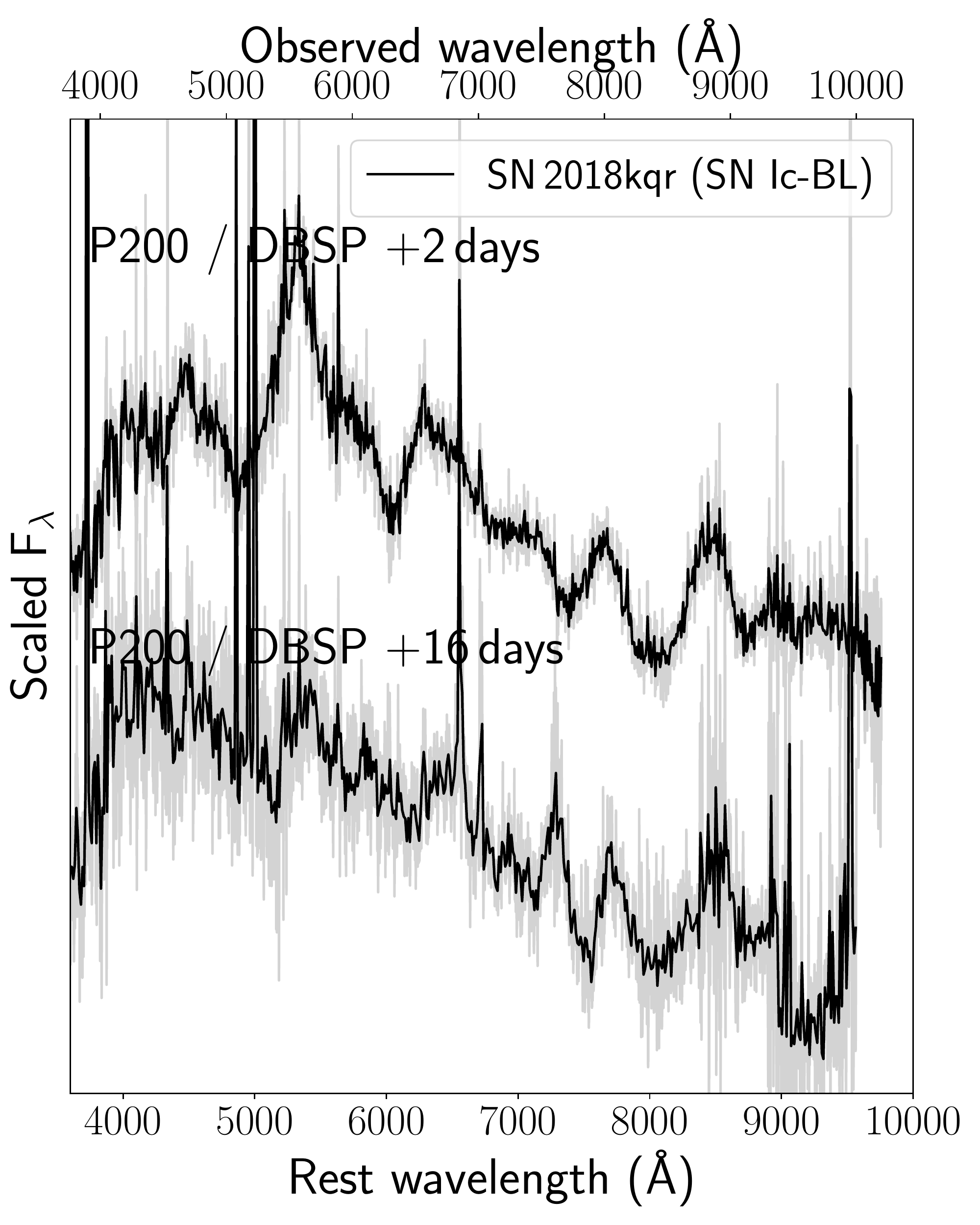}
    \caption{Continued}
\end{figure*}{}
\newpage

\begin{figure*}[]
    \ContinuedFloat
    \centering
    \includegraphics[width=0.45\textwidth]{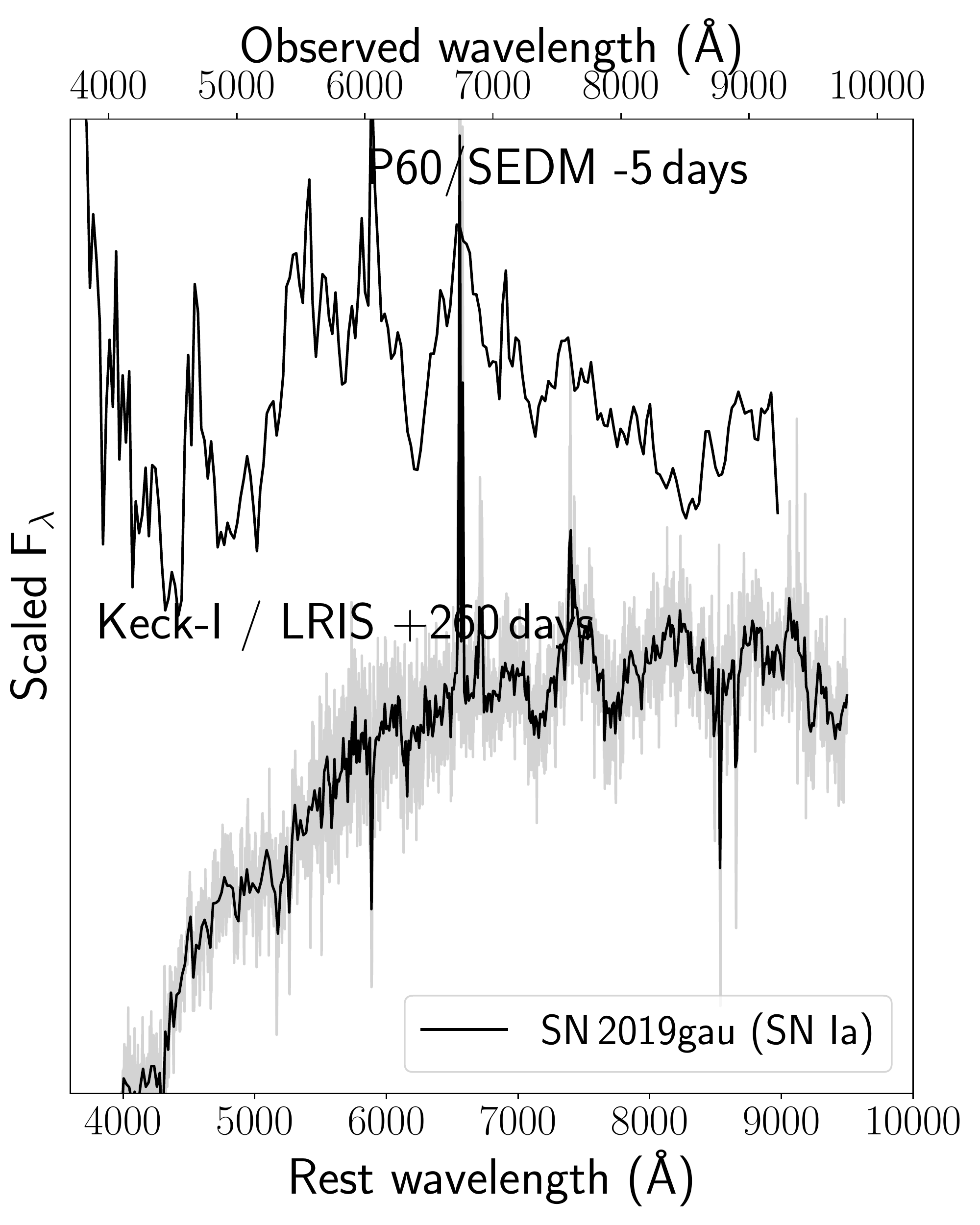}
     \includegraphics[width=0.45\textwidth]{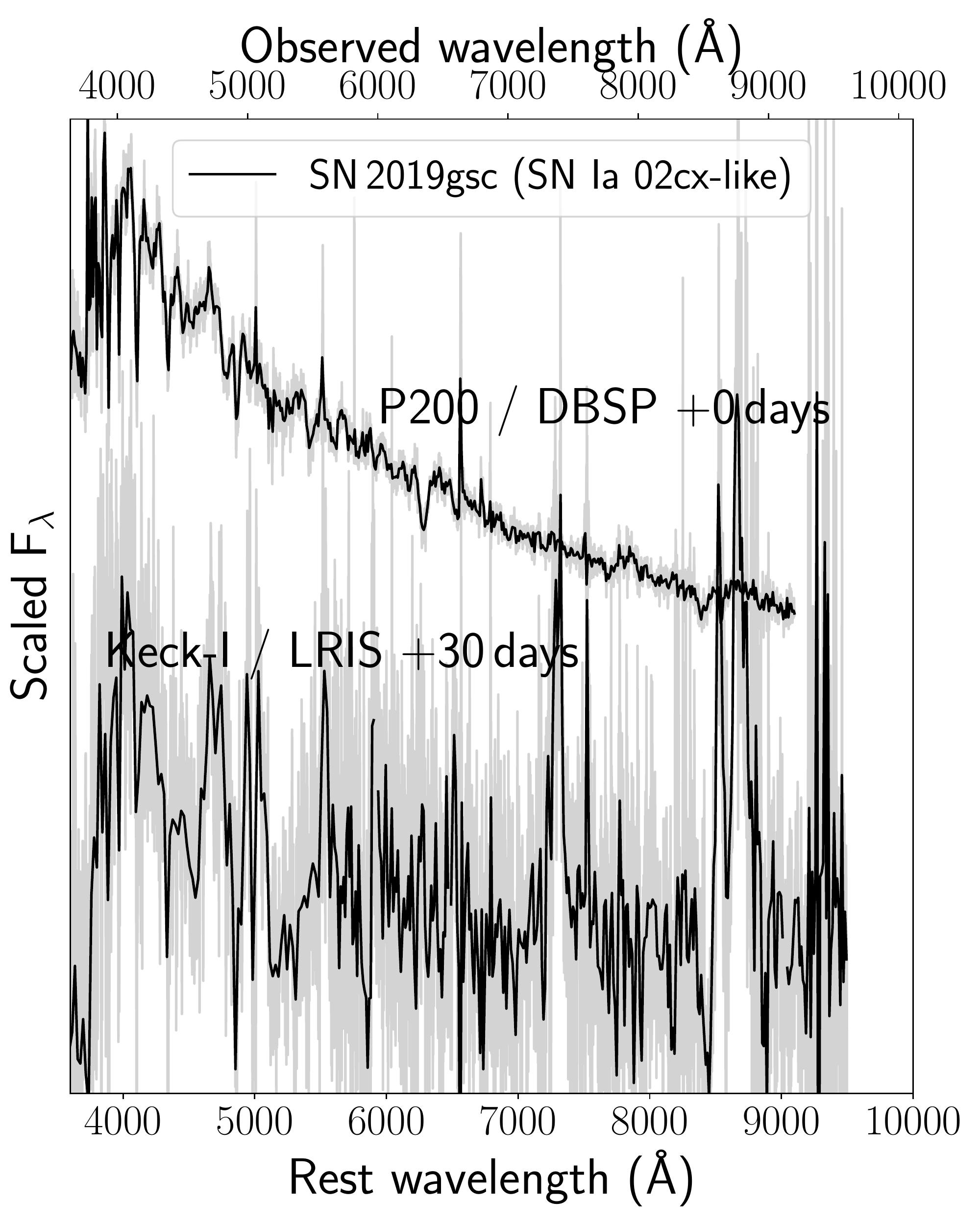}
    \includegraphics[width=0.45\textwidth]{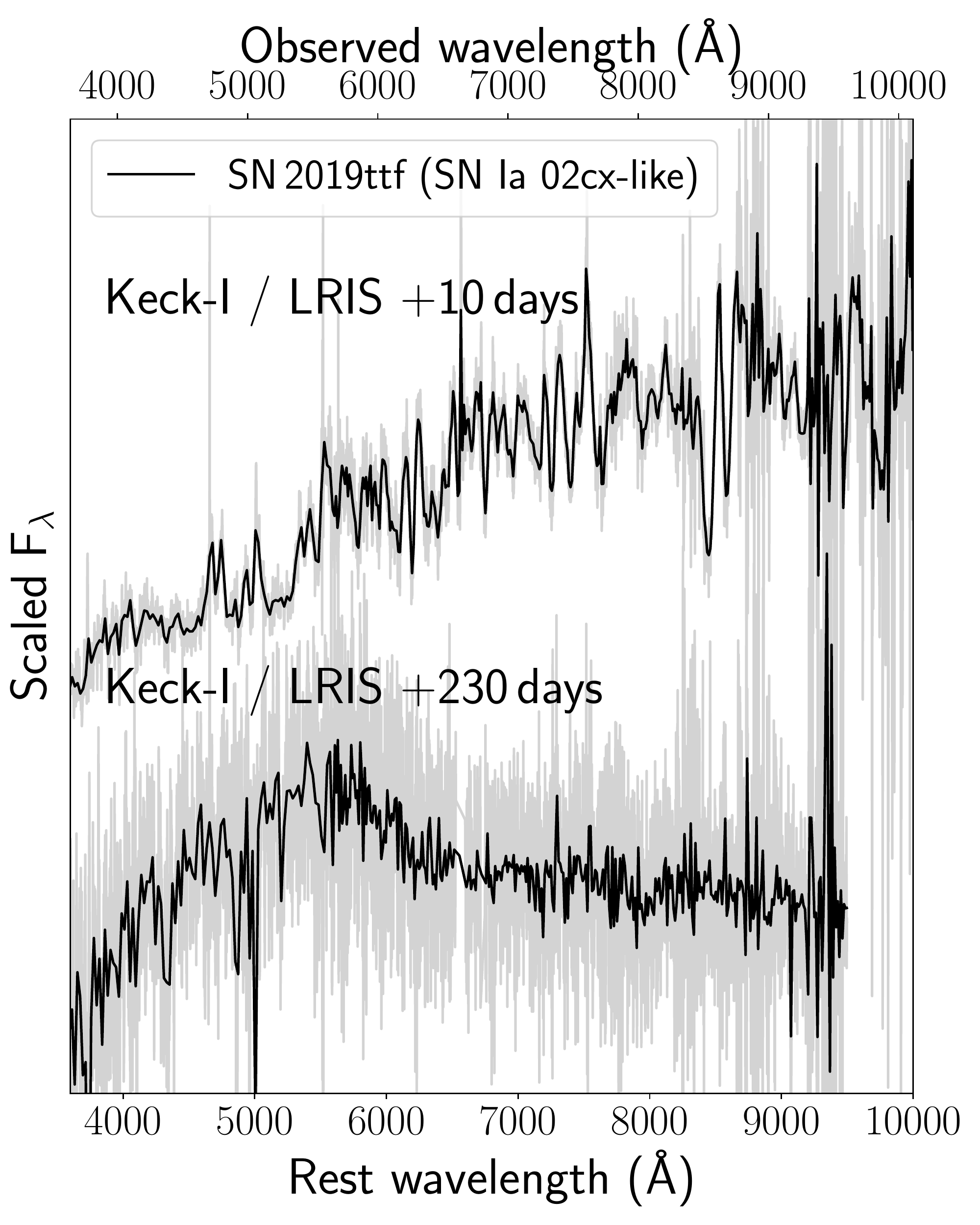}
    \caption{Continued}
\end{figure*}{}

\clearpage

\startlongtable
\begin{deluxetable*}{cccccc}
\tabletypesize{\footnotesize}
\tablecaption{Photometry of all the sources presented in this paper. The photometry has not been corrected for Galactic extinction. Phases are indicated with respect to the time from the best-fit $r$-band peak. Upper limits are at 5$\sigma$ confidence at the location of the transient. This table will be available in its entirety in machine-readable form upon publication.}
\tablehead{
\colhead{Object} &
\colhead{MJD} &
\colhead{Rest frame phase}  &
\colhead{Filter} &
\colhead{Magnitude} &
\colhead{Instrument}\\
\colhead{} &
\colhead{} &
\colhead{(days from $r$ peak)} &
\colhead{} &
\colhead{} &
}
\startdata
SN\,2018ckd & $58242.31$ & $-34.53$ & $r$ & $>20.53$ & P48+ZTF\\ 
SN\,2018ckd & $58245.34$ & $-31.57$ & $r$ & $>18.89$ & P48+ZTF\\ 
SN\,2018ckd & $58248.27$ & $-28.71$ & $r$ & $>20.95$ & P48+ZTF\\ 
SN\,2018ckd & $58255.26$ & $-21.88$ & $r$ & $>20.70$ & P48+ZTF\\ 
SN\,2018ckd & $58258.25$ & $-18.96$ & $r$ & $>20.64$ & P48+ZTF\\ 
SN\,2018ckd & $58262.30$ & $-15.01$ & $r$ & $>19.87$ & P48+ZTF\\ 
SN\,2018ckd & $58267.31$ & $-10.12$ & $r$ & $>19.65$ & P48+ZTF\\ 
SN\,2018ckd & $58270.32$ & $-7.18$ & $r$ & $20.22 \pm 0.23$ & P48+ZTF\\ 
SN\,2018ckd & $58276.19$ & $-1.45$ & $r$ & $19.19 \pm 0.03$ & P48+ZTF\\ 
SN\,2018ckd & $58279.17$ & $1.46$ & $r$ & $19.20 \pm 0.04$ & P48+ZTF\\ 
\enddata
\label{tab:photometry}
\end{deluxetable*}

\clearpage\newpage

\startlongtable
\begin{deluxetable*}{ccccccc}
\tabletypesize{\footnotesize}
\tablecaption{Log of spectroscopic observations of all objects presented in this paper. $^\dagger$ denotes spectra which did not high enough signal to noise ratio to detect features.}
\tablehead{
\colhead{Object} &
\colhead{Observation Date} &
\colhead{MJD}  &
\colhead{Phase} &
\colhead{Telescope + Instrument} &
\colhead{Range} &
\colhead{Resolution}\\
\colhead{} &
\colhead{(UTC)} &
\colhead{} &
\colhead{(days from $r$ peak)} &
\colhead{\AA} & 
\colhead{} & 
\colhead{$\lambda / \delta \lambda$}
}

\startdata
SN\,2018ckd & 2018-06-12 & 58281.3 & +3 & P200 + DBSP & 3500 - 10000 & 1000 \\
SN\,2018ckd & 2018-06-21 & 58290.2 & +12 & P200 + DBSP & 3500 - 10000 & 1000 \\
SN\,2018ckd & 2018-08-08 & 58338.3 & +57 & Keck-I + LRIS & 3500 - 10000 & 1000  \\
SN\,2018ckd & 2019-04-03 & 58576.5 & +291 & Keck-I + LRIS & 3500 - 10000$^\dagger$ & 1000 \\
SN\,2018lqo  & 2018-08-21 & 58351.2 & -1 & P200 + DBSP & 3500 - 10000 & 1000 \\
SN\,2018lqo & 2018-10-12 & 58403.3 & +49 & Keck-I + LRIS & 3500 - 10000 & 1000 \\
SN\,2018lqu & 2018-09-12 & 58373.1 & +2 & P200 + DBSP & 3500 - 10000 & 1000 \\
SN\,2018lqu & 2018-10-12 & 58403.2 & +31 & Keck-I + LRIS & 3500 - 10000 & 1000 \\
SN\,2018gwo & 2018-09-30 & 58391.8 & -12 & THO + ALPY & 3700 - 7500 & 100\\
SN\,2018gwo & 2018-10-06 & 58397.8 & -6 & THO + ALPY & 3700 - 7500 & 100\\
SN\,2018gwo & 2018-11-06 & 58428.5 & +23 & P60 + SEDM & 3800 - 9200 & 100\\
SN\,2018gwo & 2018-12-04 & 58456.6 & +51 & Keck-I + LRIS & 3500 - 10000 & 1000 \\
SN\,2018gwo & 2019-03-07 & 58549.5 & +143 & Keck-I + LRIS & 3500 - 10000 & 1000 \\
SN\,2018gwo & 2019-06-03 & 58637.3 & +230 & Keck-I + LRIS & 3500 - 10000 & 1000 \\
SN\,2018kjy & 2018-12-14 & 58466.3 & +5 & P200 + DBSP & 3500 - 10000 & 1000 \\
SN\,2018kjy  & 2019-01-04 & 58487.4 & +25 & Keck-I + LRIS & 3500 - 10000 & 1000 \\
SN\,2018kjy  & 2019-04-03 & 58576.2 & +113 & Keck-I + LRIS & 3500 - 10000 & 1000 \\
SN\,2019hty & 2019-07-01 & 58665.2 & +6 & P200 + DBSP & 3500 - 10000 & 1000 \\
SN\,2019hty  & 2019-07-02 & 58666.2 & +7 & P60 + SEDM & 3800 - 9200 & 100\\
SN\,2019hty  & 2019-08-04 & 58699.2 & +40 & P200 + DBSP & 3500 - 10000 & 1000 \\
SN\,2019ofm & 2019-08-27 & 58722.3 & -1 & P200 + DBSP & 3500 - 10000 & 1000 \\
SN\,2019ofm & 2020-02-18 & 58897.5 & +168 & Keck-I + LRIS & 3500 - 10000 & 1000 \\
SN\,2019pxu & 2019-09-24 & 58750.4 & +3 & P60 + SEDM & 3800 - 9200 & 100\\
SN\,2019pxu & 2019-10-03 & 58759.5 & +11 & P200 + DBSP & 3500 - 10000 & 1000 \\
SN\,2019pxu & 2019-10-27 & 58783.5 & +35 & Keck-I + LRIS & 3500 - 10000 & 1000 \\
SN\,2019pxu & 2020-02-18 & 58897.3 & +146 & Keck-I + LRIS & 3500 - 10000 & 1000 \\
SN\,2018dbg & 2018-08-04 & 58334.0 & +22 & P200 + DBSP & 3500 - 10000 & 1000 \\
SN\,2018fob & 2018-08-21 & 58351.0 & -8 & P200 + DBSP & 3500 - 10000 & 1000 \\
SN\,2018fob & 2018-08-31 & 58361.0 & +0 & P60 + SEDM & 3800 - 9200 & 100\\
SN\,2018fob & 2019-04-03 & 58576.0 & +209 & Keck-I + LRIS & 3500 - 10000 & 1000 \\
SN\,2018kqr & 2018-12-14 & 58466.3 & +1 & P200 + DBSP & 3500 - 10000 & 1000 \\
SN\,2018kqr & 2018-12-27 & 58479.2 & +14 & P200 + DBSP & 3500 - 10000 & 1000 \\
SN\,2019yz & 2019-02-20 & 58534.2 & +7 & NOT + ALFOSC & 3800 - 9500 & 300\\
SN\,2019yz & 2019-04-15 & 58588.2 & +61 & P60 + SEDM & 3800 - 9200 & 100 \\
SN\,2019yz & 2019-06-08 & 58642.0 & +114 & P60 + SEDM & 3800 - 9200  & 100\\
SN\,2019yz & 2019-07-02 & 58666.0 & +138 & P60 + SEDM & 3800 - 9200  & 100\\
SN\,2019yz & 2019-09-26 & 58752.2 & +224 & Keck-I + LRIS & 3500 - 10000 & 1000 \\
SN\,2019abb & 2019-01-26 & 58509.2 & -4 & P200 + DBSP & 3500 - 10000 & 1000 \\
SN\,2019abb & 2019-02-09 & 58523.0 & +9 & P60 + SEDM & 3800 - 9200  & 100\\
SN\,2019abb & 2019-02-10 & 58524.9 & +11 & NOT + ALFOSC & 3800 - 9500 & 300 \\
SN\,2019abb & 2019-04-06 & 58579.2 & +64 & Keck-I + LRIS & 3500 - 10000 & 1000 \\
SN\,2019abb & 2020-01-24 & 58872.0 & +353 & Keck-I + LRIS & 3500 - 10000 & 1000 \\
SN\,2019ape & 2019-02-12 & 58526.4 & -13 & P200 + DBSP & 3500 - 10000 & 1000 \\
SN\,2019ape & 2019-03-01 & 58543.1 & +2 & NOT + ALFOSC & 3800 - 9500  & 300\\
SN\,2019ape & 2019-12-03 & 58820.0 & +274 & Keck-I + LRIS & 3500 - 10000 & 1000 \\
SN\,2019ccm & 2019-04-06 & 58579.2 & +6 & Keck-I + LRIS & 3500 - 10000\ & 1000 \\
SN\,2019ccm & 2019-09-28 & 58754.0 & +178 & Keck-I + LRIS & 3500 - 10000 & 1000 \\
SN\,2019txl & 2019-04-06 & 58579.0 & +11 & Keck-I + LRIS & 3500 - 10000 & 1000 \\
SN\,2019txl & 2020-02-18 & 58897.0 & +318 & Keck-I + LRIS & 3500 - 10000 & 1000 \\
SN\,2019txr & 2019-06-04 & 58638.0 & +28 & Keck-I + LRIS & 3500 - 10000 & 1000 \\
SN\,2019txr & 2020-02-18 & 58897.0 & +276 & Keck-I + LRIS & 3500 - 10000 & 1000 \\
SN\,2019txt & 2019-05-13 & 58616.0 & +10 & P200 + DBSP & 3500 - 10000 & 1000 \\
SN\,2019txt & 2019-06-04 & 58638.0 & +31 & Keck-I + LRIS & 3500 - 10000 & 1000 \\
SN\,2019txt & 2020-01-24 & 58872.0 & +259 & Keck-I + LRIS & 3500 - 10000 & 1000 \\
SN\,2019gau & 2019-06-04 & 58638.9 & -4 & P60 + SEDM & 3500 - 10000 & 100\\
SN\,2019gau & 2020-02-18 & 58897.0 & +260 & Keck-I + LRIS & 3500 - 10000 & 1000 \\
SN\,2019gsc & 2019-06-04 & 58638.2 & -2 & P200 + DBSP & 3500 - 10000 & 1000 \\
SN\,2019gsc & 2019-07-04 & 58668.0 & +27 & Keck-I + LRIS & 3500 - 10000 & 1000 \\
SN\,2019ttf & 2019-07-04 & 58668.0 & +10 & Keck-I + LRIS & 3500 - 10000 & 1000 \\
SN\,2019ttf & 2020-03-22 & 58930.0 & +269 & Keck-I + LRIS & 3500 - 10000 & 1000 \\
SN\,2019mjo & 2019-08-01 & 58696.0 & +7 & P200 + DBSP & 3500 - 10000 & 1000 \\
SN\,2019mjo & 2020-01-24 & 58872.0 & +176 & Keck-I + LRIS & 3500 - 10000 & 1000 \\
SN\,2019ouq & 2019-08-04 & 58699.3 & +6 & P200 + DBSP & 3500 - 10000 & 1000 \\
SN\,2019ouq & 2020-01-24 & 58872.0 & +173 & Keck-I + LRIS & 3500 - 10000 & 1000 \\
\enddata

\label{tab:spectra}
\end{deluxetable*}{}
.
\end{document}